\documentclass[a4paper,twoside,12pt]{book}
\usepackage{epsf,epsfig,rotate}
\usepackage[french]{babel}
\usepackage{t1enc}
\usepackage{amsmath,amsthm,amsfonts,amssymb,wrapfig}
\usepackage{axodraw}
\newcommand{\beq}{\begin{equation}}
\newcommand{\eeq}{\end{equation}}

\newcommand{\beqn}{\begin{eqnarray}}
\newcommand{\eeqn}{\end{eqnarray}}

\newcommand{\ben}{\begin{eqnarray*}}
\newcommand{\een}{\end{eqnarray*}}

\def\pz{\phantom{0}}
\def\pzz{\phantom{00}}
\newcommand{\antibar}[1]{\ensuremath{\mathrm{#1\overline{#1}}}}
\newcommand{\pp}{\ensuremath{\antibar{p}}}
\def\dbar{{/\mkern-13mu{D}}}

\begin{document}

\centerline{$\ $}
\begin{flushright}
{CERN-PH-TH/2005-088}
\end{flushright}
\vspace{3cm}
\centerline{\Huge Nouvelles physiques des particules}
\vspace{1cm}
\begin{center}
{\large {\bf Julien Welzel} et {\bf David Gherson}}\\
{Institut de Physique Nucléaire de Lyon (IPNL),
Université Claude Bernard Lyon-I,\\
Villeurbanne, France}\\
\vspace{1cm}
{\large \bf John Ellis}\\
{Division de Physique Théorique, Département de Physique, CERN,\\
Genève, Suisse}\\
\vspace{1cm}
{Notes rédigées par J.W. et D.G. du cours donné par
J.E. \`a la XXXVIème école de Gif-sur-Yvette, septembre 2004, CERN.}\\
\vspace{1cm}
\underline{\large Résumé}
\end{center}
Le premier chapitre concerne le Modèle Standard de la physique des
particules et la brisure de la symétrie électrofaible. Il sert
principalement d'introduction aux chapitres suivants qui traitent de
certaines "nouvelles physiques". La supersymétrie, par exemple, est
exposée dans le second chapitre mais quelques aspects moins bien connus
comme sa brisure sont abordés plus loin. Un accent est mis sur la
phénoménologie de l'extension supersymétrique minimale du Modèle
Standard et ses perspectives aux futurs accélérateurs comme le LHC.  La
physique des neutrinos et la violation des nombres leptoniques
apparaissent au chapitre 3. Enfin, le dernier chapitre contient des
discussions sur les principales théories candidates à la description de la
physique à très haute énergie : de la Grande Unification aux théories des
cordes en passant par les dimensions supplémentaires et la supergravité.

\newpage$\ $\newpage
\section*{Avant-propos}

Ce cours a été donné par John Ellis en septembre 2004 à l'occasion de la 36ème
école d'été de Gif-sur-Yvette, intitulée "Le futur de la physique des hautes
énergies". Le public était principalement composé d'expérimentateurs et le but
de ce cours était de présenter les physiques possibles au-delà du Modèle 
Standard les plus connues et les plus étudiées. 

Il a été écrit par Julien Welzel (chapitres 1, 2, 3 et 4-section 1) et David 
Gherson (chapitre 4-sections 2, 3, 4, 5), tous deux étudiants en thèse dans le 
groupe de physique théorique
de l'Institut de Physique Nucléaire de Lyon.

Nous remercions chaleureusement John Ellis et Patrick Janot pour nous avoir
laissé l'opportunité de rédiger le manuscrit du cours et pour avoir incité et
encouragé cette pratique. Nous en tirons de grands bénéfices sur notre
compréhension de la physique au-delà du Modèle Standard. 
Nous remercions aussi sincèrement Aldo Deandrea et Maurice Kibler pour leurs 
commentaires et leurs critiques sur le cours. 

La bibliographie n'est pas exhaustive, nous avons préféré donner les
références générales qui nous ont aidé à rédiger. Cependant, il y a quand même 
un certain nombre de références plus spécifiques pour citer les sources 
des différentes valeurs, graphes,... que nous avons utilisé. 

De plus, ce cours est à prendre dans le contexte de l'école et 
complémentairement à d'autres. Ainsi, il est recommandé de consulter les cours
de D. Treille, G. Unal, A. Blondel et A. De Roeck, par exemple, si on veut avoir
de plus amples détails sur les aspects plus expérimentaux.

\begin{flushright}
J.W, D.G
\end{flushright}

\newpage
\tableofcontents
\newpage


\chapter{Nouvelle physique et brisure électrofaible} 			       %

La nécessité d'une nouvelle physique n'est pas une observation récente même si
jusqu'à il y a quelques années, le Modèle Standard (MS) de la physique des
particules n'avait pas été mis en défaut expérimentalement auprès des
accélérateurs. La physique des neutrinos avec le phénomène des oscillations de
neutrinos n'a fait que confirmer expérimentalement ce que nous
savions dejà : le MS souffre d'insuffisances et plusieurs
indices phénoménologiques ne peuvent trouver une explication qu'au-delà du MS.
Nous vivons une période charnière pour la physique des particules car
nous ne sommes qu'à quelques années des premiers résultats
du grand collisionneur de protons, le LHC, où de la nouvelle physique fera son
apparition~\footnote{Soyons optimistes !}. C'est un moment très
important pour faire un "bilan" et un récapitulatif des
possibles nouvelles physiques et des idées qui se sont développées.
\par\hfill\par

Beaucoup de cours et livres présentent et exposent les détails de la 
construction de ce qu'on appelle le Modèle Standard. Nous préfèrerons 
ici ne 
rappeler que le contenu physique et quelques détails mathématiques. Nous
ferons aussi le choix de ne pas insister sur le succès de cette théorie qui fait 
également l'objet de nombreux exposés, pour mieux présenter les 
insuffisances. 
Ainsi
il sera plus aisé de comprendre les motivations de l'élaboration d'une théorie 
plus complète qui expliquerait mieux le monde que nous observons. 
 
Dans ce chapitre introductif, nous allons donc tout d'abord exposer la 
description actuelle de la physique des particules résumée dans le Modèle 
Standard puis nous passerons en revue quelques insuffisances intrinsèques. Ainsi 
la
nécessité d'une nouvelle physique apparaitra plus clairement.

Dans un second temps nous discuterons brièvement de la place des neutrinos dans 
le MS et de l'impact de la découverte récente de leurs oscillations. 

La troisième partie, la plus longue, sera consacrée à la 
brisure électrofaible, dont l'origine est la seule inconnue du Modèle Standard. 
Nous verrons comment devrait intervenir le mécanisme de Higgs dans la génération 
des 
masses et nous 
discuterons aussi des alternatives à un boson de Higgs élémentaire que proposent
les modèles de technicouleur. Puisque nous sommes à l'aube des résultats
expérimentaux du LHC nous présenterons les possibilités de découverte du boson
de Higgs ainsi que les modèles alternatifs dans le cas où le Higgs ne paraitrait
pas dans la gamme de masses où nous l'attendons. 

Nous terminerons ce chapitre en présentant un exemple de "carte routière" des 
possibles 
nouvelles physiques et des énigmes qu'il reste à élucider dans les années à venir. 

\section{Le Modèle Standard et ses problèmes}

\subsection{Le cadre actuel : le Modèle Standard de la physique des particules}%
\subsubsection{Particules, interactions et symétries}
Avant d'aller explorer la physique des hautes énergies et les théories possibles, 
il est bon de se remémorer la vision actuelle de la physique des particules du 
Modèle Standard. 
\par\hfill\par
Aujourd'hui, en physique des particules, le monde est décrit comme étant un espace
 à 3 dimensions spatiales et 1 dimension temporelle contenant des particules 
élémentaires de la matière,
 {\it les fermions}, qui 
interagissent entre elles par l'intermédiaire de 4 forces. Ces 4 interactions 
 fondamentales -- électro-magnétique, forte, faible et gravitationelle -- se
 comprennent par l'échange d'un autre type de particules élémentaires, {\it 
les bosons de jauge}. On schématise les interactions entre particules par les diagrammes de 
 Feynman. Toutes ces interactions entre particules obéissent à des 
 symétries locales et internes et les propriétés des particules s'en 
 déduisent. Les théories qui
décrivent ces interactions fondamentales sont toutes des théories de 
jauge. 

\par\hfill\par
* L'électrodynamique quantique (QED) qui est la version relativiste et 
quantique
de l'électromagnétisme est mathématiquement construite autour de la symétrie de
jauge locale U(1) (symétrie locale de la phase des champs). Son boson de jauge
associé est le {\it photon}, de masse nulle\footnote{Ce qui est une conséquence 
directe de l'invariance de jauge. Cette dernière empêche l'introduction de 
terme de masse $m G^{\mu}G_{\mu}$ dans le
lagrangien et donc d'une masse pour les champs de jauge $G^{\mu}$.}, de spin 1, 
et il se couple à la charge 
électrique des fermions. 

\par\hfill\par
* La QCD décrit l'interaction forte et celle-ci se produit par l'échange de 
{\it gluons}, qui sont au nombre de huit\footnote{\label{nombre de bosons de jauge} 
Le 
nombre de bosons de jauge est donné par le nombre de générateurs de la symétrie de 
jauge. Pour les théories type SU(N), ce nombre est $N^2-1$.} et qui se 
couplent à la charge de "couleur" des 
fermions. Cette charge de couleur est l'analogue "fort" de la charge électrique
bien connue.  La masse des gluons est 
nulle\footnote{C'est aussi une conséquence directe de l'invariance de jauge.} et ils
 ont un spin 1. L'interaction forte se distingue entre autres
 par 2 propriétés importantes,
le {\it confinement} et la {\it liberté asymptotique}~\footnote{ La charge de
couleur du quark diminue quand on se "rapproche" du nuage de gluons autour du
quark. A la limite asymptotique (distances infiniment courtes), l'interaction de
couleur entre les quarks est donc nulle. Ceux-ci sont donc quasi-libres à très
courtes distances. C'est la \underline{liberté asymptotique}. Inversement, à des distances
grandes (environ $10^{-13}$ cm), l'interaction de couleur devient très
forte et les quarks sont \underline{confinés} en hadrons et n'existent pas à l'état libre. 
La démonstration de la liberté asymptotique a d'ailleurs valu cette année le prix 
Nobel de physique à D.~J.~Gross, F.~Wilczek et H.~D~.Politzer. En 
revanche, la démonstration 
du confinement reste toujours un défi pour les théoriciens.}.  

\par\hfill\par
* Le modèle de Glashow-Weinberg-Salam (GWS) décrit à la fois l'interaction 
faible et l'interaction
électromagnétique, réunies sous le terme d'interaction électrofaible. Mais ce
 modèle n'unifie pas vraiment les 2 interactions en une seule, il les
englobe mathématiquement dans un même formalisme, un même groupe de symétrie 
de jauge : $SU(2)_L\otimes U(1)_Y$. Dans ce modèle il y a en effet deux 
constantes de couplage différentes, conventionnellement $g$ et $g'$,
respectivement. Si on veut 
retrouver la constante de couplage électromagnétique $e$, on utilise la relation :
\beq e=\displaystyle\frac{g\,g'}{\sqrt{g'^2 + g^2}}. \eeq
Les bosons de jauge 
associés à cette symétrie sont au nombre de quatre~\footnote{
Comme dans le cas de QCD, mathématiquement, le nombre de bosons de jauge est lié
au groupe de symétrie - voir la note \ref{nombre de bosons de jauge}.}, 
trois "$W$" pour $SU(2)$ et un "$B$" pour $U(1)$. Mais les 
bosons physiques de la théorie, ceux qui s'observent dans les interactions, sont
 le photon $A$ et les bosons $W^\pm$ et $Z^0$.
Les bosons $A$ et $Z$ s'obtiennent par une rotation, dans l'espace interne, des bosons
 $W^3$ et $B$ d'un angle $\theta_W$ appelé angle électrofaible :
\beq \left(\begin{array}{c} Z^{\mu} \\ A^{\mu} \end{array}\right)=
\left(\begin{array}{cc} \cos(\theta_W) & \sin(\theta_W) 
\\ -\sin(\theta_W) & \cos(\theta_W) \end{array}\right)\left(\begin{array}{c} 
W_3^{\mu} \\ B^{\mu}\end{array}\right),
\eeq  
et nous avons la relation :
\beq \sin^2(\theta_W)=\displaystyle\frac{g'^2}{g'^2 + g^2}, \eeq
qui vaut expérimentalement $0.23120\pm0.00015$ d'après le PDG 2004~\cite{PDG04}.
C'est de ce point de vue que l'on considère l'unification des deux interactions. 
La "vraie" unification (une constante de couplage unique pour l'ensemble) 
se produirait plutôt à des énergies de l'ordre de
$10^{16}$ GeV, et incluerait aussi l'interaction forte. Nous étudierons ceci dans 
le cadre des 
théories de Grande Unification beaucoup plus tard, au dernier chapitre.  
Les trois bosons $W^\pm$ et $Z^0$ ont des masses approximatives de 80 et 91 GeV 
respectivement. 
La masse non-nulle des 
bosons faibles rend l'interaction de très courte 
portée~\footnote{Puisque $\Delta E \Delta T
\simeq \hbar$, nous trouvons que la portée $L \simeq \hbar/c M$.}, si bien 
qu'à des 
énergies faibles devant ces masses on retrouve la théorie de Fermi des 
interactions faibles.
Mais l'interaction faible a aussi une spécificité très importante 
phénoménologiquement : {\it la violation de la parité} (c'est-à-dire la non-invariance
par renversement des coordonnées spatiales : $x\to-x$) dans les interactions à 
courants 
chargés (par échange de bosons chargés $W^+$ et $ W^-$). Cette 
spécificité se traduit par le fait que seules les particules de
chiralité~\footnote{Toute particule de spin 1/2 peut être décomposée en une
partie appelée "gauche" et une partie appelée "droite" qui ont des transformations 
différentes
sous le groupe de Lorentz. La chiralité se confond
d'ailleurs avec l'hélicité, projection du spin sur l'impulsion, dans le cas de
particules de masses nulles.} gauche sont sensibles à l'interaction faible.  

\par\hfill\par
* Enfin la quatrième et dernière interaction connue mais qui n'est pas comprise 
dans le Modèle Standard, la gravitation, est
supposée véhiculée par le {\it graviton}, de masse nulle et de spin 2. Elle ne 
joue pas 
de rôle en physique des particules aux énergies qui nous sont accessibles (quelques TeV). 
L'interaction gravitationnelle est décrite classiquement par la Relativité
Générale que l'on peut considérer comme une théorie de jauge basée sur une invariance 
locale de l'espace-temps (le groupe de symétrie est alors le groupe de Poincaré).  

\par\hfill\par
Les symétries sont très utiles pour décrire les interactions fondamentales 
mais ont aussi d'importantes implications pour la classification des fermions.
Ces constituants de la matière sont classés en 2 catégories, de propriétés 
différentes (ils ont des nombres quantiques et des transformations 
de symétrie différents). Le tableau~(\ref{particules}) résume le contenu en
particules du 
Modèle Standard.
 
\begin{table}[htbp!]
\begin{center}
\begin{tabular}[t]{||l|c|c||}
\hline
{Multiplet}& {Particules}& {$SU(3)_C\otimes SU(2)_L\otimes U(1)_Y$}	\\
\hline	    
	    &{Générations}&						\\
\hline
\hline
& & \\
$L_L$ &$\left(\begin{array}{c} \nu_e \\ e^- \end{array}\right)_L$ , 
$\left(\begin{array}{c} \nu_{\mu} \\ \mu^- \end{array}\right)_L$ , 
$\left(\begin{array}{c} \nu_{\tau} \\ \tau^- \end{array}\right)_L$ &(\textbf{1},\textbf{2},-1)\\
 $E_R$	& $e^-_R$ , $\mu^-_R$ , $\tau^-_R$	&(\textbf{1},\textbf{1},-2)\\
& & \\
\hline
\hline
& & \\
 $Q_L$ &$\left(\begin{array}{c} u \\ d \end{array}\right)_L$ ,  	
 $\left(\begin{array}{c} c \\ s \end{array}\right)_L$ ,
 $\left(\begin{array}{c} t \\ b \end{array}\right)_L$
 &(\textbf{3},\textbf{2},+1/3)\\
 $U_R$ & $u_R$ , $c_R$ , $t_R$	&(\textbf{3},\textbf{1},+4/3)\\
 $D_R$ & $d_R$ , $s_R$ , $b_R$	&(\textbf{3},\textbf{1},-2/3)\\
& & \\
\hline
\end{tabular}
\end{center}
\caption[contenu du MS]{ Contenu en particules du MS. }
\label{particules}
\end{table}
 
\noindent Nous avons pour les fermions :
 
-d'un coté les {\it quarks}, sensibles aux interactions forte, faible et 
électromagnétique (et gravitationnelle). 
Ce sont des triplets de $SU(3)_C$ qui existent donc en 3 couleurs 
(conventionnellement rouge, vert ou jaune, bleu). Ils se déclinent en 3 saveurs 
($u$, $c$ et $t$) de charge électrique $+(2/3)e$ et  3 saveurs 
($d$, $s$ et $b$) de charge électrique $-(1/3)e$ ainsi qu'en 2 
chiralités (gauche et droite).  

-d'un autre coté les {\it leptons}, qui ne sont pas sensibles à l'interaction 
forte mais sont sensibles aux trois autres. Ils ne portent pas de charge de 
couleur et ont des masses généralement
moins élevées que les quarks. Nous distinguons les leptons chargés $-e$ (électron,  
muon, tau) qui existent avec les 2 chiralités gauche et droite et les neutrinos
de charge électrique nulle mais uniquement gauches\footnote{Les neutrinos droits 
n'ont pas été observés à l'heure actuelle, voir plus loin.}. 

Quarks et leptons sont aussi répartis en 3 {\it générations} ou {\it familles},
de masses de plus en plus élévées. Pour éviter les 
\textit{anomalies}~\footnote{A l'origine, les anomalies viennent de la brisure
des symétries classiques du lagrangien par les corrections quantiques à
boucles. On les observe par exemple dans la désintégration
du $\pi^0$ en 2 photons. Pour les éliminer, il faut introduire des
conditions extérieures comme des conditions sur les hypercharges ou sur les
nombres de quarks et leptons.} il faut 
que le nombre de générations soit le même entre quarks et leptons. Ce nombre est
fixé à 3 grâce à la mesure de la largeur du Z.\footnote{En mesurant la largeur de
désintégration du boson $Z^0$, c'est-à-dire l'inverse de son temps de vie, 
les expériences du
LEP ont pu déduire le nombre de saveurs de neutrinos légers. Ils sont au nombre
de $2.984\pm0.011$. D'après le MS, cela limite le nombre de doublets de leptons 
et pour éviter l'apparition d'anomalies cela limite aussi le nombre de quarks.
Il n'y a donc que 3 générations de quarks et leptons.}

Les antiparticules correspondantes possèdent les même masses, mais les charges
électriques et les chiralités sont opposées.

\par\hfill\par
Mathématiquement, les particules se distinguent par leurs propriétés de 
transformation sous les différents groupes de symétries de jauge du Modèle 
Standard, $SU(3)_C$, $SU(2)_L$ et $U(1)_Y$. A chacun de ces groupes est associé 
un nombre quantique: la couleur, l'isospin faible $T$ et sa troisième composante 
$T_3$ et l'hypercharge $Y$. La charge électrique s'obtient par la combinaison 
linéaire $Q=T_3 + Y/2$. La dernière colonne du tableau~(\ref{particules}) 
indique, pour chacun des 3 groupes du MS, la représentation à laquelle 
appartient la particule. Les nombres quantiques associés 
s'en déduisent. Par exemple, l'électron gauche se transforme dans le MS selon 
la représentation notée (\textbf{1},\textbf{2},-1). Ceci signifie :

-que c'est un singulet de couleur (charge de couleur nulle, il ne subit pas 
l'interaction forte), 

-qu'il forme avec le neutrino gauche un doublet d'isospin T=1/2 
(et son $T_3$, la projection de l'isospin faible sur un
axe, vaut -1/2 ), 

-qu'il possède une hypercharge égale à -1. 

\noindent Nous pouvons vérifier que nous obtenons bien une charge électrique
 -1 en unité de la charge électrique de l'électron, $e$.

\subsubsection{Le lagrangien du Modèle Standard }

Nous venons de décrire le contenu du MS en champs de matière et en champs de 
jauge,
nous pouvons maintenant résumer tout ceci dans le lagrangien de la théorie.
Mais il manque encore quelques élements essentiels qui n'ont pas été abordés.

Depuis Cabibbo et grâce à Glashow, Iliopoulos, Maiani puis Kobayashi 
et Maskawa, on sait maintenant que les quarks sont "mélangés". C'est un
des aspects les plus 
importants pour la phénoménologie. On distingue les {\it états de saveurs} qui 
sont les états propres
d'interaction faible et les {\it états de masse} qui sont les états propres de
propagation. Ces deux bases ne sont pas identiques, et la matrice de Cabibbo, 
Kobayashi et Maskawa (CKM) est la matrice de changement de base :

\beq \left(\begin{array}{c} d'\\s'\\b'\end{array}\right)=
V_{CKM} \left(\begin{array}{c} d\\s\\b \end{array}\right)
=\left(\begin{array}{ccc} V_{ud} & V_{us} & V_{ub} \\ V_{cd} & V_{cs} & V_{cb} 
\\ V_{td} & V_{ts} & V_{tb} \end{array}\right)
\left(\begin{array}{c} d\\s\\b \end{array}\right)
\eeq
La matrice $V_{CKM}$ s'écrit souvent sous la forme de trois rotations successives 
dans l'espace des saveurs et une phase. Cette redéfinition de la base des quarks n'a d'effet que 
sur les courants chargés (échange d'un $W$) car la matrice est unitaire 
($V\,V^{\dag}=1$). Les courants neutres ne changent que très peu la saveur~\footnote{
On utilise en général l'abréviation FCNC issue de l'anglais Flavour Changing
Neutral Current pour parler de courant neutre qui change la saveur.}.
La matrice CKM est complexe ce qui se traduit par des phases $e^{i\delta}$ dans
la définition des états propres. Mais nous pouvons montrer qu'une seule phase
est physique (observable) et reste {\it a priori} non-nulle. 
Physiquement, c'est cette phase qui serait à l'origine de la violation de la 
symétrie $CP$ dans certaines interactions faibles~\footnote{Par exemple dans
 $K\to \pi\pi$ ou dans $B^0 \to K_S J/\psi$. Si la symétrie $CP$ 
 n'est pas respectée, alors un processus réalisé avec les 
antiparticules ne donnera pas le même résultat (c'est-à-dire la même section 
efficace, par exemple) que le même processus avec des particules.}.

Le lagrangien du MS se construit en suivant deux règles essentielles : 
{\it l'invariance de jauge sous le groupe de symétrie du MS} et la {\it 
renormalisabilité}. En effet, cette dernière assure 
que les quantités mesurables calculées seront finies car les infinités 
rencontrées ne sont pas physiques. Ceci rend ce modèle très prédictif
contrairement à l'ancienne théorie des interactions faibles, la théorie de 
Fermi. Le lagrangien~(\ref{LMS}) comporte 4 parties, une pour chaque ligne; la 
première décrit {\it le 
secteur de jauge} c'est-à-dire toute la dynamique des champs de jauge 
(des bosons de jauge). Le deuxième terme est {\it le secteur de Dirac } dans 
lequel nous
retrouvons tous les champs de matière $\psi$ (les fermions) et leurs interactions 
avec les 
champs de jauge. Le troisième terme est {\it le secteur de Yukawa} qui contient
les interactions des fermions avec le champ de Higgs $\phi$ et qui donne après 
brisure
de la symétrie électrofaible toutes les masses des fermions. Enfin, la
dernière ligne correspond au {\it secteur de Higgs} qui réalise le 
{\it mécanisme de Higgs} dont nous parlerons plus loin.
\begin{eqnarray}
{\cal L} & = & - \frac{1}{4} \, F^a_{\mu \nu} F^{a\ \mu \nu}\nonumber \\
& + &i \bar{\psi}\, \dbar\, \psi + h.c. \nonumber \\
& + &\psi_i y_{ij} \psi_j \phi + h.c. \nonumber \\
& + &|D_{\mu} \phi|^2 - V(\phi).
\label{LMS}  
\end{eqnarray}
\subsection{Le Modèle Standard  et les tests de précision électrofaibles}

Même si ce cours est orienté vers les nouvelles physiques et l'après-MS, 
ne pas présenter l'extraordinaire accord du MS avec les données expérimentales ne 
serait pas lui rendre justice. C'est 
en effet un modèle très prédictif et qui a été testé et vérifié avec une précision
supérieure au pourcent, dans une large variété d'expériences, et dans un vaste 
domaine
d'énergie : de la violation de la parité dans les atomes avec une (impulsion)$^2$ 
transférée, $Q^2$, d'environ $10^{-10}$ GeV$^2$ à des collisions 
proton-antiproton à 
$Q^2 \simeq 10^5$ GeV$^2$.
Il existe des références récentes~\cite{PrecisionTests,PrecisionTests2} dans lesquelles tous ces tests sont présentés,  
nous nous contenterons de résumer l'accord de ce modèle avec les nombreux tests
de précision électrofaibles dans le tableau~(\ref{tableau EW data}). On y retrouve
par exemple les résultats de la mesure de la largeur du $Z$, de la masse du $W$ et
de celle du top. 
L'ensemble des tests forme un ensemble de contraintes très fortes. A
l'examen du tableau~(\ref{tableau EW data}) nous ne pouvons pas encore dire que 
nous avons vu de déviation significative des prédictions du MS et donc de 
nouvelle physique. Néanmoins, cet accord contraint fortement les nouvelles 
physiques possibles. 
Jusqu'à
présent le Modèle Standard est donc une description très performante de la réalité à 
des échelles d'énergies inférieures à quelques centaines de GeV. 

\begin{table}[htbp!]
\begin{center}
  \renewcommand{\arraystretch}{1.30}
\begin{tabular}{|ll||r||r|}
\hline
&Observable& Mesure$\ \ \ \ \ \ $  & fit du MS  \\
\hline
\hline
&$m_Z$ [GeV] & $91.1875\pm0.0021\pz$ &91.1873$\pz$ \\
&$\Gamma_Z$ [GeV] & $2.4952 \pm0.0023\pz$ & 2.4965$\pz$ \\
&$\sigma_h^0$ [nb]   & $41.540 \pm0.037\pzz$ &41.481$\pzz$ \\
&$R_l^0$          & $20.767 \pm0.025\pzz$ &20.739$\pzz$ \\
&$A_{FB}^{0,l}$       & $0.0171 \pm0.0010\pz$ & 0.0164$\pz$ \\
\hline
&$A_l$~(SLD)   & $0.1513 \pm0.0021\pz$ & 0.1480$\pz$ \\
\hline
&$A_l~(P_\tau)$& $0.1465 \pm0.0033\pz$ & 0.1480$\pz$ \\
\hline
&$R_b^0$       & $0.21644\pm0.00065$   & 0.21566     \\
&$R_c^0$       & $0.1718\pm0.0031\pz$  & 0.1723$\pz$ \\
&$A_{FB}^{0,b}$     & $0.0995\pm0.0017\pz$  & 0.1037$\pz$ \\
&$A_{FB}^{0,c}$     & $0.0713\pm0.0036\pz$  & 0.0742$\pz$ \\
&${\mathcal A}_b$         & $0.922\pm 0.020\pzz$  & 0.935$\pzz$ \\
&${\mathcal A}_c$         & $0.670\pm 0.026\pzz$  & 0.668$\pzz$ \\
\hline
&$\sin^2\theta^{lept}_{eff}(Q^{had}_{FB})$
     & $0.2324\pm0.0012\pz$  & 0.23140     \\
\hline
\hline
&$m_W$ [GeV]
                & $80.425\pm0.034\pzz$  &80.398$\pzz$ \\
&$\Gamma_W$ [GeV]
                & $ 2.133\pm0.069\pzz$  & 2.094$\pzz$ \\
\hline
&$m_t$ [GeV] ($\pp$)
                & $178.0\pm4.3\pzz\pzz$ &178.1$\pzz\pzz$ \\
\hline
& $\Delta\alpha^{(5)}_{had}(m_Z^2)$
                & $0.02761\pm0.00036$   & 0.02768     \\
\hline
\end{tabular}\end{center}
\caption[Overview of results]{ Ce tableau résume les mesures de précision à 
grand $Q^2$ et donne les résultats fités dans le cadre du MS. De plus amples 
détails sur la signification de certaines observables se trouvent dans la 
référence~\cite{PrecisionTests} d'où est extrait ce tableau.}
\label{tableau EW data}
\end{table}

\subsection{Les insuffisances du Modèle Standard}

Le constat expérimental précédent, en faveur du MS, confronté au constat 
théorique suivant est source d'étonnements dans la communauté des physiciens 
des particules. En effet, le MS n'explique pas, par exemple, les nombres quantiques des 
particules (charge électrique, isospin faible, couleur,...), ne prédit pas 
le spectre de masse ni n'inclut l'interaction gravitationnelle. 
Passons un peu de
temps sur chaque point pour mieux comprendre pourquoi le Modèle Standard ne 
peut pas être la description ultime de la nature.

		\subsubsection{Les paramètres libres}
Outre les nombres quantiques des particules, le MS contient au moins 19 
paramètres libres, non déterminés par la théorie et
ajustés {\it a posteriori} par l'expérience. Il n'explique donc pas l'origine de
la valeur de ces paramètres. Si nous faisons le décompte, nous avons :

- 3 constantes de couplage pour chacune des 3 interactions (le MS ne nous
 dit pas pourquoi l'interaction faible est si faible ni pourquoi 
 l'interaction forte est si forte à basse énergie, les valeurs ont été déduites
 des mesures). De façon équivalente, on peut remplacer ces trois paramètres
 libres par $\alpha_s$, $e$ et $\sin^2\theta_W$.

- 1 paramètre de QCD, $\theta_{QCD}$ qui correspond à une possible violation de
$CP$ dans les interactions fortes~\footnote{Ce paramètre est la cause de ce que
l'on appelle le "strong
$CP$ problem" car on ne comprend pas pourquoi sa valeur est si faible voire 
peut-être même nulle. En effet, les mesures
expérimentales donnent une limite supérieure sur $\theta_{QCD}$ de
$10^{-9}$~\cite{Mohapatra}.}

- les 6 masses des 6 saveurs de quarks $u$, $c$, $t$, $d$, $s$, $b$

- 3 angles de mélanges des quarks + 1 phase de violation de $CP$ faible

- les 3 masses des 3 leptons chargés e, $\mu$, $\tau$

- les 2 paramètres du potentiel de Higgs, $\lambda$ et $\mu$ (ou de façon
équivalente $M_H$ et $M_W$ ou $M_Z$).

La découverte récente des oscillations de neutrinos ajoute
encore 9 paramètres supplémentaires. De plus, nous avons aussi 
besoin  de paramètres libres supplémentaires pour générer les masses 
des neutrinos par le mécanisme de
Seesaw~\footnote{Tout ce qui est lié à la masse des neutrinos sera précisé au 
chapitre 3.}.
Ceci est un problème que la plupart des modèles rencontrent mais 
parmi les paramètres
précédents certains trouvent une explication dans les Théories de Grande
Unification (les couplages de jauge, les angles de mélanges) ou la supersymétrie 
(les paramètres de la brisure électrofaible). 

		\subsubsection{La gravitation}
Le Modèle Standard ne donne pas de description quantique de la théorie de la 
gravité. En fait, l'interaction gravitationnelle
est ajoutée, juxtaposée au MS sous sa forme actuelle c'est-à-dire la Relativité 
Générale (RG). Cela ne
pose pas de problème à l'échelle d'énergie qui nous est accessible actuellement
puisque la force gravitationelle y est considérablement plus faible que l'interaction
faible. Mais pour des énergies autour de $10^{19}$ GeV, elle est de même intensité
et nous ne pouvons plus la traiter classiquement {\it via} la RG.

De plus, certains phénomènes cosmologiques sont insuffisamment expliqués et
 il faut introduire d'autres paramètres libres, non prédits par la théorie, 
 comme la constante cosmologique $\Lambda$, au moins un paramètre 
d'inflation, l'asymétrie baryonique~\footnote{C'est-à-dire l'asymétrie 
entre le 
nombre de particules et d'antiparticules.},...

La quantification de la gravitation, {\it i.e.} sa version quantique, est depuis
longtemps une des préoccupations principales des physiciens théoriciens. La
difficulté vient notamment de sa non-renormalisabilité. Comme
nous le verrons à la fin de ce cours sur les nouvelles physiques, la théorie
des supercordes semble pouvoir être la solution et l'espoir est permis.

		\subsubsection{La matière noire}
Au début des années 90, les astronomes ont fait une découverte~\footnote{C'est
en comparant la vitesse de rotation de la galaxie avec la valeur qu'elle aurait
si elle n'était faite que de matière ordinaire.} très importante 
pour les physiciens des particules : l'existence de {\it matière noire}.
Ils ont observé que la densité de matière connue (baryonique) ne représentait que 
moins de 4$\%$ de la densité totale de l'Univers. 
20$\%$ de la densité totale viendrait d'une matière non baryonique qui n'existe pas 
dans le MS,
ce que l'on appelle la matière noire. Pire encore, le reste nous 
est totalement inconnu et on l'attribue à une sorte d'{\it énergie
noire}~\footnote{Elle est appelée ainsi car elle elle n'est pas due à la matière.}.
Tous les modèles au-delà du MS contiennent des particules nouvelles et
celles-ci, à condition d'être stables, neutres et sans interactions 
fortes, peuvent jouer le 
rôle de matière noire. La supersymétrie
offre par exemple un candidat très sérieux, le {\it neutralino}. 
Par contre, l'énergie noire est aujourd'hui un mystère total et son explication ne peut se
 trouver que dans les modèles qui incluent la gravitation.

		\subsubsection{La hiérarchie des échelles}
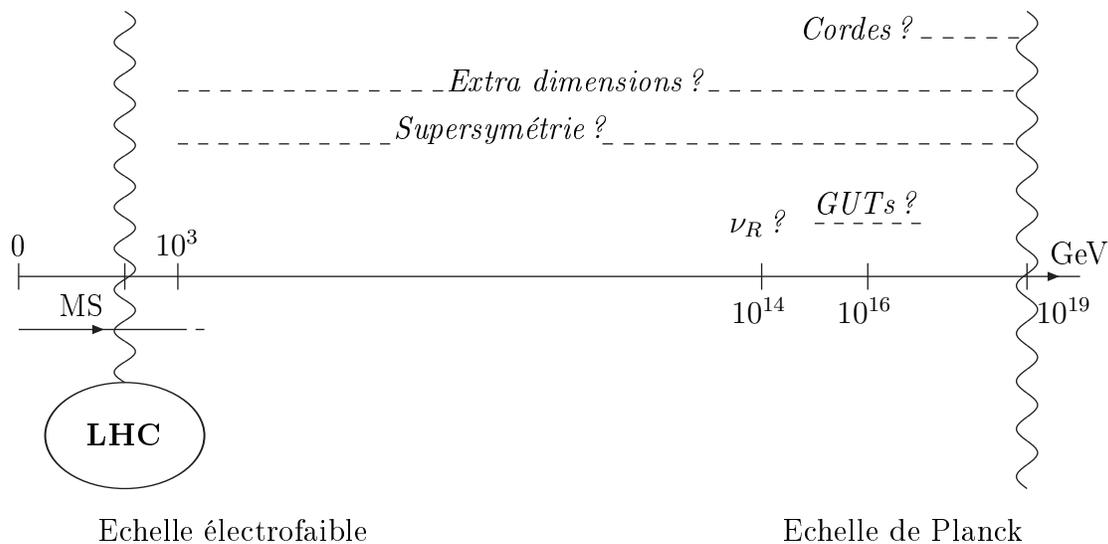
\begin{figure}[htbp!]
\begin{center}
\begin{picture}(400,200)(0,-100)
\Line(0,0)(380,0)\ArrowLine(380,0)(400,0) \Text(390,5)[lb]{GeV} 

\Photon(40,100)(40,-40){4}{9} \Text(30,-100)[lb]{Echelle électrofaible}
\Oval(40,-60)(20,30)(0) \Text(40,-60)[]{{\bf LHC}}

\Photon(380,100)(380,-80){4}{10} \Text(380,-100)[rb]{Echelle de Planck}

\ArrowLine(0,-20)(60,-20) \Text(15,-15)[lb]{MS} \DashLine(60,-20)(70,-20){4}

\DashLine(300,20)(340,20){4} \Text(320,28)[]{{\it GUTs ?}} 
\DashLine(140,50)(60,50){4} \Text(142,50)[lb]{{\it Supersymétrie ?}}
\DashLine(220,50)(375,50){4}
\DashLine(160,70)(60,70){4} \Text(162,70)[lb]{{\it Extra dimensions ?}}
\DashLine(260,70)(375,70){4} 
\DashLine(340,90)(377,90){4} \Text(338,90)[rb]{{\it Cordes ?}}
\Text(280,20)[]{$\nu_R$ \textit{?}}

\Line(0,5)(0,-5) \Text(0,8)[b]{0} 
\Line(40,5)(40,-5) 
\Line(60,5)(60,-5) \Text(60,8)[b]{$10^3$} 
\Line(380,5)(380,-5) \Text(385,-8)[lt]{$10^{19}$} 
\Line(320,5)(320,-5) \Text(320,-8)[t]{$10^{16}$} 
\Line(280,5)(280,-5) \Text(280,-8)[t]{$10^{14}$} 
\end{picture} 
\end{center}

\caption{Les différentes échelles d'énergie en physique des particules et 
les nouvelles physiques possibles. }
\label{domaine}
\end{figure}

Si nous regardons le domaine des énergies, tracé sur la figure~(\ref{domaine}), il s'étend jusqu'à $10^{19}$ 
GeV où la gravité n'est plus négligeable dans les interactions entre particules.
 Une question se pose alors, pourquoi y a-t-il 17 ordres de grandeurs qui séparent
l'échelle électrofaible de l'échelle de Planck ?! Là non plus le MS ne donne 
pas de réponse, et c'est même l'un des plus grands
défis de la physique moderne. Peu de modèles apportent une explication
satisfaisante. 
A plus petite échelle, la même observation peut être faite à propos des masses
de fermions connus. Pourquoi le top est il $4\times 10^5$ fois plus lourd que 
l'électron ? Il faut là aussi comprendre les raisons de cette autre hiérarchie
(appelée parfois hiérarchie des saveurs). L'introduction de dimensions spatiales
supplémentaires pourrait donner une réponse au problème mais ce n'est souvent
qu'une reformulation qu'elles proposent.
\par\hfill\par
Clairement, il reste des mystères qui nécessitent un autre modèle qui prendra
le relais du Modèle Standard pour la physique à plus haute énergie.

\section{Introduction au secteur des neutrinos}

Cette section a pour but d'exposer ce que la plupart des physiciens considèrent comme le premier
indice de l'existence d'une nouvelle physique au-delà du MS. Mais avant tout,
rappellons-nous les particularités des neutrinos et leur place dans le MS. 

\subsection{Les neutrinos dans le Modèle Standard }
Le neutrino fût la première fois introduit par W.Pauli pour rendre compte du spectre
continu de l'électron dans les désintégrations $\beta$ telles que : $n\to
p+e^-+\nu_e$ et c'est 25 années plus tard que F.Reines et C.Cowan Jr observèrent 
pour la première fois  
ces neutrinos~\footnote{En réalité c'était des antineutrinos électroniques.}, 
 produits par un réacteur nucléaire. Depuis les années 60, des neutrinos 
 provenant du Soleil, de l'atmosphère et même d'une supernova du Nuage de 
 Magellan ont été détectés.

Dans la th\'eorie \'electrofaible il est rang\'e avec l'électron~\footnote{Plus 
rigoureusement, c'est le neutrino électronique qui est rangé avec l'électron. 
Comme tous les fermions du MS, le neutrino existe en 3 familles : le neutrino 
électronique, muonique, tauique.} dans un doublet leptonique de chiralité 
gauche, d'isospin faible 1/2 et d'hypercharge $Y=-1$; voir
tableau~(\ref{particules}) :
\beq L=\left( \begin{array}{c} \nu_e \\ e^- \end{array}\right)_L.\eeq
Il ne subit que
l'interaction faible. Mais contrairement à l'électron ou à n'importe quel autre
lepton chargé, jusqu'à présent seul des neutrinos de chiralité gauche ont été
observés~\footnote{ Les neutrinos droits n'existent donc pas ? Si un neutrino
droit existe, étant de
chiralité droite et l'interaction faible ne concernant que les particules
gauches, il n'est pas sujet à cette interaction. Il ne lui reste que
des interactions indirectes (nous ne comptons pas la gravitation) pour se 
manifester (voir chapitre 3).} (ainsi que
des antineutrinos droits). 

La faiblesse de ses interactions avec la matière rend difficile sa détection et
les expériences qui visent à
connaitre les propriétés des neutrinos atteignent donc des tailles colossales.
Pour exemple, l'expérience SuperKamiokande au Japon est constituée d'un immense
réservoir d'eau de 45 000 tonnes. Elle est capable notamment de d\'etecter en 
temps r\'eel les \'electrons de l'eau diffus\'es 
par interaction \'elastique avec les neutrinos solaires : 
$$\nu_{\alpha}+ e^{-}\to\nu_{\alpha}+ e^{-}.$$
 
L'explication de la faiblesse des interactions se trouve dans le fait que les 
neutrinos n'interagissent que par échange de boson $W$ ou $Z$, 
c'est-à-dire :

\noindent - par courant charg\'e (\'echange d'un boson vecteur $W^{\pm}$):
 
 \beq\mathcal{L}_{cc}=\frac{-g}{\sqrt{2}}\sum_{\alpha=e, \mu, \tau}
 \nu_{L_{\alpha}}\gamma_{\mu}l_{L_{\alpha}}W^{\mu}\ +\ h.c,\eeq

\noindent - par courant neutre (\'echange d'un boson vecteur Z$^{o}$ ):
 
 \beq\mathcal{L}_{nc}=\frac{-g}{2\cos{\theta_{W}}}\sum_{\alpha=e, \mu, \tau}
 \nu_{L_{\alpha}}\gamma_{\mu}l_{L_{\alpha}}Z^{\mu}\ +\ h.c,\eeq

\noindent Les sections efficaces contiennent alors un facteur très petit:
 \beq \sigma \propto \left( \frac{g^2}{M_W^2} \right)^2 \propto G_F^2. \eeq

Nous allons maintenant discuter de la masse des neutrinos. 
Les fermions, comme l'électron, ont un terme de masse dans le lagrangien qui
s'écrit :
\beq  {\mathcal L} \supset -m^D_{\nu}\overline{f_L} f_R.  \eeq
C'est un terme qui lie un fermion gauche avec un fermion droit. La non-existence
de neutrinos droits interdit donc ce terme de masse et nous disons que le
neutrino n'a pas
de {\it masse de Dirac}.
Mais pouvons-nous écrire un terme de masse pour $\nu_L$ même sans $\nu_R$?
Oui, un terme de {\it masse de Majorana} :
\beq\label{Majorana mass} {\mathcal L} \supset -\frac{1}{2}m^M_{\nu}
\overline{(\nu_L)^c}\nu_L \eeq
avec $\nu=\nu^c=C\overline{(\nu)^T}$. $C$ est la matrice de conjugaison de
charge ($C=i\gamma^0\gamma^2$). Dans ce cas, le neutrino est sa propre 
antiparticule !

Or, il existe dans le MS une loi de conservation qui interdit ce terme.
En effet, pour expliquer la non-observation de $\mu \to e \, \gamma$, 
$\tau \to e \, \gamma$ et $\tau \to \mu \, \gamma$~\footnote{Les limites 
supérieures sur les taux de branchements c'est-à-dire sur les largeurs
partielles de désintégration sont : $BR(\mu \to e \, \gamma)<1.2\times 10^{-11}$, 
$BR(\tau \to e \, \gamma)<2.7\times 10^{-6}$, $BR(\tau \to \mu \,
\gamma)<1.1\times 10^{-6}$.} on introduit un nombre leptonique $L$ pour chaque famille
et qui se conserve. Ainsi on a les nombres leptoniques suivant pour le muon
$L_{\mu}=1,\ L_{e}=0,\ L_{\tau}=0$, l'électron $L_{\mu}=0,
\ L_{e}=1,\ L_{\tau}=0$ et par exemple la désintégration $\mu \to e \, \gamma$
ne serait pas permise à cause de la conservation de $L_{\mu}$ et $L_{e}$. Ce nombre 
leptonique est relié à une symétrie globale du lagrangien dont l'origine
profonde reste à être clarifiée.
Un terme de masse de Majorana comme l'équation~(\ref{Majorana mass}) a 
$\Delta L_{\alpha} =2$, le nombre leptonique individuel n'est pas conservé.
Il n'y a pas d'autre terme de masse possible, par conséquent \underline{le neutrino a une masse
nulle dans le Modèle Standard} ! Ceci est assez surprenant pour une particule de 
matière. Mais est-ce cohérent avec les résultats expérimentaux ?
Les observations directes sur la masse des neutrinos et les limites 
cosmologiques ne donnent malheureusement que des limites supérieures, mais il y
a un fait expérimental qui entre en désaccord direct avec des masses nulles pour les 
neutrinos : \underline{l'observation des oscillations entre saveurs de
neutrinos}. 

\subsection{Les oscillations de neutrinos : nouvelle physique}

Les oscillations de neutrinos reposent sur le même principe que le mélange CKM 
des quarks "down" : les états propres de masses, {\it i.e.} de propagation, ne 
coïncident pas avec les états propres de saveurs. 
En d'autres mots, les neutrinos produits dans les interactions faibles (états 
de saveurs) sont des combinaisons linéaires des états de propagation. Un neutrino 
de saveur purement électronique, initialement produit dans le Soleil, a donc une 
probabilité non-nulle de devenir purement de saveur muonique à l'endroit où on 
le détecte~\footnote{Dans le cas à deux saveurs, nous pouvons faire l'analogie
avec un système à deux
niveaux en mécanique quantique dans lequel la probabilité de présence oscille
entre les 2 niveaux à la pulsation de Rabi.}. 
Ainsi, les expériences comme SuperKamiokande qui peuvent faire la 
comptabilité des neutrinos électroniques provenants du Soleil observent un déficit 
important 
(plus de 50$\%$ !) par rapport au nombre de neutrinos solaires prédits par le 
modèle solaire standard~\footnote{Ce modèle est très performant et ne peut être
raisonnablement mis en doute.}. Toutes les interprétations autres que les 
oscillations,
comme une désintégration du neutrino, sont très
défavorisés~\cite{OscInterpretation, NewPhysics}.

Le phénomène d'oscillation entre saveurs nécessite une masse non-nulle pour les
neutrinos. De plus le nombre leptonique individuel n'est pas 
conservé au cours du temps. Voici pourquoi nous disons que les oscillations, et
donc les masses des neutrinos, demandent une physique au-delà du Modèle
Standard. Nous reviendrons sur ce point au chapitre 3 où les choses apparaitront 
plus clairement.
 
Il est vrai que nous pouvons toujours modifier et étendre le MS pour inclure
ce phénomène mais l'explication n'est pas naturelle. Le neutrino est visiblement
un fermion à part et l'origine de sa particuliarité doit se comprendre à partir 
d'une physique à plus haute énergie (par exemple dans les théories de Grande 
Unification que nous verrons en détail au dernier chapitre).

\section{Le secteur de Higgs : la brisure électrofaible}

\subsection{La génération des masses}
Le secteur électrofaible du MS est décrit par la symétrie de jauge
$SU(2)_L\otimes U(1)_Y$. Or
l'invariance du lagrangien sous cette symétrie ne permet pas l'introduction de
termes de masse pour les bosons ou les fermions ! A ce stade de la description,
aucune particule n'est donc massive ce qui n'est pas conforme à nos observations.  
La symétrie doit donc être brisée d'une façon ou d'une autre. Cependant, il faut
 garder cette symétrie dans 
les 
interactions c'est-à-dire dans le lagrangien. La brisure ne peut alors pas être 
{\it explicite}, elle est {\it spontanée}. 
Il existe un théorème, {\it le théorème de Goldstone} qui 
donne les deux conséquences importantes de la brisure spontanée d'une symétrie
\underline{globale} :

* Le lagrangien reste invariant mais l'état d'énergie minimale, {\it i.e.} le vide,
 n'est pas invariant. La symétrie n'est alors plus manifeste dans le spectre 
des états~\footnote{Car les états excités s'obtiennent par l'action des 
différents
générateurs sur le vide (l'état fondamental) et ne possèdent plus la symétrie.}.

* Il existe un certain nombre d'états physiques de masse nulle dont les 
propriétés sont reliées à celles des générateurs de la symétrie brisée. Ce sont 
les modes ou {\it bosons de Goldstone}. Il y en a autant que le nombre de générateurs brisés de 
la symétrie.

Quand la symétrie est approchée, par exemple la symétrie
chirale des quarks de l'interaction forte, les bosons de Goldstone ont une
masse non-nulle, mais faible, dont l'ordre de grandeur est reliée à la validité de
l'approximation. Les pions, qui ont une masse d'environ 135 MeV (donc 
beaucoup plus 
faible que les masses des autres hadrons, objets typiques de l'interaction forte), jouent
ce rôle. On les appelle les bosons de pseudo-Goldstone de la symétrie chirale de
l'interaction forte. 

Mais dans le Modèle Standard, $SU(2)_L\otimes U(1)_Y$ est une symétrie de jauge donc
\underline{locale}. Les conséquences de la brisure spontanée sont alors différentes :
certains bosons de Goldstone \underline{disparaissent} au profit de nouveaux états de
polarisation longitudinale des bosons de jauge, devenus massifs~\footnote{Nous
passons de deux états de polarisation d'un champ de jauge de spin 1 sans masse à
trois.}. Les bosons de Goldstone restants sont aussi devenus massifs. 
C'est le fameux {\it mécanisme de Higgs} de Brout, Englert, Higgs et Kibble et
c'est le seul mécanisme connu de génération des masses qui conserve la 
renormalisabilité du MS.

\par\hfill\par
Nous allons maintenant expliciter un peu plus ce mécanisme. 
Prenons le cas simple d'un champ scalaire complexe $\phi$ dans une théorie de
jauge $U(1)$. Le lagrangien est :
 \beq {\mathcal L}=D_{\mu}\phi^{\dag}D^{\mu}\phi -\frac{1}{4}
F^{\mu\nu}F_{\mu\nu} - V[\phi].\eeq
Le terme $V[\phi]$ est le potentiel effectif et
renormalisable créé par le
champ $\phi$, il peut s'écrire :
 \beq V[\phi]=-\mu^2\phi^{\dag}\phi + \lambda(\phi^{\dag}\phi)^2 \eeq
avec $\mu^2 >0$, $\lambda>0$.

Le champ $\phi$ est appelé le {\it champ de Higgs} et si nous le paramétrisons 
en fonction de 
deux champs réels, alors $\displaystyle 
\phi(x)=\frac{1}{\sqrt{2}}(\phi_1(x) +i\phi_2(x))$ et le potentiel s'écrira : 
\beq V[\phi_1,\phi_2]=\frac{\mu^2}{2}(\phi_1(x)^2+\ \phi_2(x)^2) +\frac{\lambda}{4}(\phi_1(x)^2+\
\phi_2(x)^2)^2 \eeq
\noindent Il est manifestement invariant sous une "rotation" des champs 
$\phi_1$ et $\phi_2$ et possède la symétrie $SO(2) \sim\ U(1)$~\footnote{Le 
potentiel reste invariant quand on redéfinit les champs
ainsi: $$ \left(\begin{array}{c} \phi_1^{'}\\ \phi_2^{'} \end{array}\right) 
 = \left(\begin{array}{cc} \cos\theta e^{i\delta} & \sin\theta \\ -\sin\theta
& \cos\theta e^{i\delta}\end{array}\right) \left(\begin{array}{c} \phi_1\\ \phi_2 
\end{array}\right) $$}.
Le potentiel admet une infinité de minima qui se distinguent
les uns des autres par rotation : l'état de 
vide
est dégénéré. Quand nous choisissons un minimum particulier pour le vide, en
$\phi_1=0,\ \phi_2=v/\sqrt{2}$ : 
\beq\phi_0=<0|\phi|0>= \frac{1}{\sqrt{2}}\left(\begin{array}{c} 0\\
+v \end{array}\right)\eeq 
avec 
\beq v=\sqrt{\frac{-\mu^2}{\lambda}}, \eeq
nous brisons la sym\'etrie. $\phi_0$ n'est plus 
invariant sous une transformation de jauge mais les lois
dynamiques sont préservées : $\mathcal{L}$ reste toujours 
invariant de jauge. C'est tout le principe d'une brisure
\underline{spontanée} de symétrie, par opposition à la brisure
\underline{explicite} où $\mathcal{L}$  
perd son invariance. Si nous redéfinissons le champ $\phi(x)$ comme une 
fluctuation autour de ce 
minimum particulier, en passant en coordonn\'ees polaires nous avons :
\beq \phi(x)=\frac{1}{\sqrt{2}}(v+\ \sigma(x))e^{i\pi(x)} \eeq
\noindent où $\pi(x)$ est le degré de liberté associé à la symétrie de rotation du potentiel.
 Il correspond à des modes d'excitation de masse nulle, c'est le 
 {\it boson de Goldstone} tandis que $\sigma(x)$ est le boson associé aux excitations
 radiales.
En développant la partie cinétique du lagrangien, $|D_{\mu} \phi|^2$, et en 
passant dans la jauge
unitaire~\footnote{Notre lagrangien est par construction invariant de jauge
donc nous pouvons faire ce choix particulier. Il a la particularité de faire
disparaitre les bosons de Goldstone non physiques (de masse nulle) au profit de
masse pour les bosons de jauge. Mais ce choix de jauge unitaire n'est pas toujours le plus
pratique à cause de l'existence de bosons massifs. En effet, le propagateur 
d'un boson massif s'écrit : 
\beq 
\Delta^{\mu\nu}=i\left( \frac{g^{\mu\nu}-\frac{q^{\mu}q^{\nu}}{M^2}}{q^2-M^2}\right) 
\nonumber \eeq

\noindent Il ne converge pas vers 0 quand $q^{\mu}\to\ +\infty$ (à cause du 
terme $\frac{q^{\mu}q^{\nu}}{M^2}$ ) ce qui pose problème pour la
renormalisation. Pour y rem\'edier, le {\it théorème d'équivalence} nous dit que la 
voie à suivre et de ne pas calculer les amplitudes dans la jauge unitaire et
donc de garder explicitement les bosons de Goldstone.}, nous observons 
l'apparition de termes de masse, de termes 
d'interactions, ainsi que la disparition du champ $\pi(x)$. Le champ de jauge 
a maintenant une masse $\left(\displaystyle \frac{e\, 
v}{2}\right)$~\footnote{O\`u $e$ est la constante de couplage de
la théorie que nous considérons dans cet exemple.} et le champ scalaire réel
$\sigma(x)$ une masse $\sqrt{2 \mu^2}$. 
\par\hfill\par
Comment réaliser ce mécanisme dans le MS ? Nous souhaitons briser 
$SU(2)_L\otimes U(1)_Y$ en
$U(1)_{em}$ sans toucher à $SU(3)_C$. Il nous faut donc un champ scalaire
neutre de couleur, d'isospin et hypercharge faibles non nuls et au moins une
composante de ce champ doit être nulle électriquement~\footnote{C'est pour qu'il ne se
couple pas au photon et donc qu'il laisse celui-ci sans masse.}. De plus, dans un modèle
minimal, le \underline{même} champ de Higgs brise la symétrie électrofaible et donne aussi 
une masse aux quarks et aux leptons. Dans le MS, les fermions gauches sont des 
doublets de $SU(2)_L$ et les fermions droits des singulets, donc le produit 
est un doublet de $SU(2)_L$. Le terme de masse des fermions, qui doit
être un scalaire sous $SU(2)_L\otimes U(1)_Y$~\footnote{Il doit être
invariant car le lagrangien l'est.}, ne peut s'obtenir que par couplage direct
à un champ de Higgs \underline{doublet} de $SU(2)_L$~\footnote{D'après la 
théorie des groupes, $(2 \otimes 1)\otimes 2 =3 \oplus 1$ donc on peut former 
un scalaire de $SU(2)$ avec 2 doublets et un singulet.}. Par exemple dans le 
cas des leptons,
\beq 
{\mathcal L } \supset \sum_{i,j=e,\mu,\tau}
Y^{lep}_{ij}\overline{(L_i)_L}\phi(E_j)_R
\eeq 
où $Y^{lep}$ est une constante (matrice $3\times$3 dans l'espace des saveurs)
qui caractérise le couplage au champ de Higgs. Elle est appelée {\it couplage de 
Yukawa} 
et est différente pour chaque fermion (car chaque fermion a bien une
masse distincte des autres).
Pour être en plus invariant sous $U(1)_Y$, il faut que les hypercharges vérifient
la condition~\footnote{La trace de $Y$
doit être nulle car c'est un des générateurs diagonaux du groupe de symétrie.} : 
$-Y_L+Y_{\phi}+Y_E=-(-1)+Y_{\phi}-2=0$, 
donc l'hypercharge de $\phi$ est 1. Le champ $\phi$ s'écrit alors~\footnote{En utilisant la relation $Q=
T_3 +\frac{Y}{2}$.} :

\beq \phi=\left(\begin{array}{c} \phi^+\\ \phi^0 \end{array}\right) \eeq
et il appartient à la représentation (\textbf{1},\textbf{2},+1) du MS.
Le potentiel effectif créé par $\phi$ est semblable à celui de l'exemple précédent.
Sa forme est la plus générale possible et respecte les conditions d'invariance 
sous 
$SU(2)_L\otimes U(1)_Y$ et de renormalisabilité.
\par\hfill\par
Après brisure spontanée, dans la jauge unitaire, nous remplaçons $\phi$ par sa 
valeur dans le vide : $<\phi>=\frac{1}{\sqrt{2}}\left(\begin{array}{c} 0\\
v\end{array}\right)$.

\noindent * Les fermions ont alors une masse 
$ \displaystyle M_{f}=Y_{f}\,\frac{v}{\sqrt{2}}$.

\noindent * Les bosons de jauge ont obtenu une masse : $\displaystyle
m_W=\frac{g\,v}{2}$ et 
$\displaystyle m_Z=\frac{m_W}{\cos \theta_W}$.

\noindent De $SU(2)_L\otimes U(1)_Y$ à $U(1)_{em}$, nous sommes passés de 4 
générateurs de symétrie à 1. D'aprés ce qui a été dit précédemment, 3 
générateurs sont brisés et deviennent les 3 degrés de polarisation longitudinale 
des $W^{\pm}$ et $Z^0$ qui sont maintenant massiques. Or sur les
4 champs scalaires réels~\footnote{Un doublet de champs scalaires complexes 
peut s'écrire en fonction de quatre champs scalaires réels : 
$$\phi=\left(\begin{array}{c} \phi_1+i\phi_2 \\ \phi_3 +i\phi_4 \end{array}\right)$$}
qui composent $\phi$, 3 sont les bosons de Golstones absorbés, il reste donc 
un boson de Goldstone physique, \textit{le boson de Higgs}, avec une masse :
\beq m_H^2= 2\mu^2=2\lambda v. \eeq

Si nous résumons ce qui vient d'être dit, le mécanisme de Higgs permet de briser
spontanément la symétrie électrofaible, donnant alors une masse aux bosons de jauge 
$W^{\pm}$ et $Z^0$. Les fermions obtiennent eux une masse en se couplant avec le
champ de Higgs et ce couplage est proportionnel à $\displaystyle \frac{M_f}{v}$.
De cette brisure, il reste le boson de Higgs, de spin nul et de masse 
$\sqrt{2\mu^2}$. 

\begin{figure}[htbp!]
\centerline{\epsfxsize=14cm\epsfbox{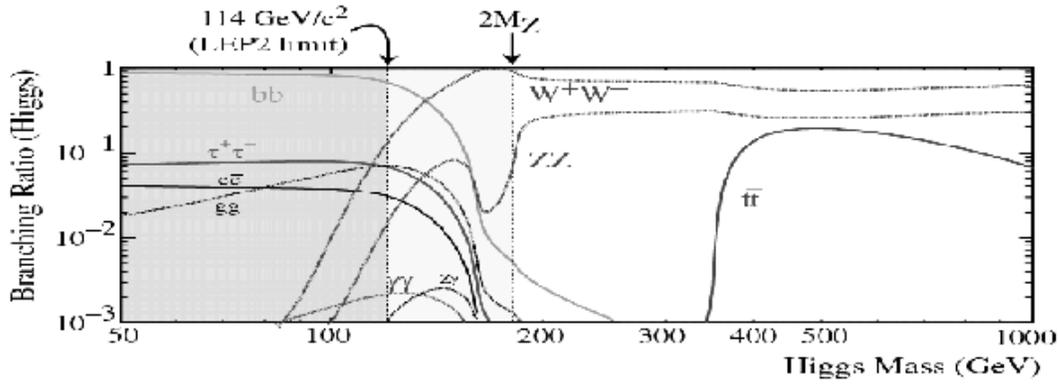}}
\caption{Quelques taux de branchements du boson de Higgs en fonction de
 sa masse.~\cite{NewPhysics}}
\label{Higgs BR}
\end{figure}

A partir de $m_W$ ou de $G_F$, reliés par la relation 
\beq 
m_W=\left(\frac{\pi\alpha/\sqrt{2}\mathrm{G}_F}{\sin\theta_W}\right)^{1/2},
\eeq
on peut déduire la valeur de $v$ :
\beq v=(\sqrt{2} G_F)^{-1/2}=246\ \mathrm{GeV}. \eeq 
Connaissant $m_W$, la masse du boson de Higgs 
est donc le seul paramètre libre de ce
secteur ce qui rend possible la prédiction de la production et des
désintégrations du Higgs en fonction de sa masse. Dans la figure~(\ref{Higgs BR}) nous montrons les
désintégrations du Higgs les plus importantes en fonction de sa masse.
Notamment, pour $m_H$ entre 100 et 150 GeV, les désintégrations en une paire 
$\bar{b} b$, $\gamma\gamma$ ou $WW$ semblent très prometteuses et les
collaborations CDF et D$\emptyset$ en font la recherche très 
active. Le LHC
pourra davantage explorer la phénoménologie du Higgs.

\subsection{Les propriétés du Higgs et les modèles Technicouleur}

Si nous calculons les {\it corrections radiatives}~\footnote{Ces corrections sont
dues aux boucles de particules virtuelles.} à la masse du Higgs dans le MS,
nous observons qu'elles divergent quadratiquement :
\beq \delta m_H^2 \propto {\mathcal O}(\frac{\alpha}{\pi})\Lambda^2 \eeq
Quand le cut-off $\Lambda$ est supérieur au TeV, les corrections n'ont plus 
de sens et nécessitent l'introduction d'une nouvelle physique. L'existence d'un scalaire fondamental dans la théorie n'est pas non plus 
conceptuellement très
satisfaisante. Pour l'instant, aucun scalaire élémentaire n'a été détecté dans 
la nature. Toutes les particules de matière sont des fermions de spin 1/2. 
L'origine du mécanisme de Higgs est inconnue et l'échelle électrofaible paraît
arbitraire. Elle implique des ajustements fins des paramètres et c'est ce qui la
rend non-naturelle. Pour remédier à tout ceci, la {\it  Technicouleur} (TC) fut introduite. 
Cette théorie repose sur
l'existence d'une brisure \underline{dynamique} (comme pour QCD) de l'interaction 
électrofaible. Le boson de Higgs est alors un condensat fortement lié de deux
fermions très lourds, les techniquarks. C'est la valeur non-nulle dans le
vide de ce condensat qui brise spontanément la symétrie électrofaible et réalise
le mécanisme de Higgs :

\beq <\phi>=<0\mid \overline{F_L}F_R\mid 0>=v. \eeq

Ce mécanisme de "condensation" s'observe en physique du solide quand un materiau
devient supraconducteur et c'est la condensation en paires de Cooper (paires
d'électrons) qui brise la
symétrie $U(1)_{em}$.
La TC donne donc une origine purement dynamique à la brisure et
l'échelle électrofaible est reliée au couplage technicouleur mobile. 
C'est une nouvelle interaction, très forte, qui se manifeste à des énergies de
l'ordre du TeV. Sa présence résoud le problème de la divergence quadratique de
$m^2_H$. Dans le modèle minimal, la symétrie de jauge sur laquelle se base la TC
 est $SU(N_{TC})$ et 
les nouveaux fermions, U et D, appartiennent à la représentation fondamentale. 
Ce sont les analogues des quarks légers de la symétrie chirale approchée de QCD.

Mais la version la plus simple, basée sur une transposition à plus haute énergie
de QCD, n'est pas compatible avec les données de précision électrofaible.
D'autres modèles non-minimaux ont été construits, comme la technicouleur étendue
(ETC), la "walking" TC, la "topcolor assisted" TC, mais nous ne les étudierons pas
dans ce cours. Le LHC devrait pouvoir découvrir ou infirmer le principe de la TC
et du Higgs composite.

\subsection{Les contraintes sur la masse du Higgs}

Les données de précision électrofaibles sont capables de nous renseigner
sur la nature du Higgs et d'une façon générale sur la nouvelle physique.
En effet, Veltman a montré la sensibilité des corrections à boucles aux
particules "invisibles", c'est-à-dire trop massives pour être produites.
Par exemple, au niveau d'une boucle, les masses du $W$ et du $Z$ sont
données par:

\beq m_W^2 \sin^2\theta_W= m_Z^2 \cos^2\theta_W\sin^2\theta_W=\frac{\pi
\alpha}{\sqrt{2}{\mathrm G_F}}(1 + \Delta r) \eeq
où $\Delta r$ est la correction radiative à calculer. Celle-ci reçoit
d'importantes contributions du quark lourd $t$:
\beq \Delta r \supset \frac{3{\mathrm G_F}}{8 \pi^2 \sqrt{2}} m_t^2 \eeq
et du boson de Higgs :
\beq \Delta r \supset \frac{\sqrt{2}{\mathrm G_F}}{16 \pi^2} m_W^2 (\frac{11}{3} \ln
\frac{m_H^2}{m_Z^2} +...),\; m_H>> m_W. \label{veltmanHiggs}\eeq

\begin{figure}[htbp!]
\centerline{\epsfxsize=8cm\epsfbox{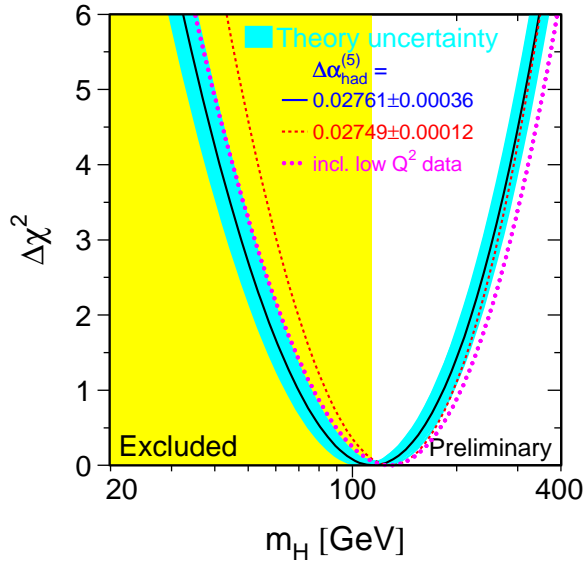}}
\caption{Contrainte sur la masse du Higgs à partir 
des données expérimentales, dans le cadre du MS. On rappelle que le 
$\chi^2$ sert à estimer la qualité du fit et que le $\chi^2$ le plus bas
correspond au meilleur fit. $\Delta\chi^2=\chi^2(m_H)-\chi^2_{min}$.~\cite{NewPhysics}}
\label{contrainte sur mH}
\end{figure}

L'influence de toute particule nécéssaire pour renormaliser le MS, par
exemple le quark $t$ et le boson de Higgs, diverge quand sa masse devient
infinie. Les tests de précision peuvent donc être sensibles à la nouvelle
physique même si les nouvelles particules sont trop lourdes pour être
produites directement. L'influence quadratique du quark $t$ a permis, il y
a plusieurs années, la prédiction de la masse du top, ce qui a été ensuite
confirmée par sa mesure a Fermilab. L'influence du Higgs au niveau d'une
boucle n'est que logarithmique, comme nous voyons dans
l'équation~(\ref{veltmanHiggs}), \`a cause de ce que l'on appelle
l'\textit{écrantage de Veltman}~\footnote{Au niveau de deux boucles,
l'influence de $M_H$ devient quadratique aussi.}. Néanmoins, en utilisant
plusieurs données différentes, on peut tout de même donner une estimation
de la masse du Higgs. En combinant toutes les données acquises par le LEP, le
Tevatron et SLD nous obtenons la figure~(\ref{contrainte sur mH}).

Les mesures expérimentales accordent au boson de Higgs une masse :
\beq m_H=114^{+69}_{-45}\ {\mathrm GeV} \eeq 
Les recherches d'observations directes du LEP donnent une limite inférieure à 
la masse du Higgs puisque celui-ci n'a pas encore été vu (région grise sur la courbe 
de~(\ref{contrainte sur mH})) : 
\beq m_H>114\ {\mathrm GeV}\ @ 95\%\ c.l. \eeq

\begin{figure}[htbp!]
\centerline{\epsfxsize=10cm\epsfbox{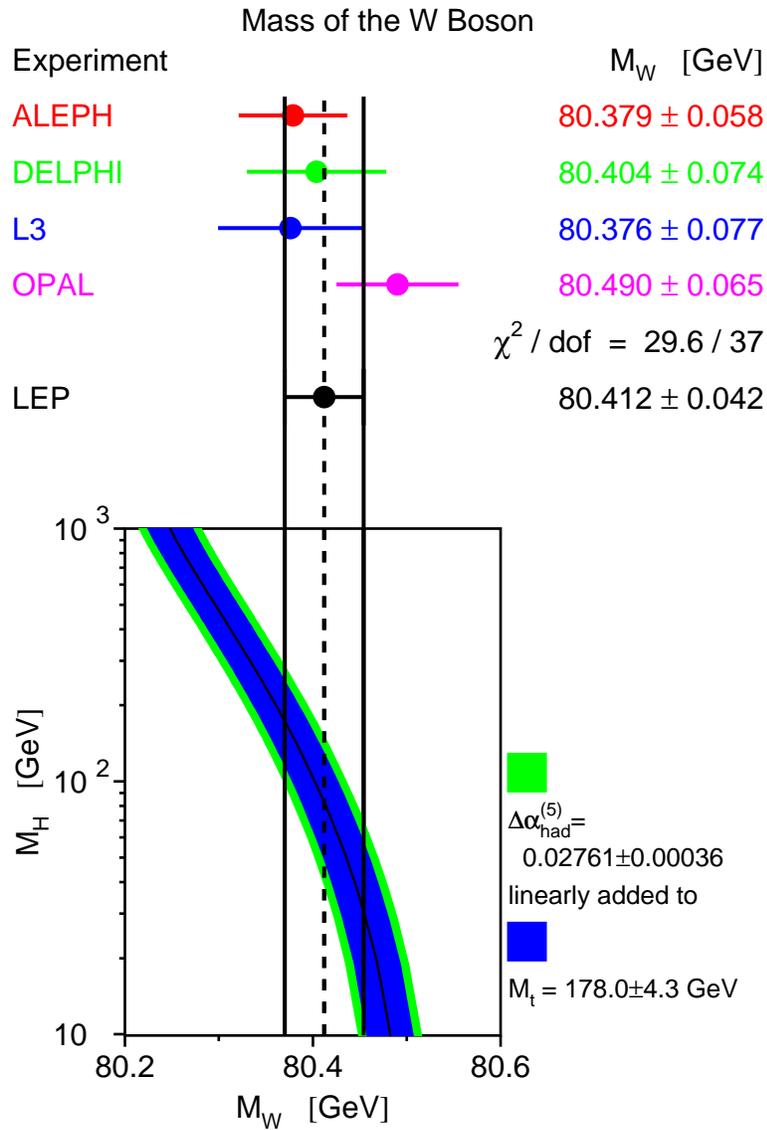}}
\caption{La masse du boson de Higgs dans le MS est tracée en fonction de la
masse $m_W$.
On y trouve la valeur expérimentale actuelle et donc la valeur favorisée de 
$m_H$ par les mesures de $m_W$.~\cite{LEPEWWG}}
\label{MWMH}
\end{figure}

De plus, nous pouvons aussi utiliser la masse du boson $W$ pour contraindre la
masse du boson de Higgs. Nous voyons là aussi, sur la 
figure~(\ref{MWMH}), que
pour être consistant avec les mesures de la masse du $W$, la masse préférée du boson de
Higgs est entre 10 et 200 GeV environ.
Les tests de précision électrofaibles donnent des indications claires en faveur
d'un boson de Higgs léger, \underline{proche de la limite directe 
actuelle de 114 GeV}. 
  
\begin{figure}[htbp!]
\centerline{\epsfxsize=10cm\epsfbox{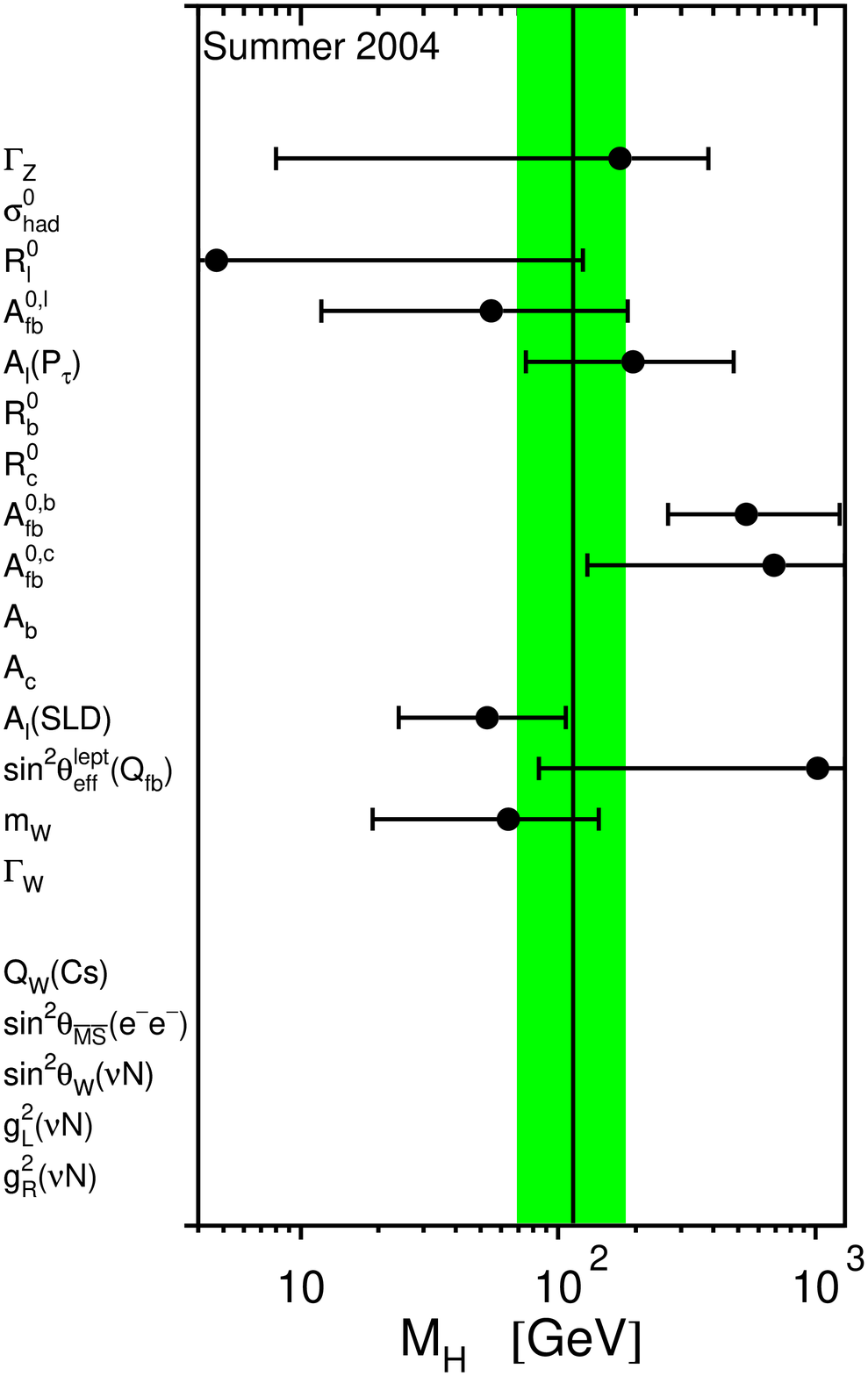}}
\caption{Masse du boson de Higgs favorisée par les différentes observables du
MS.~\cite{LEPEWWG}}
\label{MH}
\end{figure}

La figure~(\ref{MH}) donne la gamme de masse du boson du Higgs 
favorisée par les
mesures de différentes observables. Nous voyons là aussi que la masse du Higgs
est attendue aux environ de la centaine de GeV.  

\par\hfill\par
\begin{figure}[htbp!]
\centerline{\rotate[l]{\epsfxsize=8cm\epsfbox{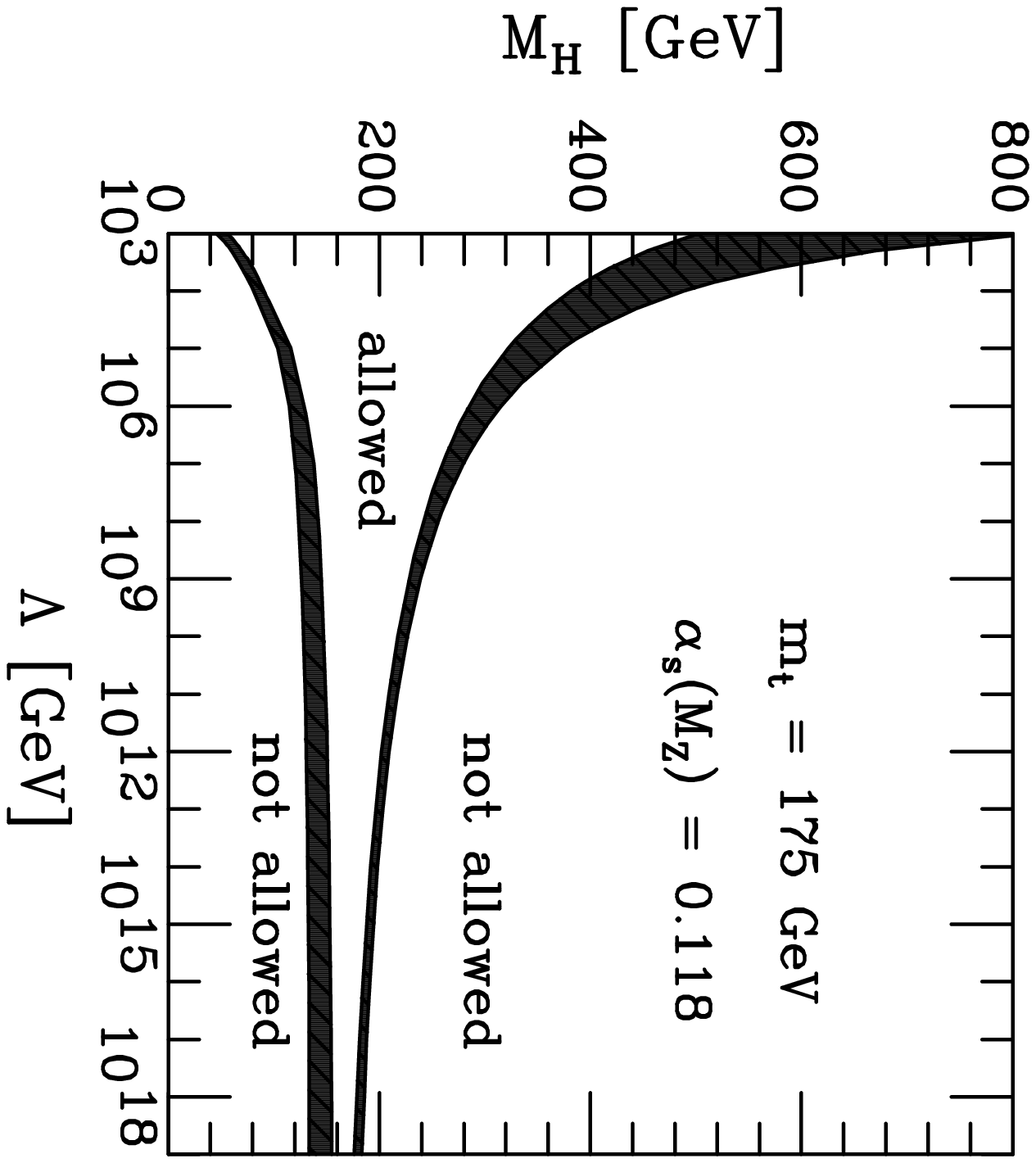}}}
\caption{Limites théoriques sur la masse du Higgs en fonction d'une échelle
d'énergie $\Lambda$. La région permise est comprise entre les deux courbes.~\cite{HiggsMassBounds}}
\label{mH limits}
\end{figure}

Mais il est aussi possible d'étudier la consistence théorique du MS en 
fonction de la masse du boson de Higgs car 2 effets s'opposent : les corrections
radiatives à l'autocouplage $\lambda$ du 
potentiel de Higgs et la stabilité du potentiel. 
En effet, quand nous prenons en compte les ordres supérieurs de la théorie des
perturbations (les diagrammes à boucles), la valeur de $\lambda$ diverge
en fonction de l'énergie. Il existe donc une énergie seuil à partir de 
laquelle le secteur de Higgs du MS cesse 
d'avoir un sens, où les corrections deviennent beaucoup plus grandes que la
valeur au niveau de l'arbre. 
De plus, si $m_H$ est trop faible, l'évolution par le groupe
de renormalisation peut rendre $\lambda$ négatif et le potentiel de Higgs est
alors instable~\footnote{Le potentiel change de forme et n'est plus limité par
le bas.}. 
Les considérations de stabilité du potentiel et 
les corrections radiatives contraignent la masse du boson de Higgs dans les
deux sens, par le bas et par le haut respectivement. La 
figure~(\ref{mH limits})
montre le "couloir" permis pour la masse du Higgs en fonction de l'échelle
d'énergie $\Lambda$. La largeur des bandes correspond à l'incertitude
théorique sur les résultats. 

\subsection{Sa détection potentielle au LHC et LC}

\begin{figure}[htbp!]
\centerline{\epsfxsize=9cm\epsfbox{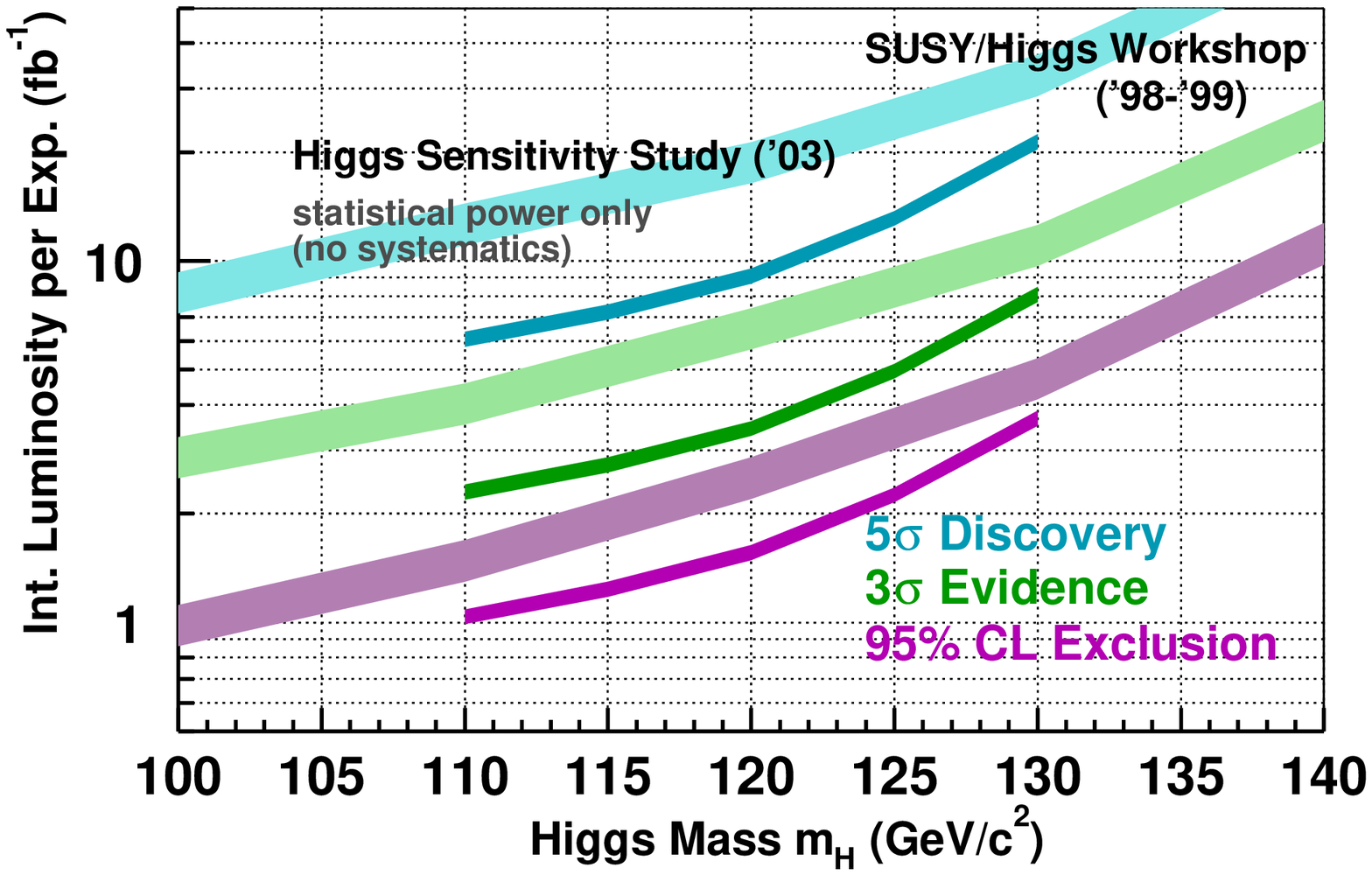}}
\caption{Sensibilité projetée du Tevatron pour la découverte du boson de Higgs
en fonction de la masse et de la luminosité intégrée. Les lignes les plus fines
sont dues au "2003 Higgs Sensitivity Study" et les plus épaisses du "1998-1999
SUSY/Higgs Workshop". Les deux études estiment que chaque expérience
aura une luminosité intégrée entre 4 et 8 $fb^{-1}$ d'ici 2009.~\cite{NewPhysics}}
\label{Tevatron Higgs}
\end{figure}

Nous allons maintenant nous tourner vers la détection du boson de Higgs dans
les accélérateurs de particules. Tout d'abord quelles sont les chances de 
l'observer au Tevatron, le collisionneur de proton-antiproton du Fermilab a
Chicago ? La 
figure~(\ref{Tevatron Higgs}) montre la sensibilité du Tevatron en 
fonction de la
masse. Un signal sera visible (une évidence à $3\sigma$) si $m_H$ est inférieure 
ou égale à 115 GeV 
mais peut être pas avant 2007, date à laquelle le LHC devrait être lui aussi 
opérationnel. L'évaluation de la sensibilité la plus récente correspond aux
lignes les plus fines.

\begin{figure}[htbp!]
\centerline{\epsfxsize=8cm\epsfbox{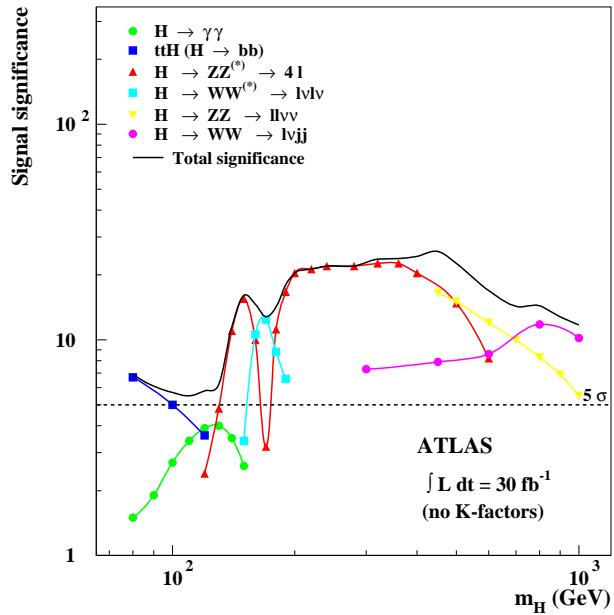}}
\caption{Sensibilité de l'experience ATLAS à la découverte d'un boson de Higgs
standard en fonction de $m_H$ et pour les désintégrations les plus importantes.~\cite{NewPhysics}}
\label{Higgs Atlas}
\end{figure}

Le LHC, avec une énergie disponible dans le centre de masse beaucoup plus
importante trouvera de façon certaine~\footnote{Dans la mesure o\`u, bien 
s\^ur, le
boson de Higgs existe.} le boson de Higgs à toutes les masses jusqu'à plusieurs 
TeV. Aprés plusieurs années de mesures, la précision
attendue sur sa masse est de l'ordre de 1 $\%$ et le LHC pourra mesurer beaucoup de 
ses propriétés ainsi que les paramètres liés à la brisure de la symétrie 
électrofaible.
Nous voyons sur la figure~(\ref{Higgs Atlas}) la sensibilité de
l'expérience ATLAS au LHC pour différentes désintégrations du Higgs.

Dans le cas d'un collisionneur linéaire $e^+$ $e^-$, les résultats seront d'une
manière générale complémentaires. L'énergie disponible dans le centre de masse
est moins élevée mais les bruits de fond sont beaucoup moins
importants et donc les incertitudes expérimentales plus faibles (ce sont des
particules fondamentales qui sont accélérées, pas des protons). 

\par\hfill\par
Globalement, deux scénarios sont possibles. Soit le Higgs est léger comme prévu,
 soit il est beaucoup plus
lourd voire inexistant. Avant de connaitre la réponse, nous pouvons déjà
réfléchir aux conséquences de ces deux types de scénario.

\subsubsection{Un Higgs léger?}
Dans ce cas, un Higgs de l'ordre de 115 GeV demande nécessairement
l'introduction d'une nouvelle physique aux alentours de $10^6$ GeV. En effet,
le potentiel effectif $V[H]$ sera "déstabilisé" par les corrections
radiatives (l'autocouplage du Higgs, $\lambda_H$, finit par 
changer de signe à cause de la renormalisation.
Le potentiel n'a alors plus sa forme de "chapeau mexicain" et le vide n'est
plus stable). Le MS n'est plus
valide à partir d'une certaine échelle et il faut introduire d'autres
particules pour modifier la renormalisation de $\lambda_H$. 

L'évolution avec l'échelle d'énergie $\mu$ des paramètres physiques dépend de 
leur
fonction béta~\footnote{La fonction béta est le coefficient d'évolution avec
l'énergie que l'on trouve dans les équations du groupe de renormalisation.}. 
Pour $\lambda_H$, en ne regardant que l'effet du quark top,
nous avons~\cite{LightHiggs} : 
\beq  \beta_{\lambda_H} \ \equiv
 \ \frac{\partial}{\partial \mu} \lambda_H
\ = \ \frac{1}{16 \pi^2} \left( 4 \lambda_H^2 \, - \, 36 \, g_t^4
 \, +\frac{27}{4} g^4 \, + \, \frac{9}{4} g^{\prime \, 4}
+ \, \frac{9}{2} g^2 g^{\prime \, 2} \right) + 2 \lambda_H \gamma_H,
 \label{beta H}
 \eeq
avec $\gamma_H$, la constante de renormalisation à une boucle du champ $H$ :
\beq
\gamma_H \ = \ \frac{1}{16\pi^2} \left( 6 g_t^2-\frac{9}{2} g^2 \ -
\frac{3}{2} g^{\prime \, 2} \right),
\label{wfr}
\eeq
 Quand nous regardons le signe devant le couplage de Yukawa du quark top, $g_t$, il
 parait clair qu'ajouter de nouveaux fermions ne résoudra pas la stabilité du
 potentiel, bien au contraire (puisque la fonction béta sera encore plus
 "négative" et $\lambda_H$ diminuera plus rapidement). La nouvelle physique 
 requise doit introduire de
 nouveaux bosons.  Dans le lagrangien nous avons alors l'ajout d'un terme
 ($\phi$ est le nouveau boson et $H$ le Higgs):
 \beq
 {\cal L} \supset\ M^2|\phi|^2 + \frac{M_0}{v^2}|H|^2\,|\phi|^2
\eeq

\begin{figure}[htbp!]
\centerline{\epsfxsize=10cm\epsfbox{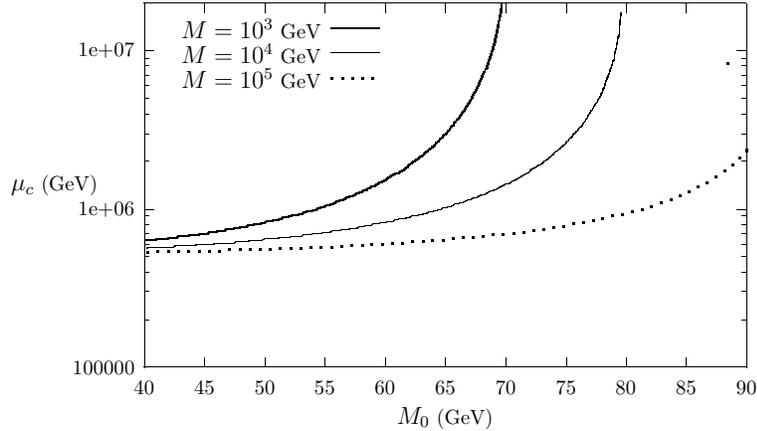}}
\caption{Echelle critique à partir de laquelle le potentiel devient instable en
fonction du couplage $M_0$ entre le boson de Higgs et le nouveau champ
scalaire, pour différents choix d'échelle M de nouvelle physique.~\cite{LightHiggs}}
\label{Light H02}
\end{figure}  

La figure~(\ref{Light H02}) donne des exemples d'échelle critique $\mu_c$,
échelle à partir de laquelle $\lambda_H$ change de signe et le potentiel
devient instable, en fonction de $M_0$. Pour une échelle de nouvelle physique M
de $10^3$ GeV, l'instabilité du potentiel est repoussée très loin.

\begin{figure}[htbp!]
\centerline{\epsfxsize=10cm\epsfbox{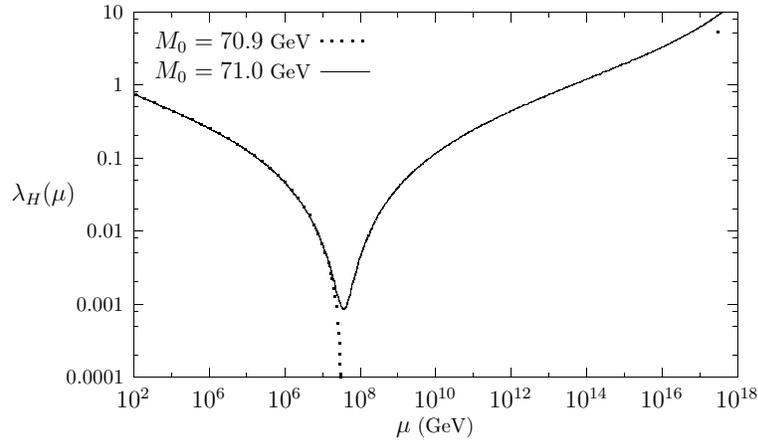}}
\caption{Exemple d'ajustement fin ("fine-tuning") nécessaire sur $M_0$ pour que 
le potentiel de
Higgs reste stable jusqu'à l'échelle de Planck. L'échelle de nouvelle physique
M est prise à 1 TeV.~\cite{LightHiggs}}
\label{Light H01}
\end{figure}

De plus, si nous augmentons la valeur de $M_0$ de façon à ce que 
l'instabilité du
potentiel n'intervienne pas avant l'échelle de Planck, $10^{19}$ GeV, un
ajustement extrèmement fin de $M_0$ est nécessaire. Si nous passons de
$M_0=70.9$ GeV à 71 GeV, $\lambda_H(\mu)$ passe de négative à 
$\mu \, \sim \, 5 \times 10^7$ GeV à un changement radical de comportement
juste aprés cette limite et devient même si grande à l'échelle de Planck que
les calculs perturbatifs deviennent incertains, figure~(\ref{Light H01}). 

\begin{figure}[htbp!]
\centerline{\epsfxsize=9cm\epsfbox{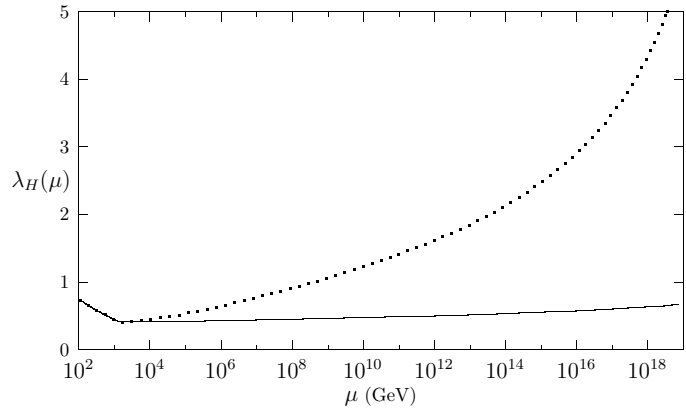}}
\caption{Apport de la supersymétrie dans l'évolution de $\lambda_H$ en fonction
de l'échelle d'énergie $\mu$. La ligne pleine correspond à la supersymétrie et
les pointillés à une théorie où les contributions des fermions Higgsinos et
Jauginos de la supersymétrie ont été enlevés.~\cite{LightHiggs}}
\label{Light H03}
\end{figure}

La supersymétrie est la seule théorie connue qui résoud de façon naturelle, et à 
tous les ordres (toutes les boucles), le problème lié à cet 
ajustement fin. 
Dans les théories supersymétriques, $\lambda_H$ est uniquement reliée aux 
couplages de jauge $SU(2)$ et $U(1)$ : 
\beq \lambda_H \ = \ \frac{3}{4}(g^2+g^{\prime \, 2}). \eeq
L'évolution, montrée sur la figure~(\ref{Light H03}), est modifiée par les liens
entre les fermions du MS (le top surtout) et les superpartenaires fermioniques et implique
 une remarquable
stabilité de $\lambda_H$, donc du vide électrofaible.

Ainsi, un boson de Higgs léger, d'une masse d'environ 115 GeV pointerait de
manière très forte l'existence de la supersymétrie comme nouvelle physique 
au-delà du MS.

\subsubsection{Un Higgs lourd ou pas de Higgs?}
Si nous n'observons pas de boson de Higgs avec une masse inférieure à 130 GeV, où
nous sommes-nous trompés? Voici quatres pistes que nous ne
détaillerons cependant pas pour éviter d'alourdir le cours.
\par\hfill\par

Si le boson de Higgs n'a pas la masse attendue, peut-être n'avons nous pas
correctement interprété les données de précision
électrofaibles~\cite{Interpretation}. 
En effet, les résultats 
de l'expérience NuTeV (collisions inélastiques $\nu-N$) montrent un léger 
désaccord de quelques écarts standards par
rapport au fit dans le MS. De plus, il y a aussi un désaccord sur les valeurs
favorisées de la masse du Higgs entre les asymétries leptoniques et hadroniques
de la masse du Z. Si ces désaccords s'avèrent confirmés c'est que nous ne 
comprenons pas tout à fait les effets hadroniques ou alors qu'il y a de la 
nouvelle physique derrière ces anomalies. Et dans ce cas, il n'y a pas de
prédiction vraiment solide en faveur d'un boson de Higgs léger puisque les
analyses se basent sur le fit dans le cadre strict du MS.
\par\hfill\par

Deuxième point, le MS est une théorie \underline{effective}. Nous pouvons donc 
nous attendre à ce que les
effets de nouvelle physique se fassent sentir {\it via} l'introduction dans le
lagrangien de termes \underline{non-renormalisables}. Nous avons alors un lagrangien 
de cette
forme :
\beq
{\mathcal L}_{eff} = \underbrace{{\mathcal L}_{MS}}_{{\mathrm renormalisable}} 
+ \sum_i \underbrace{\frac{c_i}{\Lambda_i^P}{\mathcal
O}_i^{4+P}}_{{\mathrm non-renormalisable}}, 
\eeq
où ${\mathcal O}_i^{4+P}$ sont des opérateurs de dimension de masse à la
puissance ($4+P$) traduisant les interactions supplémentaires. Le lagrangien 
étant de dimension de (masse)$^4$,
ces opérateurs sont supprimés par une échelle de masse $\Lambda_i$ élevée à la
puissance $P$.

\begin{figure}[htbp!]
\centerline{\epsfxsize=12cm\epsfbox{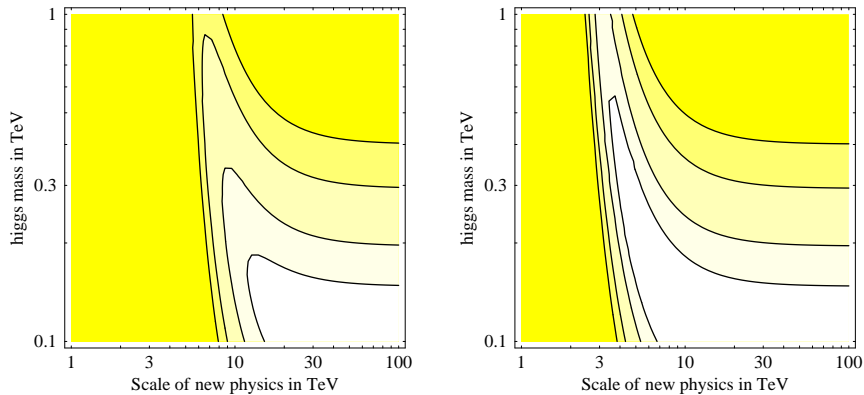}}
\caption{Couloirs vers un boson de Higgs lourd quand sont inclus des operateurs
non renormalisables dans le fit global du MS. La masse du Higgs est donnée en
TeV et en fonction de l'échelle d'apparition d'une nouvelle physique en TeV.~\cite{Operators} }
\label{operateurs}
\end{figure}

Le fit global du MS suggère, pour un Higgs léger, que les coefficients des
nouvelles interactions sont faibles : $\Lambda_i\approx 10$ TeV pour des $c_i$
de l'ordre de $\pm 1$. Mais il reste possible malgré tout qu'une nouvelle
physique à cette énergie rende $m_H$ grande. La figure~(\ref{operateurs}) nous le 
montre : il
existe un étroit couloir pour une valeur élevée de $m_H$ encore compatible avec
les données actuelles.

\par\hfill\par
 Les modèles "Little Higgs" reposent sur l'annulation des divergences
 quadratiques de la masse du Higgs et des divergences dues aux boucles de bosons
 de jauge par l'introduction de nouvelles particules : un
 pseudo-quark~\footnote{On l'appelle ainsi car son introduction a pour but de
 contrer l'effet du quark top. Cependant, il a un spin 1.} top T
 lourd ainsi que d'autres bosons de jauge $W^{'}$ et des bosons de Higgs. Le 
 spectre de masse des  particules est~\cite{LittleHiggs}, en plus d'un
 boson de Higgs relativement léger : $M_H \approx 200$ GeV,
\begin{equation}
M_T< 2\ {\mathrm TeV}\left( \frac{m_H}{200 GeV}\right)^2, \;
M_{W^{'}}< 6\ {\mathrm TeV}\left( \frac{m_H}{200 GeV}\right)^2, \;
M_{H^{++}}< 10\ {\mathrm TeV}.
\end{equation}
 Pour obtenir de nouveaux bosons, le MS est inclu dans un groupe de jauge plus
 large, brisé spontanément.
 Ces modèles, bien qu'incomplets, resolvent partiellement le problème de la 
 hiérarchie des masses et constituent des alternatives intéressantes à la 
 supersymétrie pour des énergies inférieures à 10 TeV environ~\footnote{Au del\`a de cette échelle d'énergie, il faudrait 
enrichir le modèle "Little Higgs" pour avoir une théorie plus 
complète.}. De plus, leurs
 conséquences phénoménologiques seront observables au LHC ce qui rend l'étude
 de ces modèles attractive. 
 \par\hfill\par
 
 La dernière option, plus radicale, est qu'il n'existe pas de boson de Higgs du
 tout. Des modèles sans Higgs ("Higgsless")~\cite{Higgsless} ont déjà été construits mais 
 prédisent tous une
 diffusion WW forte à l'échelle du TeV. En effet, le résultat du calcul des sections
 efficaces des processus de type $f\bar{f}\to W^+
 W^-$ et $W^+W^-\to W^+W^-$ (diffusion WW) diverge si il n'y a pas de particule
 de spin 0 pour annuler les divergences dues aux échanges de $\nu$, $\gamma$ et
 $Z^0$. 
 Dans le cadre quadridimensionel standard, c'est {\it a priori} incompatible avec
 les données de haute précision. Mais en ajoutant une dimension spatiale 
 supplémentaire et en postulant
 des conditions aux bords qui brisent la symétrie électrofaible~\footnote{Sans
 rentrer dans le détail, puisque le boson de Higgs n'existe pas, il nous faut
 trouver un autre moyen de briser la symétrie électrofaible. Ceci est fait en
 "jouant" sur la géométrie de la dimension supplémentaire {\it via} des
 conditions aux bords.}, on décale la
 prédiction d'une diffusion forte WW vers 10 TeV. L'observation de cette
 diffusion est donc un test crucial des modèles Higgsless, 
qui éprouvent toujours quelques difficultés avec les mesures de haute 
précision.

\section{Conclusions et carte de route}

Le MS est une construction mathématique dotée d'un fort pouvoir prédictif malgré
le grand nombre de paramètres libres. Dans ce modèle, il reste néanmoins à 
connaitre de façon certaine la brisure de symétrie électrofaible. Le responsable
présumé de cette brisure, le boson de Higgs, n'a en effet pas encore été observé
expérimentalement. Mais mis à part ceci, le MS est en extraordinaire accord avec les données 
expérimentales accessibles actuellement (énergies aux environs de 100 GeV). Nous 
savons cependant que de la nouvelle physique doit apparaître à l'échelle du TeV
et au-delà. Le MS ne peut pas donner de réponses à 
toutes les questions et les mystères de la physique des particules, un nouveau
modèle doit prendre le relais. 
Dans ce contexte le MS est l'approximation à basse énergie d'une 
description plus fondamentale mais incontournable aux énergies supérieures. Il
est alors vu comme une théorie {\it effective}. 

Parmis la liste des questions en suspens que nous avons vu à la section 1.1.3,
certaines peuvent se regrouper. Globalement, il reste 3 grands problèmes
qu'une "théorie du tout" doit résoudre : 

-{\bf Le problème de l'unification}:
Pourquoi observons nous 4 interactions et comment pouvons nous les unifier ? 
Existe t'il un groupe de symétrie qui peut toutes les englober? Les Théories de 
Grande Unification (G.U.T en anglais) existent-elles ? 

-{\bf Le problème de la masse}:
Qu'elle est l'origine de la masse des particules ? Sont elles dues au boson de
Higgs? Pourquoi sont elles si petites devant la masse de Planck ?

-{\bf Le problème des saveurs}: 
Pourquoi y-a-t'il autant de quarks et de leptons? Pourquoi y-a-t'il ce mélange
observé entre les saveurs? Existe-t'il des symétries supplémentaires entre les
différentes saveurs?  

\par\hfill\par
De toutes les possibilités de nouvelle physique étudiées, une en particulier 
parait incontournable à bon nombre de physiciens...


\chapter{Supersymétrie}							       %

Comme nous venons d'en discuter, le Modèle Standard (MS) est une description 
valable des phénomènes physiques à des énergies inférieures à quelques centaines 
de GeV. Au-delà, la supersymétrie pourrait jouer un rôle important et 
nous nous rapprochons de plus en plus de sa possible découverte. En effet, le 
LHC va sonder les énergies de l'ordre du TeV et nous verrons dans ce chapitre 
que c'est à cette énergie que la supersymétrie est sensée faire son apparition. 
\par\hfill\par
Ce cours est principalement orienté vers la présentation et la discussion des
modèles supersymétriques. Nous tenterons de discuter clairement de ce que peut apporter la 
supersymétrie en physique des particules. 
Nous commencerons par introduire la supersymétrie de façon historique et nous 
donnerons, sans trop de détails, les principales motivations de son utilisation.
Une section plus formelle suivra, où nous présenterons la structure générale 
d'une théorie physique avec supersymétrie.
Nous poursuivrons le cours par une partie également théorique mais appliquée à 
la physique des particules à basse énergie (autour du TeV). Nous y trouverons 
les différents modèles, comme le Modèle Standard Supersymétrique Minimal (MSSM),
qui servent de base à l'analyse de la phénoménologie. 
C'est tout naturellement que nous aborderons ensuite l'aspect expérimental, avec
d'abord les premières contraintes expérimentales puis les scénario possibles de
la détection de la supersymétrie. 

\section{Histoire et motivations}

\subsection{Qu'est-ce que la supersymétrie}

Comme son nom l'indique, la supersymétrie est tout simplement une nouvelle 
symétrie et l'effet d'une transformation de supersymétrie est de 
transformer un état \underline{bosonique} en un état \underline{fermionique} et 
vice-versa, avec $\Delta S=\pm 1/2$ où S est le spin. Si nous appelons Q le 
générateur de la supersymétrie qui réalise la transformation, alors très schématiquement 
\beqn
&Q|Boson> &= |Fermion> \\
&Q|Fermion> &= |Boson>.
\eeqn

\subsection{L'introduction de la supersymétrie}

Mais ceci n'est que l'aspect "physique des particules" de la supersymétrie qui
est une symétrie tout à fait générale. Plus formellement, la supersymétrie est 
en fait la seule et dernière extension possible du groupe de Poincaré des 
symétries d'espace-temps et c'est tout d'abord ainsi qu'elle fut découverte. 

En effet, à l'origine, on cherchait à combiner les symétries 
\underline{externes} (d'espace-temps, comme les translations)
et \underline{internes} (surtout globales) c'est-à-dire à 
étendre le groupe de Poincaré par des transformations internes. Il y a eu 
plusieurs essais dans les années 1960 mais en 1967, Coleman et Mandula 
montrèrent de façon formelle qu'il est impossible de combiner les deux types de 
symétries. C'est leur fameux {\it 
théorème no-go}. En fait, il était sous entendu que c'était impossible en 
utilisant des générateurs \underline{bosoniques} (donc de spin entier) habituels.
Mais en 1971, Golfand et Likhtman réussirent l'extension du groupe de Poincaré en
utilisant des charges \underline{fermioniques}, donc de spin demi-entier. C'est la
supersymétrie. La même année, Neveu, Schwarz puis Ramond proposèrent des modèles
supersymétriques à 2 dimensions pour obtenir des cordes supersymétriques qui
expliqueraient l'origine des baryons. 
Quelques années plus tard, en 1973, Volkov et Akulov appliquèrent la 
supersymétrie aux neutrinos (ils voulaient en faire le fermion de Goldstone) mais 
il fut montré un peu plus tard que leur théorie, à 4 dimensions,
du neutrino ne décrivait pas correctement les interactions de basse énergie 
observées expérimentalement. 
La même année, Wess et Zumino proposèrent la première théorie des champs 
supersymétrique à 4 dimensions de vrai intérêt du point de vue 
phénoménologique. 
Puis, ensemble avec Iliopoulos et Ferrara, ils montrèrent que la supersymétrie 
permettait de supprimer beaucoup de divergences des théories des champs usuelles. 
Ceci a rendu la supersymétrie très attractive et pendant un temps donné, elle 
fut beaucoup utilisée pour tenter d'unifier les bosons et les fermions. Par 
exemple pour unifier les particules de matière de spin 1/2 avec les particules 
de jauge de spin 1, ou les particules de matière et les champs de Higgs, dans 
les même supermultiplets.
Puis en 1976, deux groupes découvrirent que la supersymétrie locale (la
transformation de supersymétrie dépend alors des coordonnées dans 
l'espace-temps) incluait une description de la gravitation. C'est ce que l'on a 
appellé la {\it supergravité}.

Depuis, la phénoménologie de la supersymétrie a été énormément étudiée et les 
théories basées sur la supersymétrie se sont imposées comme les candidates les 
plus sérieuses pour la physique au-delà du MS.

\subsection{Pourquoi la supersymétrie}

Pourquoi introduire la supersymétrie en physique des particules ?
Qu'est-ce qui la rend si attractive pour les physiciens des particules ? 
Les motivations de son introduction en physique des particules sont 
principalement d'ordre \underline{physique} avant d'être esthétiques. 
\par\hfill\par
\begin{figure}
\centerline{\psfig{figure=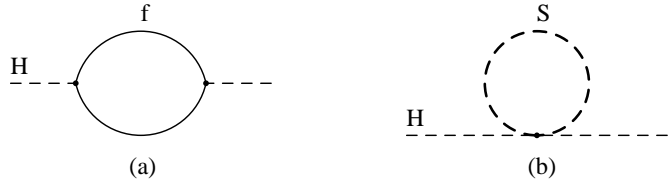,height=1in}}
\caption{Corrections quantiques à une boucle à la masse-carrée du Higgs. (a) boucle
de fermion, (b) boucle de boson scalaire.
\label{fig:higgscorr1}}
\end{figure}

$\bullet$ En effet, la supersymétrie apporte quelque chose de très
important dans les théories des champs. Dans le cours précédent, nous avons
discuté des corrections radiatives à la masse du Higgs. Quand nous calculons
la contribution d'une boucle fermionique comme celle de la
figure~(\ref{fig:higgscorr1} a), nous obtenons~\footnote{Pour le calcul, nous 
avons pris la forme usuelle du couplage de Yukawa du 
boson de Higgs scalaire au fermion : $y_f H\overline{\psi}\psi$.}:
\beq
\Delta m_H^2=-\frac{y_f}{16\pi^2}[2\Lambda^2 + 6m_f^2 \ln(\Lambda/m_f)+...],
\eeq   
où $\Lambda$ est un cut-off ultraviolet utilisé pour restreindre les impulsions 
dans la boucle et qui est de l'ordre de l'échelle de la nouvelle physique au
delà du MS. Nous voyons que la masse du Higgs diverge et que si nous supposons
le MS valable jusqu'à l'échelle de Planck, $M_p\simeq 10^{19}$ GeV, alors
$\Lambda=M_p$ et cette
correction est $10^{30}$ fois plus forte que la valeur raisonnable de la masse
du Higgs au carré, ($10^2$ GeV)$^2$ !

Cette constation est la même si nous considérons plutôt une boucle d'un champ
scalaire S, figure~(\ref{fig:higgscorr1} b). 
\beq
\Delta m_H^2=\frac{\lambda_S}{16\pi^2}[\Lambda^2 - 2m_S^2 \ln(\Lambda/m_S)+...],
\eeq 
où $\lambda_S$ est son couplage avec le boson de Higgs.
Le problème qui vient d'être énoncé est le problème de la hiérarchie vu au
chapitre 1. 
Que nous apporte la supersymétrie dans ce cas ? Si nous regardons de plus près les 2 
équations précédentes, les deux contributions divergentes (leur terme 
$\propto \Lambda^2$) s'annulent si pour chaque fermion de notre théorie entrant
dans la boucle nous avons aussi 2 scalaires avec $\lambda_S=y_f^2$ . Nous 
verrons juste après que c'est exactement ce que la supersymétrie se propose 
d'apporter ! Et de plus, à tous les ordres (c'est-à-dire quand nous considérons des
corrections à plusieurs boucles) ceci est réalisé. Il nous reste alors une
divergence logarithmique mais qui n'induit pas de problèmes d'ajustements fins. 
A ce jour, la supersymétrie fournit la résolution du problème de la hiérarchie la plus 
naturelle et la plus efficace. 
\par\hfill\par
$\bullet$  De plus, nous avons vu à la fin
du chapitre précédent que les données expérimentales penchent en faveur d'un
Higgs relativement léger et qu'un Higgs léger demande, pour stabiliser le vide électrofaible, 
que la physique à plus
haute énergie partage beaucoup de choses en commun avec la supersymétrie.
Inversement, les calculs faits à partir du Modèle Standard Supersymétrique
Minimal donnent la prédiction $m_H\leq 130$ GeV qui est donc compatible avec les
ajustements expérimentaux vus au chapitre précédent. 
\begin{figure}
\centerline{\psfig{figure=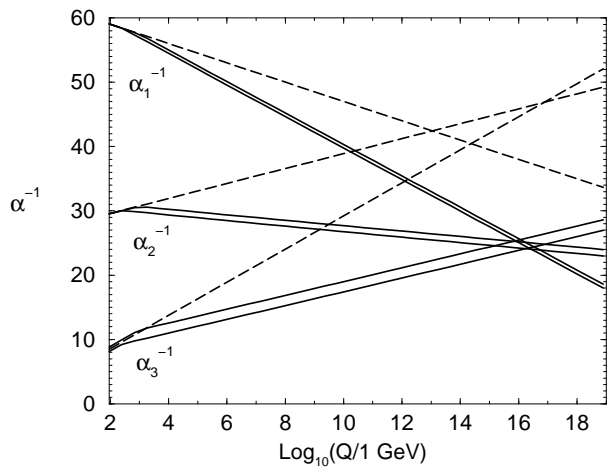,height=2.7in}}
\caption{Evolution, en fonction du logarithme de l'énergie, des couplages de jauge inversés, $\alpha_a^{-1}(Q)=4\pi/g_a^2$, 
dans le MS (tirets) et dans le MSSM (lignes pleines).
\label{fig:gaugeunification}}
\end{figure}

\par\hfill\par
$\bullet$ Les deux premiers arguments ne concernent que le Higgs. Considérons
maintenant les constantes de couplages qui caractérisent chaque force
fondamentale. Si nous faisons évoluer les 3 couplages du Modèle Standard en 
fonction de l'énergie, nous observons qu'ils tendent tous à 
se croiser à la même échelle. Enfin presque, c'est quand nous ajoutons la
supersymétrie dans les calculs d'évolution que les couplages se croisent quasi 
exactement au même point (autour de $2\times 10^{16}$ GeV). Personne n'est obligé de 
croire en une "Grande Unification" qui repose sur cette unification des couplages, 
mais il est très étonant d'observer que la supersymétrie réalise l'unification 
aussi précisément. C'est tout de même un argument fort en faveur 
de la supersymétrie car les théories de Grande Unification ont beaucoup de caractéristiques 
et de prédictions intéressantes pour 
la physique de basse énergie (elles expliquent par exemple les nombres
quantiques des particules vus dans le tableau (1.1) du chapitre précédent).
\par\hfill\par
$\bullet$  Il y a aussi un argument de ce type en faveur de l'existence de
la supersymétrie dans la nature. Les théories de cordes, qui fournissent à
l'heure actuelle une des seules solutions au problème de la gravitation
quantique (avec la gravité quantique à boucles), ne peuvent se passer de la
supersymétrie sans souffrir d'inconsistences physiques et mathématiques. 
Là aussi la supersymétrie semble un ingrédient essentiel pour construire des 
théories cohérentes à haute énergie.
\par\hfill\par
$\bullet$ Enfin, la supersymétrie peut aussi jouer un rôle dans l'explication de
la matière noire. C'est une des seules théories qui possèdent dans son spectre 
de particules des candidats sérieux. Ils ont pour nom, neutralino ou gravitino, 
et la supersymétrie prédit, dans un cadre plausible, leur stabilité. Ainsi, 
une fois créés, il ne se
désintègrent pas et ne peuvent pas être observés directement. Leur nombre peut donc
contribuer à la masse manquante sous la forme de matière noire.  

\par\hfill\par 
Cette discussion démontre que la supersymétrie est une symétrie trés séduisante 
théoriquement, mais 
surtout trés \underline{utile}. Nous n'avons encore aucune preuve de son 
existence mais au vu de ce qu'elle apporte en physique des particules, il est 
difficile d'y rester insensible. 
Nous n'avons donné que des arguments qualitatifs dans cette partie, nous allons 
maintenant entrer beaucoup plus en détail.

\section{La structure d'une théorie supersymétrique}

\subsection{Interlude: "spinorologie"}

Avant de démarrer une présentation plus théorique de la supersymétrie nous
allons dans ce brève interlude donner nos notations et conventions sur ce 
qui concerne les spineurs. 

\noindent $\bullet$ {\it Un spineur de Weyl} décrit une particule de spin 1/2 et 
de chiralité donnée.
C'est un spineur à 2 composantes. Nous nous efforcerons le plus souvent de les 
noter par des lettres minuscules grecques avec un indice. Par exemple 
$\psi_{\alpha}$, $\xi_{\beta}$,... où $\alpha,\ \beta,...= 1,2$. Un spineur 
$\psi_{\alpha}$ ou $\psi_L$ est par convention de chiralité gauche, le spineur 
droit est noté $\overline{\psi}^{\dot{\alpha}}$ ou $\psi_R$. A noter que :
\beqn
&(\psi_{\alpha})^*=\overline{\psi}_{\dot{\alpha}}&
\\
&(\overline{\psi}^{\dot{\alpha}})^*=\psi^{\alpha}&
\eeqn
et que les matrices
$\varepsilon_{\alpha\beta}=\varepsilon_{\dot{\alpha}\dot{\beta}}=i\sigma_2$ et 
$\varepsilon^{\alpha\beta}=\varepsilon^{\dot{\alpha}\dot{\beta}}=-i\sigma_2$
permettent de monter et descendre à volonté les indices spinoriels $\alpha$ et 
$\beta$. 

\noindent $\bullet$ {\it Un spineur de Dirac} se construit avec 2 spineurs de 
Weyl et réunit les deux chiralités d'une particule donnée. C'est donc un 
spineur à 4 composantes.
Nous les noterons par des lettres majuscules grecques : $\Psi$, $\chi$,
$\Phi$,...
En termes de ses spineurs de Weyl, nous avons :
\beq \Psi=\left(\begin{array}{c} \psi_L \\ \psi_R \end{array}\right)=
\left(\begin{array}{c} \psi_{\alpha} \\ \overline{\eta}^{\dot{\alpha}} 
\end{array}\right)
\eeq
Les opérateurs de projection $P_{R,L}=\frac{1}{2}(1\pm\gamma_5)$ permettent de 
sélectionner l'une ou l'autre chiralité : $\Psi_{R,L}=P_{R,L}\Psi$.

\noindent $\bullet$ {\it Un spineur conjugué de charge} est un spineur auquel 
l'opérateur de
conjugaison de charge a été appliqué. Il décrit la même particule mais sa charge
électrique est opposée.
\beq \Psi^c=C \overline{\Psi}^T = 
\left(\begin{array}{c} \eta_{\alpha} \\ \overline{\psi}^{\dot{\alpha}} 
\end{array}\right)
\eeq
où la matrice de conjugaison de charge $C$ peut s'écrire :
\beq C=i\gamma^0\gamma^2 \eeq

\noindent $\bullet$ {\it Un spineur de Majorana} se construit avec un seul 
spineur de Weyl mais l'englobe dans une notation à 4 composantes. Il est égal à
son conjugué de charge, $\Psi_M=\Psi^c_M$.
\beq \Psi_M=\left(\begin{array}{c} \psi_L \\ -i\sigma_2(\psi_L)^* \end{array}\right)=
\left(\begin{array}{c} \psi_{\alpha} \\ \overline{\psi}^{\dot{\alpha}} 
\end{array}\right)
\eeq

\noindent $\bullet$ {\bf {\it La représentation des matrices $\gamma$}} choisie 
est la représentaion de Weyl dans
laquelle :
\beq \gamma^{\mu}=\left( \begin{array}{cc} 0 & \sigma^{\mu} \\
\overline{\sigma}^{\mu} & 0 \end{array}\right) \eeq
avec $\sigma^{\mu}=(\mathbf{1}_2,\sigma^{i}),\ \overline{\sigma}^{\mu}
=(\mathbf{1}_2,-\sigma^{i})$ où
les $\sigma_{i}$ sont les matrices de Pauli, et
$\gamma_5=i\gamma^0\gamma^1\gamma^2\gamma^3=\mathrm{diag}(-\mathbf{1}_2,\mathbf{1}_2)$.

Nous avons aussi $\{\gamma^{\mu},\gamma^{\nu}\}=2\eta_{\mu\nu}$ où
$\eta_{\mu\nu}=\mathrm{diag}(+1,-1,-1,-1)$ est la métrique de Minkowski utilisée pour monter 
et descendre les indices de Lorentz.

\subsection{L'algèbre et les supermultiplets}

Comme il a été dit précédemment, la supersymétrie combine des transformations 
spatio-temporelles du
groupe de Poincaré et des transformations de symétrie interne 
(d'un groupe de jauge donné). La supersymétrie introduit des nouveaux générateurs 
$Q_{\alpha}$, fermioniques, donc \underline{anticommutant} entre eux (ce sont des spineurs): deux 
opérations de supersymétrie ne commutent donc pas entre elles. {\it A priori}, rien 
n'interdit l'introduction de plusieurs générateurs, mais dans la version simple
de la supersymétrie il n'y a qu'un seul nouveau couple de générateurs, $Q_{\alpha}$ 
\underline{et}~\footnote{Car les Q et $\overline{Q}$ 
sont des spineurs donc des objets complexes qui se transforment de façons
différentes sous le groupe de Lorentz.} $\bar{Q}^{\dot{\alpha}}$. C'est 
la 
"supersymétrie $\mathcal{N}=1$" et c'est la seule dont nous parlerons. Si la
raison est ici pédagogique, dans la section suivante nous donnerons les raisons
physiques de ce choix.

La connaissance de l'alg\`ebre de la supersymétrie (et d'une quelconque
symétrie) se r\'esume \`a la
connaissance des relations de commutation de tous ses 
générateurs (son algèbre de Lie). Nous avons, en plus des relations de 
commutation isues de l'alg\`ebre de Poincaré, celles qui font 
intervenir les générateurs $Q_{\alpha}$ et $\bar{Q}^{\dot{\alpha}}$:
\begin{eqnarray}
&[P^{\mu},Q_{\alpha}]&=0=[P^{\mu},\bar{Q}^{\dot{\alpha}}] \label{com1}
\\
&\lbrace Q_{\alpha},\bar{Q}_{\dot{\beta}}\rbrace &
=2(\sigma_{\mu})_{\alpha\dot{\beta}}P^{\mu}  \label{com2}
\\
&\lbrace Q_{\alpha},Q_{\beta}\rbrace & =
\lbrace \bar{Q}^{\dot{\alpha}},\bar{Q}^{\dot{\beta}}\rbrace=0 \label{com3}
\\
&\lbrace M_{\mu\nu}, Q_{\alpha}\rbrace & =
\frac{1}{2}(\sigma_{\mu\nu})_{\alpha}^{\beta}Q_{\beta} \label{com4}
\\
&\lbrace M_{\mu\nu}, \bar{Q}_{\dot{\alpha}}\rbrace & =
\frac{1}{2}(\overline{\sigma}_{\mu\nu})^{\dot{\beta}}_{\dot{\alpha}}\bar{Q}_{\dot{\beta}}
\label{com5}
\end{eqnarray}

Quelle est la signification des $Q_{\alpha}$ ? Tout d'abord, $Q$ est une charge
au sens du théorème de Noether c'est-à-dire la charge conservée dans la
symétrie. Elle commute donc avec l'Hamiltonien du système, {\it cf}~(\ref{com1}). 
Elle possède un spin 1/2 et peut alors s'écrire sous la 
forme d'un spineur de Weyl~\footnote{Elle possède donc 2 composantes 
complexes.}. On pourrait aussi écrire $Q$ sous la forme d'un spineur de Majorana 
à 4 composantes. De plus, si nous regardons le commutateur~(\ref{com2}), nous avons 
schématiquement $\lbrace Q,\bar{Q} \rbrace \sim P$ c'est-à-dire que $Q^2$ est 
une translation d'espace-temps. \underline{$Q$ peut donc se voir comme la "racine carré"
d'une translation dans l'espace-temps}.

\par\hfill\par
Nous allons maintenant étudier les représentations 
irréductibles de cet algèbre (\textit{les supermultiplets})
et en détailler le contenu. En effet, nous voulons appliquer la supersymétrie à
la physique des particules, il nous faut donc savoir comment ranger nos
particules et quelles seront leurs propriétés de transformation.
Dans le groupe de Poincaré, il y a 2 éléments \underline{invariants} de Casimir : 
l'opérateur de spin $W^2=W^{\mu}W_{\mu}$, avec
$W^{\mu}=\frac{1}{2}\epsilon^{\mu\nu\rho\sigma}P_{\nu}M_{\rho\sigma}$
le vecteur de Pauli-Lubanski, et l'opérateur de masse $P^2=P^{\mu}P_{\mu}$, où 
$P^{\mu}$ est la quadri-impulsion. Dans un multiplet du groupe de Poincaré, les
particules ont la même masse et le même spin. Mais dans l'algèbre supersymétrique,
$W^2$ n'est plus un invariant de l'algèbre. Nous avons 
\begin{eqnarray}
&[P^2,Q_{\alpha}]&=0
\\
&[W^2,Q_{\alpha}]&\not=0
\end{eqnarray}
et donc, \underline{dans un supermultiplet les particules ont la même masse mais des 
spins différents}. Nous pouvons tout de même corriger $W$ pour obtenir un nouvel
invariant dont les valeurs propres sont sous la forme $2j(j+1)m^4$ avec 
$j=0,\frac{1}{2},1,...$ le nombre quantique de \textit{superspin}. Ce nouveau
$W$
est invariant donc chaque représentation irréductible peut être caractérisée par
un couple $[m,\ j]$ et le lien entre le spin $S$ et $j$ est déduit de la 
relation : $M_S=M_j,\ M_j+ \frac{1}{2},\ M_j- \frac{1}{2},\ M_j$.
Dans un même supermultiplet, on aura donc des particules de \underline{même masse}
et de même \underline{superspin}.
De plus, une propriété importante est qu'\underline{il y 
a égalité dans un supermultiplet entre le nombre de degrés de liberté 
bosoniques et} \underline{fermioniques : $n_B=n_F$}.

\par\hfill\par
Nous pouvons maintenant construire les différentes représentations :

\noindent $\triangleright$ La repr\'esentation fondamentale $[m,\ 0]$ est appel\'ee 
\textit{supermultiplet chiral (ou scalaire)}. La valeur $j=0$  implique 
$M_S=0,+\frac{1}{2},-\frac{1}{2},0$ donc ce supermultiplet contient 2 champs
scalaires réels réunis sous la forme d'un champ scalaire complexe (le sfermion),
 $\phi$ et un champ fermionique de Weyl (de spin 1/2), $\psi$. Ces champs ont la même masse. 
Pour que la supersymétrie soit préservée dans les boucles où les particules ne
sont pas sur leur couche de masse (c'est-à-dire $P^2\not=M^2$) il faut que les 
degr\'es de libert\'e fermioniques et bosoniques soient aussi équilibrés dans 
ce cas (\textit{off-shell}). En effet, {\it off-shell} un fermion de Weyl possède 4 
degrés de liberté de spin  au lieu de 2 {\it on-shell}. Il faut donc ajouter au 
contenu de cette représentation un autre champ scalaire complexe mais qui ne 
se propage pas (on dit qu'il est \textit{auxiliaire} et il n'a pas de 
terme cinétique). {\it On-shell}, nous utiliserons l'équation du mouvement $F=F^*=0$ 
pour l'éliminer. Le contenu total du supermultiplet chiral est donc 
\begin{equation} \Psi=(\phi,\ \psi_{\alpha},\ F). \end{equation}

$\triangleright$ La seconde représentation que nous allons utiliser 
dans la suite est le \textit{supermultiplet vecteur (ou de
jauge)} $[m,\ 1/2]$. Son contenu en champ est obtenu de la même façon : 
un fermion de Weyl~\footnote{Où de façon équivalente un fermion de 
Majorana.} 
(le jaugino), $\lambda^a_{\alpha}$, 
un boson de jauge (de masse nulle), $A_a^{\mu}$, et comme 
pour le supermultiplet chiral, un champ scalaire réel auxiliaire, $D^a$. 
\begin{equation} \Phi=(\lambda^a_{\alpha} ,\ A^a_{\mu},\ D^a), 
\end{equation}
où $a$ est un indice adjoint du groupe de jauge :

C'est dans ces deux représentations que les particules du MS et leurs
superpartenaires seront rangés. Nous allons maintenant construire avec ces 
deux représentations une théorie des champs supersymétrique.

\subsection{Théorie des champs supersymétrique}

Avant de discuter de modèles en particulier et surtout de l'extension
supersymétrique minimale du MS (le MSSM) nous allons d'abord présenter, sans 
démonstrations trop lourdes, la structure générale d'une théorie des champs avec
supersymétrie. Pour commencer progressivement, nous allons introduire le modèle
de Wess et Zumino sans interaction, juste pour voir comment concrètement se
transforment les champs. Puis nous introduirons les interactions et ceci nous
conduira à la notion nouvelle de {\it superpotentiel}. Enfin, nous discuterons 
des champs de jauge dans une théorie supersymétrique. A l'issue de cette
section, nous devrions posséder suffisamment de bagage théorique pour
comprendre comment le MSSM se présente et en étudier les aspects concrets comme
les prédictions expérimentales.

\subsubsection{Le lagrangien libre globalement supersymétrique}

L'action la plus simple que l'on peut construire avec le supermultiplet chiral 
est celle du mod\`ele de Wess-Zumino, sans masse et sans
interaction. Dans le cas {\it on-shell} (sans $F$), qui est suffisant quand il
n'y a pas d'interaction, nous avons simplement un fermion $\psi$ et un boson 
scalaire $\phi$ :
\beqn 
&S=\int d^4x\ (\mathcal{L}_{scalaire}+\mathcal{L}_{fermion})& 
\\
&\mathcal{L}_{scalaire}=-\partial^{\mu}\phi\,\partial_{\mu}\phi^{*} & 
\\
&\mathcal{L}_{fermion}= -i\psi^{\dag}\bar{\sigma}^{\mu}\,\partial_{\mu}\psi&
\eeqn
Si on applique une transformation supersymétrique globale de paramètre 
$\epsilon_{\alpha}$, fermion de Weyl infinitésimal indépendant 
des coordonnées d'espace-temps ($\partial^{\mu}\epsilon_{\alpha}=0$ ), sur le champ 
scalaire $\phi$, le résultat doit être proportionnel au champ fermionique $\psi$ :   
\beq 
\delta\phi= \epsilon^{\alpha}\psi_{\alpha} \ \ \mathrm{et}\ \ 
\delta\phi^*= \bar{\epsilon}_{\dot{\alpha}}\,\bar{\psi}^{\dot{\alpha}}
\eeq
\beq 
\Rightarrow \delta \mathcal{L}_{scalaire} = -\epsilon^{\alpha}\,
(\partial^{\mu}\psi_{\alpha})\,\partial_{\mu}\phi^*
-\partial^{\mu}\phi\,\bar{\epsilon}_{\dot{\alpha}}\,
(\partial_{\mu}\bar{\psi}^{\dot{\alpha}}) 
\eeq
A noter que le fermion infinitésimal $\epsilon_{\alpha}$ a la dimension d'une
masse à la puissance -$\frac{1}{2}$ contrairement à un fermion de Weyl usuel
qui a la dimension (masse)$^{3/2}$ :
\beq 
[\phi]=1,\ \ [\psi]=\frac{3}{2}, \ \ [\epsilon]=-\frac{1}{2}. 
\eeq 
La théorie physique est invariante si l'action est invariante (principe de
moindre action). Mais pour que l'action reste invariante sous une 
transformation de supersymétrie, il faut que la somme 
$\delta L_{scalaire}+\delta L_{fermion}$ soit nulle à une 
divergence totale près, qui ne contribuera pas à l'action. En considérant aussi
la dimension des 3 champs dont nous disposons, nous voyons que les champs 
fermioniques se transforment comme:
\beq 
\delta\psi_{\alpha}=i(\sigma^{\mu}\epsilon^{\dag})_{\alpha}\,
\partial_{\mu}\phi \ \ \mathrm{et}\ \ \delta\bar{\psi}^{\dot{\alpha}}=
- i(\epsilon\,\sigma^{\mu})^{\dot{\alpha}}\,\partial_{\mu}\phi^{*} 
\eeq
Mais est-ce que cette transformation correspond bien à une transformation de
supersymétrie? Pour s'en convaincre il suffit de partir, soit d'un fermion 
$\psi$ soit d'un boson $\phi$, et de leur appliquer 2 fois ces transformations.
Nous avons la chaîne suivante :
\beq \phi\to\psi\to\partial\phi,\ \ \psi\to\partial\phi\to\partial\psi, \eeq
c'est-à-dire que dans les deux cas l'action de 2 transformations
supersymétriques successives 
est équivalent à la dérivation donc à l'opérateur impulsion $P$ (car
$P^{\mu}\sim i\,\partial^{\mu}$). Nous retrouvons le
résultat de la section précédente, $Q^2\sim P$, et donc nos
transformations réalisent bien l'algèbre supersymétrique.

Dans le cas {\it off-shell}, l'action $S$ est modifiée par l'ajout d'un terme
comportant le champ $F$:
\beqn 
&S=\int d^4x\ (\mathcal{L}_{scalaire}+\mathcal{L}_{fermion}+\mathcal{L}_{aux}),&
\\
&\mathcal{L}_{aux}=F^*\,F,&
\eeqn
et les transformations de supersymétrie des champs $\psi$ et $\phi$ s'en
trouvent modifiées. Pour le champ scalaire $F$, sa transformation doit faire 
intervenir le champ $\psi$. Remarquons que la dimension du champ $F$ est une masse
au carré. Avec les lois de transformation suivantes,
\beq 
\delta F= i\,\bar{\epsilon}^{\dot{\alpha}}
\,(\overline{\sigma}^{\mu})_{\dot{\alpha}}^{\beta}
\,\partial_{\mu}\psi_{\beta} \ \ \mathrm{et}\ \ 
\delta F^*=-i\,\partial_{\mu}\bar{\psi}^{\dot{\beta}}
\,(\bar{\sigma}^{\mu})_{\dot{\beta}}^{\alpha}
\,\epsilon_{\alpha},
\eeq
la variation du terme $\mathcal{L}_{aux}$ donne :
\beq 
\delta\mathcal{L}_{aux}=i\,\bar{\epsilon}
\,(\overline{\sigma}^{\mu})
\,\partial_{\mu}\psi\,F^* - i\,\partial_{\mu}\bar{\psi}
\,(\bar{\sigma}^{\mu})\,\epsilon\,F.
\eeq
Cette variation s'annulent bien {\it on-shell} avec l'équation du mouvement
$F=F^*=0$. Pour compenser cette variation dans le cas {\it off-shell}, la loi 
de transformation de $\psi$ devient :
\beq 
\delta\psi_{\alpha}=i(\sigma^{\mu}\bar{\epsilon})_{\alpha}\,
\partial_{\mu}\phi+\epsilon_{\alpha}F \ \ \mathrm{et}\ \ 
\delta\bar{\psi}^{\dot{\alpha}}=-
i(\epsilon\,\sigma^{\mu})^{\dot{\alpha}}\,\partial_{\mu}\phi^{*}+
\bar{\epsilon}^{\dot{\alpha}} F^*.
\eeq
Les transformations de $\phi$ sont inchangées.
Nous pouvons vérifier que $\delta S=0$ sans faire référence aux équations du
mouvement et donc la supersymétrie est aussi réalisée {\it off-shell} avec ces
lois de transformation.
\par\hfill\par
Nous avons vu que le champ $F$ était auxiliaire et ne servait qu'à l'utilisation 
{\it off-shell} de la supersymétrie. En fait, il a aussi un autre rôle. En effet, 
nous sommes partis dans l'explication de la
supersymétrie mais il ne faut pas oublier qu'aux échelles d'énergies accessibles
actuellement nous n'avons pas observé de supersymétrie. Si la supersymétrie
existe dans la nature, elle a dû forcément être brisée à un moment ou un autre.
Le champ auxiliaire $F$, mais aussi le champ auxiliaire $D$, servent à briser 
la supersymétrie. Nous verrons cet aspect au dernier chapitre. Après cette
section formelle, nous oublierons ces champs auxiliaires en utilisant à chaque
fois les équations du mouvement pour les éliminer des équations.

\subsubsection{Les interactions du multiplet chiral}

Nous allons maintenant ajouter à notre théorie la possibilité de termes 
d'interaction entre ces deux types de champs qui composent les supermultiplets
chiraux.
Nous n'allons pas le démontrer mais le terme d'interaction le plus général, 
invariant sous les transformations de supersymétrie et renormalisable, que nous allons 
ajouter dans le lagrangien libre vu précédemment, s'écrit sous la forme :
\beq 
\mathcal{L}_{int}=-\frac{1}{2}W^{ij}(\phi)\psi_i\psi_j + V(\phi,\ \phi^*) + c.c. \label{LW} 
\eeq
La quantité $W^{ij}$ ne doit dépendre que des champs $\phi$, pour assurer 
que la variation lors de la transformation
de supersymétrie du premier terme de $\mathcal{L}_{int}$ puisse être
compensée par la variation d'un autre terme. Pour la m\^eme raison, 
$W^{ij}$ doit être
complétement symétrique. La seule forme possible pour $W^{ij}$ est alors :
\beq 
W^{ij}=\frac{\partial^2 W(\phi)}{\partial\phi_i\,\partial\phi_j},
\eeq
où on définit le {\it superpotentiel} $W$ que l'on écrit sous la forme :
\beq W=\frac{1}{2}m^{ij}\phi_i\phi_j+\frac{1}{6}y^{ijk}\phi_i\phi_j\phi_k
\label{superpotentiel}\eeq
dans le cadre d'une théorie renormalisable.

Qu'est-ce que ce superpotentiel ? Tout d'abord, sa dimension est celle d'une 
masse au cube. Il fait intervenir la matrice symétrique de masse $m^{ij}$ des 
fermions~\footnote{La supersymétrie assure que c'est aussi la matrice de masse 
des bosons scalaires associés.} et la matrice totalement symétrique 
des couplages de Yukawa $y^{ijk}$ entre un scalaire et 2 fermions. \underline{Il 
résume donc toutes les interactions qui ne sont pas de jauge} (c'est-à-dire avec les 
bosons de jauge). C'est, de plus, une fonction analytique des champs complexes 
$\phi_i$ c'est-à-dire qu'il est fonction de $\phi_i$ mais pas du complexe 
conjugué $\phi_i^*$. Ceci est très important pour la suite.

De même, en imposant que $\mathcal{L}_{int}$ soit invariant sous transformation
de supersymétrie, on détermine la forme du potentiel $V$.
En présence d'interactions, donc d'un superpotentiel non-nul, les équations du
mouvement des champs auxiliaires $F^i$ sont :
\beq F_i=-\frac{\partial W(\phi)}{\partial\phi^i}=-W^*_i,\ \ \ \ \ 
F^{*i}=-\frac{\partial W(\phi)}{\partial\phi_i}=-W^i. \eeq
Nous pouvons les utiliser pour écrire le lagrangien sans les champs $F$,
comme dit plus haut.
Le {\it potentiel scalaire} $V$ de la théorie est :
\beq V=W^*_i W^i=F_iF^{*i}, \eeq
qui est automatiquement non-négatif puisque c'est la somme de carrés. 
Si nous employons la forme générale~(\ref{superpotentiel}) du superpotentiel, nous
avons alors le lagrangien général pour un supermultiplet chiral en interaction :
\beq 
\mathcal{L}=
-\partial^{\mu}\phi\,\partial_{\mu}\phi^{*}
-i\psi^{\dag}\bar{\sigma}^{\mu}\,\partial_{\mu}\psi
-\frac{1}{2}m^{ij}\psi_i\psi_j-\frac{1}{2}m^{*}_{ij}\psi^{\dag i}\psi^{\dag j}
-V-\frac{1}{2}y^{ijk}\phi_i\psi_j\psi_k
-\frac{1}{2}y^{*}_{ijk}\phi^{*i}\psi^{\dag j}\psi^{\dag k}.
\eeq

\subsubsection{Théorie de jauge supersymétrique}

Le MS, qui est la théorie qui nous intéresse et que nous voulons
"supersymétriser", a, outre des champs fermioniques chiraux (les quarks, les
leptons), des champs de jauge de spin 1 (bosons $W$, $Z$, gluons,...). Dans la
section dédiée à l'algèbre supersymétrique nous avons vu que le supermultiplet
vecteur pouvait accueillir de tels champs de jauge. Voyons donc comment se
comporte un tel supermultiplet, sans et avec interaction.

Le supermultiplet vecteur contient un boson de jauge $A^{\mu}_a$, de masse nulle,
 un fermion de Weyl, le jaugino $\lambda_a$, également de masse nulle, ainsi 
qu'un champ scalaire réel auxiliaire $D_a$ qui est l'analogue du champ $F$ 
précédent. Ce supermultiplet est dans la représentation adjointe du groupe de
jauge (de constantes de structure $f^{abc}$). La forme du lagrangien est 
complètement déterminée par la condition d'invariance de jauge et la renormalisabilité :
\beq 
\mathcal{L}_{jauge}=
-\frac{1}{4}F_{\mu\nu}^a F^{a \mu\nu}
-i\lambda^{a\dag}\bar{\sigma}^{\mu}D_{\mu}\lambda^{a}
+\frac{1}{2}D^aD^a, \label{Ljauge}
\eeq
où les dérivées covariantes de jauge $D_{\mu}$ et $F_{\mu\nu}^a$ sont donnés par
:
\beqn
&F_{\mu\nu}^a&=\partial_{\mu}A^a_{\nu}-\partial_{\nu}A^a_{\mu}
-gf^{abc}A^b_{\mu}A^c_{\nu}, \\
&D_{\mu}\lambda^{a}&=\partial_{\mu}\lambda^{a}-gf^{abc}A^b_{\mu},
\eeqn
comme habituellement pour une théorie de jauge.
Ce lagrangien est déjà supersymétrique et les transformations de supersymétrie
de paramètre $\epsilon$ pour les champs du supermultiplet vecteur sont :
\beqn
&\delta A^a_{\mu}&=
\frac{1}{\sqrt{2}}\left(\epsilon^{\dag}\bar{\sigma}^{\mu}\lambda^{a}
+\lambda^{a\dag}\bar{\sigma}^{\mu}\epsilon\right), \\
&\delta \lambda^{a}_{\alpha}&=
-\frac{i}{2\sqrt{2}}(\sigma^{\mu}\bar{\sigma}^{\nu}\epsilon)_{\alpha}F_{\mu\nu}^a
+\frac{1}{\sqrt{2}}\epsilon_{\alpha}D^a, \\
&\delta D^a&=
\frac{i}{\sqrt{2}}\left(\epsilon^{\dag}\bar{\sigma}^{\mu}D_{\mu}\lambda^{a}
-D_{\mu}\lambda^{a\dag}\bar{\sigma}^{\mu}\epsilon\right).
\eeqn
Sans interaction avec aucun supermultiplet chiral, l'équation du mouvement 
pour le champ 
$D^a$ est simplement
$D^a=0$ que nous obtenons directement du Lagrangien~(\ref{Ljauge}). Il n'a pas
de terme cinétique et ne se propage donc pas.
 
Dans le MS, les champs de jauge interagissent avec les fermions chiraux. Dans
notre version supersymétrique il nous faut donc considérer les interactions
entre le supermultiplet chiral et le supermultiplet vecteur. Comme dans le cas
non-supersymétrique, les dérivées usuelles $\partial^{\mu}$ des fermions sont 
maintenant à remplacer par des dérivées covariantes de jauge $D^{\mu}$. 
De plus, le lagrangien doit comporter des termes supplémentaires qui traduisent les 
interactions entre supermultiplets chiraux et vecteurs. Les lois de transformation
supersymétriques du supermultiplet chiral changent pour prendre en 
compte la variation des nouveaux termes. 
L'équation du mouvement pour $D^a$ est alors (les
$T^a$ sont les générateurs du groupe de jauge selon lesquels les
supermultiplets chiraux se transforment et $g$ est la constante de couplage) :
\beq D^a=-g(\phi^{*}T^a\phi), \eeq
et le potentiel scalaire complet est :
\beq 
V=F_iF^{*i}+\frac{1}{2}\sum_aD^aD^a=
W^*_iW^i+\frac{1}{2}\sum_ag^2(\phi^{*}T^a\phi)^2. \label{pot}
\eeq 
Ce potentiel scalaire est automatiquement non-négatif et s'avère très 
important 
pour la brisure de symétrie. Nous parlons de termes $F$ et de termes $D$ pour faire référence respectivement au 
premier et deuxième terme du potentiel. Ce potentiel est complètement déterminé
par les couplages de Yukawa ({\it via} le terme $F$) et par les interactions de
jauge ({\it via} le terme $D$).

\section{Les modèles supersymétriques à basse énergie}

La section précédente était assez abstraite, nous allons tout de suite appliquer
les différents résultats obtenus. Le Modèle Standard fonctionne très bien, nous
l'avons vu. Nous allons donc juste lui offrir une promotion en le
"supersymétrisant" et en conservant toutes ses caractéristiques. Le modèle
minimal que nous pouvons obtenir est le MSSM (Minimal Supersymmetric Standard
Model). Nous présenterons son contenu en particules (la nomenclature des
nouvelles particules), nous expliquerons comment la symétrie électrofaible peut 
se briser, et nous décrirons la brisure effective de supersymétrie (brisure dite
"douce"). Nous 
aborderons juste après les prédictions typiques du MSSM. Enfin, nous parlerons 
aussi des variantes possibles du MSSM car la nature a pu très bien choisir une 
voie un peu plus complexe que ce modèle minimal.

\subsection{Les modèles $\mathcal{N}\geq 2$}

Comme nous l'avons déjà dit dans la section 1.2.2, {\it a priori} le nombre de générateurs 
supersymétriques $Q_{\alpha}$ que nous pouvons introduire peut être supérieur ou
égal à 1 (nous parlerons de supersymétrie $\mathcal{N}\geq 1$). Après tout, les théories
supersymétries $\mathcal{N}\geq 2$ possèdent d'avantages de symétries et de ce fait il se
trouve qu'elles ont moins de divergences ce qui les rend très intéressantes.
En effet, dans le cas $\mathcal{N}= 2$ il n'y a qu'un nombre fini de
diagrammes qui divergent et dans le cas $\mathcal{N}=4$ il n'y en a plus du
tout ! Une théorie supersymétrique $\mathcal{N}=4$ est intrinsèquement
\underline{finie}. Tout
naturellement, nous aimerions donc construire un modèle $\mathcal{N}=4$
englobant le MS. 
Malheureusement, \underline{à basse énergie} (autour du TeV), \underline{les modèles $\mathcal{N}\geq 2$ 
ne sont pas réalistes}. Ils ne permettent pas la violation de la parité que nous
observons dans les interactions faibles. En effet, un supermultiplet d'une
théorie supersymétrique $\mathcal{N}\geq 2$ possède toujours les 2 hélicités opposées à la 
fois donc particules "gauches" et "droites" siègent dans le même supermultiplet.
Ce qui implique qu'elles ont les mêmes interactions (car elles sont dans la même
représentation du groupe de jauge). C'est malheureusement une conclusion 
contraire aux
observations expérimentales qui nous disent par exemple que l'électron "gauche" 
(qui fait
partie d'un doublet dans le MS) n'a pas la même interaction avec les bosons $W$
que l'électron "droit" (qui est un singulet d'isospin faible nul et qui ne 
"ressent" pas l'interaction faible). Les modèles $\mathcal{N}\geq 2$ ne peuvent
donc pas décrire la physique des particules à basse énergie.

\subsection{La zoologie du Modèle Standard Supersymétrique Minimal}

La sous-section précédente nous a enseigné que le cas minimal $\mathcal{N}=1$
était aussi le seul cas réaliste à basse énergie pour englober le MS. Les
supermultipets dont nous disposons sont : 

$\bullet$ le supermultiplet chiral qui comprend un fermion de spin 1/2 et un
boson de spin 0,

$\bullet$ le supermultiplet vecteur qui comprend un boson de spin 1 et un
fermion de spin 1/2.

\par\hfill\par
Pouvons nous ranger toutes nos particules du MS dans ces multiplets ? Autrement
dit, pouvons nous associer les quarks et leptons aux bosons W, Z, au photon, 
etc ? 

Malheureusement, cela poserait des problèmes pour la conservation des nombres
quantiques. En effet, les bosons de jauge et les fermions n'ont pas les mêmes 
propriétés de transformation sous les groupes de jauge donc possèdent des nombres 
quantiques différents. La supersymétrie ne modifie pas ces nombres quantiques, 
on ne peut donc pas associer un boson de jauge à un fermion
connu ou inversement. Cela poserait aussi des problèmes pour la conservation 
d'autres nombres comme le nombre leptonique car les bosons de jauge que nous 
connaissons ont un nombre leptonique nul contrairement aux leptons. 
Il nous faut donc inventer des (super)partenaires à toutes les
particules connues ! Le tableau suivant, (\ref{sparticules}), donne à chaque 
particule connue le nom, le spin et l'abréviation de son spartenaire.

\begin{table}[htbp!]
\begin{center}
\begin{tabular}[t]{||l||c|c||}
\hline
{Particule}& {Spartenaire}& {Spin}	\\
\hline
\hline
& & \\
quarks q & squarks $\tilde{q}$ & 0 \\
$\to$ top t & stop $\tilde{t}$ & \\
$\to$ bottom b & sbottom $\tilde{b}$ & \\
...& & \\
leptons l & sleptons $\tilde{l}$ & 0 \\
$\to$ neutrino $\nu_e$ & sneutrino $\tilde{\nu_e}$ & \\
$\to$ muon $\mu$ & smuon $\tilde{\mu}$ & \\
...& & \\
\hline
\hline
& & \\
bosons de jauge & jauginos & 1/2\\
$\to$ photon $\gamma$ & photino $\tilde{\gamma}$ & \\
$\to$ boson $Z$ & zino $\tilde{Z}$ & \\
$\to$ boson $B$ & bino $\tilde{B}$ & \\
$\to$ boson $W$ & wino $\tilde{W}$ & \\
$\to$ gluon $g$ & gluino $\tilde{g}$ & \\
\hline
\hline
& & \\
bosons de Higgs $H_i^{\pm,0}$ & higgsinos $\tilde{H}_i^{\pm,0}$& 1/2 \\
& & \\
\hline
\end{tabular}
\end{center}
\caption[contenu du MSSM]{ Les particules du MSSM. }
\label{sparticules}
\end{table}
 
Avant de passer aux sections suivantes, nous allons formuler plusieurs
remarques. 
Tout d'abord nous avons aussi, outre les nouveaux spartenaires, au moins deux
doublets de bosons de Higgs. Pourquoi a-t-il fallu aussi ajouter des 
bosons de Higgs ? Dans l'étude des théories de champs supersymétriques nous 
avons introduit la notion de superpotentiel. Celui-ci résume toutes les 
interactions possibles des particules
(mais qui ne font pas intervenir les bosons de jauge) donc en particulier les
interactions de Yukawa avec les champs de Higgs. Ce superpotentiel ne peut pas
être fonction de champs complexes conjugués. Or dans le MS, pour donner une
masse aux quarks type "up" nous utilisons un terme $QU^cH^{*}$. Comme
en supersymétrie ce genre de terme est interdit nous devons utiliser un
\underline{nouveau} 
champ de Higgs, d'hypercharge $Y=-1$, et le coupler simplement (sans 
conjugaison complexe) : $QU^cH_u$. 
Ce nouveau champ, après brisure électrofaible nous laissera donc d'autres 
bosons de Higgs dont certains seront chargés (voir plus loin). Ce nouveau
doublet de Higgs est aussi nécesaire pour annuler les possibles anomalies.

Deuxièmement, nous savons bien que dans le MS un fermion droit subit un 
traitement
différent d'un fermion gauche. Ils auront en supersymétrie chacun un supermultiplet avec
chacun un spartenaire. Par exemple $q_L\to \tilde{q}_L$ et $q_R\to \tilde{q}_R$.
Ces deux squarks sont bien différents et pour les identifier nous laisserons
l'indice de chiralité $L$ ou $R$ tout en sachant qu'il n'a pas de sens physique
pour une particule scalaire (spin 0 donc une seule hélicité $\lambda=0$).

Troisièmement, pourquoi avons nous fait le choix d'avoir des spartenaires 
de spin
inférieur ? {\it A priori} nous aurions pu associer à tous les fermions du MS des
spartenaires de spin 1 et aux bosons de jauge de spin 1 des spartenaires de spin
3/2. Cependant, introduire une particule de spin 1 signifie introduire une
nouvelle interaction et implique un modèle non-minimal. De plus, introduire des
particules de spin >1 rend la théorie non-renormalisable~\footnote{En analysant
en théorie des champs ce qui rend les diagrammes divergents, on aboutit à une
condition pour qu'un terme du lagrangien soit renormalisable: 
$\Delta=4-d-\sum_i n_i(s_i+1)\geq 0$ où $d$ est le nombre de dérivées, $n_i$ est
le nombre de champs du type $i$ dans le terme d'interaction et $s_i$ leur spin.
Si le spin est trop élevé on tombe inévitablement sur des termes
non-renormalisables.}.

Enfin, les $\tilde{\gamma},\ \tilde{Z},\ \tilde{W},\ \tilde{H}$ ne
s'observent pas directement. En effet, ils se mélangent et donc n'apparaissent
expérimentalement que des combinaisons de ces jauginos et higgsinos : celles-ci
ont pour nom les {\it neutralinos} et les {\it charginos} :
 
$\bullet$ Les neutralinos $\tilde{N}_{1,2,3,4}^0$~\footnote{Notés aussi 
dans la littérature $\tilde{\chi}_{1,2,3,4}^0$.} sont de charge électrique 
nulle et 
mélangent en particulier les fermions $\tilde{B}$, $\tilde{W}^0$, $\tilde{H}_u^0$ et
$\tilde{H}_d^0$. 

$\bullet$ Les charginos $\tilde{C}_{1,2}^{\pm}$~\footnote{Notés aussi 
souvent 
$\tilde{\chi}_{1,2}^{\pm}$.} sont chargés électriquement et mélangent les 
$\tilde{W}^{\pm}$ et les $\tilde{H}^{\pm}$.

\noindent Ces mélanges sont dûs au fait que les jauginos et higgsinos
possèdent les même nombres quantiques et ne sont pas distinguables séparément.

\subsection{Le modèle}

Le MSSM est l'extension supersymétrique minimale du MS. Les quarks
et les leptons sont alors mis dans des superchamps chiraux avec leurs 
superpartenaires et ces
superchamps forment des supermultiplets chargés sous $SU(3)_C$, $SU(2)_L$ et 
$U(1)_Y$ de la même façon que les multiplets du MS. Les bosons de jauge sont
quant à eux placés avec leurs superpartenaires fermioniques dans des superchamps vecteurs. 

Le superpotentiel le plus général, mais minimal, du MSSM est alors :
\beq \mathcal{W}=\mathcal{Y}_u\bar{U}QH_u + \mathcal{Y}_d\bar{D} Q H_d +\mathcal{Y}_e\bar{E}
L H_d +\mu H_u H_d. \eeq
La notation $\bar{\psi}$ signifie que les champs du supermultiplet sont des
champs conjugués de charge $\psi^c$. Les
champs conjugués apparaissent car nous avons choisi de ne travailler qu'avec
des champs gauches. Les champs droits s'obtiennent justement par cette opération
de conjugaison. De plus, les indices de $SU(2)$ ont été supprimés pour ne pas
alourdir l'expression. Nous avons en fait
$\mu(H_u)_{\alpha}(H_d)_{\beta}\epsilon^{\alpha\beta}$, 
$(\mathcal{Y}_u)^{ij}\bar{U}_{ai}Q^a_{j\alpha}(H_u)_{\beta}\epsilon^{\alpha\beta}$,
...

Les $\mathcal{Y}$ sont les matrices de Yukawa, $3\times3$ dans l'espace des
saveurs et sont sans dimensions. Elles donnent les masses des quarks et leptons
ainsi que les angles et phases de CKM après la brisure électrofaible.

Les deux champs de Higgs, $H_u$ et $H_d$, ont été introduit pour respecter la condition
d'analycité du superpotentiel~\footnote{C'est-à-dire l'absence de champs
complexes conjugués $\phi^*$.} et donner une masse aux particules "up" et "down"
ainsi que pour la condition d'annulation des anomalies. Mais une fois que nous avons 
deux superchamps de Higgs, un terme qui couple les deux peut {\it a priori} 
exister. Ce terme
cependant (le couplage $\mu$) donne naissance au problème "$\mu$" :
phénoménologiquement, il est de l'ordre du TeV alors que dans le MSSM rien ne le
force à être aussi bas. Dans un modèle plus fondamental, $\mu$ pourrait être lié à
l'échelle de brisure de la supersymétrie.

Une fois que ce superpotentiel a été écrit, nous pouvons trouver toutes les
interactions possibles (mais non de jauge) entre les particules et écrire le
lagrangien d'interaction, grâce à la
formule~(\ref{LW}), ainsi que le potentiel effectif de la théorie,
formule~(\ref{pot}).

\subsection{La brisure douce de la supersymétrie}

Il reste cependant à introduire dans le modèle la brisure de la supersymétrie.
Mais le mécanisme et l'échelle réelle de la brisure sont encore inconnues. 
Ce que nous pouvons faire c'est \underline{paramétriser} à basse énergie cette 
brisure. Ceci se fait en ajoutant des termes au lagrangien qui brise
explicitement la supersymétrie. La forme générale de ce lagrangien de brisure 
$\mathcal{L}_{soft}$ est :
\beq \mathcal{L} \supset
\mathcal{L}_{soft}=-\frac{1}{2}(M_{\lambda}\lambda^a\lambda^a +\ c.c) - m_{ij}^2
\phi_j^{*}\phi_i +
(\frac{1}{2}b_{ij}\phi_i\phi_j+\frac{1}{6}a_{ijk}\phi_i\phi_j\phi_k+\ c.c)
\eeq
Il brise bien la supersymétrie car seuls les scalaires ($\phi_i$) et les
jauginos ($\lambda^a$) ont un terme de masse. La brisure, bien
qu'\underline{explicite}~\footnote{Par opposition à spontanée.}, est
dite "douce" ("soft" en anglais) car on peut montrer qu'elle n'introduit pas de
divergences quadratiques. Tous les modèles de brisure, qu'ils aient leur origine dans
les théories des cordes ou de supergravité (voir chapitre 4), conduisent à basse
énergie à cette forme de $\mathcal{L}_{soft}$.
Avec les superchamps du MSSM, $\mathcal{L}_{soft}$ s'écrit : 
\beqn -\mathcal{L}_{soft} &=&\frac{1}{2}(M_3 \tilde{g}\tilde{g} + M_2\tilde{W}\tilde{W}+M_1\tilde{B}\tilde{B}+\ c.c) \nonumber \\ 
&+&  \tilde{Q}^{\dag}m_Q^2\tilde{Q} + \bar{\tilde{U}}^{\dag}m_{\bar{U}}^2\bar{\tilde{U}} +\bar{\tilde{D}}^{\dag}m_{D}^2\bar{\tilde{D}} +\bar{\tilde{L}}^{\dag}m_{L}^2\bar{\tilde{L}} +\bar{\tilde{E}}^{\dag}m_{\bar{E}}^2\bar{\tilde{E}} \nonumber \\ 
&+& (\bar{\tilde{U}}^{\dag}a_U \tilde{Q}H_u - \bar{\tilde{D}}^{\dag}a_D\tilde{Q}H_d -\bar{\tilde{E}}^{\dag}a_E\tilde{L}H_d +\ c.c) \nonumber \\ 
&+& m_{H_u}^2 H_u^{*}H_u +m_{H_d}^2 H_d^{*}H_d + (b H_uH_d +\ c.c) 
\label{superpotentielMSSM}\eeqn 

Les masses $M_3$, $M_2$, $M_1$ des jauginos sont en général complexes, 
ce qui introduit 6 paramètres. 
Les $m_Q$, $m_L$, $m_{\bar{u}}$,..., sont les matrices de masse des squarks et 
sleptons, hermitiennes et de taille $3\times3$
dans l'espace des saveurs, ce qui fait 45 paramètres inconnus.
Les couplages $a_U$, $a_D$,..., sont des couplages trilinéaires, $3\times3$ et
complexes donc caractérisés par 54 paramètres.
Enfin, les couplages bilinéaires des Higgs introduisent 4 paramètres. 
En tout, $\mathcal{L}_{soft}$ contient 109 paramètres inconnus ! La
supersymétrie introduit donc beaucoup de paramètres mais en contrepartie fait intervenir très
peu de principes. Ce nombre de paramètres "soft" peut cependant 
être diminué en redéfinissant les champs grâce à des symétries ou des 
hypothèses supplémentaires. La mesure des paramètres de brisure douce 
permettra de tester les modèles de plus haute énergie.  

\subsection{La brisure électrofaible et les bosons de Higgs supersymétriques}

\subsubsection{Le potentiel scalaire}
Le secteur de Higgs du MSSM contient 2 doublets complexes : 
\beq
H_u=\left(\begin{array}{c} H_u^0 \\ H_u^- \end{array}\right),
\ H_d=\left(\begin{array}{c} H_d^+ \\ H_d^0 \end{array}\right).
\eeq
La brisure électrofaible est donc un peu plus complexe que dans le cas du Modèle
Standard (à 1 seul doublet). 
Au niveau classique ("arbre"), le potentiel scalaire effectif s'écrit, 
après 
plusieurs
simplifications que nous ne détaillerons pas :
\beqn 
V&=&(|\mu|^2+m_{H_u}^2)|H_u^0|^2 + (|\mu|^2+m_{H_d}^2)|H_d^0|^2 - b(H_u^0 H_d^0
+c.c) \nonumber \\
&&+\frac{1}{8}(g_2^2+g_1^2)(|H_u^0|^2-|H_d^0|^2)^2. 
\eeqn
Les termes proportionnels à $|\mu|^2$ tirent leur origine des termes $F$ et le
terme proportionnel aux couplages de jauge $(g_1,\ g_2)$ des termes $D$. Les
autres termes viennent de $\mathcal{L}_{soft}$ (en omettant les autres scalaires
qui ne joueront aucun rôle ici).  
Une brisure spontanée de la symétrie électrofaible ne peut exister avec cette
forme de potentiel que si le paramètre $b$ vérifie :
\beq
b^2>(|\mu|^2+m_{H_u}^2)(|\mu|^2+m_{H_d}^2), \label{EWSBb1}
\eeq
et de plus, pour que le potentiel soit limité inférieurement, il faut :
\beq
2b<2|\mu|^2+m_{H_u}^2+m_{H_d}^2. \label{EWSBb2}
\eeq
Ces conditions ne sont valables qu'au niveau "arbre".
Quand la brisure de la symétrie a lieu, les deux champs $H_u^0$ et $H_d^0$
développent une {\it v.e.v} :
\beq
<H_u^0>=v_u,\ <H_d^0>=v_d,
\eeq
qui sont reliées à celle du MS par :
\beq
v^2=v_u^2+v_d^2 =\frac{2m_Z^2}{(g_2^2+g_1^2)}.
\eeq
On définit aussi le paramètre $\tan\beta$ par :
\beq
\tan\beta=\frac{v_u}{v_d},\ 0<\beta<\frac{\pi}{2}.
\eeq 
Au minimum du potentiel, 
\beq
\frac{\partial V}{\partial H_u^0}=\frac{\partial V}{\partial H_d^0}=0,
\eeq 
nous avons alors  les deux relations :
\beqn
|\mu|^2+m_{H_u}^2 & = & b\tan\beta -\frac{m_Z}{2}\cos2\beta, \\
|\mu|^2+m_{H_d}^2 & = & b\cot\beta +\frac{m_Z}{2}\cos2\beta.
\eeqn
Elles sont très importantes car lient une quantité mesurable, $m_Z$, à des
paramètres de brisure douce. Nous pouvons aussi noter que la phase de $\mu$
n'est pas déterminée. 

\subsubsection{Les bosons de Higgs supersymétriques}

Les deux doublets complexes comptent 8 degrés de liberté donc 8 scalaires réels.
Le mécanisme de Higgs, quand la brisure électrofaible se réalise, en utilise 3
pour donner une masse aux 2 bosons $W$ et au $Z$. Il reste alors 5 bosons de
Higgs qui restent dans le spectre : 

$\hookrightarrow$ 2 sont neutres et pairs sous une transformation de $CP$, $h^0$ et
$H^0$,

$\hookrightarrow$ un est neutre et impair sous $CP$, $A^0$,

$\hookrightarrow$ les 2 derniers sont chargés, $H^{\pm}$.

\par\hfill\par
Les masses de ces bosons de Higgs supersymétriques sont au niveau de l'arbre : 
\beqn
m^2_{h^0, H^0} &=& \frac{1}{2}\left(m_{A^0}^2+m_Z^2 \mp 
\sqrt{(m_{A^0}^2+m_Z^2)^2 -4m_{A^0}^2m_Z^2\cos^22\beta}\,\right), \\
m_{A^0}^2 &=& \frac{2b}{\sin2\beta}, \\
m_{H^{\pm}}^2 &=& m_{A^0}^2 + m_W^2, 
\eeqn
et de plus la masse du $h^0$ est bornée supérieurement par :
\beq
m_{h^0}<|\cos2\beta|m_Z.
\eeq
En particulier, cette dernière relation nous donne $m_{h^0}<m_Z$. 

\begin{figure}[htbp!]
\centerline{\epsfxsize=10cm\epsfbox{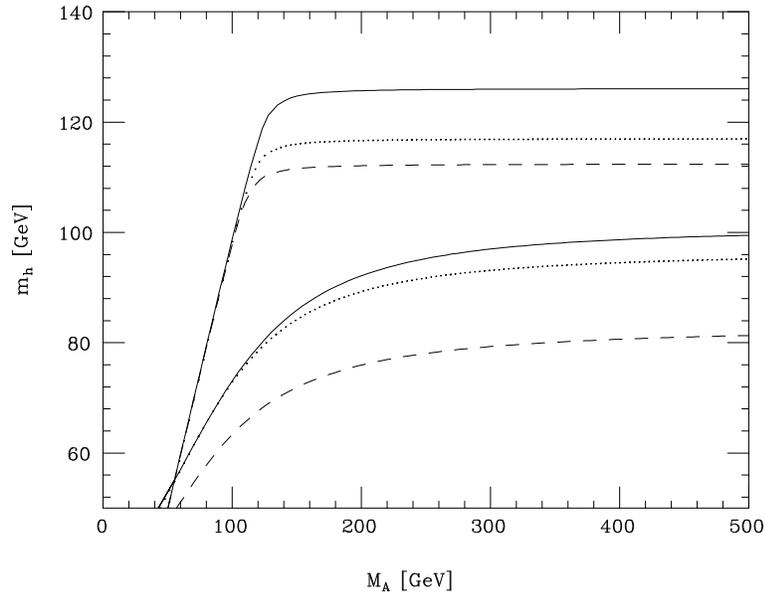}}
\caption{Masse du boson de Higgs supersymétrique le plus léger en fonction de
$m_A$ et pour différentes valeurs de $\tan\beta$.~\cite{JE-SUSY}}
\label{mHSUSY}
\end{figure}

Or ces relations ne sont valables \underline{qu'à l'arbre}. Nous avons déjà vu que la masse 
des scalaires de Higgs subissent des corrections radiatives non-négligeables au 
niveau d'une boucle. On a pour la masse de $h$ au carré :
\beq
\Delta m_h^2=\frac{3m_t^4}{4\pi^2 v^2}\ln\left(\frac{m_{\tilde{t}_1}m_{\tilde{t}_2}}{m_t^2}\right) 
+ \frac{3m_t^4}{8\pi^2
v^2}\mathrm{f}(m_{\tilde{t}_1}^2,m_{\tilde{t}_2}^2,\mu,\tan\beta),  
\eeq 
où $m_{\tilde{t}_{1,2}}$ sont les masses physiques des stops (qui sont des
mélanges des états $\tilde{t}_R$ et $\tilde{t}_L$) et $\mathrm{f}(m_{\tilde{t}_1}^2,m_{\tilde{t}_2}^2,\mu,\tan\beta)$ 
est une fonction que l'on peut retrouver explicitement dans~\cite{JE-SUSY}.
La correction $\Delta m_h^2$ dépend \underline{quartiquement} de la masse du 
top, ce qui la rend importante.  
En prenant en compte cette correction, la masse du boson de Higgs 
supersymétrique le plus léger, $h$, pour des masses de sparticules autour d'un 
TeV, est :
\beq
m_h \lesssim 130\ \mathrm{GeV}.
\eeq
Dans la figure~(\ref{mHSUSY}), nous pouvons voir $m_h$ en fonction de $m_{A^0}$ 
pour différentes valeurs de $\tan\beta$. 
Ce boson de Higgs supersymétrique le plus léger a donc toutes les caractéristiques
nécessaires pour être le boson de Higgs favorisé par les ajustements des données
électrofaibles ! Ceci constitue une prédiction importante de la supersymétrie.
Il y en a d'autres et nous allons en voir tout de suite un aperçu. 

\subsection{La $R$-parité et la matière noire}

\subsubsection{La $R$-parité}

Purement sur des considérations d'invariance de jauge, d'invariance de Lorentz et de
renormalisabilité, nous pouvons introduire dans le
superpotentiel~(\ref{superpotentielMSSM}) d'autres termes qui n'ont pas de
correspondance avec le MS et qui ne
conservent ni le nombre baryonique, ni le nombre leptonique~\footnote{La 
conservation de
$B$ et $L$ dans le MS est accidentelle mais n'est {\it a priori} pas 
obligatoire dans ses extensions (supersymétriques ou pas). En effet, dans le MS, il 
n'existe pas de
termes renormalisables qui violent $L$ ou $B$, ce qui n'est forcément le 
cas dans d'autres théories. 
Cependant, dans le MS, il y a des possibles interactions 
non renormalisables qui violent $L$ ou $B$.}. Ces termes sont :
\beq \mathcal{W}_{RPV}=\lambda_{ijk}L_iL_jE_k + \lambda'_{ijk} L_iQ_j\bar{D}_k +
\lambda''_{ijk} \bar{U}_i\bar{D}_j\bar{D}_k, \label{WRPV}\eeq
où $\lambda,\lambda',\lambda''$ sont des couplages inconnus sans dimensions. 

Cependant, une combinaison des deuxième et troisième termes pourrait 
conduire à une désintégration rapide du
proton, alors que le proton est très stable~\footnote{Sa durée de vie 
excède $10^{33}$
ans !}. Phénoménologiquement, il faut pouvoir
s'assurer que ces termes soient supprimés~\cite{RPVSUSY}:
\begin{equation}
|\lambda' \lambda''| \; < \; {\cal O}(10^{-9}). 
\end{equation}
C'est ce qui se passe quand on
postule une nouvelle symétrie, la $R$-parité. Elle est définie comme suit : 
\beq R_p=(-1)^{3(B-L)+2S} \eeq
où $S$ est le spin. C'est un nombre quantique multiplicatif, les particules du MS
sont alors paires et les sparticules impaires. Si cette $R$-parité est conservée, 
elle implique plusieurs choses importantes :

$\bullet$ les sparticules sont produites par paires, par exemple :
$\bar{p}\,p\to\tilde{q}\,\tilde{g}\,X$, $e^+\,e^-\to \tilde{\mu}^+\,
\tilde{\mu}^-$,

$\bullet$ une sparticule se désintègre en une autre sparticule (ou en un nombre
impair), par exemple : $\tilde{q}\to q\, \tilde{g}$, $\tilde{\mu}\to\mu\,
\tilde{\gamma}$,

$\bullet$ la sparticule la plus légère, la LSP, est stable. 
\subsubsection{La LSP et la matière noire}

Dans le MSSM, la $R$-parité est conservée et la sparticule la plus légère est 
stable donc ne peut pas se désintégrer. La LSP a un fort interêt car pourrait 
constituer la majorité de la matière
noire froide favorisée par les modèles cosmologiques de formation des structures
(galaxies,...). En effet, au fur et à mesure du refroidissement de l'Univers (son
expansion) les particules se sont désintégrées, ont formé des baryons, atomes
etc, sauf les neutrinos et la LSP. Cette dernière nous parait "invisible" car
elle n'interagit que très peu avec la matière et elle  
contribue alors à la densité relique de matière noire dont les bornes sont,
d'après les mesures récentes de WMAP~\cite{BDEGOP}, $0.094<\Omega_{MN} h^2<0.129$ 
(à 2$\sigma$).
L'existence d'une LSP est une prédiction très importante mais sa contribution totale à la
densité de matière noire dépend des paramètres du MSSM.   

Les particules candidates sont le neutralino le plus léger qui est de spin 1/2 
et le gravitino qui est de spin 3/2. Le sneutrino le plus léger a déjà été 
exclu par les recherches directes au
LEP. Toutes sont des particules neutres car si elles étaient chargées, elles 
auraient dû être détectées par les mesures d'isotopes lourds anormaux 
car une LSP aurait pu se lier au noyau. En
effet, si la LSP est chargée électriquement ou colorée, le nombre de ces 
isotopes lourds anormaux par rapport au nombre d'isotopes normaux devrait 
être supérieur à $10^{-6}$. Expérimentalement, ce rapport est inférieur à
$10^{-15}$ voir même $10^{-30}$ !
Pour l'instant, même si le neutralino est plus "à la mode", la nature de la LSP
diffère grandement selon les scénario et les paramètres de la supersymétrie.
 
\subsection{Les variantes du modèle minimal}

Le MSSM n'est pas le seul modèle qui a été construit et ce n'est bien sûr pas le
seul à pouvoir être réalisé dans la nature. Dans les analyses et études de la
phénoménologie au-delà du Modèle Standard, on parle souvent de violation de la
$R$-parité et du NMSSM. Nous allons les présenter 
brièvement dans cette section.

\subsubsection{La violation de la $R$-parité}

\begin{figure}[htbp!]
\begin{center}
\begin{tabular}{c c}
\mbox{\epsfig{file=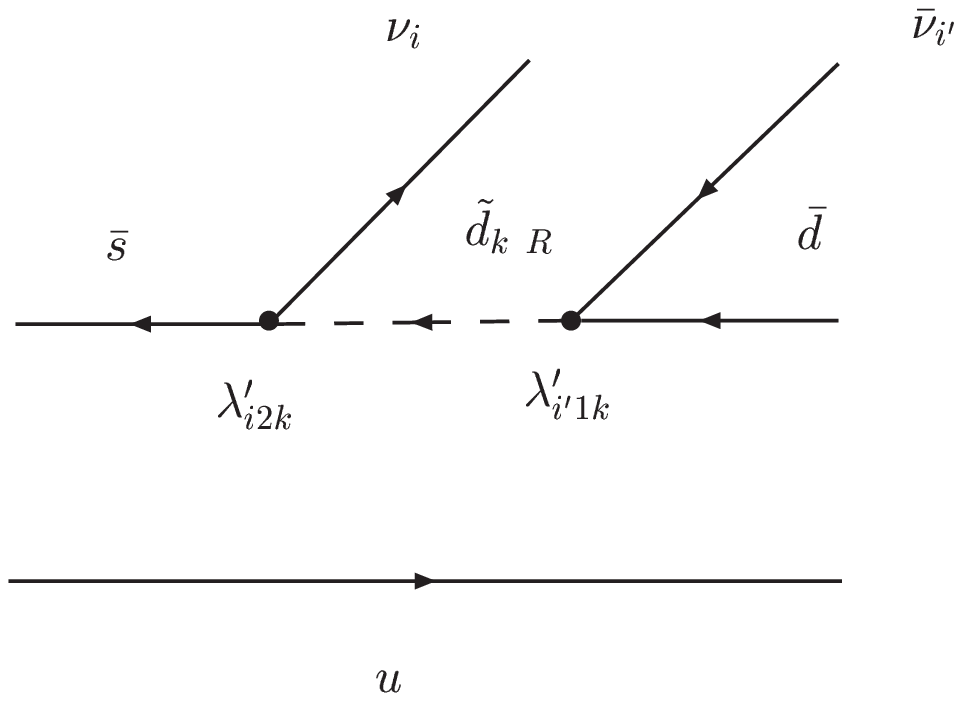,width=7cm}} &
\mbox{\epsfig{file=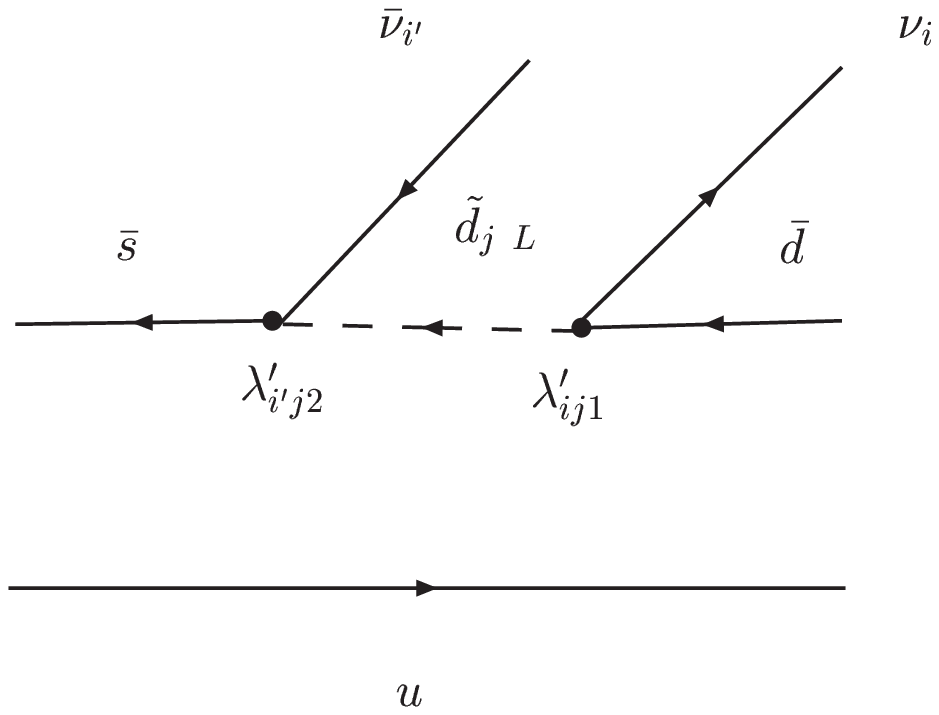,width=7cm}} \\
\end{tabular}
\end{center}

\caption{Les deux contributions possibles, au niveau de l'arbre, et qui violent
la $R$-parité, au processus rare $K^+\to\pi^+\nu\bar{\nu}$~\cite{Kpinunubar}.
 Dans les deux cas on a échange d'un squark de type "down".}

\label{kpinunu}
\end{figure}

Dans le cas ou la $R$-parité est violée, les termes de 
$\mathcal{W}_{RPV}$, eq (\ref{WRPV}), donneront des contributions à tous les 
processus. Par exemple, $b\to s\, \gamma$, les oscillations $B-\bar{B}$  ou la 
désintégration rare $K^+\to\pi^+\nu\bar{\nu}$ (voir
figure~\ref{kpinunu}). 

Les paramètres $\lambda$ sont en tout au nombre de 45 (9+27+9) et les bornes
supérieures sont typiquement de l'ordre de 
$(10^{-2}-10^{-1})\times m_{\tilde{q}}/(100$ GeV). Ils sont cependant 
difficiles à contraindre car
ils apparaissent sous forme de produits, $|\lambda_{ijk}\lambda'^*_{ijk}|$ par
exemple, et nous ne
pouvons les isoler que sous certaines hypothèses~\footnote{Par exemple, supposer 
qu'il existe
une hiérarchie entre les couplages qui violent $B$ ou $L$ et une hiérarchie 
selon les différentes générations de quarks et leptons. Ceci semble raisonnable 
car les limites
obtenues à partir des données expérimentales suggèrent qu'il ne peut y avoir
violation de $B$ et $L$ simultanément.}. La 
phénoménologie de la 
violation et de la conservation de la $R$-parité sont très différentes et très 
riches, leur recherche est donc déjà en cours au Tevatron et aussi prévue aux 
futurs collisionneurs. 

\subsubsection{Le NMSSM}

Le NMSSM ("next-to-minimal-supersymmetric-standard-model") est la plus 
simple 
extension du MSSM. Dans ce modèle, seul le contenu
en particules est modifié car on ajoute un nouveau supermultiplet chiral $S$,
singulet de jauge. Le superpotentiel est alors :
\beq \mathcal{W}_{NMSSM}=\frac{1}{6}k S^3 +\frac{1}{2}\mu_S S^2 + \lambda S H_u
H_d +\mathcal{W}_{MSSM} \label{NMSSM}.\eeq 
Le supermultiplet $S$ contient à la fois un fermion chiral et son partenaire scalaire. Le principal interêt du NMSSM est de proposer une solution au problème
"$\mu$". En effet, en supposant que la partie scalaire de $S$ développe une 
valeur dans le vide non-nulle $\langle S \rangle$, le terme 
dans~(\ref{NMSSM}) donne un $\mu$
effectif :
$\mu_{eff}=\lambda \langle S \rangle$. Or $S$ apparait aussi dans 
$\mathcal{L}_{soft}$ et sa
\textit{v.e.v}
est naturellement de l'ordre de $m_{soft}$, $\mathcal{O}(1)$ TeV, la masse typique des scalaires et
jauginos. Ainsi, la valeur effective de $\mu$
est de l'ordre de $m_{soft}$ plutôt que d'être un paramètre libre et 
indépendant de la brisure de supersymétrie. 

Phénoménologiquement le NMSSM diffère du MSSM parce qu'il permet un boson de
Higgs léger plus lourd. De plus, le scalaire $S$ peut {\it a priori} se mélanger
avec les autres scalaires du MSSM et former 5 états de neutralinos. Ainsi, les
signatures expérimentales des sparticules peuvent changer radicalement.

\section{Les premières contraintes expérimentales}

Comme nous avons vu, le secteur de la brisure douce du MSSM compte une centaine de paramètres.
L'analyse des données et l'extraction de valeurs expérimentales pour ces
paramètres est alors difficile. Une hypothèse simplificatrice, à la base du
modèle CMSSM (constrained-minimal-supersymmetric-standard-model), est de
supposer l'universalité à une certaine échelle : 

$\bullet$ des masses de tous les jauginos : $M_3=M_2=M_1=m_{1/2}$,

$\bullet$ des masses des scalaires : $m_Q^2=m_{\bar{u}}^2=...=m_0^2 \mathbf{1}$
et $m_{H_u}^2=m_{H_d}^2=m_0^2$,

$\bullet$ et que les couplages trilinéaires soient reliés par un paramètre
universel $A_0$
: $a_u=A_0y_u$, $a_d=A_0y_d$, $a_e=A_0y_e$.

\noindent Ainsi, le passage du MSSM au CMSSM fait passer de plus d'une centaine
de paramètres à seulement 5 paramètres ! 
Cette hypothèse est très pratique d'un point de vue phénoménologique, bien que
discutable d'un point de vue purement théorique. Le CMSSM et la simplification
de $\mathcal{L}_{soft}$ sont en fait inspirés des modèles de supergravité où la
brisure de supersymétrie se fait par la médiation de la gravité (chapitre 4).

Pour ces raisons nous nous placerons dans ce modèle jusqu'à la fin du chapitre
pour discuter des contraintes expérimentales.

\subsection{Les accélérateurs}

Les premières contraintes expérimentales sur les modèles supersymétriques viennent du fait 
qu'aucune sparticule 
n'a été découverte directement, ni au LEP ni au Tevatron. Ceci implique que les 
masses des superpartenaires soient supérieures à une centaine de GeV environ, selon
les cas. 

Les contraintes sur les sparticules sont aussi obtenues {\it via} la limite 
expérimentale sur la masse du Higgs.
\beq m_H\geq114\ {\mathrm GeV}. \eeq 
En effet, la masse du Higgs est très 
sensible à la masse des particules (celle du stop et du top surtout) 
circulant dans les boucles, comme nous avons vu précédemment.

Ces deux types de contraintes sont résumées sur la
figure~(\ref{paraspacewmap}).  La limite sur $m_H$ donne une limite
sur la masse des jauginos $m_{1/2}$ qui entre dans le calcul de la masse
du stop.

De plus, les désintégrations $b\to s\,\gamma$ ont un taux de branchement
en accord avec le MS. Les boucles de sparticules (de charginos et de Higgs
chargés principalement) ne doivent donc contribuer que très peu, ce qui contraint 
aussi leurs masses. Cette contrainte venant de $b\to s,\,\gamma$ est aussi tracée
sur la figure~(\ref{paraspacewmap}). 

Enfin, le moment magnétique anormal du muon ($g_{\mu}-2$) peut être utilisé comme 
autre résultat expérimental contraignant l'espace des paramètres, et cette
contrainte a été reportée dans la figure~(\ref{paraspacewmap}).
Mais l'observation d'une réelle déviation de ($g_{\mu}-2$) par rapport à la 
prédiction du MS est sujette à débats actuellement dans la communauté des physiciens, la
contrainte précédente est donc à prendre avec prudence.

\subsection{La cosmologie}

\begin{figure}[htbp!]

\begin{center}\begin{tabular}{c c}
\mbox{\epsfig{file=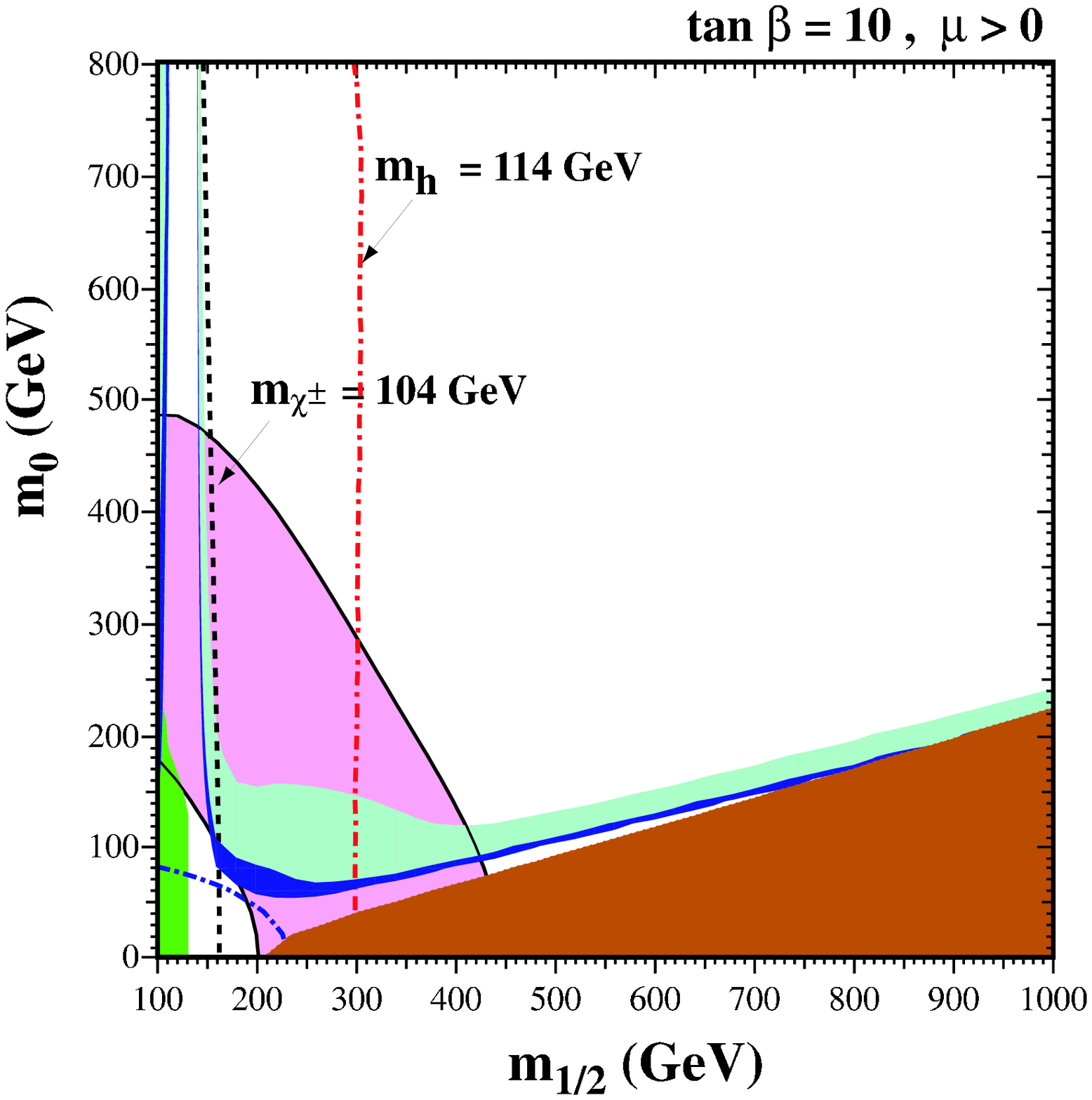,height=7.5cm,width=7.5cm}} &
\mbox{\epsfig{file=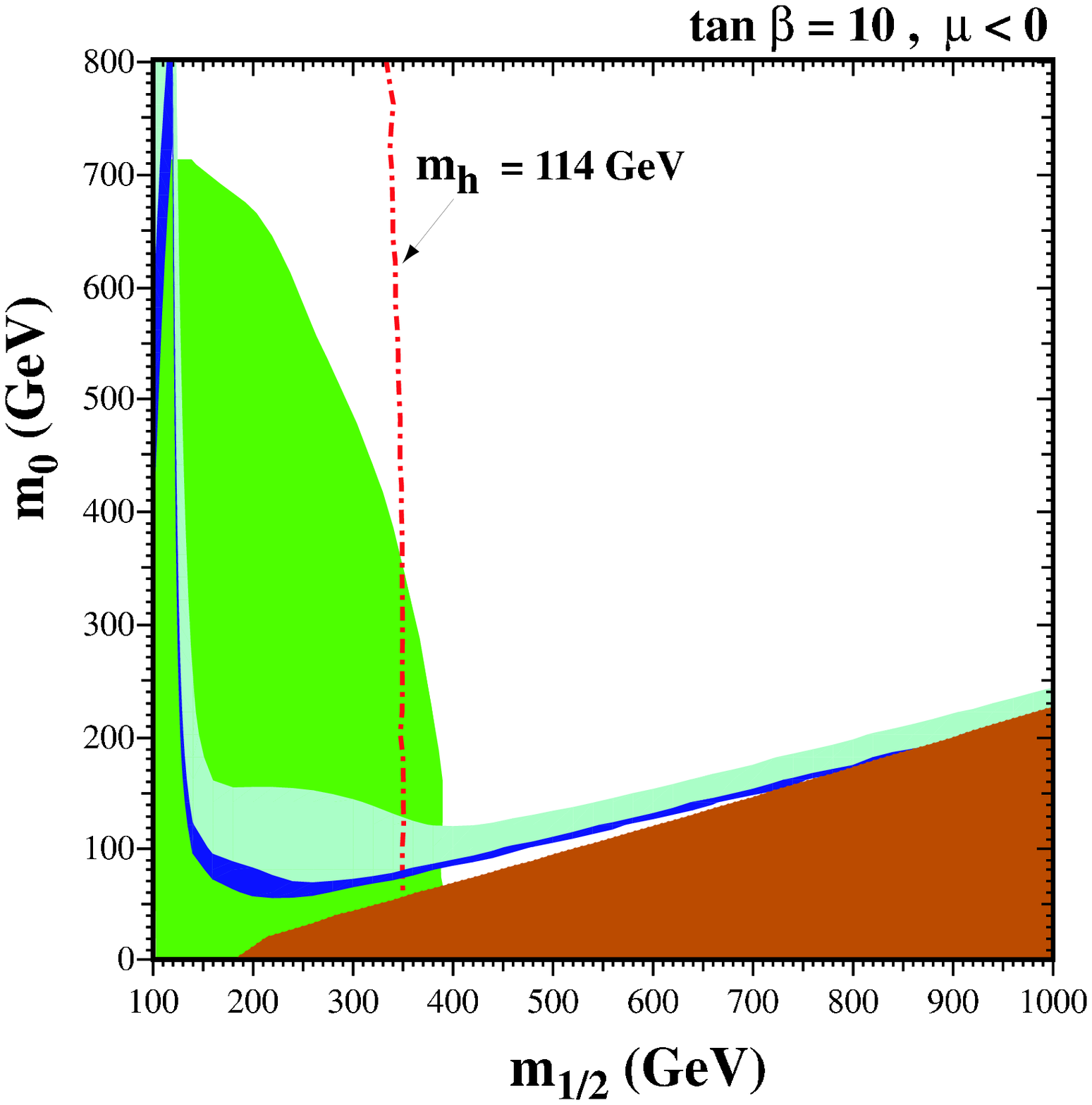,height=7.5cm,width=7.5cm}} \\
\mbox{\epsfig{file=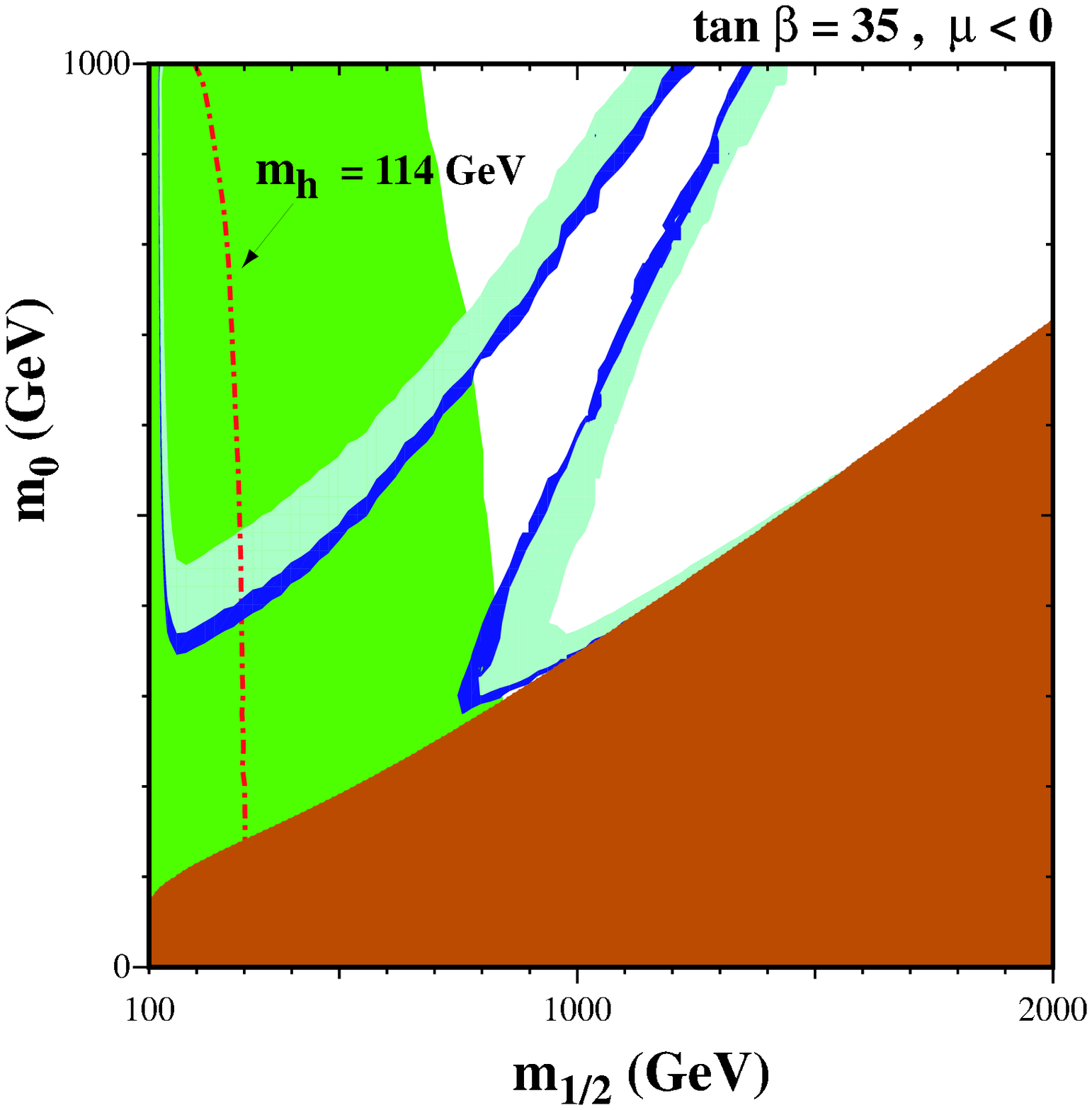,height=7.5cm,width=7.5cm}} &
\mbox{\epsfig{file=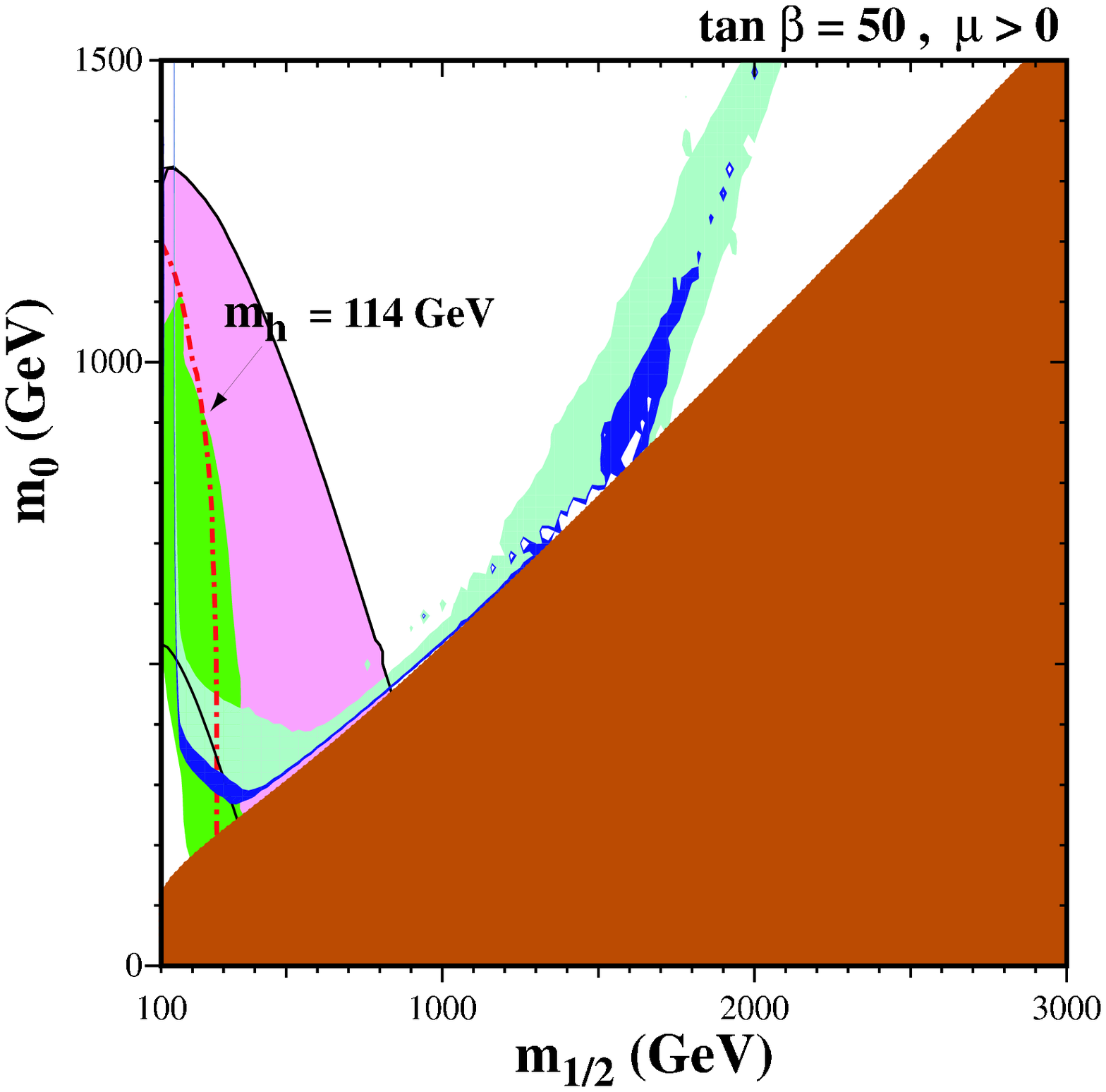,height=7.5cm,width=7.5cm}} \\
\end{tabular}
\end{center}

\caption{Contraintes expérimentales appliquées au plan ($m_{1/2},m_0$) pour
différentes valeurs de $\tan\beta$, et les 2 signes de $\mu$.
La région à gauche des tirets noirs est
exclue par la contrainte du LEP sur $m_{\chi^{\pm}}$ et la région à gauche des
tirets rouges est exclue par la recherche directe du boson de Higgs au LEP.
La région exclue par les mesures de $b\to s,\,\gamma$ est en vert
c'est-à-dire la large région à gauche.
La région rose en bas à gauche correspond à la région permise par $g_{\mu}-2$
et les fines bandes foncées à celle par WMAP.
La région foncée en bas à droite est interdite parce que l\`a la LSP a une charge
électrique.~\cite{BDEGOP}}

\label{paraspacewmap}

\end{figure}

La cosmologie joue un rôle de plus en plus important en physique des particules.
Quand la $R$-parité est conservée, la supersymétrie apporte un candidat, la LSP,
pour la matière noire froide. Pour respecter les mesures cosmologiques de WMAP
sur la densité totale de matière noire $\Omega_{MN}h^2$, la contribution de la LSP 
$\Omega_{LSP}h^2 $ ne peut pas excéder 0.129~\cite{BDEGOP}.
Les figures~(\ref{paraspacewmap}) incluent aussi les régions permises de la densité 
de matière noire d'origine supersymétrique par WMAP. De plus, 
la LSP ne doit pas être chargée, et cette condition est aussi prise en compte,
la région triangulaire foncée est la région exclue.

\section{La détection de la supersymétrie aux collisionneurs}

\subsection{Les benchmarks}

\begin{figure}[htbp!]
\begin{center}
\begin{tabular}{c c}
\mbox{\epsfig{file=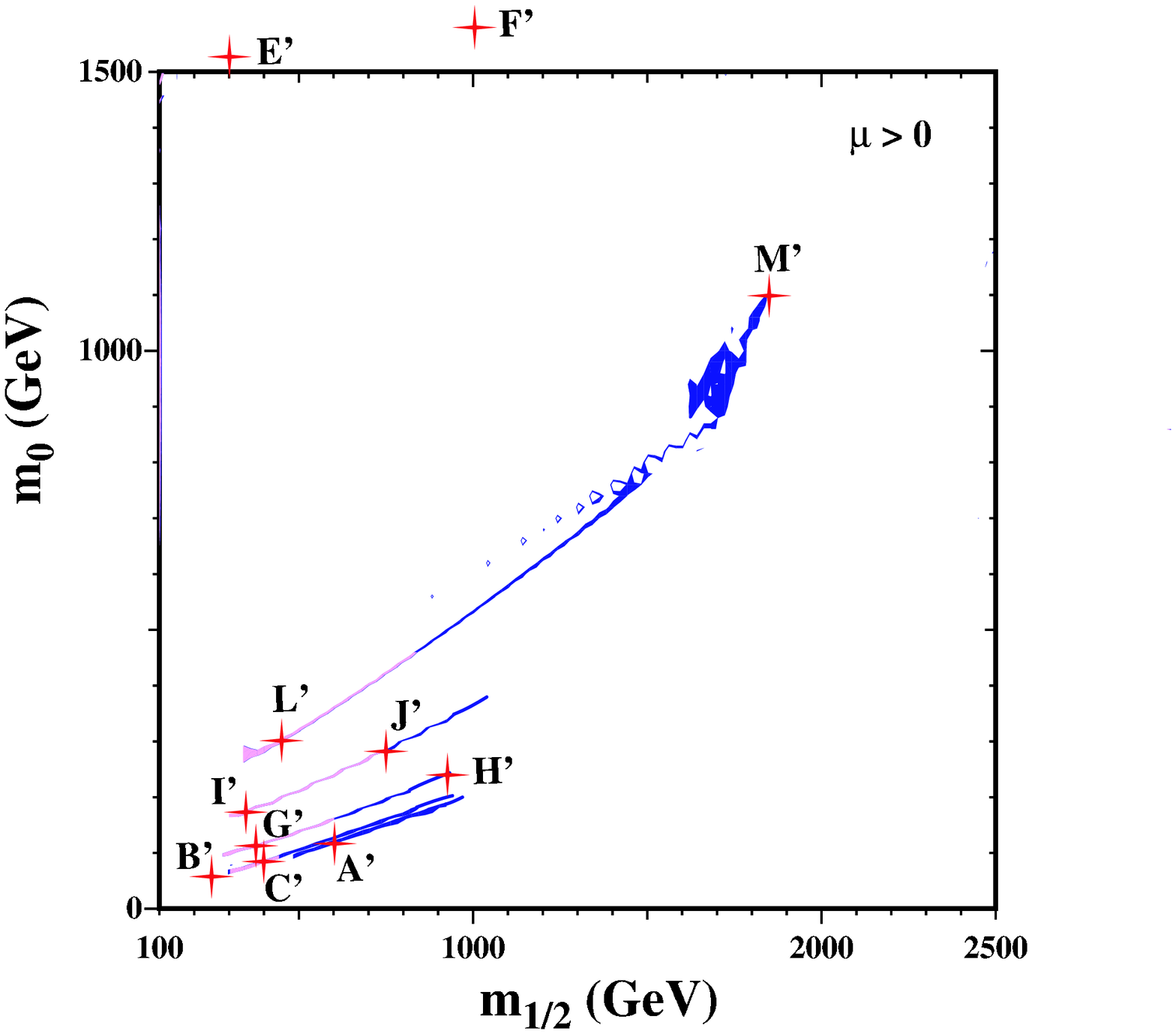,height=7.5cm}} &
\mbox{\epsfig{file=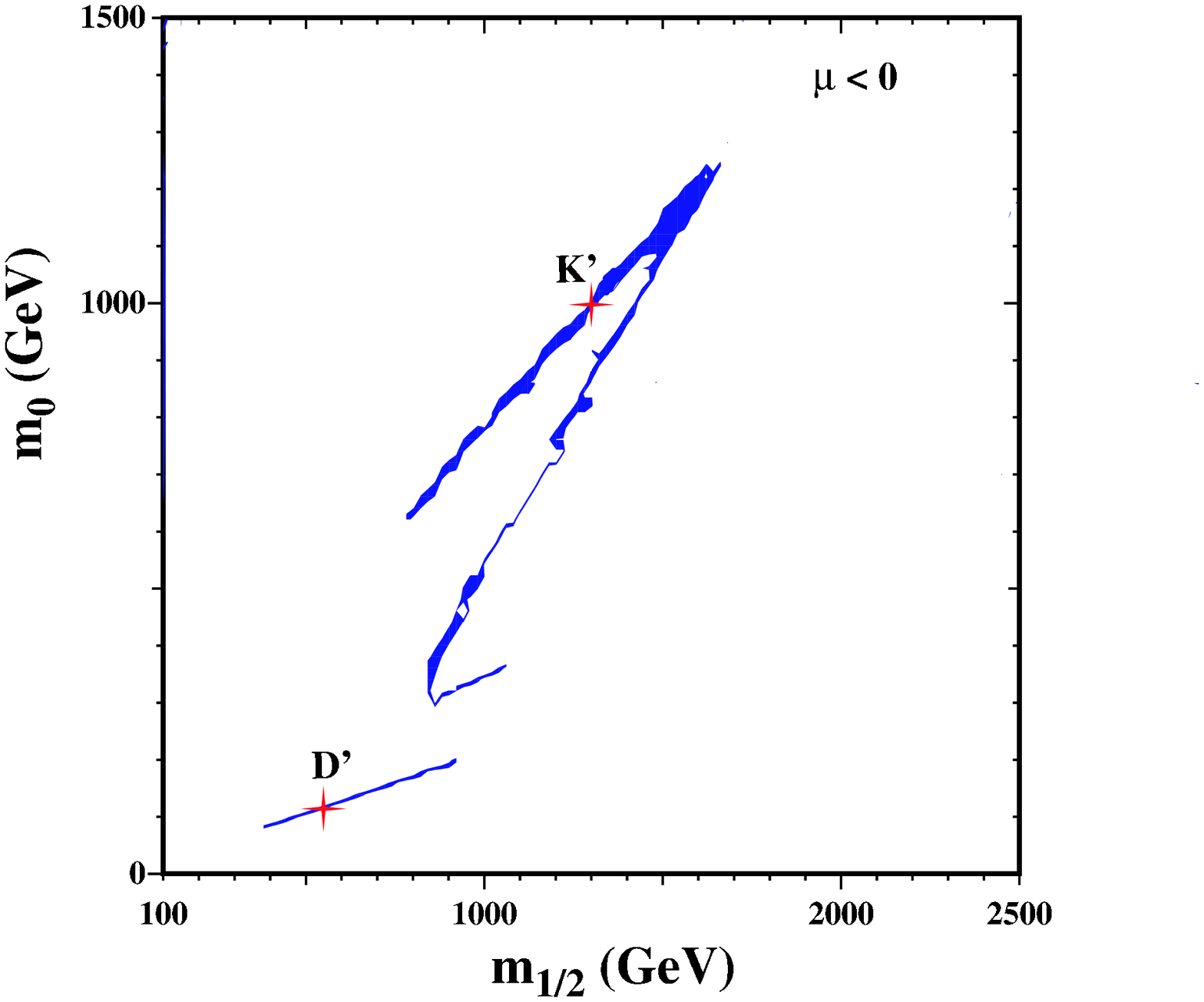,height=7.5cm}} \\
\end{tabular}
\end{center}

\caption{Lignes compatibles avec les résultats de WMAP, dans le plan
($m_{1/2},m_0$). Dans la première figure  $\mu>0$ et $\tan\beta=$5, 10, 20, 35, 50,
et dans la seconde, $\mu<0$ et $\tan\beta=$10, 35. Les courbes à $\tan\beta$
élevé sont les plus hautes.~\cite{BDEGOP}}

\label{benchWMAPstrips}
\end{figure}

Pour étudier les possibilités de découverte de la supersymétrie aux futurs
collisionneurs, il est très utile de faire appel à des "benchmarks", c'est-à-dire
à des modèles tests dont les paramètres sont fixés. Ceux ci
servent de repères et permettent de focaliser la discussion. Ces points de
l'espace des paramètres doivent être bien choisis de manière à illustrer les
différentes possibilités. Les régions permises par WMAP et les autres mesures
tracent des sortes de bandes fines dans l'espace des paramètres du CMSSM. La
supersymétrie devrait donc se trouver sur un point de l'une de ces bandes. Une
dizaine de points significatifs ont été choisi, nommés par une lettre de
l'alphabet (figure~(\ref{benchWMAPstrips})). Certains points correspondent à de petites
masses de sparticules, ce qui est donc favorable à une détection directe,
d'autres sont éparpillés sur les lignes de co-annihilation et 2 points ont été
choisis dans des régions où les LSP peuvent s'annihiler rapidement (points K et
M). 2 points, les "points focus" (E et F) sont des points très sensibles à la masse du top.

\subsection{Les perspectives au LHC et au LC}

\begin{figure}[htbp!]
\begin{center}
\mbox{\epsfig{file=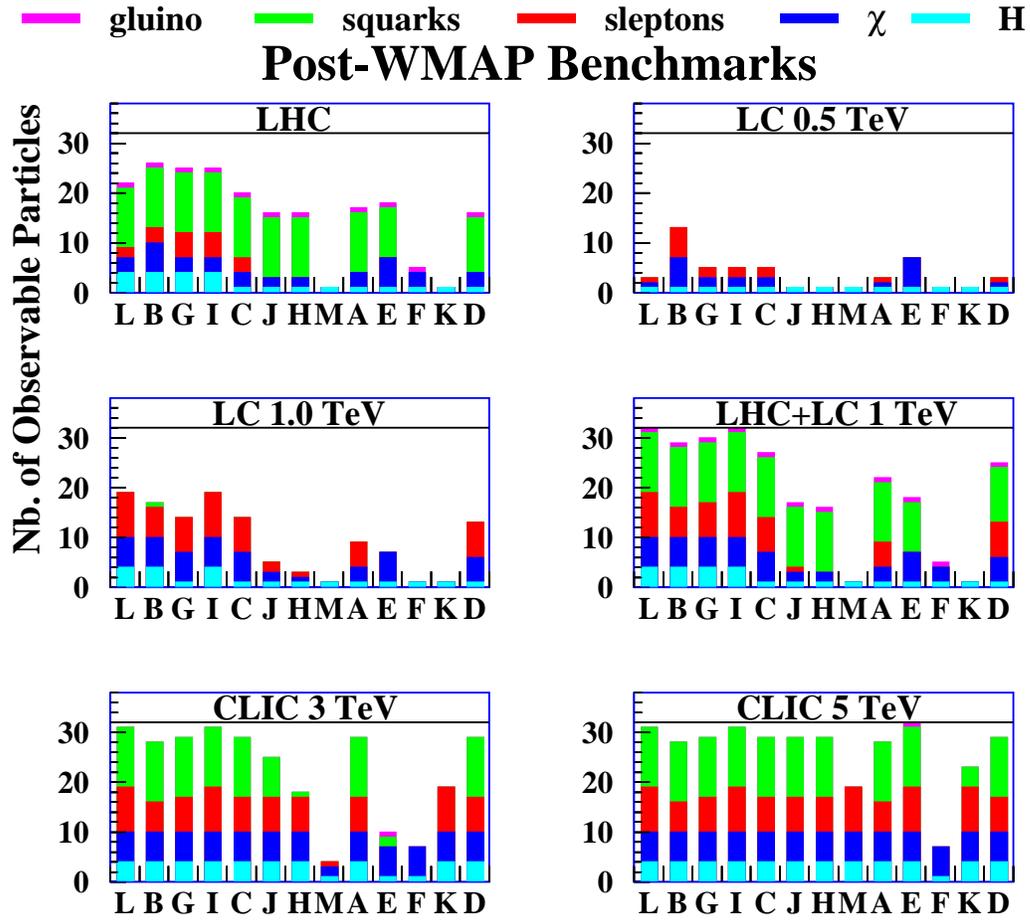,width=14.0cm}}
\end{center}

\caption{Nombre de types de particules du MSSM qui peuvent être détectables par les divers
accélerateurs, en fonction des différents benchmarks.~\cite{BDEGOP}}

\label{benchWMAPresults}
\end{figure}

Bien entendu, la possibilité de découverte dépend des caractéristiques du
collisionneur. Sachant que les masses des sparticules peuvent aller de plusieurs
centaines de GeV à quelques TeV, le Tevatron n'a qu'une mince chance de découvrir la
supersymétrie. En revanche, le LHC et le LC ont toutes les chances de la
découvrir car leurs caractéristiques permettent de couvrir une large région de
l'espace des paramètres. Les résultats simulés des benchmarks sont donnés sur la
figure~(\ref{benchWMAPresults}) qui indique le type de particules que nous pouvons
espérer
observer avec le LHC, le LC et la combinaison LHC+LC ou un collisionneur linéaire plus
puissant, le CLIC. Au LHC, la supersymétrie semble donc "facile" à voir dans la
plupart des benchmarks. Dans ce cas, le nombre important de désintégrations en
cascade permettra d'observer plusieurs sparticules à la fois. Les mesures d'états
finaux supersymétriques permettront d'obtenir les masses des sparticules visibles
avec une précision d'environ $10\%$. De plus, l'observation d'un
grand nombre d'événements avec beaucoup d'énergie transverse manquante pourra
indiquer la présence de supersymétrie avec conservation de la $R$-parité. Toutes
ces mesures du spectre donneront accés aux paramètres fondamentaux des modèles
comme le CMSSM qui pourront être utilisés pour calculer certaines des
observables de basse énergie.

\subsubsection{Exemple d'utilisation des résultats}

\begin{figure}[htbp!]
\begin{center}
\mbox{\epsfig{file=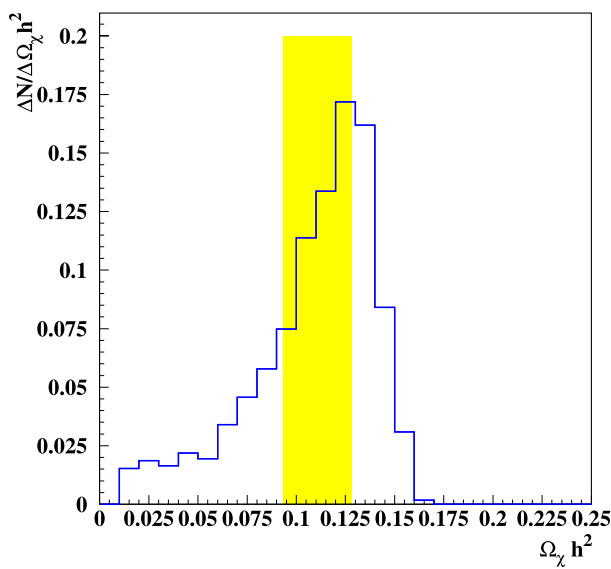,width=8.0cm}}
\end{center}

\caption{Précision estimée à laquelle la densité $\Omega_{LSP}h^2$ (ici notée
$\Omega_{\chi}h^2$) peut être
prédite grâce aux résultats du LHC dans le cas du benchmark "B".~\cite{BDEGOP}}

\label{CDM}
\end{figure}

Dans les cas les plus favorables, on pourra estimer les paramètres du MSSM au LHC
(et peut-être aussi avec un collisionneur linéaire ?) et
calculer par exemple la densité de matière noire. Juste pour donner un exemple,
le spectre de masse suivant, correspondant au benchmark "B", pourra être
mesurable au LHC avec les précisions indiquées~\cite{BDEGOP} (les masses sont en
GeV) :
\beqn
&m_{\tilde{g}}= 595.1 \pm 8.0 \nonumber\\
&m_{\tilde{q}_L}=540.3 \pm 8.8,\  m_{\tilde{q}_R}=520.4 \pm 11.8 \nonumber\\
&m_{\tilde{b}_1}=491.9 \pm 7.5,\  m_{\tilde{b}_2}=524.5 \pm 7.9 \nonumber\\
&m_{\tilde{l}_L}=202.3 \pm 5.0,\  m_{\tilde{l}_R}=143.1 \pm 4.8,\
m_{\tilde{\tau}_1}=132.5 \pm 6.3 \nonumber\\
&m_{\tilde{\chi}}=96.2 \pm 4.8,\ m_{\tilde{\chi}_2}=176.9 \pm 4.7,\
m_{\tilde{\chi}_4}=377.9 \pm 5.1.
\eeqn
En supposant le signe de $\mu$ connu, en posant~\footnote{$A_0$ n'a de toute
façon que peu d'impact sur
$\Omega_{LSP}h^2$.} $A_0=0$ on trouve :
\beq m_0=(103\pm 8)\ \mathrm{GeV},\ \ m_{1/2}=(240\pm  3)\ \mathrm{GeV},\ \
\tan\beta=10.8\pm 2. \eeq
Ce qui donne (voir figure~\ref{CDM}) :
\beq \Omega_{LSP}h^2=0.11^{+0.02}_{-0.03}. \eeq
Dans le calcul de la densité de LSP, la limite expérimentale de WMAP ne
sera pas forcément saturée. Il pourra toujours y avoir d'autres contributions
non négligeables à $\Omega_{MN}h^2$ qu'il nous faudra trouver. De plus, le CMSSM
n'est pas forcément le modèle le plus réaliste. Cependant, c'est un des modèles
les plus simples pour étudier la physique au-delà du MS mesurable dans les
collisionneurs à venir.

\section{Conclusions}

La supersymétrie est pour l'instant hypothétique. Seulement, beaucoup
d'arguments sont en faveur de son existence. Ces arguments sont essentiellement
théoriques mais la supersymétrie est aussi \underline{suggérée
expérimentalement}. D'un côté elle résout le problème de la hiérarchie des
échelles, et est essentielle pour les théories des cordes. De l'autre, la
supersymétrie ajoutée au MS réalise à partir des données du LEP l'unification des couplages de jauge à haute
énergie. Elle favorise aussi un boson de Higgs léger qui est préféré
expérimentalement. De plus, sa phénoménologie est très riche : elle introduit
une foule de nouvelles particules avec lesquelles de nouvelles interactions
sont possibles. Elle
contribue aux processus rares violant la saveur leptonique et la symétrie
$CP$. Le concept de $R$-parité induit aussi une riche phénoménologie comme par
exemple une candidate pour la matière noire (la LSP).
Tout ceci l'a rendue extrèmement populaire. Les considérations de stabilité de
la masse du Higgs envers les corrections radiatives suggèrent des masses de
l'ordre du TeV pour les spartenaires. Ainsi, elle est très attendue au LHC !

La supersymétrie séduit car elle permet d'unifier les bosons et les fermions.
C'est aussi la dernière symétrie de l'espace-temps à 4D à ne pas être (encore)
observée dans la nature, et elle peut inclure naturellement la gravitation (théories de
supergravité, voir chapite 4).
Cependant, il est vrai qu'elle introduit beaucoup de nouvelles particules, dont
des scalaires et ceux-ci n'ont encore pas été observés dans la nature. De plus,
la supersymétrie seule n'explique pas tout : les masses, et la physique des saveurs
d'une manière générale, ne sont pas expliquées. Peut être faudrait-il ajouter
une Grande Unification, des dimensions supplémentaires ?
Il reste encore beaucoup de choses à étudier dans la supersymétrie, et
notament sa \underline{brisure}. Expérimentalement aussi, la découverte de la
supersymétrie, du modèle (MSSM, NMSSM,...) et de tous ses paramètres est un
défi pour les années à venir. Dans quelques années nous serons enfin
capable de tester l'idée de la $R$-parité, de la LSP, nous pourrons voir les
connexions entre la supersymétrie et la cosmologie mais surtout nous devrions
enfin connaître le mécanisme et l'origine de la brisure électrofaible.


%
\chapter{Physique des neutrinos {\it et al.}}                                  %
%

Nous avons déjà discuté des neutrinos et de leur impact sur le Modèle Standard.
La physique des neutrinos est riche et les conséquences du mélange des
neutrinos sont nombreuses. Ce chapitre aborde donc des sujets très divers mais
tous liés : les oscillations de neutrinos, la violation de $CP$ leptonique,
la violation des nombres leptoniques, la leptogenèse et l'inflation
cosmologique. Nous verrons ainsi l'importance des spécificités du secteur des
neutrinos sur la physique des particules et la cosmologie.

\section{Les Masses et oscillations des neutrinos}

\subsection{L'échelle de masse des neutrinos}

Dans le Modèle Standard, nous connaissons maintenant plus ou moins
précisemment les masses de tous les quarks, de tous les leptons chargés mais pas
du tout celles des neutrinos. A vrai dire, on a même longtemps cru qu'ils n'avaient
pas de masse. De plus, jusqu'à présent nous n'avons que des limites
supérieures sur les masses des 3 neutrinos.
De la désintégration $\beta$ du Tritium nous savons que~\cite{JE-NU} :
\beq m_{\nu_e} \leq 2.5\ \mathrm{eV}. \eeq
L'expérience KATRIN devrait pouvoir sonder des masses de l'ordre de 0.5 eV.
Ensuite, des mesures de la désintégration $\pi\to\nu\mu$, nous avons :
\beq m_{\nu_{\mu}} \leq 190\ \mathrm{KeV} \eeq
et il y a des projets pour baisser cette limite d'un facteur $\sim 20$.
Finalement, des mesures de $\tau\to n\pi\nu$, nous avons :
\beq m_{\nu_{\mu}} \leq 18.2\ \mathrm{MeV} \eeq
qui devrait dans le futur être améliorée jusqu'à $\sim$5 MeV.
Les masses des neutrinos sont donc plus faibles que celles de la plupart des 
fermions du MS. 
Mais les limites astrophysiques sont beaucoup plus contraignantes que ces limites
obtenues en laboratoire. Les données extraites de l'observation des
structures à grande échelle dans l'univers (galaxies, amas de galaxies,...) 
peuvent être utilisées pour obtenir une limite supérieure de 1.8 eV sur la
\underline{somme} des masses des neutrinos. Cette limite a été récemment 
améliorée par WMAP en : 
\beq \sum_{i}m_{\nu_i} < 0.7\ \mathrm{eV}. \eeq
Dans le cas où la limite est saturée et où les neutrinos sont dégénérés en
masse, celà donne une masse individuelle d'environ 0.23 eV, 2000 fois plus
petite que celle de l'électron ! Même si l'extraction de la masse des neutrinos
à partir d'observations astronomiques dépend du modèle et de notre connaissance
actuelle de la cosmologie, cette limite est robuste.

Une autre manière intéressante d'obtenir des limites sur les masses des
neutrinos  
vient de la recherche de la désintégration double-$\beta$ sans neutrinos ($\beta \beta
_{0\nu}$). Sa mesure contraint la somme des masses de Majorana des
neutrinos gauches, pondérées par leurs couplages à l'électron :
 \beq <m_{\nu}>_{ee} = |\sum_{i}m_{\nu_i} U_{ei}^2 |\leq 0.35\ \mathrm{eV} \eeq
Les futures expériences ont l'intention d'améliorer cette limite jusqu'à $\sim
0.01$ eV.
\par\hfill\par
\begin{figure}[htbp!]
\centerline{\epsfxsize=6cm\epsfbox{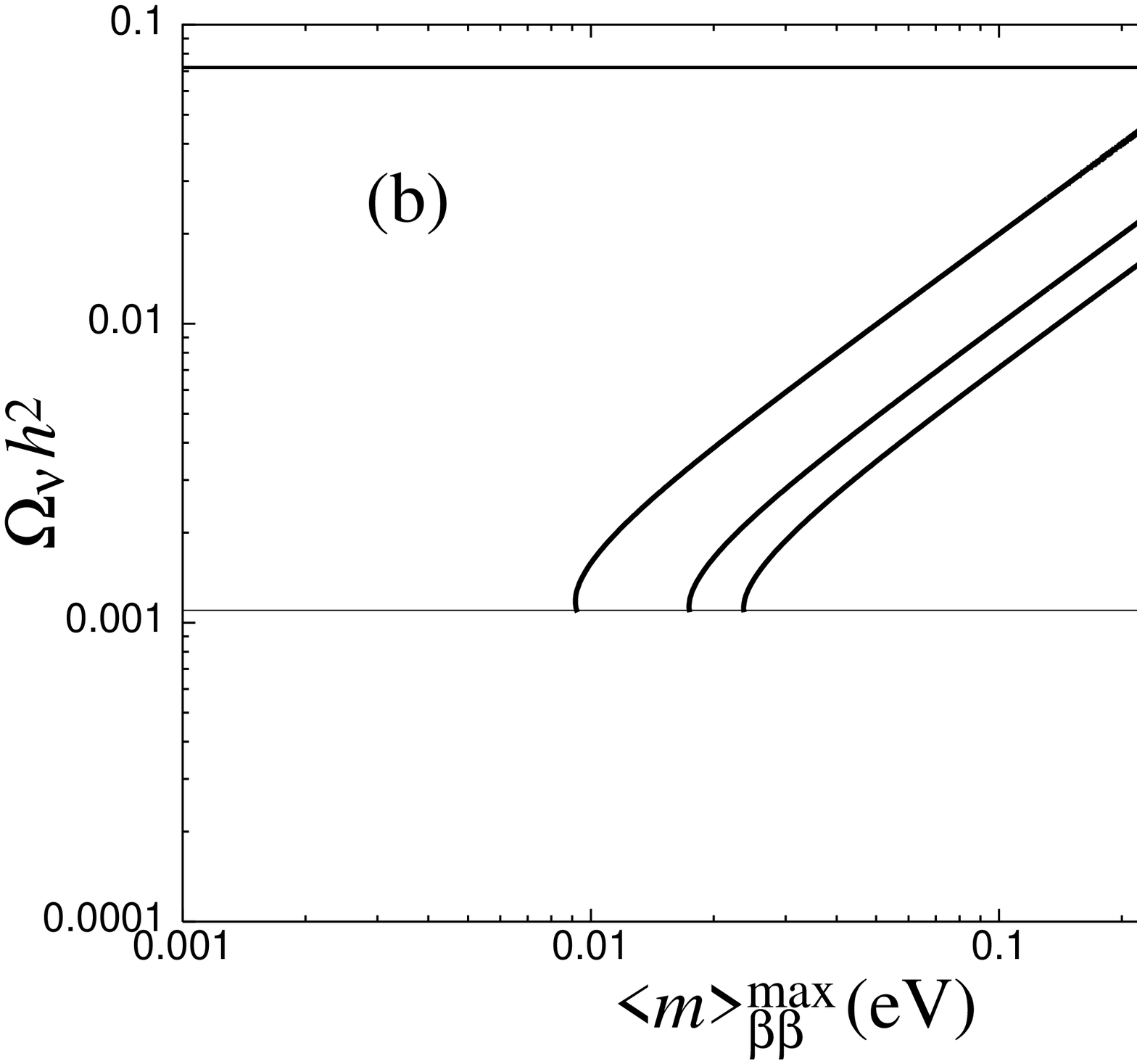}}
\caption{Corrélation entre $\Omega_{\nu} h^2$ et $<m_{\nu}>_{ee}$ en prenant en
compte l'intervalle permis par les oscillations de neutrinos.~\cite{Correlation}}
\label{WMAP02}
\end{figure}

Sur la figure~(\ref{WMAP02}) nous pouvons voir la corrélation entre $\Omega_{\nu} h^2$ 
et $<m_{\nu}>_{ee}$, les données des oscillations et de WMAP permettent aussi de
contraindre la somme des masses de Majorana des
neutrinos gauches. On a
$0.001\lesssim\Omega_{\nu} h^2\lesssim0.07$ donc on s'attend d'après la figure 
à avoir à peu près 0.01 eV $\lesssim <m_{\nu}>_{ee} \lesssim$ 1 eV.

\subsection{Les oscillations des neutrinos}

\subsubsection{La Matrice MNS}

\begin{figure}[htbp!]
\centerline{\epsfxsize=12cm\epsfbox{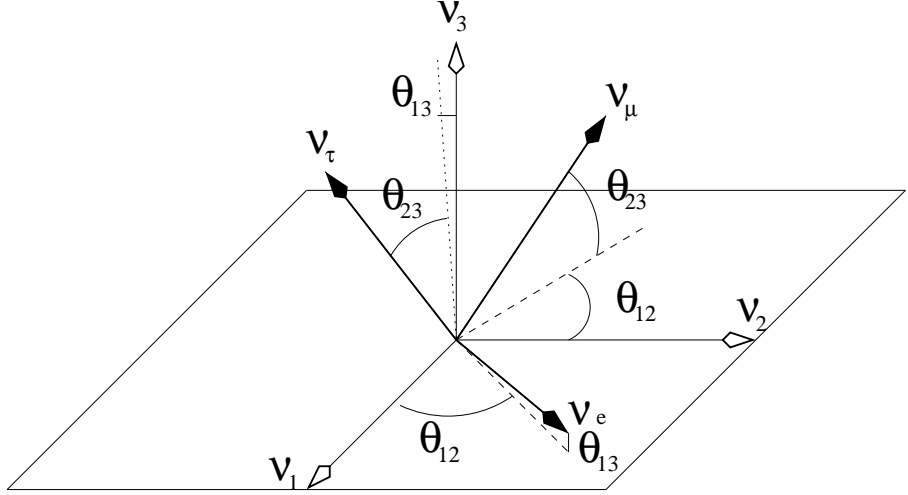}}
\caption{Relation entre les états de saveurs $\nu_e$, $\nu_{\mu}$, $\nu_{\tau}$ 
et les états propres de masse $\nu_1$, $\nu_2$, $\nu_3$ en fonction des 3 
angles d'Euler $\theta_{12}$, $\theta_{23}$, $\theta_{13}$.}
\label{angles}
\end{figure}

Nous avons évoqué au chapitre 1 le phénomène des oscillations entre les saveurs
de neutrinos. Celui-ci vient de la différence entre les états propres de masses de
ceux de saveurs. Il existe une matrice unitaire qui relie ces deux types d'états, 
la matrice MNS, telle que :
\beq 
|\nu_{\alpha}> = \sum_{i=1,2,3} \mathcal{U}_{(MNS)\, \alpha\, i}\, |\nu_{i}>. 
\eeq
$\mathcal{U}_{(MNS)}$ est la matrice de rotation dans l'espace des saveurs
$\alpha$ (voir figure~(\ref{angles})) et peut se paramétriser en fonction de 3 
angles et 3 phases :
\beq 
\mathcal{U}=
\left(\begin{array}{ccc} 1&0&0 \\ 0&c_{23}& s_{23}\\ 0&-s_{23}&c_{23}\end{array}\right)
\left(\begin{array}{ccc} c_{13} &0&s_{13}e^{-i\delta} \\ 0&1&0 \\ -s_{13}e^{-i\delta}&0& c_{13}\end{array}\right)
\left(\begin{array}{ccc} c_{12}&s_{12}&0 \\ -s_{12}& c_{12}&0 \\ 0&0&1 \end{array}\right)
\left(\begin{array}{ccc} e^{-i\phi_1}&0&0 \\ 0&e^{-i\phi_2}&0 \\ 0&0&1 \end{array}\right)
\eeq
où $c_{ij}=\cos{\theta_{ij}}$ et $s_{ij}=\sin{\theta_{ij}}$.

La première matrice est mesurable par les expériences sur les neutrinos
atmosphériques~\footnote{C'est-à-dire des neutrinos $\nu_{\mu}$ produits dans l'atmospère terrestre par
les rayons cosmiques puis détectés sur terre.} comme par exemple
SuperKamiokande (SK). L'angle de mélange de ce secteur est \`a peu 
pr\`es maximal, 
$\theta_{23} \sim 45^0$ et $\Delta m_{23}^2 \simeq 2 \times 10^{-3}$
eV$^2$~\cite{PDG04}.

La seconde matrice est encore inconnue expérimentalement. Elle est accessible
depuis les expériences basées sur la production de neutrinos par des réacteurs
nucléaires et par des accélérateurs. Jusqu'à présent, les expériences comme Chooz n'ont établi que des limites
supérieures sur $\theta_{13}$ et il n'y aucune information sur la phase 
$\delta$ de violation de $CP$ (CPV). $\theta_{13}\lesssim 15^0 $~\cite{PDG04}.

La troisième matrice est celle du secteur des neutrinos
solaires~\footnote{Ils sont appelés ainsi car ces
neutrinos sont produits dans les réactions nucléaires du coeur du Soleil et
nous parviennent sur Terre.}. Les valeurs
favorisées par SK et SNO sont $\theta_{12}\sim 32^0$ et $\Delta m_{12}^2\simeq
7 \times 10^{-5}$ eV$^2$~\cite{PDG04}.

Enfin, la dernière matrice contient les 2 phases de Majorana possibles si les
neutrinos gauches sont de type Majorana (c'est-à-dire égaux aux antineutrinos). Elles
sont, en principe, observables dans les expériences sur la désintégration 
$\beta \beta _{0\nu}$ mais pas dans les oscillations de neutrinos. D'autres
quantités observables peuvent être sensibles indirectement à ces phases, nous
en discuterons un peu plus tard.

Si, pour avoir une idée de se qui se passe physiquement, nous prenons le cas
simple à 2 saveurs, le calcul de la propagation de ces neutrinos dans le vide et 
de la probabilité de transition donnent :
\beq 
P(\nu_{\alpha} \to \nu_{\beta})= \sin^22\theta\sin^2(1.27\Delta m^2L/E).
\eeq
A des longueurs L différentes (ou des temps différents), les
probabilités d'observer un neutrino de saveur $\beta$ sont différentes. La
probabilité oscille en fonction de $\Delta m^2_{ij}=m_i^2-m_j^2$ et de l'angle 
de mélange $\theta$ entre les 2 saveurs (à $L/E$ fixé). 
\underline{Les expériences basées sur les} \underline{oscillations entre saveurs des neutrinos n'ont
donc accés qu'aux différences de masses-carrées}  \underline{ainsi qu'aux angles 
de mélanges de la matrice MNS} (et à la phase $\delta$).

\subsubsection{Les résultats expérimentaux et le bilan}

Avant SK et KamLAND, seuls des déficits dans les flux de neutrinos 
venant des sources astrophysiques avaient
été observés. Mais depuis, des oscillations ont été vues dans les graphes.
De plus, deux types d'expériences, SK/SNO et KamLAND, ont démontré que les
neutrinos solaires se comportent de la même façon que les neutrinos issus
des réacteurs, et SK et K2K ont démontré que les neutrinos atmosphériques
se comportent de la même façon que les neutrinos issus d'un accélérateur.
Même si le signal n'est pas suffisamment clair pour parler de découverte
ou de confirmation des oscillations de neutrinos, les autres explications
qui prédisent des déficits mais pas d'oscillations sont à plus de
$\sim3\sigma$ du meilleur ajustement. L'hypothèse des
oscillations semble donc très solide et vraisemblable. Les expériences
MINOS et OPERA devraient faire beaucoup mieux et dans peu de temps nous
pourrons avoir une confirmation nette des oscillations.
 
\begin{figure}[htbp!]
\centerline{\epsfxsize=12cm\epsfbox{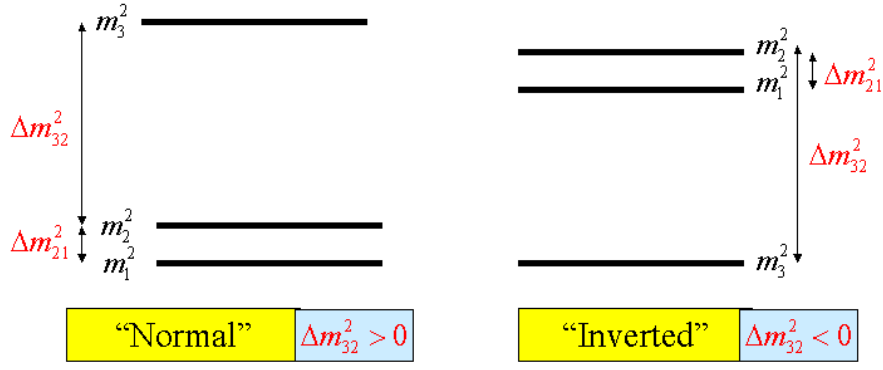}}
\caption{Hiérarchies possibles entre les 3 masses propres $m_1$, $m_2$, $m_3$
compatibles avec les 2 $\Delta m^2_{ij}$ mesurés.}
\label{hierarchie}
\end{figure}

A l'heure actuelle, nous pouvons dire être arrivé au "consensus" suivant sur
notre connaissance des neutrinos :

$\bullet$ il y a 3 neutrinos légers,

$\bullet$ leurs masses propres sont rangées dans l'ordre hiérarchique (cf
figure~(\ref{hierarchie})),

$\bullet$ le mélange est presque bimaximal :
\beq 
\mathcal{U}_{(MNS)}\simeq
\left(\begin{array}{ccc} 
\frac{1}{\sqrt{2}}&-\frac{1}{\sqrt{2}}&0 \\ 
\frac{1}{2}&\frac{1}{2}&-\frac{1}{\sqrt{2}}\\ 
\frac{1}{2}&\frac{1}{2}&\frac{1}{\sqrt{2}}
\end{array}\right)
\eeq

$\bullet$ les phases $\delta$, $\phi_{1,2}$ sont complètement inconnues,

$\bullet$ la nature de Dirac ou de Majorana des neutrinos est aussi inconnue,

$\bullet$ les moments dipolaires des neutrinos sont sans doute faibles,

$\bullet$ les temps de vie des neutrinos sont sans doute longs.

Cependant, nous pouvons nous poser la question de l'existence d'autres
neutrinos légers mais stériles~\footnote{L'expérience LSND aurait vu un 
troisième
$\Delta m^2$ complètement différent des deux autres, ce qui s'expliquerait par l'existence
d'au moins un autre neutrino qui n'aurait pas été détecté avant.}
 (sans interactions faibles), ce qui est à
confirmer par l'expérience MiniBoone. Les neutrinos peuvent aussi
avoir une hiérarchie inversée, rien n'est encore sûr de ce côté. De plus, du
côté des angles de la matrice MNS, il reste à connaître $\theta_ {13}$. J-PARC,
Double-Chooz et d'autres projets utilisant des réacteurs nucléaires devraient
dans un futur proche pouvoir diminuer la limite actuelle. Ensuite, l'angle
"solaire" $\theta_ {12}$ n'est pas vraiment maximal, il reste à savoir
pourquoi. Finalement, la violation de $CP$ dans le secteur leptonique reste à
déterminer.

\subsubsection{La violation de $CP$ leptonique}

\begin{figure}[htbp!]
\centerline{\epsfxsize=12cm\epsfbox{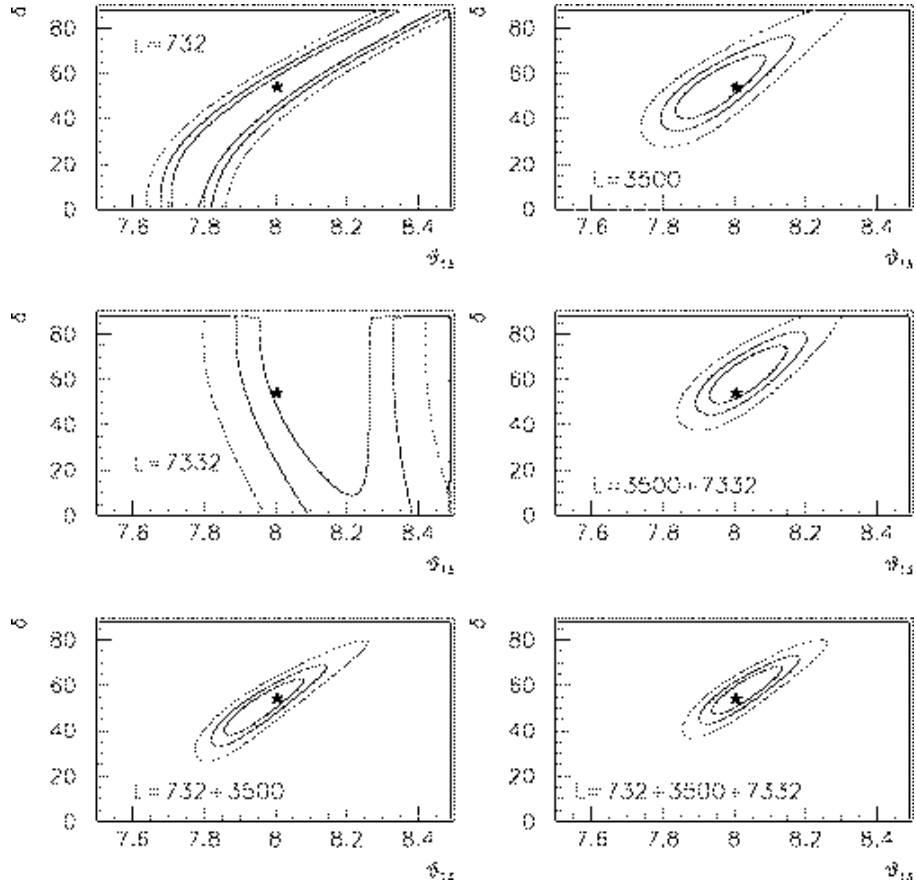}}
\caption{Ajustements combinés de $\theta_{13}$ et $\delta$ en utilisant les
informations de différentes installations situées à différentes distances de la
source.~\cite{CPfit}}
\label{CPVnu02}
\end{figure}

Avant de passer aux masses des neutrinos, nous allons encore dire quelques mots
sur la violation de $CP$ leptonique. Pour pouvoir déterminer expérimentalement
l'angle $\delta$, il est possible d'utiliser l'observable suivante :
\beqn
\mathcal{P}(\nu_e\to\nu_{\mu})-\mathcal{P}(\bar{\nu}_e\to\bar{\nu}_{\mu}) &=&
16 s_{12} c_{12} s_{13} c_{12}^2 s_{23} c_{23} \sin\delta \times \nonumber \\
&&\sin\left(\frac{\Delta m_{12}^2}{4E}L\right) \sin\left(\frac{\Delta
m_{13}^2}{4E}L\right) \sin\left(\frac{\Delta m_{23}^2}{4E}L\right)
\label{CPobs}\eeqn
c'est-à-dire la différence entre les probabilités d'oscillations de
$\nu_e\to\nu_{\mu}$ et $\bar{\nu}_e\to\bar{\nu}_{\mu}$. Mais l'extraction de
$\sin\delta$ ne sera possible que si $\theta_
{13}$ est suffisamment grand.
Si $\theta_{13}$ est vraiment faible, l'extraction de $\sin\delta$ risque d'être très
difficile à obtenir et la violation de $CP$ à établir. En revanche, la
connaissance de $\delta$ pourra entre autres nous
donner des indications indirectes pour la
leptogenèse, un des possibles mécanismes responsables de l'asymétrie
baryonique~\footnote{C'est-à-dire le fait que le nombre de baryons soit si grand devant celui
des antibaryons, {\it cf} la fin du chapitre.}.
En général, le maximum de l'asymétrie~(\ref{CPobs}) est à quelques centaines ou
 milliers de kilomètres de la source, ce qui nécessite de très longue installations
 ("Long baseline"). De plus, pour obtenir une mesure claire et sans ambigüités
possibles dans
l'interprétation des résultats, il faudra combiner les informations pour
différentes distances comme nous le montre la figure~(\ref{CPVnu02}).

\begin{figure}[htbp!]
\begin{center}
\begin{tabular}{c c}
\mbox{\epsfig{file=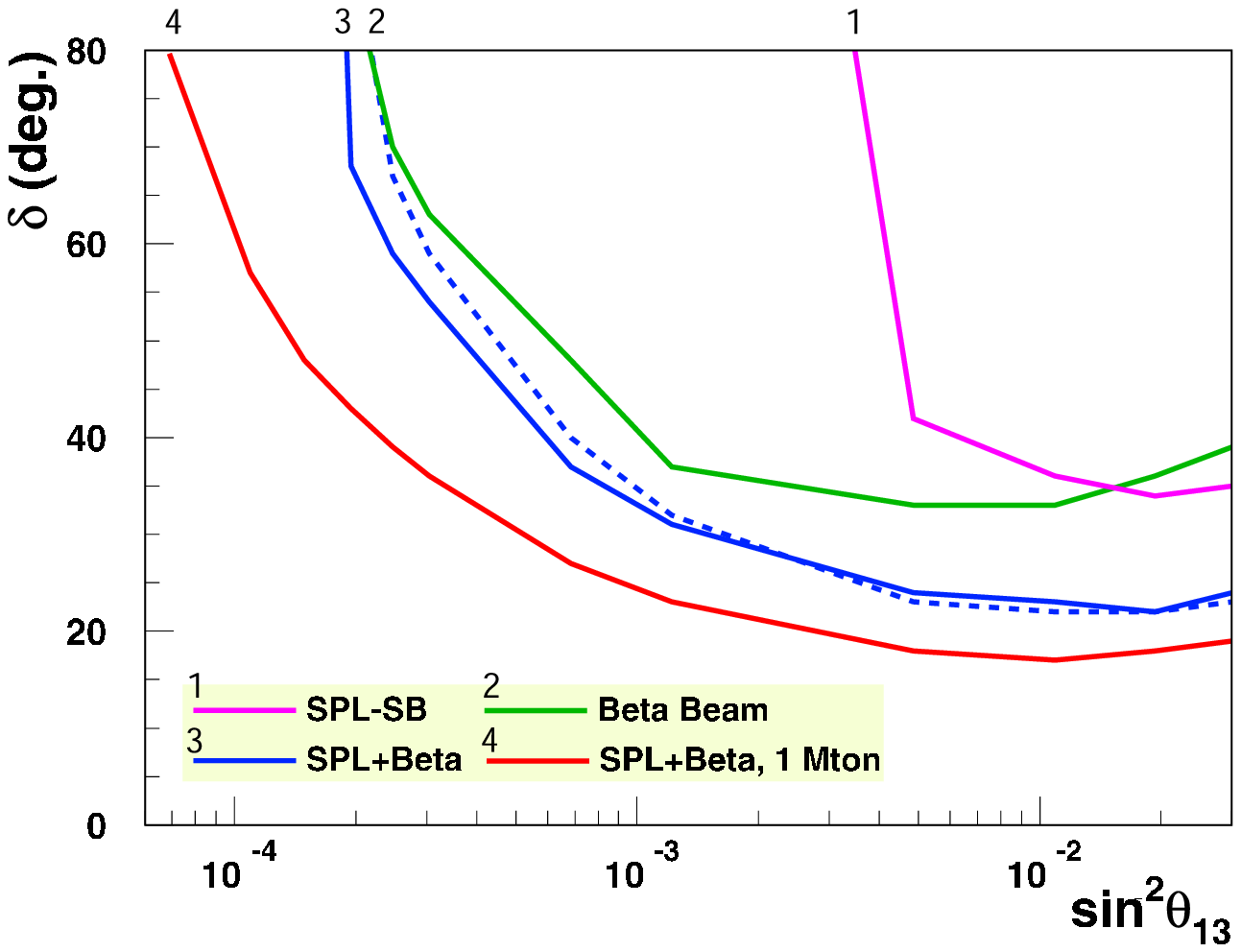,width=8.0cm}} &
\mbox{\epsfig{file=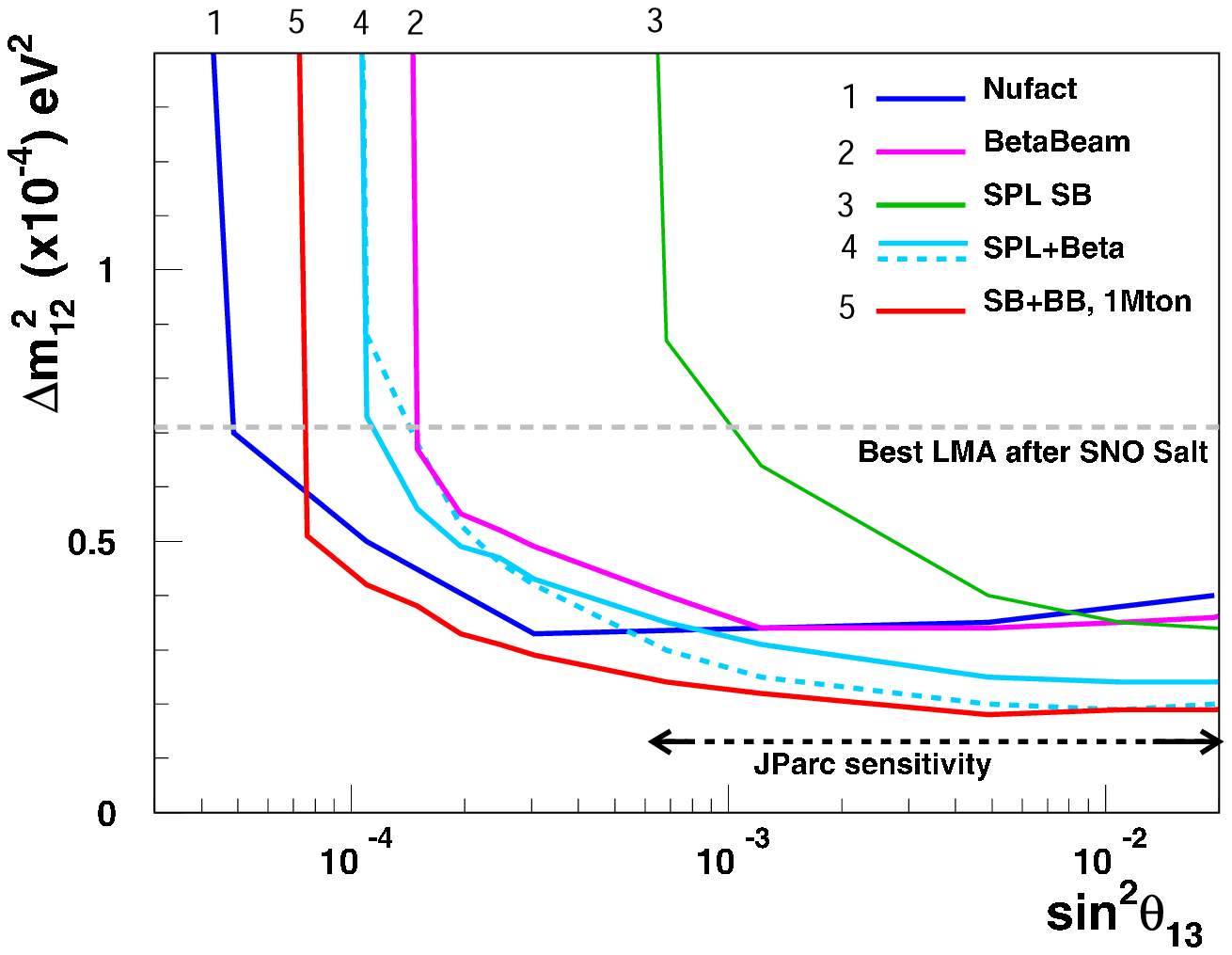,width=8.0cm}} \\
\end{tabular}
\end{center}
\caption{Simulation des limites expérimentales possibles sur $\theta_{13}$ et
$\delta$, selon différents projets.~\cite{CPex}}
\label{CPVnu0304}
\end{figure}

Avec quels types d'expériences pouvons nous obtenir les mesures de
$\theta_{13}$ et $\delta$ ?
Entre autre, et sans rentrer dans les détails, avec des
{\it usines à neutrinos} mais aussi avec des
{\it beta-beams} (faisceaux béta). Le premier projet est basé sur le
stockage de muons qui se désintègrent en $\nu_\mu$ et ${\bar \nu}_e$. Dans
le second,
les faisceaux de neutrinos ou d'anti-neutrinos sont créés
par un stockage de noyaux instables qui se désintègrent en créant un neutrino
ou anti-neutrino dans l'état final. Par exemple, $^6He: n\to p\,e^-\,\bar{\nu}_e$
et $^{18}Ne: p\to n\,e^+\,\nu_e$. Dans ce cas, le faisceau créé est très pur.
Une simulation des limites possibles sur $\Delta m^2_{12}$ et $\delta$ en fonction de
$\theta_{13}$ est donnée dans la figure~(\ref{CPVnu0304}). En particulier, l'angle
$\theta_{13}$ pourra être mesuré si $\sin^2\theta_{13}\gtrsim 10^{-4}$.

\subsection{Les masses des neutrinos}

\subsubsection{Au-delà du Modèle Standard}

Les neutrinos ont une masse, mais dans le cadre strict du MS ceci n'est pas
possible. Cependant, puisque le MS n'est qu'une
description effective à basse énergie de la physique, nous pouvons tout à fait
ajouter les termes non-renormalisables qui génèrent des masses aux
neutrinos, sans même ajouter d'autres champs. Par exemple, un terme
non-renormalisable de la forme~\footnote{C'est le terme non-renormalisable le
plus simple. Il en existe bien sûr d'autres mais ils font intervenir des
puissances de $1/M$ plus grandes et des opérateurs plus complexes.} :
\beq \mathcal{L} \supset \frac{1}{M}\nu H. \bar{\nu} \bar{H} \label{eqNR}\eeq
où M est une échelle d'énergie caractéristique de la nouvelle physique
responsable de ce terme. Ce terme, à la brisure électrofaible, va générer un
terme de masse aux neutrinos :
\beq
\sim\underbrace{\frac{\langle 0|H|0 \rangle^2}{M}}_{m_{\nu}}\nu.\bar{\nu} \\
\eeq
Or, une nouvelle interaction comme~(\ref{eqNR}) semble non naturelle et peu
fondamentale, il faut comprendre \underline{l'origine} du terme et celle de
l'échelle de masse $M$.

Numériquement, en prenant $m_{\nu}\sim\sqrt{\Delta
m_{23}^2}\sim 5.10^{-2}$ eV et $\langle 0|H|0 \rangle\sim
200$ GeV, nous trouvons que $M\sim 10^{15}$ GeV. L'échelle de masse de la
physique qui fourni le terme effectif~(\ref{eqNR}) est donc très haute,
comparable à l'échelle de Grande Unification.

\subsubsection{Le mécanisme de Seesaw}

Pour expliquer la masse faible des neutrinos à partir d'une haute échelle de
masse, nous pouvons utiliser le {\it mécanisme de Seesaw} (mécanisme de la balançoire).
C'est un mécanisme minimal, c'est-à-dire sans nouvelles interactions de jauge,
renormalisable,
et qui requiert seulement l'introduction de neutrinos lourds singulets de l'interaction
faible, donc droits. Ils sont notés $N_i$. Ces nouveaux neutrinos droits vont pouvoir se
coupler avec les neutrinos gauches {\it via } un couplage de Yukawa et conduire
à un terme de masse de Dirac à la brisure électrofaible. Cette masse de Dirac
$M_D$,
puisque intimement liée à la brisure électrofaible, est donc naturellement de
l'ordre de $m_W$. De plus, les neutrinos droits peuvent eux-même former un
terme de masse de Majorana, de masse $M$ qui peut être beaucoup plus élevée que
$m_W$. Par exemple, elle peut être de l'ordre de l'échelle de la Grande
Unification, $M\sim\mathcal{O}(M_{GUT})$. En effet, ce terme de masse
n'a rien à voir avec la brisure électrofaible donc ne nécessite pas d'y être
relié. Si nous combinons ces deux types de masse, nous obtenons la matrice de
masse dite de Seesaw :
\beq
\left(\nu_L,\ N\right)\, \underbrace{\left(\begin{array}{cc} 0 & M_D \\
M_D^{\mathrm{T}} & M \end{array}\right)}_{M_{seesaw}}
\left(\begin{array}{c} \nu_L \\ N \end{array}\right),
\eeq
ce qui nous donne, en diagonalisant, la matrice de masse \underline{effective}
des neutrinos gauches :
\beq \mathcal{M}_{\nu}=M_D \frac{1}{M} M_D^{\mathrm{T}} << M_{q,l},
\label{seesaw} \eeq
où $M_D$ et $M$ sont des matrices $3\times3$ dans l'espace des saveurs.

Afin de retrouver avec 3 masses de neutrinos non-nulles les 2 différences de
masses-carrées observées, nous avons besoin d'introduire au moins 3 neutrinos
droits $N_i$.
Même si dans ce modèle nous n'avons pas besoin d'une Grande Unification, 3
neutrinos droits apparaissent naturellement dans certains modèles ($SO(10)$ en
particulier). En résumé, nous avons donc des masses effectives de neutrinos
gauches \underline{faibles et inversement proportionnelles} \underline{aux masses de
neutrinos singulets lourds}. Le terme~(\ref{eqNR}) s'obtient par échange de ces neutrinos
lourds.

\par\hfill\par
Il existe d'autres mécanismes de Seesaw, non-minimaux, dans lesquels sont ajoutés des
champs supplémentaires. Le mécanisme précédent est souvent appelé le {\it mécanisme
de Seesaw de type I} et n'est pas la seule possibilité d'obtenir un terme effectif à
basse énergie comme~(\ref{eqNR}). Dans le {\it mécanisme de Seesaw de type II}, on
ajoute des Higgs
lourds triplets de $SU(2)_L$ qui sont échangés. Dans le {\it double Seesaw}, ce sont 
d'autres singulets de neutrinos (en plus des 3 du mécanisme type I) qu'on ajoute,
mais qui ne se couplent pas aux neutrinos gauches. Le mécanisme se produit 
grâce au couplage avec les neutrinos droits. Ces mécanismes sont un peu 
complexes mais reposent cependant sur le même principe. Nous continuerons donc
l'exposé dans le cadre du modèle seesaw de type I.

\subsubsection{Les alternatives}

Le mécanisme de Seesaw n'est pas le seul à pouvoir fournir des masses très
faibles aux neutrinos, il existe d'autres alternatives, par exemple dans le
cas d'une théorie supersymétrique avec violation de la $R$-parité.
En effet, comme le nombre leptonique n'est pas conservé, il n'est pas étonnant
que la supersymétrie sans $R$-parité puisse apporter une explication aux masses
et mélanges des neutrinos. Il y a plusieurs sources possibles aux masses des
neutrinos. Au niveau de l'arbre nous pouvons avoir un terme de masse de la 
forme :
\beq
(M_{\nu})_{ij}=\frac{\mu_i\mu_j}{m_{SUSY}} \label{treenu}
\eeq
où $\mu_i$ est le couplage bilinéaire entre les champs leptoniques et le champ
de Higgs de type "up".

Au niveau d'une boucle il y a beaucoup de possibilités, par exemple une boucle de
slepton-lepton qui donnerait :
\beq
(M_{\nu})_{ij}=\frac{1}{8\pi^2}\lambda_{ink}\lambda_{jkn}\frac{m_{l_n}m_{l_k}}{m_{SUSY}}
\label{loopnu},\eeq
ou bien une boucle de quark-squark, ou encore une boucle de sneutrino-neutralino.
L'importance relative de ces contributions dépend du modèle. Nous pouvons
facilement établir une hiérarchie entre les masses, par exemple en générant une
masse à l'arbre grâce au terme~(\ref{treenu}) et les deux autres par des termes du
genre~(\ref{loopnu}). Nous n'irons cependant pas plus loin car c'est le
mécanisme de Seesaw qui sera utilisé par la suite. 

\section{Au-delà : la violation des nombres leptoniques}

\subsection{Les nombres leptoniques}

Les oscillations entre saveurs de neutrinos remettent en cause la conservation
de la saveur leptonique qui est accidentelle dans le MS. Cela n'est pas si
choquant car après tout, pourquoi n'y aurait-il pas de violations du/des 
nombre(s) leptonique(s) ?

\underline{Expérimentalement}, c'est vrai, aucune violation de la saveur leptonique
(LFV) n'a été observée chez les leptons chargés. Par exemple les taux de 
branchements de
$\mu\longrightarrow e,\,\gamma$ et $\tau\longrightarrow\mu,\,\gamma$ sont
respectivement inférieurs à $1.2\times 10^{-12}$ et 
$1.1\times 10^{-6}$~\cite{PDG04}. 

Mais \underline{théoriquement}, il n'y a pas de raison vraiment forte pour
justifier la conservation de $L$, le nombre leptonique total, ou $L_e$,
$L_{\mu}$ et $L_{\tau}$, les nombres leptoniques individuels. En physique
des particules, les seuls nombres quantiques conservés sont associés à des
symétries de jauge exactes. De plus, les particules de masses nulles sont
aussi associées à de telles symétries~\footnote{La masse nulle du photon
est associée à la symétrie de jauge $U(1)$ de l'électromagnétisme, et
celles des gluons à la symétrie de jauge de couleur $SU(3)_C$ de QCD.}.
Pourtant, il n'y a pas de symétrie de jauge exacte associée au(x) 
nombre(s) leptonique(s).
Chez les neutrinos, les oscillations entre les différentes saveurs
montrent bien que les nombres leptoniques individuels ne se conservent pas
et que les neutrinos ont des masses non-nulles. Au-delà du Modèle Standard
en général et dans les modèles supersymétriques en particulier, nous nous
attendons à trouver des LFV dans certains processus~\footnote{La question
de la conservation/violation du nombre leptonique \underline{total},
$L_e+L_{\mu}+L_{\tau}$, reste cependant ouverte...}. Dans cette section,
nous discuterons du lien étroit entre le secteur des neutrinos et la LFV.

\subsection{Au-delà du mécanisme de Seesaw}

\subsubsection{Le comptage des paramètres}

Dans le mécanisme de Seesaw minimal, le lagrangien du secteur des neutrinos 
contient :
\beq 
- {\cal L}_\nu \supset \underbrace{\left(Y_{\nu}\right)_{ij} H 
\bar{N}_i 
\left(\begin{array}{c}
\nu \\ L \end{array}\right)_j}_{\mathrm{``masse\ de\ Dirac"\,/\,couplage\ de\ Yukawa}} + 
\underbrace{\frac{1}{2} \bar{N}_i \mathcal{M}_{ij}\bar{N}_j}_{\mathrm{``masse\ de\
Majorana"}}
\eeq
Le mécanisme de Seesaw minimal implique 18 paramètres physiques. D'un coté les 
9 paramètres observables à basse énergie : les 3 masses légères, les 3 angles de 
mélange et les 3 phases violant $CP$ de la matrice de masse effective à basse
énergie (4 de ces paramètres sont déjà connus expérimentalement, $\Delta m_{12}^2$, $\Delta
m_{23}^2$, $\theta_{12}$, $\theta_{23}$). D'un autre côté, la matrice de masse des 
neutrinos singulets lourds a aussi 9 degrés de liberté, incluant {\it a
priori} 3 masses lourdes, 3 angles réels et 3 phases CPV. 
Ceci nous fait un total de 6 phases : la phase de Dirac 
$\delta$, les 2 phases $\phi_{1,2}$ qui affectent la double désintégration beta
sans neutrinos, et 3 phases "lourdes" qui contrôlent la leptogenèse.
Comment pouvons-nous avoir
accès à ces 18 paramètres ? Les oscillations de neutrinos n'étant pas
suffisantes, quelles sont alors les autres observables que nous pouvons
étudier pour déterminer tous les paramètres ?

\subsubsection{La paramétrisation}

A haute énergie, le mécanisme de Seesaw se paramétrise par $(Y_{\nu})$ la matrice 
de couplage de Yukawa des neutrinos droits aux neutrinos gauches (ou par la
matrice de masse de Dirac $M_D$) et par $(M_N)$ la matrice de masse de Majorana
des neutrinos singulets lourds. A basse énergie, nous avons accès à la matrice de
masse effective $\mathcal{M}_{\nu}$ qui est reliée à $(Y_{\nu})$ et $(M_N)$
par l'équation~(\ref{seesaw}). Les 9 paramètres additionnels nécessaires pour paramétriser 
complètement le mécanisme de Seesaw peuvent former une matrice $3\times 3$ hermitienne 
$\mathcal{H}$. Nous avons alors schématiquement :
\beq  \underbrace{(Y_{\nu},\,M_N)}_{hautes\ energies} \longrightarrow 
\underbrace{(\mathcal{M}_{\nu},\,\mathcal{H})}_{basses\ energies}. \eeq

Reste à savoir comment obtenir des informations sur $\mathcal{H}$ à partir de
mesures d'observables à basses énergies et comment relier $\mathcal{H}$ à $(Y_{\nu})$ et
$(M_N)$. 

Il se trouve que dans le modèle supersymétrique seesaw minimal, les processus 
violant le nombre leptonique permettent justement l'étude indirecte des 
paramètres du secteur des neutrinos car ils donnent accès à cette matrice 
$\mathcal{H}$.

\subsection{Les processus violant la saveur leptonique}

Dans un scénario seesaw minimal et supersymétrique, certains processus qui
violent la saveur leptonique peuvent être induits par l'interaction de Yukawa 
$\left(Y_{\nu}\right)_{ij} H_u \bar{N}_i L_j $. En effet, si la supersymétrie 
se brise à des énergies plus grandes que l'échelle de masse des neutrinos lourds, 
des termes non-diagonaux des matrices de masse des sleptons permettant la violation 
de $L$ et de $CP$ seront 
induites radiativement grâce aux couplages de Yukawa des neutrinos légers, et
ceci même si les paramètres d'origine de la brisure sont 
indépendants de la saveur~\footnote{Si on suppose que les termes softs sont
réels et universels à l'échelle de l'unification, comme suggéré par l'absence de
grandes contributions supersymétriques aux processus LFV et CPV, les violations de
$CP$ et des nombres leptoniques dans le secteur des sleptons sont entièremement
induites par les effets de renormalisation des paramètres de couplage des
neutrinos.}.  C'est la renormalisation de la masse "douce" des sleptons qui va 
créer le mélange des saveurs. A l'ordre d'une boucle et quand $M_{GUT}>>M_{N_k}$, 
la renormalisation des paramètres de brisure de supersymétrie à basse énergie se trouve
être proportionnelle à :
\beq
\sum_k(Y_{\nu}^*)_{ki}(Y_{\nu})_{kj}\ln\left(\frac{M_{GUT}}{M_{N_k}}\right),
\label{Hij}\eeq
où $M_{GUT}$ est l'échelle où on impose les conditions initiales sur les
paramètres de brisure de supersymétrie (universalité des masses scalaires à $M_{GUT}$
...). Les $M_{N_k}$ sont les masses de neutrinos lourds. Or~(\ref{Hij}) peut
exactement jouer le rôle de la matrice $\mathcal{H}$~\cite{Param}.
En particulier, nous avons les renormalisations suivantes :
\beqn
(\delta m_{\tilde{L}}^2)_{ij} &\simeq& -\frac{1}{8\pi^2}(3m_0^2 + A_0^2)
\mathcal{H}_{ij}, \nonumber\\
(\delta A_l)_{ij} &\simeq&
-\frac{1}{8\pi^2}A_0Y_{l_i}\mathcal{H}_{ij},
\label{RN-SUSYX}
\eeqn
Dans ce cas, la seule combinaison des angles et des phases qui viole $CP$,
analogue à l'invariant de Jarlskog dans le MS~\footnote{Il traduit
l'amplitude de la violation de $CP$ dans le secteur des quarks.} est:
\beq
J_{\tilde{L}}=\mathrm{Im}[(m_{\tilde{L}}^2)_{12}(m_{\tilde{L}}^2)_{23}(m_{\tilde{L}}^2)_{31}]
\eeq
et elle ne dépend que d'une seule phase.

\noindent \underline{Remarque :} si, à $M_{GUT}$, les valeurs initiales des paramètres de brisure
douce ne sont pas indépendantes de la saveur, il y aura des sources
additionnelles aux processus changeant la saveur en dehors de celles discutées
ici. Seconde remarque, si les neutrinos lourds ne sont pas dégénérés, il y a
d'autres contributions à $J_{\tilde{L}}$ qui dépendront alors de 3
phases~\cite{CPV}.
\par\hfill\par
Tout ceci peut conduire à des taux de branchements observables pour les
processus LFV comme $\mu\to e\gamma$, $\tau\to e\gamma$, $\tau\to \mu\gamma$,
la conversion $\mu-e$ dans les noyaux,
$\mu\to 3e$ et $\tau\to 3 l$. En général, comme $(Y_{\nu})$ est complexe, elle
conduit à une violation de $CP$ dans les oscillations de neutrinos, dans les processus
rares LFV ainsi que dans les moments dipolaires électriques. Cette violation de
$CP$ leptonique est primordiale pour que l'asymétrie baryonique observée dans
l'Univers ait une origine dans la leptogenèse. Nous allons voir des exemples
numériques de tous ces cas.

\subsubsection{La LFV chez les leptons chargés}

\begin{figure}[htbp!]
\begin{center}
\begin{tabular}{c c}
\mbox{\epsfig{file=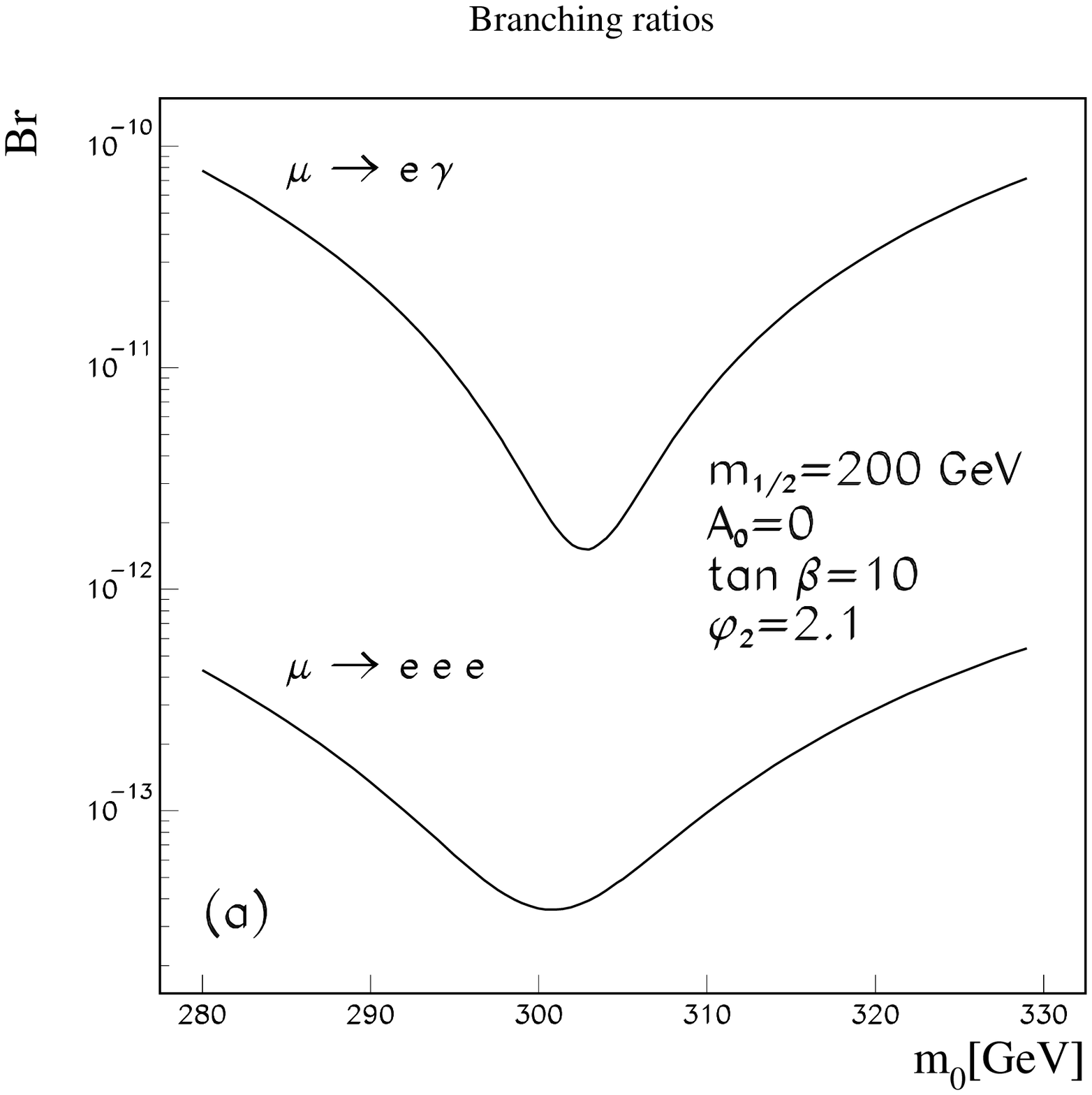,width=7.0cm}} &
\mbox{\epsfig{file=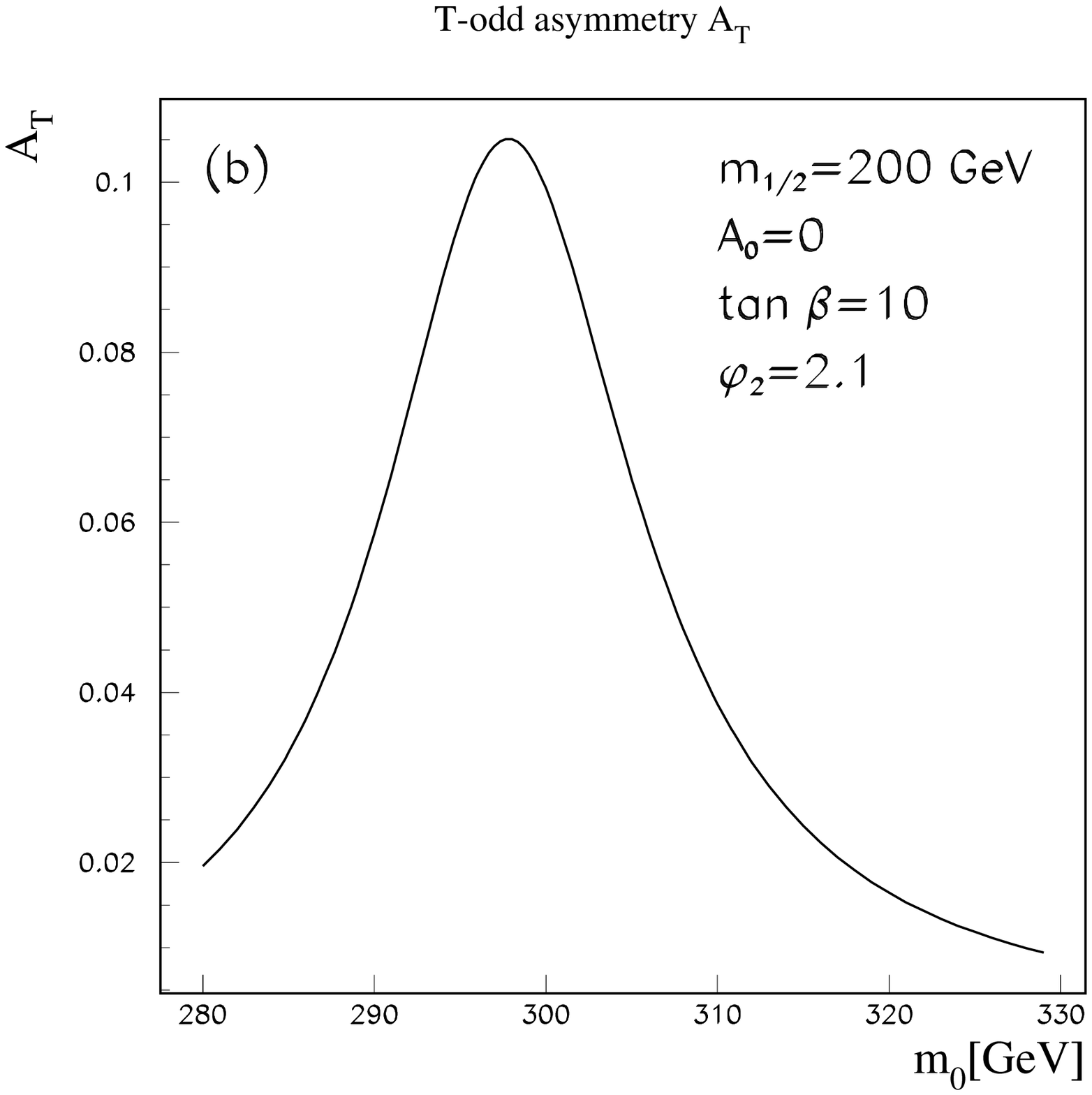,width=7.0cm}} \\
\end{tabular}
\end{center}
\caption{(a) Taux de branchement de $\mu^+\to e^+\,\gamma$ et $\mu\to e^+\,e^+\,e^-$ et (b) asymétrie
de $\mu\to e^+\,e^+\,e^-$ qui viole
T, en fonction de $m_0$ la masse "douce" commune des scalaires et pour un certain
choix des paramètres de neutrinos.~\cite{CPV}}
\label{mu-1}

\end{figure}

\begin{figure}[htbp!]
\begin{center}
\begin{tabular}{c c}
\mbox{\epsfig{file=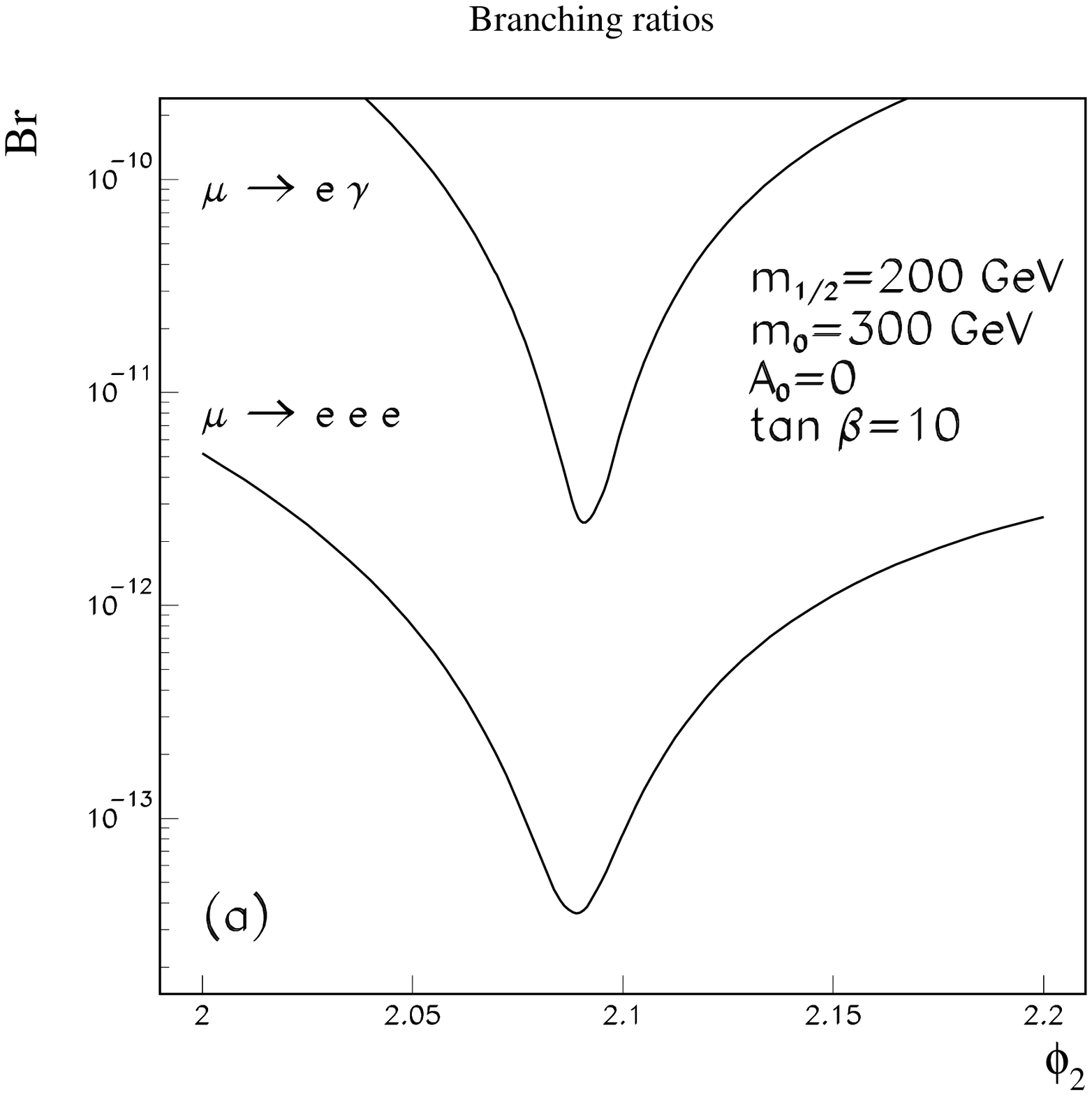,width=7.0cm}} &
\mbox{\epsfig{file=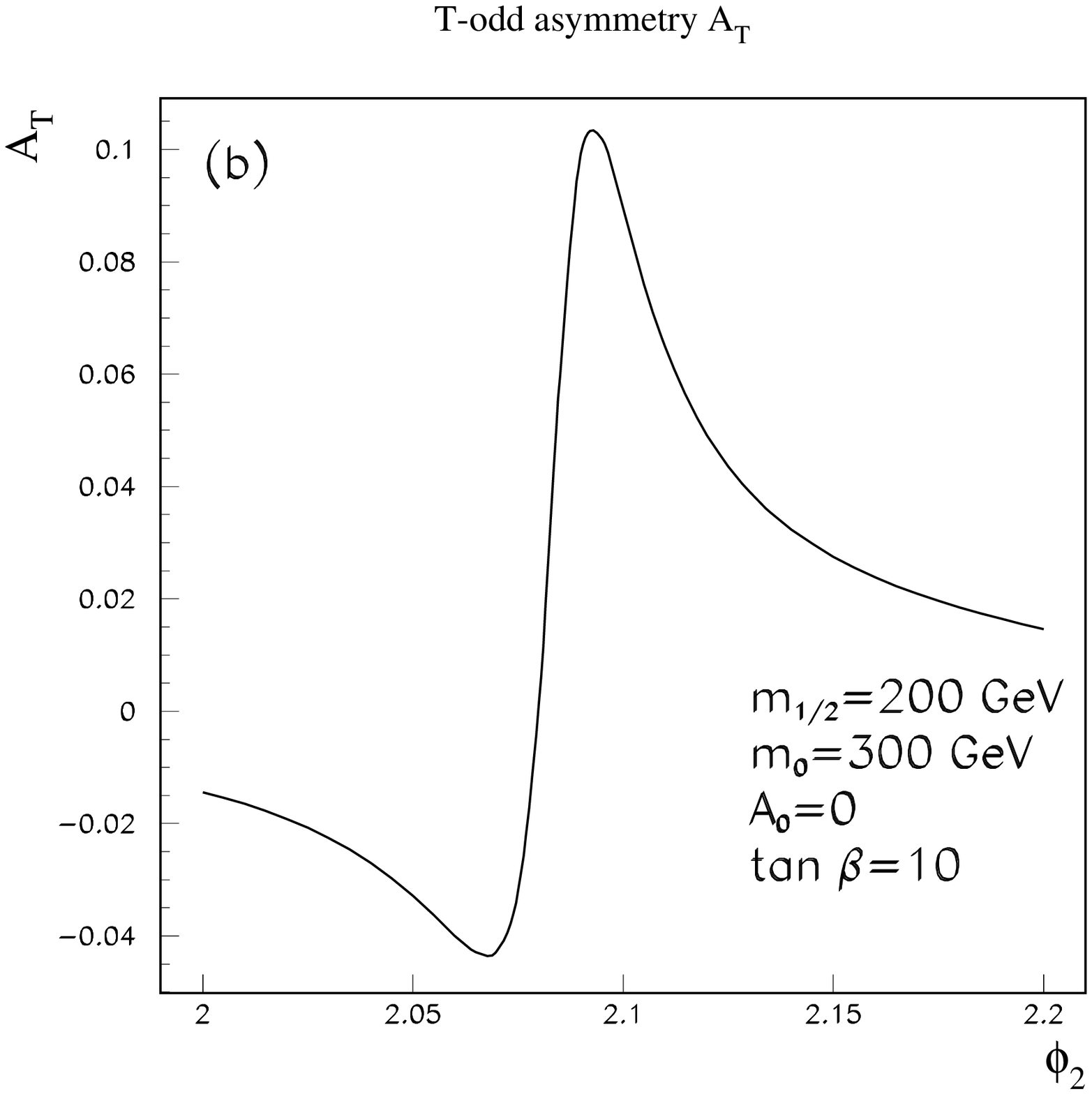,width=7.0cm}} \\
\end{tabular}
\end{center}
\caption{(a) Taux de branchement de $\mu^+\to e^+\,\gamma$ et $\mu\to e^+\,e^+\,e^-$ et (b) asymétrie
de $\mu\to e^+\,e^+\,e^-$ qui viole
T, en fonction de la phase de la phase de Majorana $\phi_2$ et pour un certain
choix des paramètres de neutrinos.~\cite{CPV}}
\label{mu-1bis}
\end{figure}

Nous supposons dans les exemples numériques suivants que les seules sources de
LFV et CPV sont dûes aux interactions avec les neutrinos singulets lourds.
Si nous calculons cette
désintégration dans le modèle seesaw \underline{supersymétrique} et minimal,
(avec un choix
particulier des paramètres des neutrinos) et pour une grande partie de
l'espace des paramètres supersymétriques, la prédiction est au-dessus de la limite expérimentale.
Mais quand la prédiction du taux de branchement est minimale ({\it i.e.} $m_0
\simeq 300$ GeV, figure~(\ref{mu-1}); $\phi_2 \simeq 2.1$,
figure~(\ref{mu-1bis})) l'asymétrie $\mathcal{A}_{T}$ de $\mu\to 3e$
qui viole $T$ (donc de façon équivalente $CP$) est maximale (elle peut être de
10$\%$ environ) : il y a anti-corrélation entre $\mathcal{A}_{T}$ et $Br(\mu\to
e\,\gamma)$. Le taux de branchement $\mu\to 3e$ est donc dans ce cas comparable à celui de $\mu\to
e\,\gamma$. Ceci est en fait dû à des annulations
possibles dans les différents diagrammes pingouins photoniques.
Ce résultat est commun à d'autres choix sur les paramètres des neutrinos,
nous n'entrerons donc pas dans les détails.
Ainsi, il est en principe possible de mesurer expérimentalement la
violation de $CP$ dans
le secteur des leptons chargés avec la désintégration $\mu\to 3e$.

\begin{figure}[htbp!]
\begin{center}\begin{tabular}{c c}
\mbox{\epsfig{file=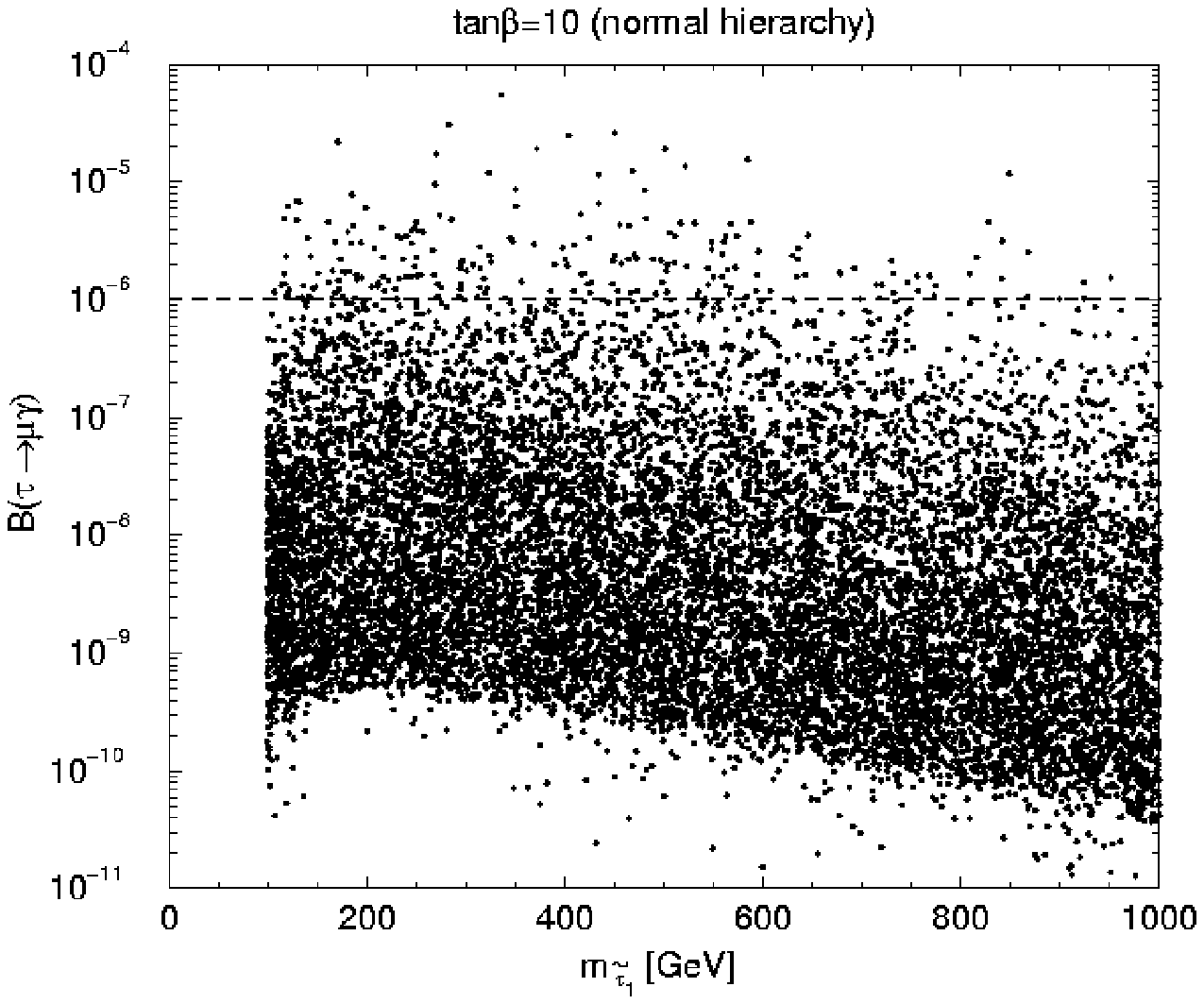,width=7.0cm}} &
\mbox{\epsfig{file=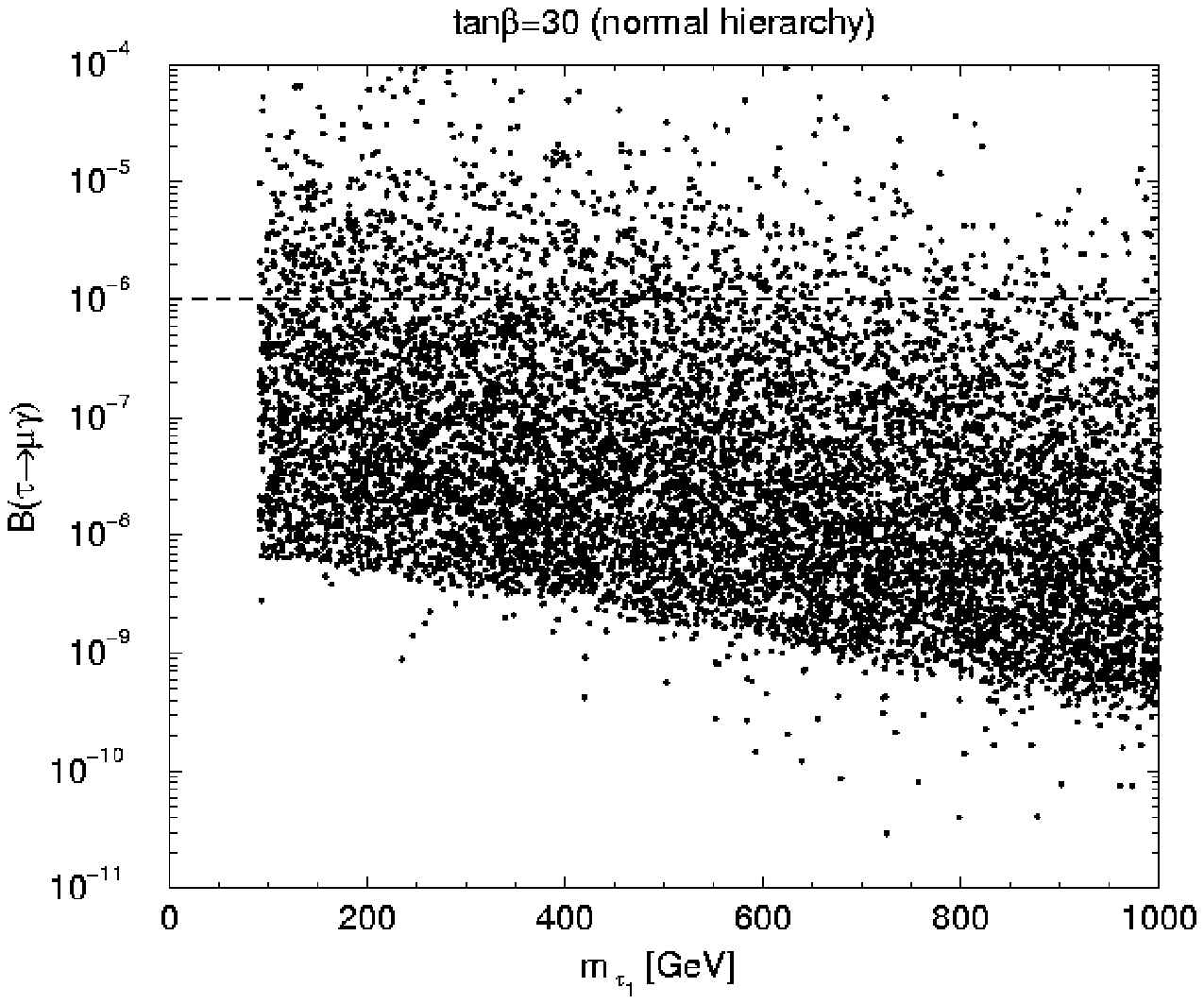,width=7.0cm}} \\
\mbox{\epsfig{file=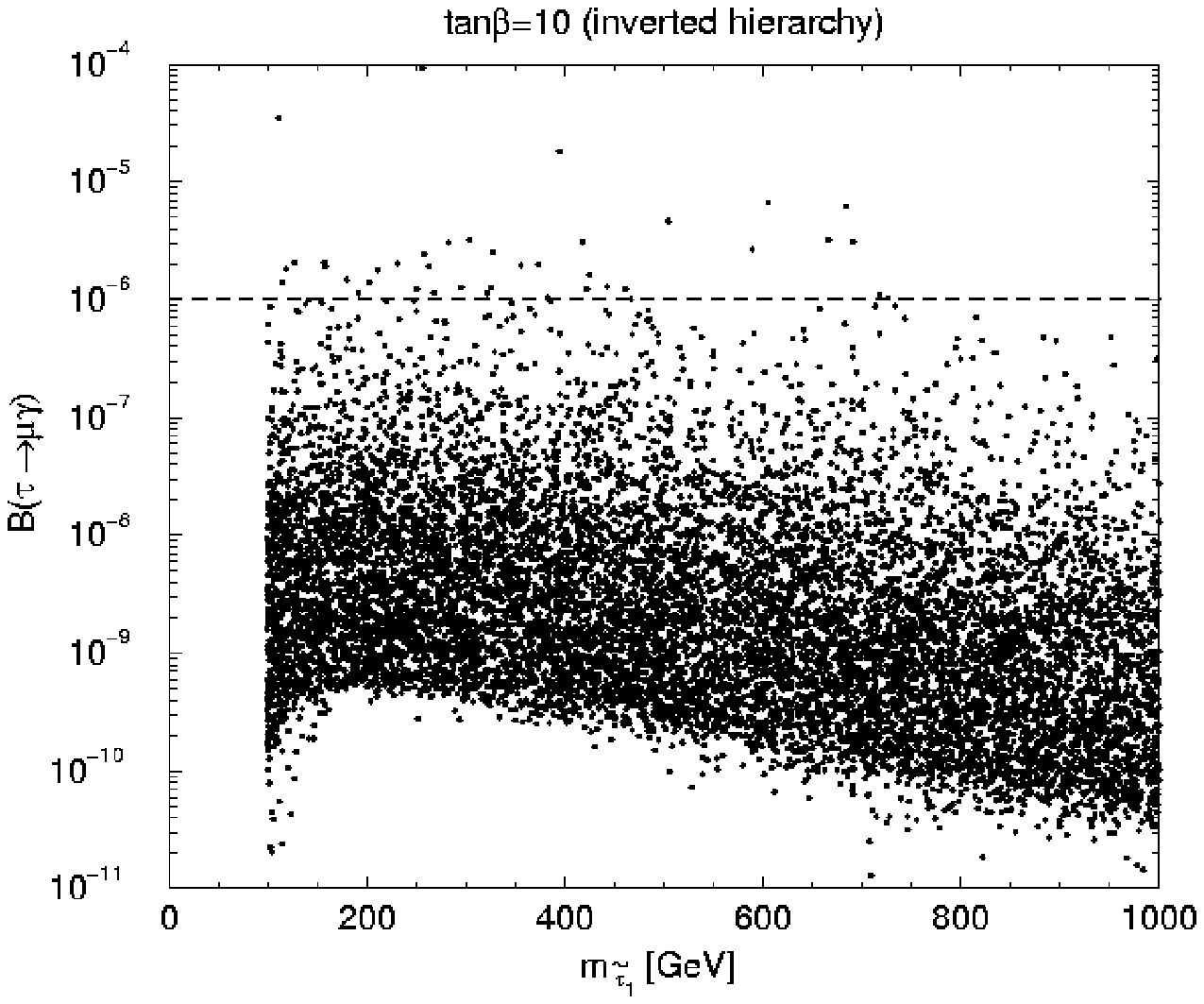,width=7.0cm}} &
\mbox{\epsfig{file=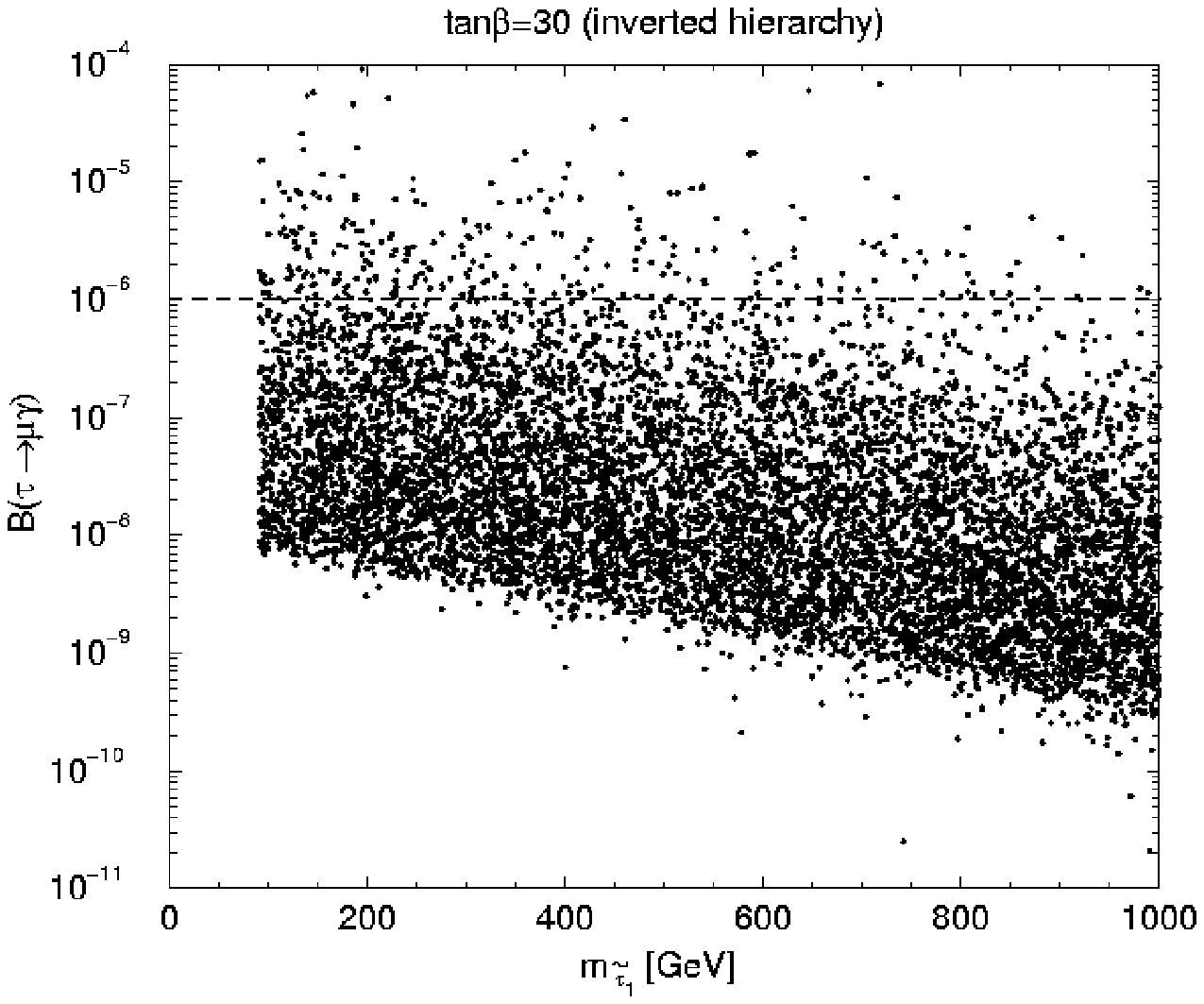,width=7.0cm}} \\
\end{tabular}
\end{center}
\caption{Taux de branchement de $\tau \to \mu\,\gamma$ en fonction de la masse
du stau le plus léger, pour un certain
choix des paramètres de neutrinos (angles de mélange). Les quatres graphes correspondent à
différents choix de $\tan\beta$ et de hiérarchie entre les masses des neutrinos
légers.~\cite{Param}}
\label{brtau}
\end{figure}

La violation de la saveur leptonique peut aussi se voir dans les
désintégrations de taus, par exemple $\tau\to
l\, \gamma$ ou $\tau\to 3l$. Mais pour $\tau \to \mu\,\gamma$ et
$\tau \to e\,\gamma$, il n'y a pas de telles annulations
dans la région de paramètres considérée dans les figures~(\ref{brtau}).
Dans ce scénario, la plupart des points se situent un peu en dessous des
limites expérimentales actuelles et donc les taux de branchements pourraient, en
principe, avoir des chances d'être observables au LHC ou à Babar et Belle.

\subsubsection{Les moments dipolaires électriques leptoniques}

Les violations de $CP$ et des saveurs leptoniques ont aussi un impact sur les
moments dipolaires électriques (EDM) des leptons. Ceux-ci dépendent en principe
de phases qui apparaissent aussi dans la leptogenèse.  Les valeurs numériques de
ces EDM sont augmentés de plusieurs ordres de grandeur quand les neutrinos
lourds sont non-dégénérés. Or dans la plupart des modèles de masse des neutrinos phénologiquement
viables, les neutrinos lourds ne sont pas dégénérés.

L'EDM d'un lepton $l$ est défini comme étant le coefficient $d_l$ de
l'interaction suivante :
\beq
\mathcal{L}=-\frac{i}{2}d_l \bar{l}\sigma_{\mu\nu}\gamma_5 l F^{\mu\nu}.
\eeq
Dans le MSSM, $d_l$ reçoit des contributions des boucles de neutralinos et de
charginos et dépend fortement de $A_0$. Les contributions dominantes à $d_e$ et
$d_{\mu}$ sont dues aux termes suivants~\footnote{Ce sont en fait les même
termes que l'équation~(\ref{RN-SUSYX}) mais dans le cas de neutrinos lourds non dégénérés.}
de la renormalisation des termes de brisure douce:
\beqn
(\delta m_{\tilde{L}}^2)_{ij} &\simeq& \frac{18}{(4\pi)^4}(m_0^2 + A_0^2)
\{Y_{\nu}^{\dag}LY_{\nu},\,Y_{\nu}^{\dag}Y_{\nu}\}_{ij}
\ln\left(\frac{M_{GUT}}{M_{N}}\right), \nonumber\\
(\delta A_l)_{ij} &\simeq&
-\frac{1}{(4\pi)^4}A_0Y_e\left(11\{Y_{\nu}^{\dag}LY_{\nu},\,Y_{\nu}^{\dag}Y_{\nu}\}
+
7[Y_{\nu}^{\dag}LY_{\nu},\,Y_{\nu}^{\dag}Y_{\nu}]\right)_{ij}
\ln\left(\frac{M_{GUT}}{M_{N}}\right)
\eeqn
$L_{ij}=ln\left(\frac{\bar{M}_{N}}{M_{N_i}}\right)_{ij}$ et $\bar{M}_{N}=\
^3\sqrt{M_{N_1}M_{N_2}M_{N_3}}$. $\bar{M}_{N}$ est la
masse moyenne des neutrinos singulets lourds (qui ne sont pas dégénérés dans ce
cas).

\begin{figure}[htbp!]
\begin{center}
\begin{tabular}{c c}
\mbox{\epsfig{file=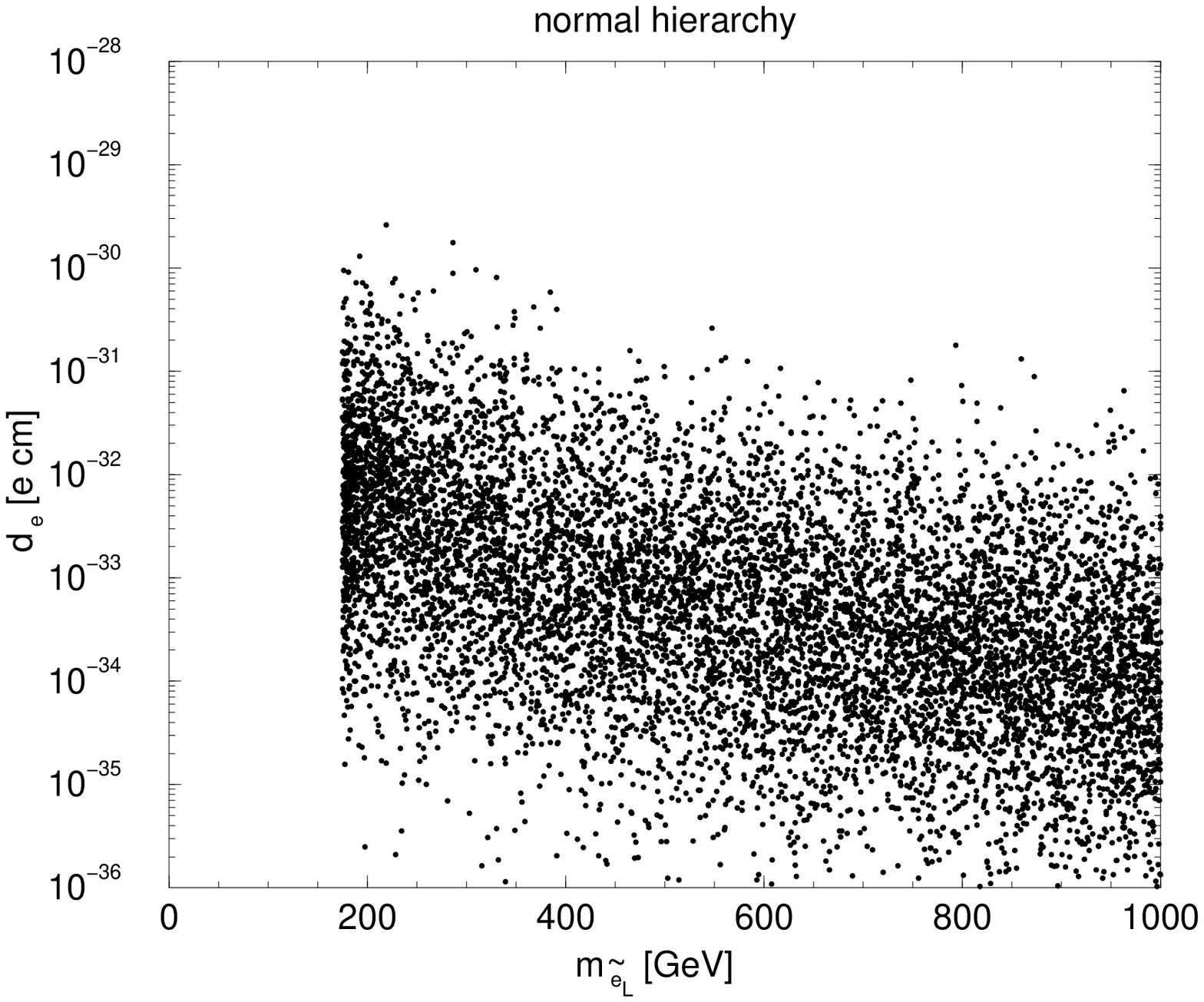,width=6.0cm}} &
\mbox{\epsfig{file=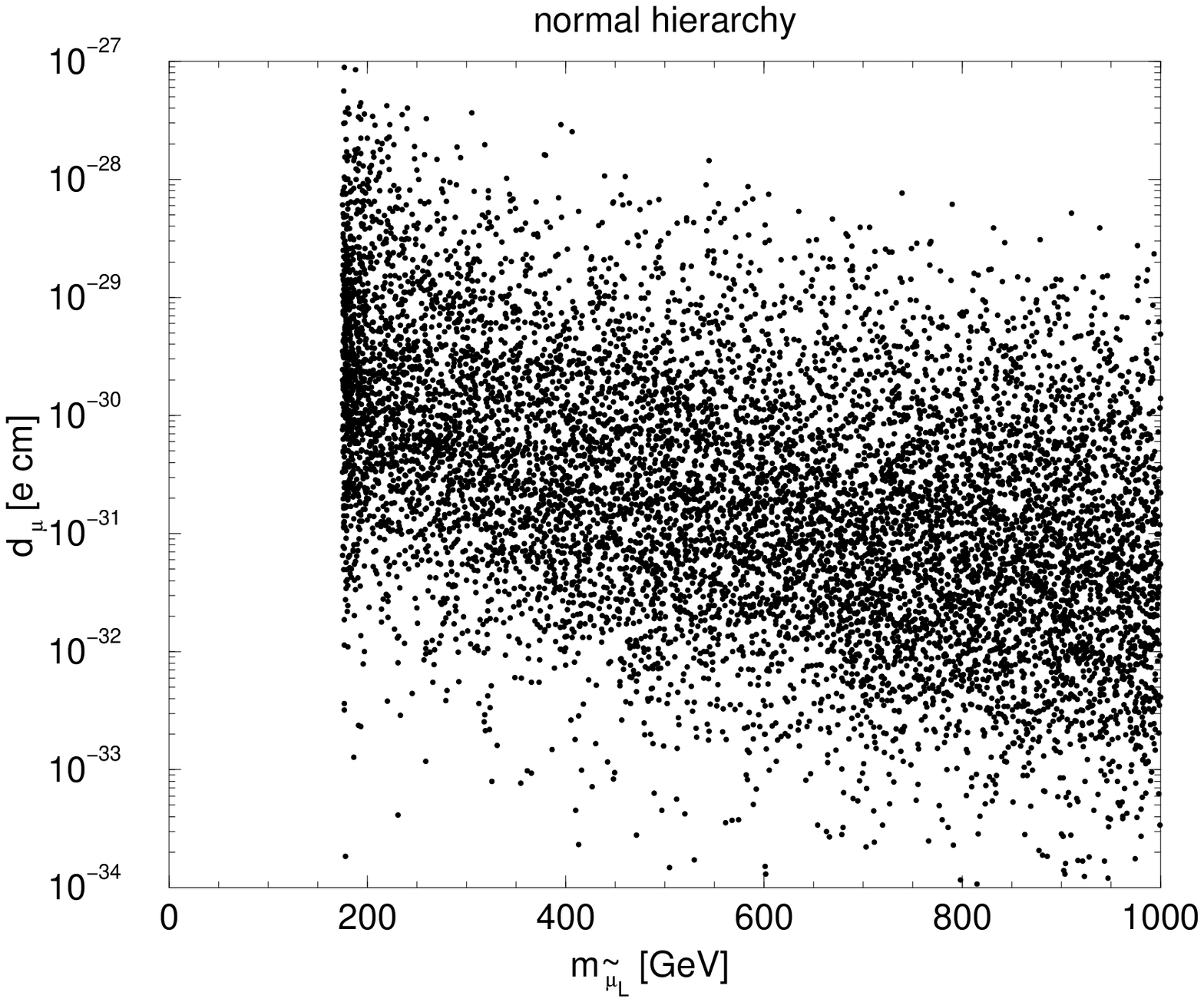,width=6.0cm}} \\
\end{tabular}
\end{center}
\caption{EDM de l'électron (a) et du muon (b) en fonction de la masse
du selectron gauche (a) et du smuon gauche (b). Chaque point correspond à un
certain choix des paramètres inconnus. Dans l'exemple donné, la hiérarchie des
neutrinos légers est "normale".~\cite{Param}}
\label{de}
\end{figure}

Les prédictions des EDM leptoniques sont illustrées numériquement dans la
figure~(\ref{de}). Les EDM du muon et de l'électron peuvent atteindre $10^{-(27-28)}\
e$.cm et $10^{-(29-30)}\ e$.cm respectivement.

Les limites supérieures
actuelles sont~\cite{EDM} :
\beqn
d_e & < & 4.3\times 10^{-27}\ \mathrm{e}.\mathrm{cm}, \nonumber \\
d_{\mu} & = & (3.7\pm 3.4)\times 10^{-19}\ \mathrm{e}.\mathrm{cm},\nonumber \\
|d_{\tau}| & < & 3.1 \times 10^{-16}\ \mathrm{e}.\mathrm{cm}.
\eeqn
Les expériences futures à BNL et auprès d'usines à neutrinos projettent
respectivement de sonder l'EDM du muon jusqu'à $10^{-24}\ e$.cm et 
$10^{-26}\ e$.cm. La mesure de l'EDM de
l'électron devrait atteindre les environs de $10^{-33}\ e$.cm. Ces expériences 
ont donc la possibilité de tester le modèle seesaw supersymétrique 
minimal {\it via} ses prédictions sur $d_e$ ($d_{\mu}$ risque malheureusement 
d'être difficile à mesurer).

\subsection{La désintégration de sparticules}

Une autre façon de mesurer la violation du/des nombres leptoniques est
d'observer la désintégration de sparticules. Par exemple, $\chi_2\to\chi_1
+(e\mu)$ où $\chi_1$ est la LSP et  $\chi_2$ la particule la plus légère après 
$\chi_1$ (NLSP). En particulier, les désintégrations de sparticules qui violent $L_{\tau}$ sont
les plus intéressantes car les effets de la renormalisation des paramètres de
brisure douce sont plus importants dans le cas du $\tau$ (le couplage de Yukawa
du $\tau$ est plus grand). Cela rend les effets de violation du nombre
leptonique $L_{\tau}$ potentiellement grands pour les désintégrations comme
$\chi_2\to\chi_1 +\tau\mu$ ou $\chi_2\to\chi_1 +\tau e$. Les recherches au
LHC de ces désintégrations pourraient donc être complémentaires à celles de 
$\tau\to l\,\gamma$.

\begin{figure}[htbp!]
\centerline{\includegraphics*[width=9cm]{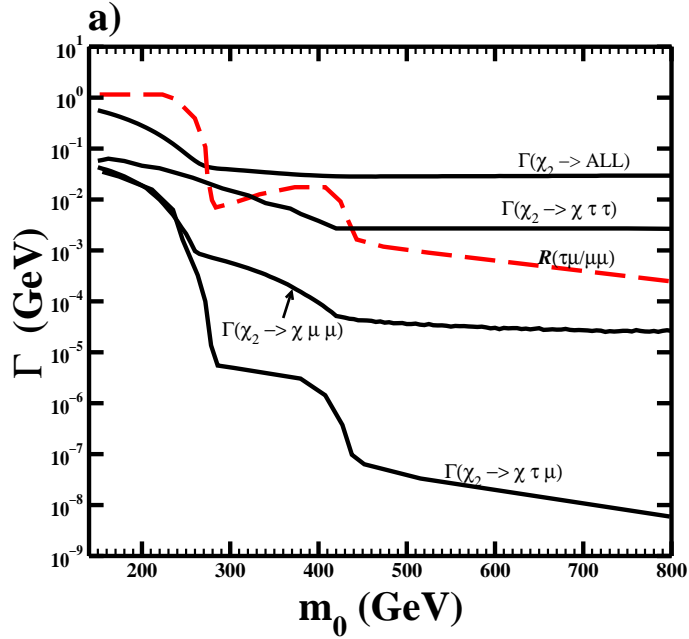}}
\caption{Comparaison des largeurs de désintégrations entre quelques modes de 
désintégration de $\chi_2$ en fonction de $m_0$.~\cite{TauLFV}}
\label{widthLFV1}
\end{figure}

Dans la figure suivante,~(\ref{widthLFV1}), les lignes en traits pleins dénotent les
largeurs de désintégration de $\chi_2$ en différents modes, dans le cas
particulier du CMSSM~\footnote{En fait, une non-universalité des masses
scalaires a été introduite, $(m_0^2)_{LL}=diag(m_0^2,m_0^2,x\times m_0^2)$ avec
$x\sim 0.9$. De plus, l'angle de mélange dans la matrice de Yukawa entre la 
2ème et 3ème génération de lepton chargés, $\phi$, a été pris égal à $\pi/6$ 
dans les 2 exemples.} avec les paramètres $\tan\beta=10$, $\mu>0$, $m_{1/2}=600$
GeV. Le rapport des taux de branchements $R(\tau\mu/\mu\mu)\equiv
\Gamma(\chi_2\to\chi_1+\tau\mu)/\Gamma(\chi_2\to\chi_1+\mu\mu)$ est représenté 
par les tirets. Nous voyons que pour $m_0<270$ GeV, $R(\tau\mu/\mu\mu)$ 
est de l'ordre de 1 donc $\chi_2\to\chi_1+(\tau\mu)$ peut être comparable à 
$\chi_2\to\chi_1+(\mu\mu)$. La violation du nombre leptonique peut donc être
importante. 

\begin{figure}[htbp!]
\centerline{\includegraphics*[width=9cm]{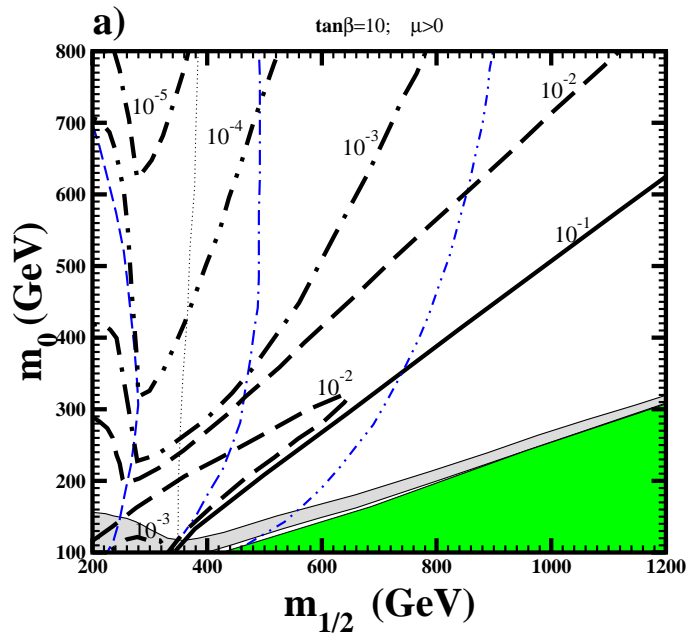}}
\caption{Contours du rapport $R(\tau\mu/\mu\mu)$ (traits noirs) et de
$BR(\tau\to\mu\,\gamma)$ (lignes pointillées bleues) dans le plan $(m_0,
m_{1/2})$ pour $\tan\beta=10$ et $\mu>0$.~\cite{TauLFV}}
\label{widthLFV2}
\end{figure}

De plus, sur la figure~(\ref{widthLFV2}) les contours des 
taux de branchements sont tracés pour le même choix de $\tan\beta$ et $\mu$ 
mais dans le plan $(m_0, m_{1/2})$.

\section{Les neutrinos et la cosmologie}

\subsection{L'asymétrie baryonique}

La physique des neutrinos peut aussi jouer un rôle en cosmologie,
particulièrement pour répondre à la question de l'origine de la matière. En
effet, la quasi-totalité de ce que nous observons dans l'Univers est fait de
matière, nous sommes fait de matière, mais pas d'antimatière. Pourquoi cette
asymétrie ? 
\par\hfill\par
La densité baryonique de l'Univers est : 
\beq
\eta=\frac{n_B}{n_{\gamma}}=(1.5-6.3)\times 10^{-10},
\eeq
où $n_B$ et $n_{\gamma}$ sont respectivement le nombre de baryons et le nombre
de photons. L'asymétrie baryonique cosmologique est alors
$Y_B=(n_B-n_{\bar{B}})/s$, où $s$ est la densité d'entropie, et vaut environ 
$\eta/7$. Cette asymétrie peut s'expliquer~\footnote{La
possibilité que nous ne vivions que dans une région dans l'Univers constituée
de matière et que l'antimatière serait localisé hors de notre horizon est très
défavorisée.} par un très
faible déséquilibre dans l'Univers primordial en faveur de la matière, causé par
les interactions. Par la suite, l'antimatière s'est alors annihilée avec la 
matière et le reste a formé un peu plus tard les baryons dont nous sommes faits. Mais 
une asymétrie matière-antimatière ne peut être dynamiquement générée dans un 
univers en expansion que si les interactions entre
particules et l'évolution cosmologique satisfont \underline{les 3 conditions de
Sakharov :} 
\begin{itemize}
\item[$\bullet$] violation du nombre baryonique ($\Rightarrow$ les interactions
changent alors le nombre de quarks) 
\item[$\bullet$] violation de $C$ et $CP$ ($\Rightarrow$ les interactions sont donc
différentes pour la matière et l'antimatière)
\item[$\bullet$] déviation de l'équilibre thermique, par exemple lors d'une
transition de phase ($\Rightarrow$ évite que les produits de désintégration ne puisse
recréer la particule initiale et donc permet de cumuler le déséquilibre)
\end{itemize}
La question de l'asymétrie baryonique permet donc d'établir un lien entre le
modèle standard cosmologique et le modèle standard de la physique des
particules.
A noter qu'une asymétrie générée à une époque de l'Univers peut être 
effacée par
la suite si elle n'est pas protégée, ce qu'il faut éviter. Cependant, nous ne
discuterons pas de ce genre de détails (bien que ce soit important).

\subsection{La leptogenèse}

Le mécanisme de génération de cette asymétrie baryonique est appelée la {\it
baryogenèse}. Il existe dejà plusieurs scénario viables. Mais ce dont nous
allons discuter ici est la {\it leptogenèse}, dans laquelle l'asymétrie 
baryonique est obtenue {\it via} une asymétrie \underline{leptonique} dans la 
désintégration de neutrinos singulets lourds (les neutrinos droits de Majorana),
ce qui est possible si le nombre leptonique n'est pas conservé. La
désintégration $N\to Higgs+l$ se fait avec un taux de branchement différent de
$N\to \overline{Higgs}+\bar{l}$. L'asymétrie entre le nombre de leptons et le
nombre d'antileptons est convertie en une asymétrie entre le nombre de quarks et 
d'antiquarks par des interactions électrofaibles \underline{non-perturbatives}
(les sphalérons) qui donneraient~\cite{Lepto}:
\beq Y_B=\frac{C}{C-1}Y_L, \eeq
avec $C=8/23$ dans le MSSM.

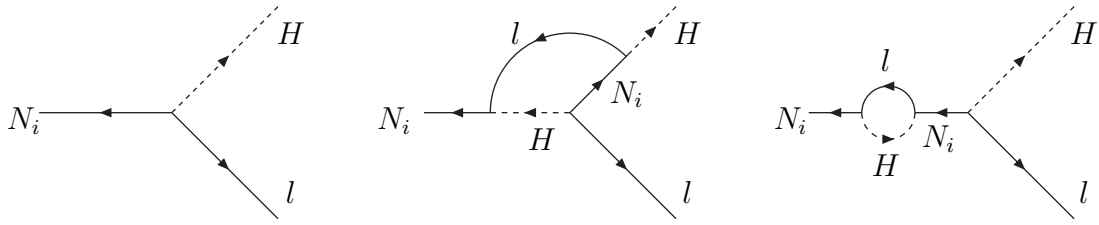
\begin{figure}
\begin{center}
\begin{picture}(350,100)(-30,-30)
\ArrowLine(0,0)(-50,0)
\ArrowLine(0,0)(40,-40)
\DashArrowLine(0,0)(40,40){2}
\Text(45,30)[]{$H$}
\Text(45,-30)[]{$l$}
\Text(-56,-3)[]{$N_i$}

\DashArrowLine(150,0)(120,0){2}
\ArrowLine(120,0)(95,0)
\ArrowLine(150,0)(190,-40)
\ArrowLine(150,0)(172,22)
\DashArrowLine(172,22)(190,40){2}
\ArrowArc(150,0)(30,45,180)
\Text(195,30)[]{$H$}
\Text(195,-30)[]{$l$}
\Text(84,-3)[]{$N_i$}
\Text(140,-10)[]{$H$}
\Text(130,+30)[]{$l$}
\Text(172,+7)[]{$N_i$}

\ArrowLine(300,0)(340,-40)
\DashArrowLine(300,0)(340,40){2}
\ArrowLine(300,0)(280,0)
\ArrowArc(270,0)(10,0,-180)
\DashArrowArc(270,0)(10,180,0){2}
\ArrowLine(260,0)(240,0)
\Text(345,30)[]{$H$}
\Text(345,-30)[]{$l$}
\Text(234,-3)[]{$N_i$}
\Text(270,+20)[]{$l$}
\Text(270,-20)[]{$H$}
\Text(290,-10)[]{$N_i$}
\end{picture}
\end{center}
\caption{ Diagrammes de Feynman à l'arbre et à une boucle qui contribuent à la désintégration
des neutrinos lourds $N_j$.}
\label{Nudecay}
\end{figure}

Le taux de désintégration total d'un neutrino singulet lourd $N_i$ peut s'écrire
ainsi (sans sommation sur i) :
\beq
\Gamma_i = \frac{1}{8 \pi} \left( Y_{\nu} Y_{\nu}^{\dag}\right)_{ii}M_i.
\eeq
Les diagrammes à une boucle qui impliquent l'échange d'un neutrino lourd,
figure~(\ref{Nudecay}), peuvent générer une asymétrie $CP$ dans la désintégration
des neutrinos lourds $N_i$.
Cette asymétrie s'écrit sous la forme suivante :
\beq
\epsilon_{ij}=\frac{1}{8 \pi} \frac{1}{\left( Y_{\nu} Y_{\nu}^{\dag}\right)_{ii}}
\mathrm{Im}[\left( Y_{\nu} Y_{\nu}^{\dag}\right)_{ij}]^2\times
f\left(\frac{M_j}{M_i}\right), \label{epsilon}
\eeq
où $f\left(\frac{M_j}{M_i}\right)$ est une fonction cinématique connue et
calculable et
\beq \epsilon=\frac{\Gamma(N\to H+l)-\Gamma(N\to
\bar{H}+\bar{l})}{\Gamma(N\to H+l)+\Gamma(N\to \bar{H}+\bar{l})}<<1. \eeq

\begin{figure}[htbp!]
\begin{center}
\includegraphics*[width=7.5cm]{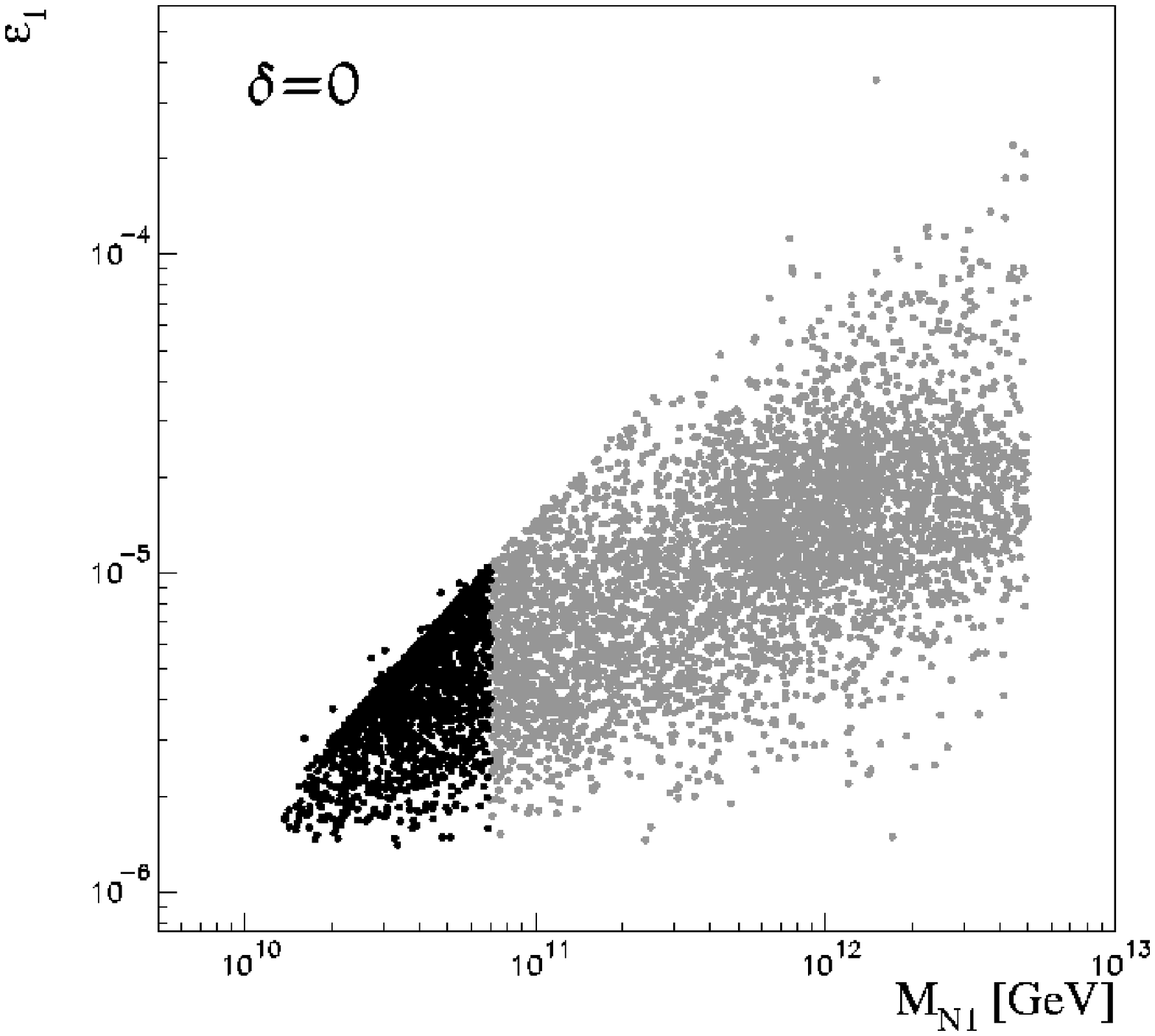}
\includegraphics*[width=7.5cm]{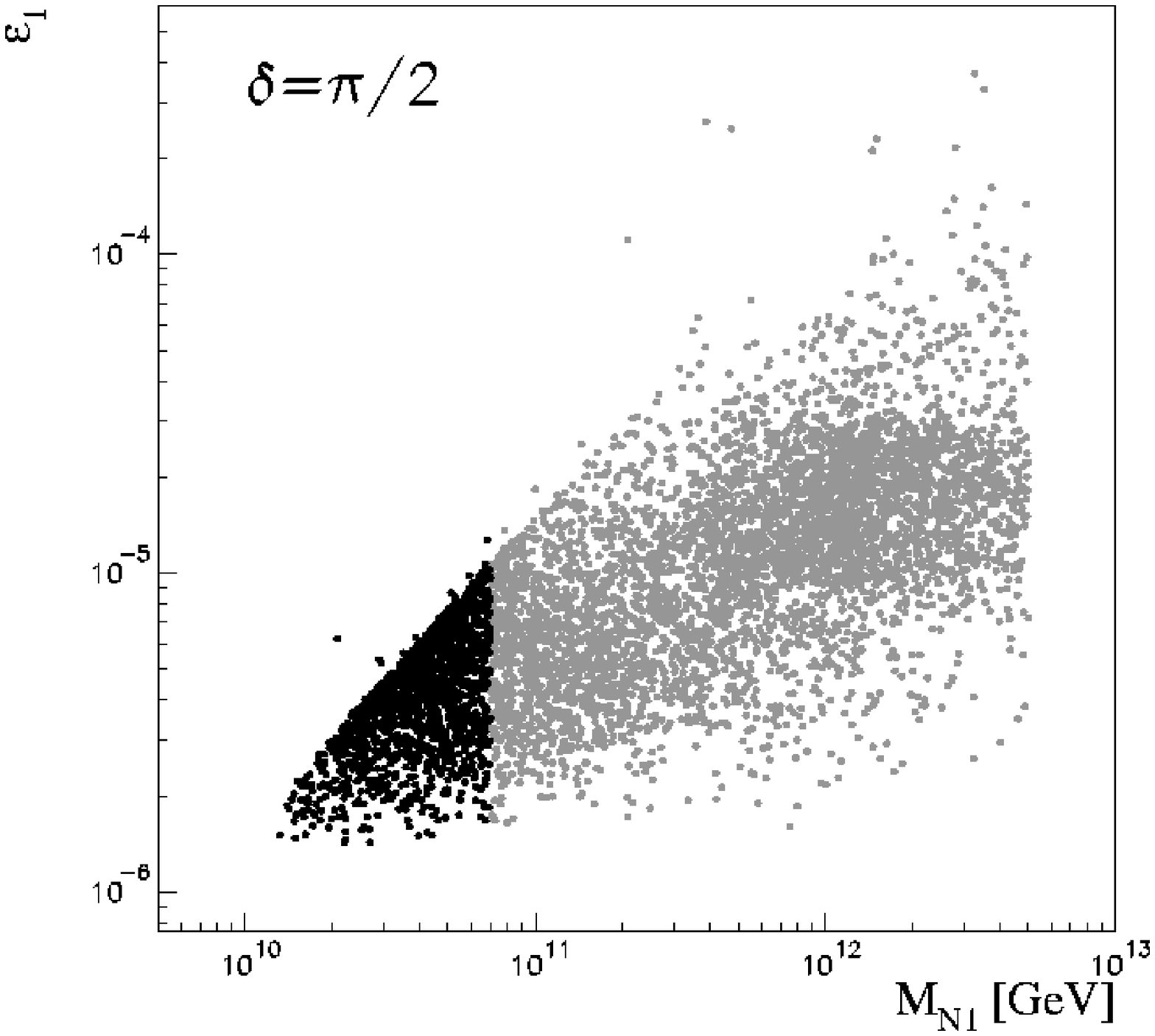}
\end{center}
\caption{"scatter plots" de l'asymétrie $\epsilon_1$ en fonction de la masse
lourde $M_{N_{1}}$ la plus légère, pour les 2 choix $\delta=0$ et
$\delta=\pi/2$.~\cite{Lepto}}
\label{lepto}
\end{figure}

Nous voyons dans~(\ref{epsilon}) que la leptogenèse est proportionnelle au produit
$\left( Y_{\nu} Y_{\nu}^{\dag}\right)$ qui dépend de 9 paramètres réels et
de 3 phases CPV mais pas celles de basse énergie ($\delta,\ \phi_{1,2}$).
L'existence d'une asymétrie leptonique ne requiert donc pas que $\delta$ soit
non-nulle. Sur la figure~(\ref{lepto}), nous voyons que les cas $\delta=0$ et
$\delta=\pi/2$ sont indistinguables.
\par\hfill\par
Comment avoir accès à la leptogenèse ? Il est possible de formuler une
"stratégie" pour calculer la leptogenèse en terme d'observables mesurables en
laboratoire :
\begin{itemize}
\item[$\bullet$] mesurer la phase $\delta$ des oscillations de neutrinos et les
phases de Majorana $\phi_{1,2}$,
\item[$\bullet$] mesurer les observables reliées à la renormalisation des
paramètres de brisure douce de la supersymétrie qui sont fonctions de $\delta$,
$\phi_{1,2}$ et des phases de la leptogenèse,
\item[$\bullet$] extraire les effets connus de $\delta$ et $\phi_{1,2}$ pour
isoler les paramètres de la leptogenèse.
\end{itemize}

\begin{figure}[htbp!]
\begin{center}
\includegraphics*[width=7.5cm]{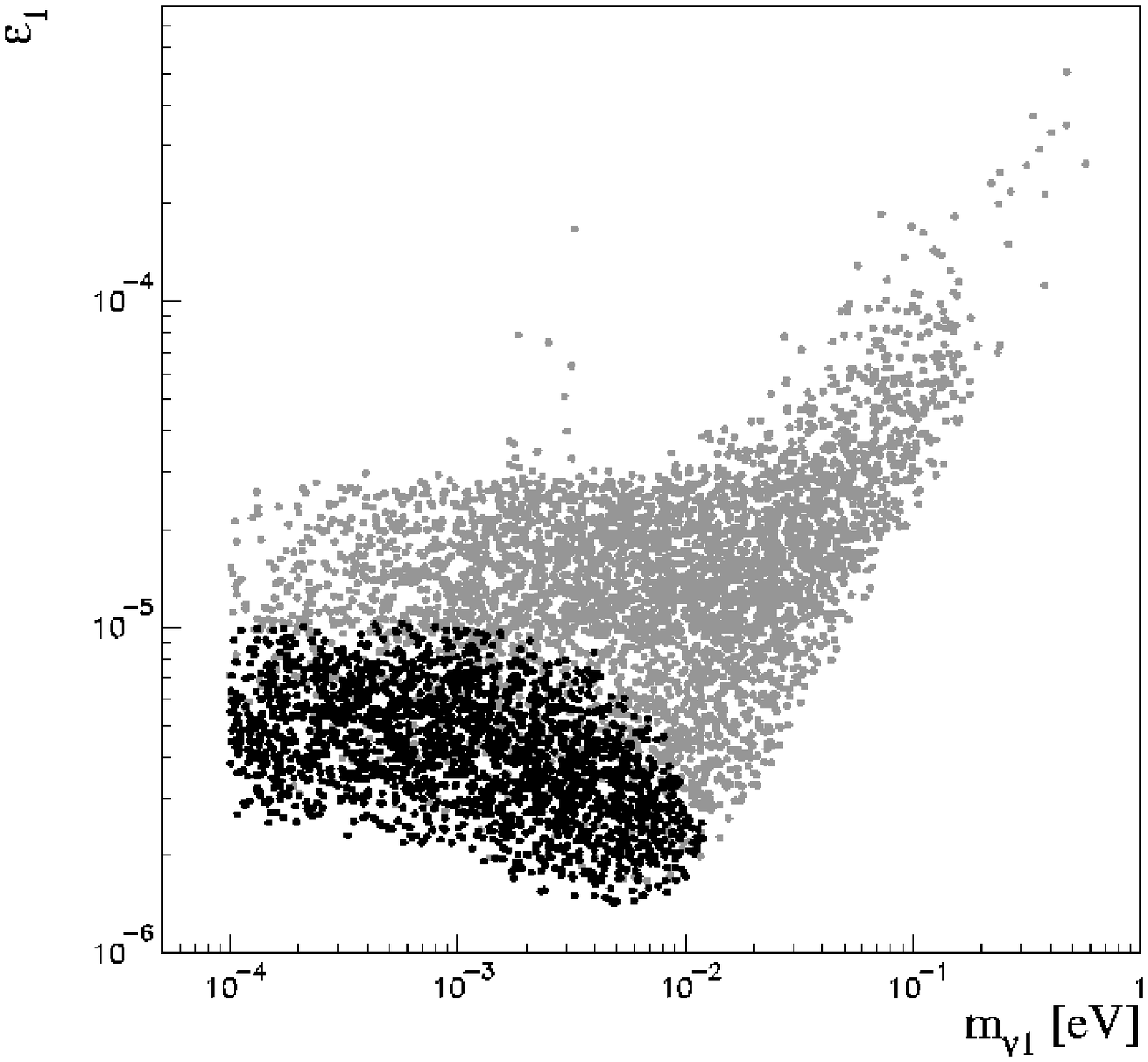}
\includegraphics*[width=7.5cm]{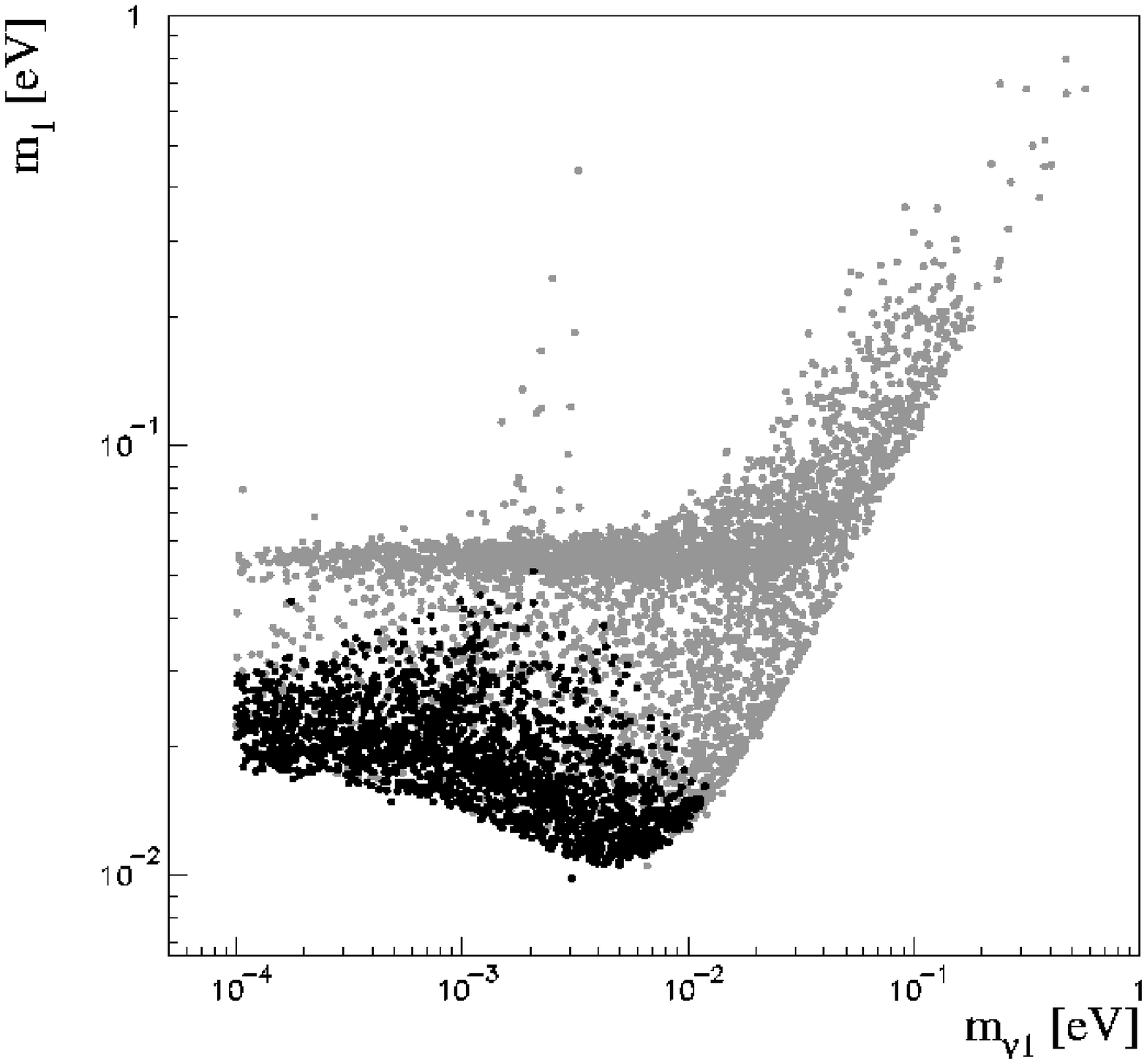}
\end{center}
\caption{"scatter plots" de l'asymétrie $\epsilon_1$ et de la masse effective
$m_1=(Y_{\nu}Y_{\nu}^{\dag})_{11}\frac{v^2\sin^2\beta}{M_{N_1}}$ en
fonction de la masse légère $m_{\nu_{1}}$ la plus légère. La
contrainte sur l'asymétrie baryonique a été prise en compte.~\cite{Lepto}}
\label{lepto2}
\end{figure}

\begin{figure}[htbp!]
\begin{center}
\includegraphics*[width=7.5cm]{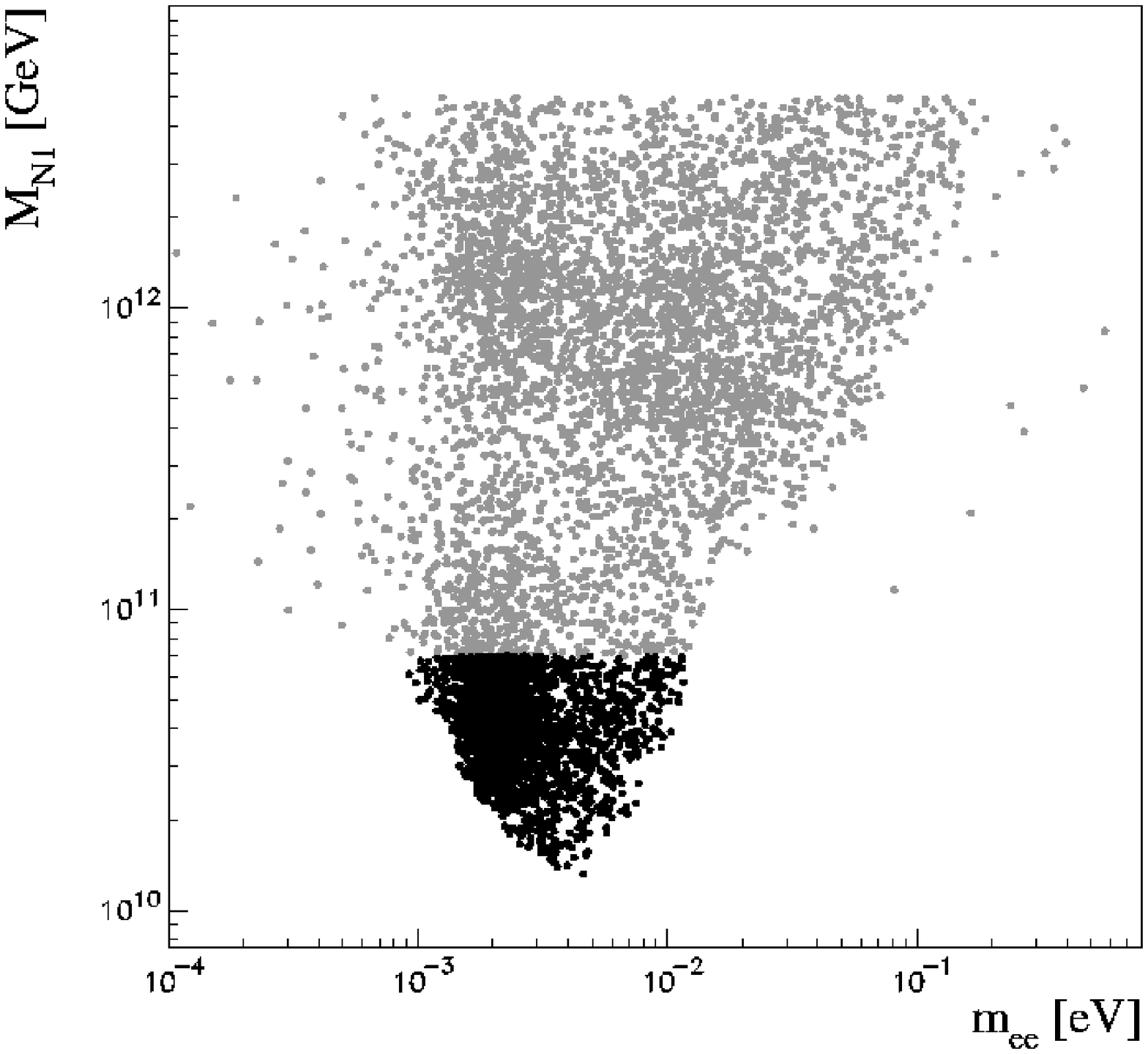}
\includegraphics*[width=7.5cm]{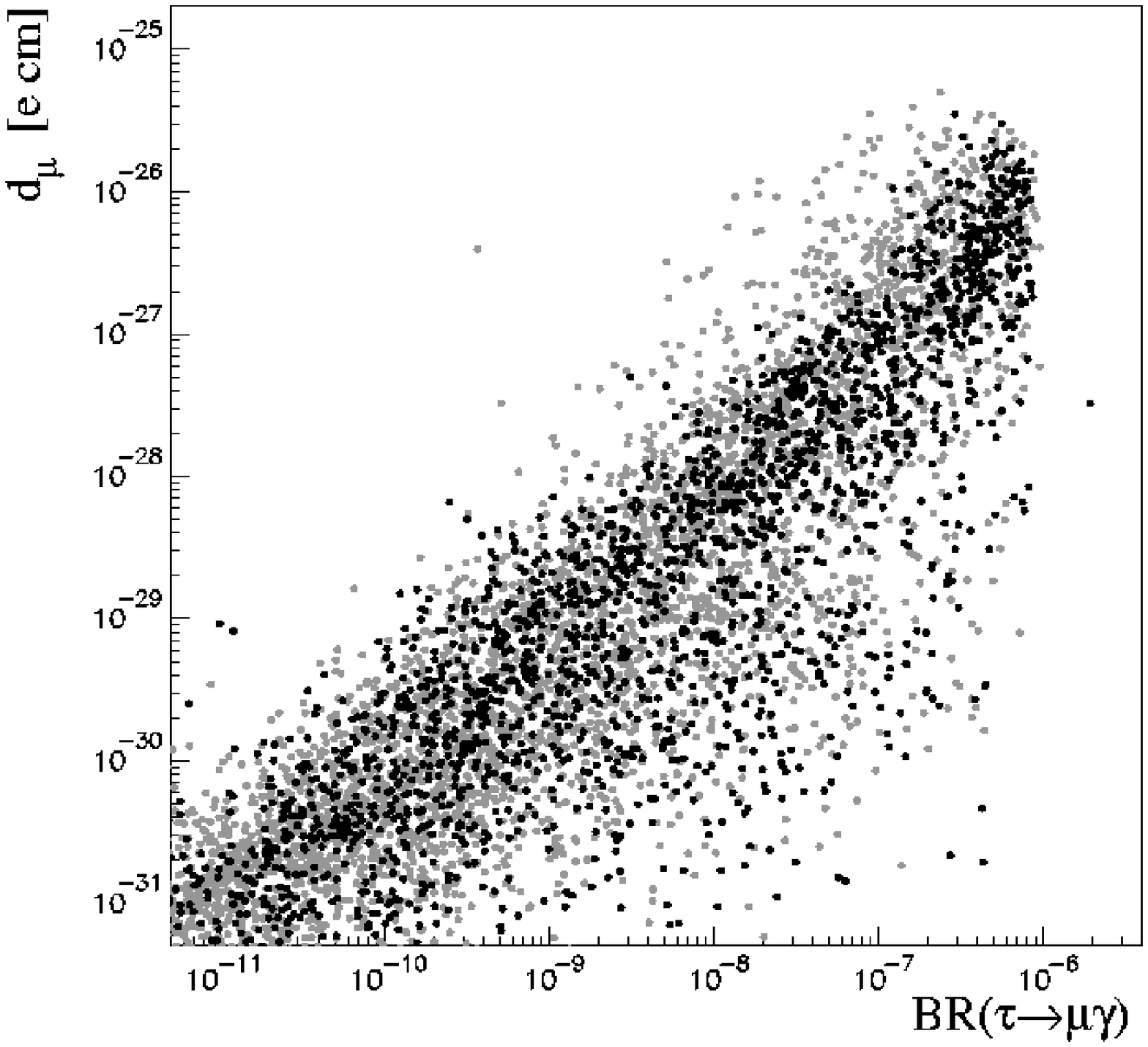}
\end{center}
\caption{"scatter plots" de $M_{N_1}$ en fonction du paramètre de $\beta\beta_{0\nu}$
$m_{ee}$, et du moment dipolaire électrique du muon en fonction du taux de
branchement de $\tau\to\mu\gamma$. La
contrainte sur l'asymétrie baryonique a aussi été prise en compte
ici.~\cite{Lepto}}
\label{lepto3}
\end{figure}

A l'heure actuelle il nous manque les informations sur les 2 premières étapes.
Nous pouvons juste explorer l'espace des paramètres. Pour chaque ($Y_{\nu}$) et
($M_N$), il est possible de calculer l'observable $m_{ee}$ de la double
désintégration
beta sans neutrinos, les EDM, les processus LFV et l'asymétrie $\epsilon_i$. Il
est aussi possible d'observer les corrélations entre tous ces paramètres et les
dépendances aux masses légères et lourdes des neutrinos. Les figures suivantes,
(\ref{lepto2}) et~(\ref{lepto3}), montrent qu'il est possible
d'obtenir le bon nombre de baryons à partir de la physique des neutrinos et de la
physique de basse énergie, en
observant les moments dipolaires électriques des leptons chargés, les désintégrations
rares $\tau\to e(\mu) \gamma$, etc. De
plus, il est possible d'obtenir des renseignements sur l'espace des paramètres
de haute énergie comme les masses $M_N$. La limite $Y_B\gtrsim 3\times 10^{-11}$
implique une limite sur la masse du neutrino lourd le plus léger :
\beqn
M_{N_1} &\gtrsim& 10^{10}\ \mathrm{GeV}\ (hierarchie\ normale) \nonumber\\
M_{N_1} &\gtrsim& 10^{11}\ \mathrm{GeV}\ (hierarchie\ inversee)
\eeqn
Mais ceci peut potentiellement entrer en conflit avec la
limite inférieure non-triviale dûe à une possible surproduction de gravitinos.
Cette surproduction modifierait alors les abondances des éléments produits
au
cours de la nucléosynthèse primordiale (He, Li,...).

La leptogenèse est un scénario important à prendre en compte et montre tout
l'intérêt de considérer les liens entre la physique des neutrinos et la
cosmologie.

\newpage
\subsection{L'inflation sneutrinique}

Un peu plus en marge du cours, mais néanmoins intéressante, est la possibilité
que {\it l'inflaton}, champ scalaire responsable de l'inflation de l'Univers, soit
incarné par l'un des partenaires supersymétriques des neutrinos singulets lourds
(voir par exemple la référence~\cite{JE-Sneutrino}).

\subsubsection{Un brève interlude sur l'inflation}
Nous allons tout d'abord commencer par quelques mots sur la cosmologie et
l'inflation.
\par\hfill\par
A l'origine de l'hypothèse de l'inflation il y a les problèmes auxquels sont
confrontés les cosmologistes : pourquoi l'Univers semble si homogène ? Pourquoi
est-il si vieux ou si vaste ? Pourquoi est-il si plat ? Pourquoi l'entropie de
l'Univers est-elle si élevée ? Pour remédier à celà, nous pouvons introduire l'idée d'une période
inflationnaire c'est-à-dire l'idée qu'à une époque de son histoire, \underline{l'expansion de
l'Univers fut quasi-exponentielle}. Cette expansion exponentielle agrandit alors
considérablement les dimensions de l'Univers et réduit la densité d'énergie (et
par conséquent la température globale).
Ceci est rendu possible par l'introduction d'un champ scalaire,
l'inflaton, qui, au tout début de l'Univers, contenait la quasi-totalité de
l'énergie.

\begin{figure}
\centerline{\includegraphics*[width=14cm]{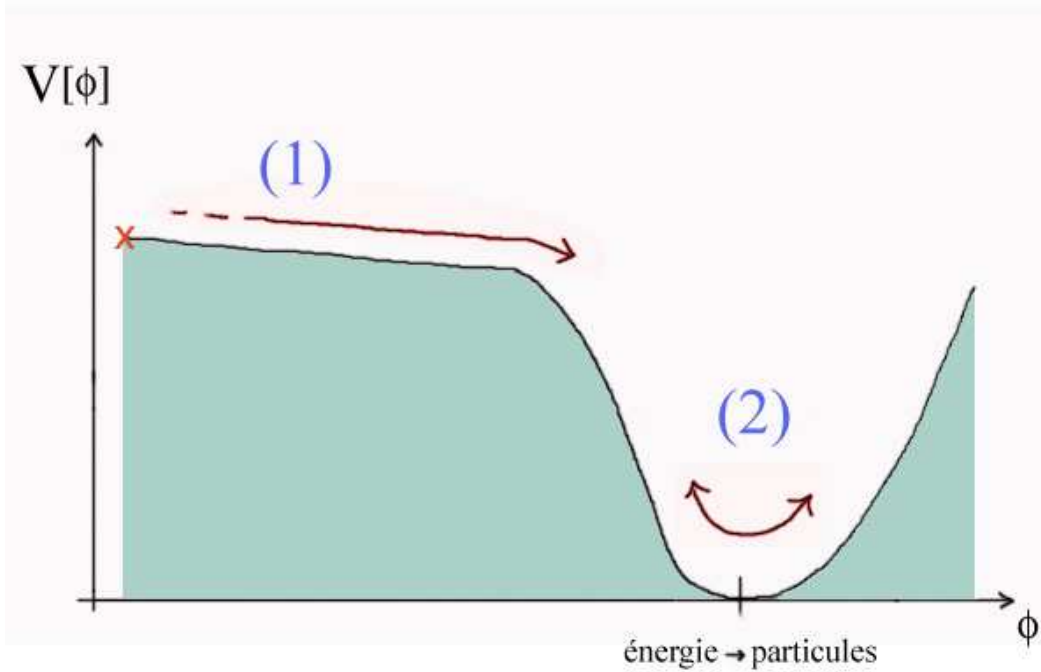}}
\caption{Forme typique du potentiel scalaire en fonction de la valeur de
l'inflaton.}
\label{potentiel-infl}
\end{figure}

La figure~(\ref{potentiel-infl}) montre une forme possible
du potentiel scalaire de l'inflaton et les différentes étapes de l'inflation. A
la première étape, le potentiel de l'inflation part d'une valeur initiale et
"roule" jusqu'au puit : l'Univers subit une expansion inflationnaire
exponentielle. Ensuite, quand l'inflaton arrive dans le puit du potentiel,
il y
passe un certain temps en oscillant jusqu'à se stabiliser. Ceci correspond au
moment où l'inflaton se désintègre en matière par la conversion de
l'énergie
initialement sous forme d'inflaton en photons et autres particules
légères. Ainsi l'univers est réchauffé jusqu'\`a une température de
réchauffement $T_{RH}$.

L'inflation donne aussi une origine aux petites fluctuations de densité observées
par WMAP (mais déjà vues par COBE), qui sont indépendantes de l'échelle de la
structure : elles viendraient de fluctuations quantiques de la valeur du
potentiel pendant l'inflation.

Il existe plusieurs
modèles d'inflation, mais l'idée présentée reste la même. Cependant, personne 
ne sait ce qu'est l'inflaton et à quelle particule il peut être identifié.  

\subsubsection{Le sneutrino, un candidat potentiel}

\begin{figure}[htbp!]
\begin{center}
\includegraphics*[width=7.5cm]{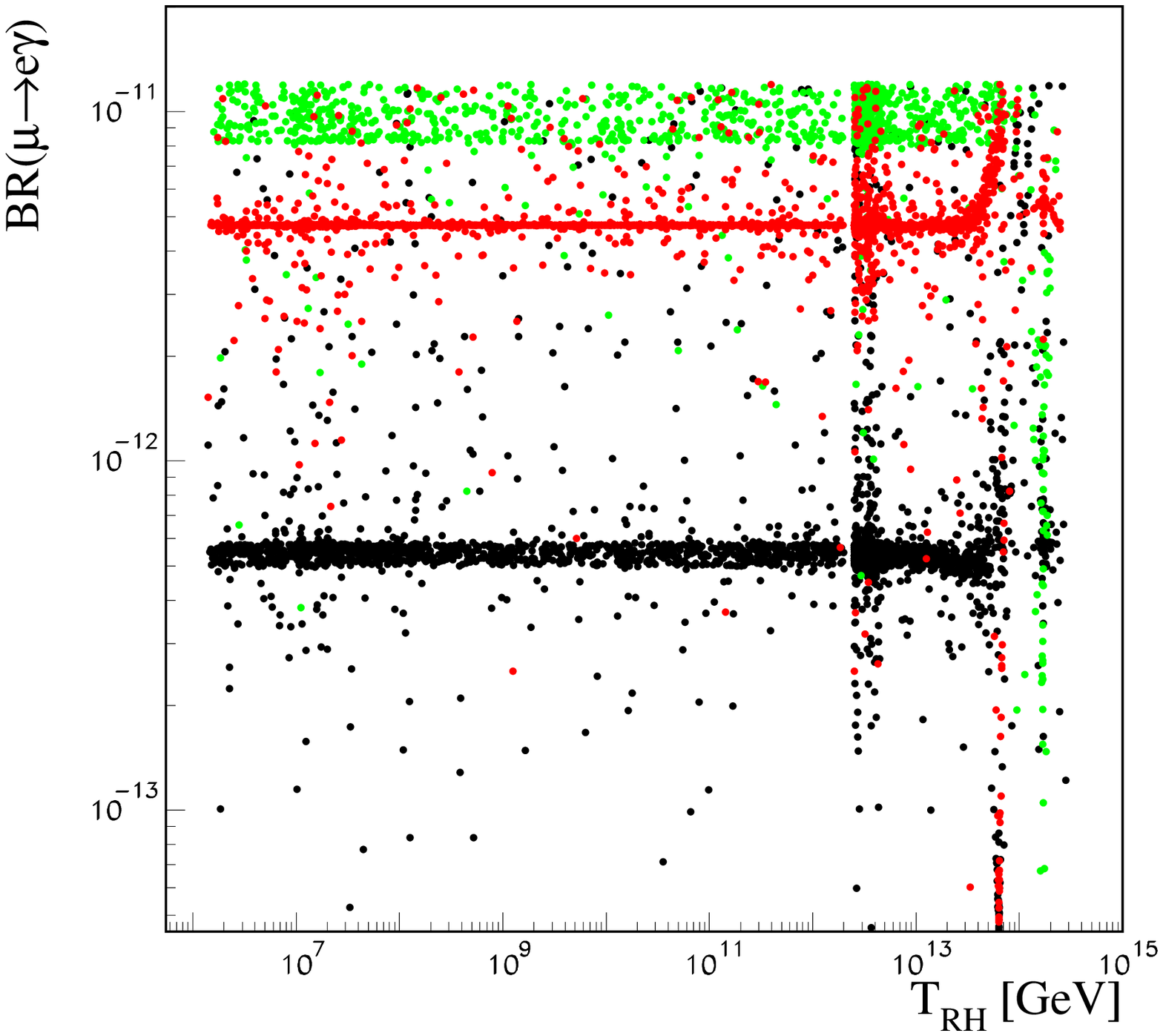}
\includegraphics*[width=7.5cm]{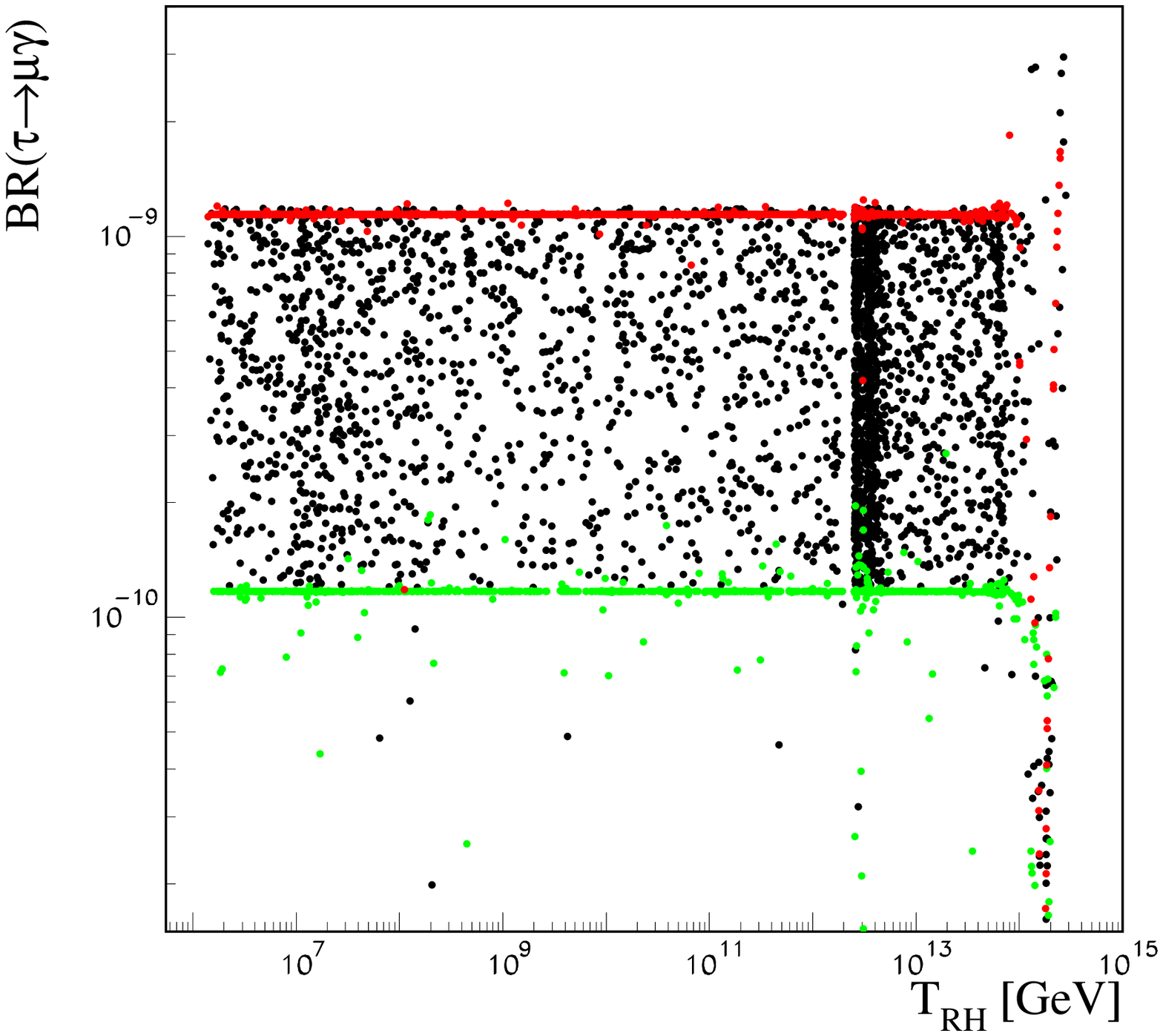}
\end{center}
\caption{$BR(\mu\to e\gamma)$ et $BR(\tau\to\mu\gamma$) en fonction de la
température de réchauffement. Les points noirs correspondent aux valeurs :
$\sin\theta_{13}=0.0$, $M_{N_2}=10^{14}$ GeV et $5\times 10^{14}$ GeV $<M_{N_3}<
5\times 10^{15}$ GeV. Les points rouges à : $\sin\theta_{13}=0.0$,
$M_{N_2}=5\times 10^{14}$ GeV et $M_{N_3}=5\times 10^{15}$ GeV. Enfin les points
verts (les points les plus clairs, en haut de la première figure et en bas de la
seconde) correspondent à : $\sin\theta_{13}=0.1$,
$M_{N_2}=10^{14}$ GeV et $M_{N_3}=5\times 10^{14}$ GeV.~\cite{JE-Sneutrino}}
\label{Infl-LFV}
\end{figure}

Dans le cas particulier d'un modèle d'inflation chaotique~\footnote{C'est un
scénario inflationnaire dans lequel il n'y a pas de structure privilégiée dans
le potentiel $V(\phi)$. Il peut être une simple puissance, $V\sim \phi^n$ ou
exponentiel, $V\sim e^{\alpha\phi}$. Une région donnée de l'Univers est supposée
commencer avec une valeur particulière de $\phi$, donc de $V$, qui décroit
ensuite de façon monotone et lente vers zéro (c'est le "slow-roll").} avec un
potentiel de la forme $V=\frac{1}{2}m_I^2\phi^2$, les données de WMAP donnent pour 
la masse de l'inflaton~\cite{Sneutrino} : 
\beq
m_I\simeq 2\times 10^{13} \ \mathrm{GeV}.
\eeq
Or cette masse correspond plus ou moins à l'intervalle de masse considéré pour les
neutrinos singulets lourds (et {\it a priori} aussi pour les sneutrinos associés)
: $M_{N_i}\sim 10^{10-15} \ \mathrm{GeV}$. De plus, l'inflaton est sensé
ne pas avoir d'interaction de jauge, ce qui peut très bien être le cas des
sneutrinos lourds. Si un de ces sneutrinos est l'inflaton, alors le problème
cosmologique de l'inflation sera connecté au reste de la physique des
particules. De plus, cette hypothèse peut être utilisée pour donner des prédictions sur les
désintégrations violant le nombre leptonique car elle contraint
significativement les paramètres du modèle seesaw supersymétrique minimal. 
Dans 
la figure~(\ref{Infl-LFV}),
nous voyons entre autres que le processus $\mu\to e\gamma$ est très proche de la
limite actuelle et dans les possibilités observationnelles des futures
expériences.  Nous observons aussi que les taux de branchement de $\mu\to e\gamma$ 
et $\tau\to\mu\gamma$ semblent indépendants de la température de réchauffement (pour
$T_{RH}<10^{12}$ GeV) et que $BR(\mu\to e\gamma)$ est très sensible à
$\sin\theta_{13}$ et $M_{N_3}$. 

\newpage
\section{Conclusions}

\begin{figure}[htbp!]
\vspace{1cm}
\begin{center}
\begin{picture}(400,300)(-200,-150)
\Oval(0,0)(35,70)(0)
\Text(-25,13)[lb]{ ${\bf Y_\nu}$  ,  ${\bf M_{N_i}}$}
\Text(-50,-2)[lb]{{\bf $\Rightarrow$ 18 paramètres}}
\Text(-30,-15)[lb]{{\bf physiques}}
\EBox(-80,90)(90,150)
\Text(-60,135)[lb]{{\bf Mécanisme de Seesaw :}}
\Text(-8,117)[lb]{${\bf {\cal M}_\nu}$}
\Text(-65,100)[lb]{{\bf $\Rightarrow$ 9 paramètres effectifs}}
\EBox(-200,-140)(-60,-80)
\Text(-167,-95)[lb]{{\bf Leptogenèse :}}
\Text(-165,-113)[lb]{ ${\bf Y_\nu Y_\nu^\dagger}$ , ${\bf M_{N_i}}$}
\Text(-175,-130)[lb]{{\bf $\Rightarrow$ 12 paramètres}}
\EBox(60,-140)(200,-80)
\Text(80,-95)[lb]{{\bf Renormalisation :}}
\Text(95,-113)[lb]{${\bf Y_\nu^\dagger L Y_\nu}$ , ${\bf M_{N_i}}$}
\Text(80,-130)[lb]{{\bf $\Rightarrow$ 16 paramètres}}
\LongArrow(0,35)(0,87)
\LongArrow(-45,-25)(-130,-77)
\LongArrow(45,-25)(130,-77)
\end{picture} 
\end{center}

\caption{La reconstruction des paramètres du mécanisme de Seesaw par les
différentes observables.}
\label{Seesaw}
\end{figure}
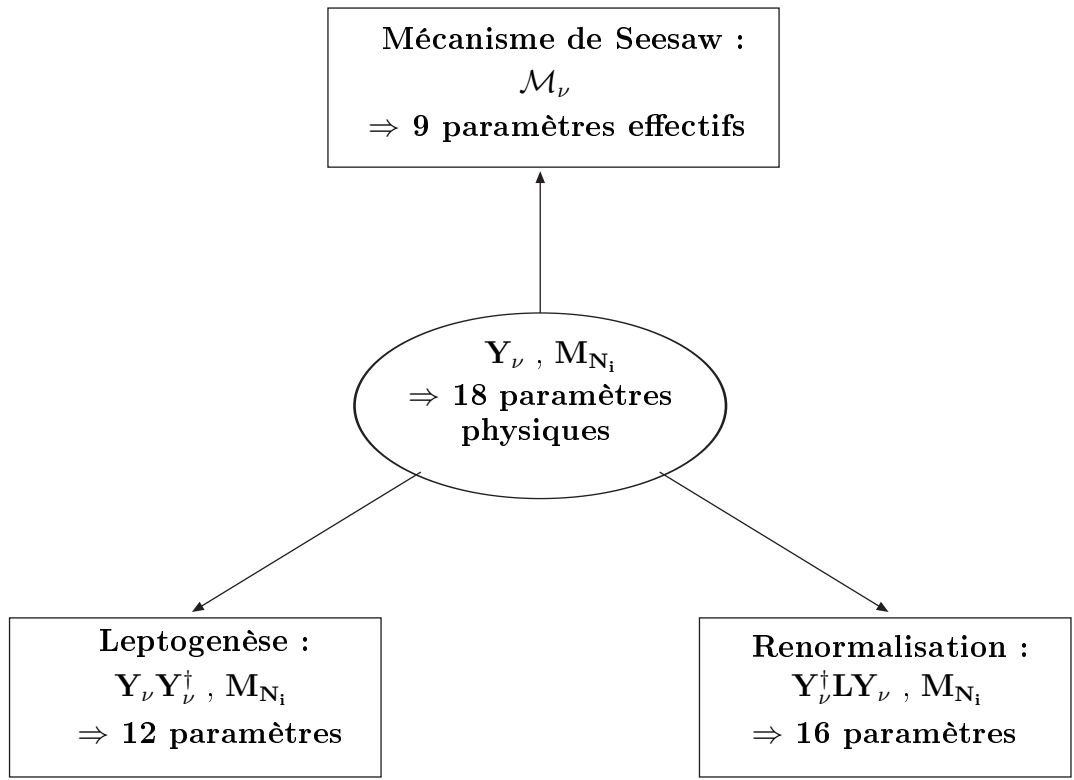

Le mélange des neutrinos observé {\it via} les oscillations peut avoir de
nombreuses conséquences, notamment dans un cadre supersymétrique où la masse si
faible des neutrinos tire son origine du mécanisme de Seesaw. Par exemple, les taux
de branchement des 
processus violant la saveur leptonique et les moments dipolaires électriques 
leptoniques calculés dans ce
modèle seesaw supersymétrique minimal montrent que la violation des nombres 
leptoniques et la violation de $CP$ leptonique peuvent avoir lieu de manière 
significative. Les futures expériences devraient avoir
la possibilité de mesurer pour la première fois ces violations. Ceci donne alors
de grandes chances de pouvoir reconstruire les paramètres de haute énergie de la
physique des neutrinos dans ce modèle, inaccessibles depuis les expériences qui
observent les oscillations des neutrinos (figure~(\ref{Seesaw})).  
De plus, la physique des neutrinos permet aussi de résoudre certains problèmes 
de nature cosmologique grâce aux hypothèses telles que la leptogenèse et
l'inflation sleptonique. 

\chapter{Le grand "Au-delà"}

\section{La Grande Unification}

Les théories de jauge (surtout non-abéliennes, appelées théories de
Yang-Mills) semblent offrir un cadre unique pour décrire
la physique des particules. Dans le MS, il y a 3 couplages différents (si la
gravitation
n'est pas comptée) associés aux trois groupes de jauge $SU(3)_C$, $SU(2)_L$ et 
$U(1)_Y$
et certains paramètres sont ajustés arbitrairement pour rendre compte des
diverses observations expérimentales. Un prolongement logique du MS est de
considérer l'existence à plus haute énergie d'une symétrie plus grande. Celle-ci
pourrait permettre de relier les différents paramètres et d'unifier les 3 
couplages~\footnote{On note les couplages ainsi : $g_1$ pour le couplage de
$U(1)$, $g_2$ pour $SU(2)$ et $g_3$ pour $SU(3)$. Dans les chapitres précédents
on avait $g$ pour $g_2$ et $g'$ pour $g_Y$.} $g_3$, $g_2$, $g_Y$. La philosophie de la Grande Unification (GU) est 
justement de chercher un groupe de jauge G simple qui comprend  
$SU(3)_C \otimes SU(2)_L\otimes U(1)_Y$. Ainsi, les 3 forces fondamentales ( 
électromagnétique, faible, forte) se trouveraient unifiées en une force 
"électronucléaire" de couplage unique $g_{GU}$. Ceci semble du
point de vue le plus courant en physique des particules plus 
"satisfaisant" car 
cela va dans le sens d'une physique plus "simple" aux échelles de longueurs plus 
petites. Cependant, les effets de l'interaction gravitationnelle sont supposés 
être toujours 
négligeables, au moins en première approximation. Si l'échelle d'unification 
$M_{GU}$ se trouve 
être significativement plus petite que la masse de Planck alors cette 
supposition est justifiée. Il se trouve que les estimations typiques, basées sur 
une extrapolation à très haute énergie de la physique connue (c'est-à-dire le MS),
donnent une échelle de l'ordre de $10^{16}$ GeV, environ mille fois plus petite 
que 
l'échelle de Planck $M_{Pl}=\mathcal{O}(10^{19})$ GeV. 

Le fait d'avoir un unique 
groupe de jauge pour décrire la physique à cette échelle implique aussi des 
relations tout à fait inédites entre les particules, ainsi que de nouveaux 
bosons de jauge. En 
effet, la symétrie change donc l'organisation des particules (les multiplets) 
change. Certains indices comme la quantification de la charge ({\it i.e.}
l'existence de charges électriques fractionnaires) ou la compensation 
des anomalies~\footnote{C'est-à-dire le fait que les anomalies dues aux leptons s'annulent
exactement avec celles dues aux quarks.} font aussi penser à une organisation 
plus simple que celle du MS.
Bien sûr, à basse énergie, nous devons retrouver le Modèle Standard, et dans ces
Théories de Grande Unification (GUT) nous devrons donc étudier la brisure du 
groupe de jauge G en $SU(3)_C \otimes SU(2)_L\otimes U(1)_Y$. 

Cette section commence donc par une présentation des équations d'évolution des 
3 couplages et de leur unification. Nous exposerons ensuite quelques exemples de
modèles dont le "prototype" basé sur le groupe $SU(5)$ permet d'aborder 
beaucoup de propriétés des GUT sans alourdir la discussion. Puis nous
étudierons les prédictions typiques de ces modèles comme la désintégration du
proton et les relations entre les masses des quarks et leptons. Nous terminerons
en discutant des avantages, des
problèmes et des perspectives des modèles basés sur l'idée d'une Grande Unification
à haute énergie.

\subsection{Les équations d'évolution des couplages $g_i$}

\subsubsection{L'unification des couplages}

\begin{figure}[htbp!]
\centerline{\psfig{figure=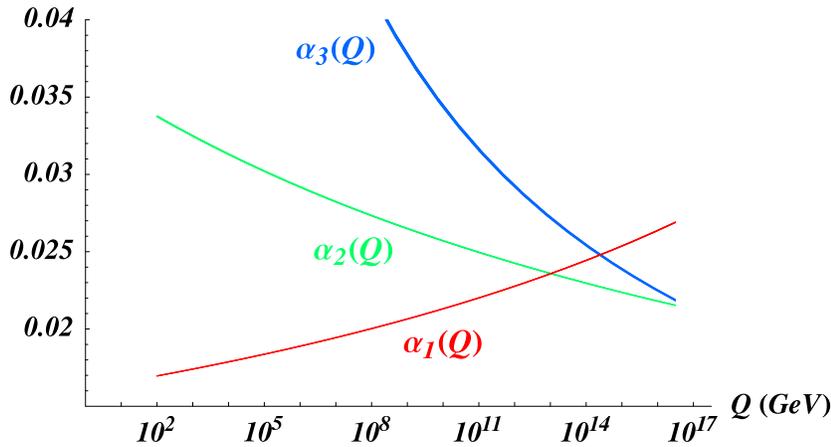,height=2.7in}}
\caption{Evolution des couplages de jauge en fonction de
l'énergie, dans le MS. La largeur des courbes correspond à l'erreur
expérimentale.~\cite{Masina}}
\label{fig:gaugeunificationMS}
\end{figure}

L'un des résultats les plus importants de la renormalisation est l'évolution 
des constantes de couplage des forces en fonction de l'échelle d'énergie :
$g_i\equiv g_i(\mu)$.
En 
supposant que le MS reste valable jusqu'à des énergies de l'ordre de la masse de 
Planck nous pouvons résoudre analytiquement ces équations du groupe de 
renormalisation (RGE) et 
observer l'évolution des 3 couplages des 3 interactions fondamentales du MS. Le 
résultat, surprenant, initialement obtenu dans les années 70 est que les 
couplages évoluent 
logarithmiquement (nous justifierons un peu plus tard cette remarque) jusqu'à se 
croiser au même point (\`a peu pr\`es) et donc atteindre une valeur 
commune. Cette observation 
légitimise la philosophie de la Grande Unification. 
En réalité, depuis la première fois où le calcul a été fait, la précison 
expérimentale a grandement augmenté et les couplages ne se croisent plus tout-à-fait 
au même point. La figure~(\ref{fig:gaugeunificationMS}) montre clairement ce 
"rendez-vous" manqué. 

\begin{figure}[htbp!]
\centerline{\psfig{figure=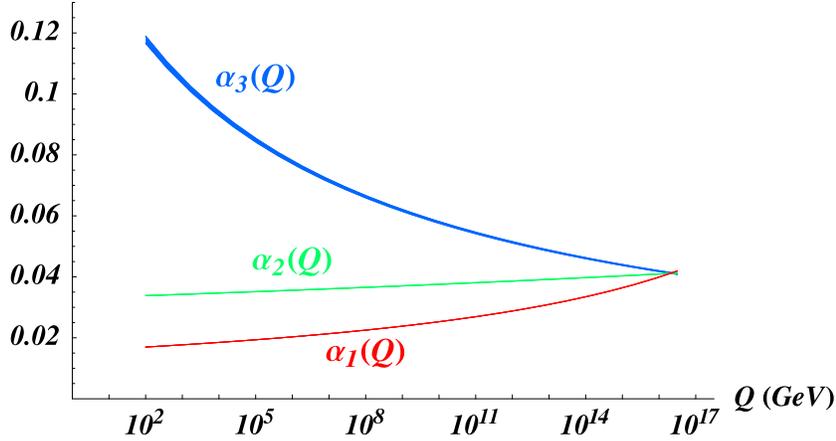,height=2.7in}}
\caption{Evolution des couplages de jauge en fonction de
l'énergie, dans le MSSM.~\cite{Masina}}
\label{fig:gaugeunificationMSSM}
\end{figure}

Cela veut-il dire que nous devons oublier la GU ? Non, simplement parce que 
nous ne connaissons 
pas toute la physique au-delà du MS et que pour obtenir la 
figure~(\ref{fig:gaugeunificationMS}) nous avons 
fait une supposition qui n'était sans doute pas justifiée : le MS est valable jusqu'à $M_{Pl}$. 
Au chapitre 1 nous nous sommes justement efforcés de montrer que le MS ne
pouvait 
pas être valable au-delà d'une échelle d'énergie, de l'ordre du TeV. Il y a de la 
nouvelle physique aux alentours de cette échelle et celle-ci peut changer 
drastiquement les équations d'évolution des constantes de couplage. De plus, au 
chapitre 2 nous avons vu que la supersymétrie était une candidate très sérieuse
pour cette nouvelle physique 
et celle-ci introduit un grand nombre de nouvelles particules qui 
interviennent dans la renormalisation des couplages. Le calcul des RGE dans le 
cadre du MSSM donne alors la figure~(\ref{fig:gaugeunificationMSSM}). Le résultat 
est très surprenant lui aussi
puisque nous retrouvons l'unification des couplages ! La supersymétrie semble 
donc un cadre approprié pour la GU. 

Pour "mesurer la qualité" de cette unification, on 
adopte souvent un point de vue inverse en partant de l'unification de $\alpha_1$ 
et  $\alpha_2$, en supposant l'unification avec $\alpha_3$ exacte et au même point, 
puis en calculant alors l'évolution de $\alpha_S$ vers les basses
énergies~\footnote{En effet, les 3 couplages de basse énergie sont maintenant
fonction de 2 paramètres indépendants, $\alpha_{GU}$ et $M_{GU}$, il y a donc
\underline{une prédiction} possible.}. Nous
pouvons alors comparer la valeur de $\alpha_3$ (à $M_Z$ par
exemple) avec la valeur expérimentale et vérifier la compatibilité. Dans le MSSM, 
pour une unification à $2\times 10^{16}$ GeV, nous trouvons alors la
valeur~\cite{HN} : 
\beq 
\alpha_3(M_Z)^{MSSM}=0.130\pm0.004+\Delta_{SUSY}+\Delta_{GUT}, 
\label{alpha_S}
\eeq
où $\Delta_{SUSY}$ et $\Delta_{GUT}$ sont des corrections qui dépendent du modèle. 
Les données expérimentales donnent la valeur suivante : 
\beq 
\alpha_3(M_Z)^{exp}=0.117\pm0.002,
\eeq
qui est en assez bon accord avec la prédiction. Sans les corrections
$\Delta_{SUSY}$ et $\Delta_{GUT}$, l'accord ne serait pas parfait, ce qui 
voudrait dire que l'unification à haute énergie des 3 couplages ne serait pas 
tout à fait exacte. Mais il ne faut pas oublier que dans le calcul des RGE 
nous nous sommes arrêtés à une certaine précision (à un ordre donnée de la 
théorie des perturbations). 
La correction $\Delta_{SUSY}$ est un effet de seuil à basse énergie,
venant de l'échelle de brisure effective (douce) de la supersymétrie.
En plus, si effectivement il y a une Grande Unification et donc des nouvelles 
particules aux hautes \'energies, celles-ci devaient se faire "sentir" quand 
nous nous rapprochons de l'échelle $M_{GU}$. Plus précisemment, aux échelles 
d'énergie proches de $M_{GU}$, les boucles virtuelles de ces bosons ont un 
effet non négligeable, car les masses de ces bosons ne sont plus négligeables 
devant l'échelle d'énergie où l'on se trouve. On parle d'effet de seuil GUT, 
représenté par la quantité $\Delta_{GUT}$ de l'équation~\ref{alpha_S}. 
Malheureusement, il dépend de la physique à l'échelle $M_{GU}$, et pour 
calculer quantitativement sa valeur il faut se placer dans un modèle donné. 
On parle de "sensibilité à la théorie UV"~\footnote{UV pour ultra-violette, 
c'est-à-dire de plus haute énergie.} de l'unification des couplages. 

Une remarque importante : par rapport au MS, dans un modèle GUT le couplage de 
l'hypercharge a subit une modification d'un facteur $\sqrt{5/3}$ ; c'est en réalité la 
relation $g_1=\sqrt{\frac{5}{3}}g_{Y}=g_2=g_3=g_{GU}$ que nous avons à 
l'échelle $M_{GU}$. Nous verrons un peu plus tard pourquoi ce facteur a 
été introduit. Maintenant, nous allons être un peu plus technique et donner ces 
RGE dans les cas discutés à savoir le MS et le MSSM.

\subsubsection{Les RGE}

Les équations du groupe de renormalisation pour les trois couplages des trois
interactions non-gravitationnelles se présentent à une boucle sous la forme :
\beq \frac{{\mathrm d}g_i}{{\mathrm dt}}=\beta_i(g_i)\equiv b_i
\frac{g_i^3}{16\pi^2},\ \ \ i=1,2,3, \eeq
où $t=\ln(\mu)$ avec $\mu$ une échelle d'énergie. Nous utiliserons plus souvent 
$\alpha_i$ à la place des $g_i$ où $\displaystyle\alpha_i=\frac{g_i^2}{4\pi^2}$. 
Les RGE s'écrivent alors :
\beq \frac{{\mathrm d}\alpha_i}{{\mathrm dt}}= b_i
\frac{\alpha_i^2}{2\pi},\ \ \ i=1,2,3. \eeq
Le tableau~\ref{bi} donne les différentes valeurs de $b_i$ dans le cas du MS (un seul
doublet de Higgs et 3 générations) et du MSSM. Les expressions où le nombre de doublets de Higgs
et de familles de quarks sont gardés explicites peuvent se trouver
dans~\cite{JE-Beyond}.

\begin{table}[htbp!]
\begin{center}
\begin{tabular}[t]{|c||c|c|c|}
\hline
& {$\ b_1\ $}& {$\ b_2\ $}& {$\ b_3\ $}	\\
\hline
\hline
& & &\\
MS & $\displaystyle\frac{41}{10}$ & $-\displaystyle\frac{19}{6}$ & -7 \\
& & &\\
\hline
\hline
& & &\\
MSSM & $\displaystyle\frac{33}{5}$ & 1 & -3\\
& & &\\
\hline
\end{tabular}
\end{center}
\caption[fonctions b]{ Les valeurs de $b_1,\ b_2,\ b_3$ dans le MS et le MSSM.}
\label{bi}
\end{table}

Ces équations différentielles peuvent se résoudre analytiquement et les 
solutions s'écrivent ainsi : 
\beq \alpha^{-1}_i(\mu)=\alpha^{-1}_i(M_{GU})+
\frac{b_i}{2\pi}\ln\left(\frac{M_{GU}}{\mu}\right) \eeq
Avec ces solutions, nous pouvons donc maintenant vérifier à quelle précision 
et à quelle échelle les couplages s'unifient, dans le MS et le MSSM. En général, on
utilise la prédiction de $\sin^2\theta_W$ ou celle de $\alpha_S$ à $m_Z$. Nous 
n'introduisons aucunes corrections d\^ues aux effets de seuil d'un 
modèle GUT particulier ou aux effets de seuil à basse énergie et nous supposons 
la relation 
$\alpha_3=\alpha_2=\alpha_1=\alpha_{GU}$ à $M_{GU}$. En prenant comme échelle
$\mu$ l'échelle $m_Z$,
\beq \sin^2\theta_W(m_Z)=\frac{g_{Y}^2(m_Z)}{g^2(m_Z)+g_{Y}^2(m_Z)}, \eeq
et en utilisant les relations~\footnote{Le facteur 5/3 vient de la normalisation
du générateur de la charge électrique dans les modèles GU. Dans le modèle
standard, $Q=T_3+Y/2$. Dans les GUTs,
$Q=\frac{1}{2}(T^{11}+\sqrt{\frac{5}{3}}T^{12})$ où $T^{11}$ et $T^{12}$ sont
les deux générateurs de $SU(2)_L\otimes U(1)_Y$ du groupe d'unification.
L'isopin et l'hypercharge font partie d'un seul groupe simple $G$ et ne sont pas
indépendants. L'hypercharge est alors redéfinie par un facteur $\sqrt{\frac{5}{3}}$ 
que l'on répercute sur le couplage $g_Y$.}
\beq \alpha_1(\mu)=\frac{5}{3}\frac{g_{Y}^2(\mu)}{4\pi}, \ \ \
\alpha_2(\mu)=\frac{g^2(\mu)}{4\pi}, \eeq
et la combinaison particulière :
\beq \alpha^{-1}_2(\mu)-\alpha^{-1}_1(\mu)=
\frac{b_2-b_1}{2\pi}\ln\left(\frac{M_{GU}}{\mu}\right), \eeq
on trouve dans le MS:
\beq 
\sin^2\theta_W(m_Z)\simeq
\frac{3}{8}\left(1-\frac{55\alpha_1(m_Z)}{24\pi}\ln\left(\frac{M_{GU}}{m_Z}\right)\right),
\eeq
et dans le MSSM :
\beq 
\sin^2\theta_W(m_Z)\simeq
\frac{3}{8}\left(1-\frac{7\alpha_1(m_Z)}{4\pi}\ln\left(\frac{M_{GU}}{m_Z}\right)\right).
\eeq
A noter que dans le cas du MSSM, nous ne connaissons pas les masses des
sparticules. En les mettant toutes à 1 TeV on retrouve approximativement la
valeur de $\sin^2\theta_W$ observée expérimentalement. 

Pour examiner ces prédictions avec plus de détails nous pouvons étudier
les équations du groupe de renormalisation jusqu'à l'ordre de 2 boucles. Dans le
MSSM, l'unification est cependant stable car les corrections à 2 boucles sont
faibles devant celles à 1 boucle, même si cela reste moins vrai pour $\alpha_3$. 
Le résultat est une unification encore plus précise c'est-à-dire une prédiction
de $ \sin^2\theta_W(m_Z)$ plus proche de la valeur expérimentale. On peut
retrouver les expressions à 2 boucles et plus de détails sur les effets de seuil
des masses des sparticules dans~\cite{JE-Beyond}.

\subsection{Les modèles Grand Unifiés}

\subsubsection{Le choix du groupe d'unification}
Quels sont les groupes susceptibles de nous intéresser pour construire une
théorie Grande Unifiée?

Tout d'abord, ces groupes sont des groupes dits "de Lie" suffisamment "grands" 
pour englober entièrement celui du Modèle Standard. Ce dernier 
est de rang 4, c'est-à-dire que le nombre de générateurs de la symétrie 
simultanément diagonalisables~\footnote{Chacun est associé à un
nombre quantique, une "charge", et les états de particules observés sont indexés par ceux-ci. A
noter que ces générateurs diagonaux sont de trace nulle.} est égal à 4 : 
$SU(3)_C$ en contient 2, $SU(2)_L$ 
un seul et $U(1)_Y$ aussi. Il nous faut donc trouver dans la classification des 
groupes de Lie, faite par Cartan, un groupe de rang supérieur ou égal à 4. 
Enfin, ils doivent comporter des représentations 
complexes pour que les fermions de chiralité différentes puissent être dans des 
représentations différentes. 

Les deux possibilités de rang 4 sont: $SU(5)$ et $SU(3)\otimes SU(3)$. Mais  
$SU(3)\otimes SU(3)$ ne permet pas simultanément aux leptons d'avoir une charge électrique 
entière et aux quarks d'avoir une charge électrique fractionnaire. Le groupe 
$SU(5)$ est donc le groupe le plus simple possible capable d'englober le Modèle 
Standard. Les autres groupes possibles et couramment utilisés mais de rangs 
supérieurs à 4 sont $SO(10)$, de rang 5, et le groupe exceptionnel $E_6$, de
rang 6.
Nous le voyons, les considérations physiques précédentes contraignent fortement 
la liste des groupes de Lie compatibles. A titre d'exemple et pour 
bien comprendre comment peut changer la physique quand on change la symétrie du 
MS nous allons étudier certains aspects du groupe $SU(5)$ puis plus brièvement
ceux du groupe $SO(10)$.

\subsubsection{Le groupe $SU(5)$}
Tout comme dans le Modèle Standard, il nous faut "ranger" nos particules dans 
les "casiers" disponibles de $SU(5)$. Dans $SU(5)$ nous disposons~\footnote{Une 
façon simple de les trouver et de "jouer" avec les tableaux d'Young.} 
d'une représentation spinorielle fondamentale de dimension 5 et une 
représentation spinorielle antisymétrique de dimension 10 pour les fermions 
d'une 
génération. Mais pour répartir ces quinze fermions ($3\times2\times2=12$ quarks + 2 leptons chargés + 
1 neutrino), il y a quelques règles à
respecter. Tout d'abord, nous pouvons exprimer les représentations de $SU(5)$ 
en terme des représentations de $SU(3)\otimes SU(2)$ : 
\beqn
\bf{5} &=& (\bf{3},\bf{1}) + (\bf{1},\bf{\bar{2}}) \label{rep 5 de SU(5)}\\
\bf{\bar{5}} &=& (\bf{\bar{3}},\bf{1}) + (\bf{1},\bf{2}) \\
\bf{10} &=& (\bf{\bar{3}},\bf{1}) +(\bf{3},\bf{2})+ (\bf{1},\bf{1}).
\eeqn
Par exemple (\ref{rep 5 de SU(5)}) veut dire que dans la représentation {\bf 5} de 
$SU(5)$, nous
devons mettre un triplet de couleur, singulet de $SU(2)$, et un singulet de couleur,
anti-doublet de $SU(2)$. De plus, il faut que la somme des charges dans chacun 
de ces 2 multiplets soit nulle\footnote{La charge électrique est une combinaison
linéaire des générateurs diagonaux (dans le MS, on a $Q=T_3 +Y$), qui sont de
trace nulle, donc est de trace nulle aussi.}. La seule combinaison possible qui 
soit en accord avec le MS~\footnote{Le choix des signes est fait en accord avec la 
définition
de la conjugaison de charge et de façon à obtenir les bonnes interactions du MS.} 
est :
\beq {\bf \bar{5}} : (\psi_i)_L=\left(\begin{array}{c} \bar{d}_1 \\ \bar{d}_2 \\
\bar{d}_3 \\ e^- \\ -\nu_e \end{array}\right)_L, \eeq
et donc pour le reste des fermions de la premi\`ere g\'en\'eration:
\beq {\bf 10} : (\chi^{ij})_L=\frac{1}{\sqrt{2}}\left(
\begin{array}{ccccc} 
0 & \bar{u}_3 & -\bar{u}_2 & u_1 & d_1 \\ 
-\bar{u}_3 & 0 & \bar{u}_1 & u_2 & d_2 \\ 
u_2 & -\bar{u}_1 & 0 & u_3 & d_3 \\ 
-u_1 & -u_2 & -u_3 & 0 & e^+ \\ 
-d_1 & -d_2 & -d_3 & -e^+ & 0 
\end{array}\right)_L , \eeq
o\`u nous négligeons les éventuels mélanges entre les familles.
Comme une famille entière gauche entre dans ($\bf{10}+\bf{\bar{5}}$), nous devons donc 
répliquer la classification précédente pour les 3 familles
et en conséquence nous ne pouvons pas donner d'explication de la 
présence de 3 familles et de leurs différences dans $SU(5)$.

\par\hfill\par
Nous venons de parler des fermions, nous allons maintenant aborder les bosons de
jauge. Les groupes de type $SU(N)$ ont $N^2-1$ générateurs de symétries donc 
$SU(5)$ a 24 bosons de jauge ($SU(3)_C$ a 8 gluons, $SU(2)$ a 2 $W$ et un $Z$ etc).
Parmis ces 24 bosons de jauge, 12 correspondent au MS et 12 sont nouveaux. Ils
appartiennent toujours à la représentation adjointe du groupe de jauge, et celle-ci
est de dimension 24 pour $SU(5)$. Si comme pour les fermions nous décomposons la
représentation adjointe en représentations de $SU(3)\otimes SU(2)$, et que nous
identifions les bosons déjà connus nous obtenons :
\beq
{\bf 24} =\underbrace{({\bf 3},{\bf 2},\frac{5}{3}) \oplus ({\bf \bar{3}},{\bf 2},-\frac{5}{3})
}_{nouveaux\ bosons}  \oplus \underbrace{({\bf 8},{\bf 1},0)}_{gluons\ G_i} \oplus 
\underbrace{({\bf 1},{\bf 3},0)}_{W_i} \oplus \underbrace{({\bf 1},{\bf
1},0)}_{B}.
\eeq
Le 3ème chiffre dans les parenthèses est l'hypercharge du sous-multiplet.
Les nouveaux bosons sont appelés les \textit{leptoquarks} $X$ et $Y$, sont chargés
électriquement~\footnote{Respectivement 4/3 et 2/3.}, sont colorés et ont 
un isospin 1/2. Il peut donc y avoir des
interactions directes entre quarks et leptons, nous en verrons les conséquences
juste après.

En notation matricielle, 
\beq A=\sum_{a=1}^{24} T_a A^a=\left(
\begin{array}{ccccc}
G_i&G_i&G_i& \bar{X} & \bar{Y} \\
G_i& G_i &G_i& \bar{X} & \bar{Y} \\
G_i&G_i&G_i& \bar{X} & \bar{Y} \\
X & X & X &W_i&W_i\\
Y & Y & Y &W_i& W_i\\
\end{array}\right),
\eeq
où les $T_a$ sont les générateurs de SU(5) représentés par des matrices
$5\times5$ (l'équivalent pour SU(5) des matrices de Pauli de SU(2)). Le choix de
la base étant libre, nous avons choisi de les écrire telles que $SU(3)_C$
agisse sur les 3 premières lignes et colonnes et $SU(2)_L$ sur les 2 dernières.
D'où les deux blocs avec les gluons et les W. Le boson $B$ se situe sur la
diagonale.

\par\hfill\par
Puisque $SU(5)$ n'est qu'une symétrie valable à très haute énergie, elle se brise
à plus basse énergie pour donner le MS. De plus, les multiplets fermioniques
étant ceux de $SU(5)$ nous devons adapter le secteur de Higgs et trouver les
bons multiplets qui brisent la symétrie électrofaible.
Pour briser $SU(5)$ nous pouvons choisir (c'est le choix le plus simple) un
multiplet adjoint $\Phi$, de dimension {\bf 24}, de bosons de Higgs. La
\textit{v.e.v} de
$\Phi$ qui préserve la couleur (3 premières lignes et colonnes), l'isospin (2
dernières) et l'hypercharge (la diagonale) s'écrit :
\beq
<0|\Phi|0>=\frac{v_{\Phi}}{2}\left( 
\begin{array}{ccccc}
2&0&0&0&0 \\
0&2&0&0&0 \\
0&0&2&0&0 \\
0&0&0&-3&0 \\
0&0&0&0&-3 \\
\end{array}\right).
\eeq
Cette première étape rend les bosons $X$ et $Y$ massifs en laissant les bosons du MS
de masses nulles.
\beq m_X^2=m_Y^2=\frac{25}{8}g_5^2 v_{\Phi} \eeq
La \textit{v.e.v} $v_{\Phi}$ doit donc être de l'ordre de $10^{15-16}$ GeV ($\sim M_{GU}$).
La brisure électrofaible se
fait dans $SU(5)$ de façon minimale par un multiplet {\bf 5} qui contient un
triplet de Higgs colorés et le doublet de Higgs du MS : 
\beq H_j[\bf{5}]= (H_C,\ H). \eeq
La \textit{v.e.v} la plus simple qui brise seulement $SU(2)_L\otimes U(1)_Y$ est de la 
forme :
\beq <0|H|0>=v_{H}\left( 
\begin{array}{c}
0 \\
0 \\
0 \\
0 \\
1 \\
\end{array}\right).
\eeq
Les bosons électrofaibles sont alors rendus massifs et de l'ordre de $v_H$, 
\beq m_W^2 = g_2^2 v_H^2, \eeq
et les fermions acquièrent une masse de l'ordre de $v_H$ grâce aux couplages 
de Yukawa.

Le potentiel effectif est donc fonction
des champs $H$ et $\Phi$. En toute rigueur, outre une partie responsable de la
première étape et une autre partie responsable de la brisure électrofaible (de
la même forme que le potentiel du MS), nous devons introduire un terme qui
mélange les deux champs. Le traitement des brisures est donc un peu plus 
complexe.
Mais puisque les 2 échelles de brisure sont séparées par $10^{13}$ GeV, elle
peuvent être considérées raisonnablement indépendantes. Cependant, cette
hiérachie entre les masses des Higgs requiert un ajustement extrêmement fin des
paramètres du potentiel, de 13 ordres de grandeurs, ce qui paraît peu naturel.
En effet, comme nous l'avons discuté au premier chapitre, la masse du Higgs $H$,
va recevoir des corrections venant des boucles de particules de masse $\sim
M_{GU}$. Pour que $m_H$ reste de l'ordre de 100 GeV, il faut des annulations
accidentelles extrêmement précises entre des contributions de l'ordre 
de $M_{GU}$. Ce problème peut être résolu par l'apport de la
supersymétrie qui justifie ces annulations.

Autre nouveauté de ce secteur de Higgs : l'existence de Higgs colorés. Leurs
échanges  
conduisent à de nouvelles interactions qui n'existaient pas dans le MS qui
peuvent se révéler très importantes (voir plus loin). 
De plus, d'autres multiplets de Higgs peuvent exister si nous voulons briser 
$SU(5)$ en 
$SU(3)_C \otimes SU(2)_L\otimes U(1)_Y$. Le secteur de Higgs peut être beaucoup 
plus compliqué dans les GUTs que dans le MS et peut faire intervenir des
représentations plus grandes.

\subsubsection{Une autre possibilité : $SO(10)$}

Une autre possibilité très souvent étudiée est le groupe d'unification $SO(10)$.
C'est un groupe de rang 5,   
\beq SU(5)\subset SO(10). \eeq
L'avantage principal de $SO(10)$ sur $SU(5)$ est qu'il possède une représentation
spinorielle fondamentale de dimension 16 ce qui permet de ranger
\underline{tous} les 
fermions d'une génération \underline{et} un neutrino droit, en une seule fois . Ceci en fait un groupe "naturel" du point de vue de la physique
des neutrinos et du mécanisme de Seesaw. Dans $SU(5)$, les neutrinos droits sont à 
rajouter "à la main" comme singulets. Mais la structure des interactions de jauge
$SO(10)$ n'a aucune incidence sur les neutrinos, $SO(10)$ est juste plus
adapté si on croit au Seesaw. En termes des représentations de $SU(5)$, 
\beq {\bf 16} = {\bf 10} \oplus {\bf \bar{5}} \oplus {\bf 1}. \eeq
Dans $SO(10)$, le nombre de bosons de jauge s'elèvent à 45, ce qui nous fait 33
bosons supplémentaires par rapport au MS et donc beaucoup d'interactions
possibles.
De plus, la brisure de $SO(10)$ est plus complexe car se fait en deux étapes :
en passant par $SU(5)\otimes U(1)$ ou par $SU(4)\otimes SU(2)_L \otimes SU(2)_R$
puis intervient la brisure de $SU(2)\otimes U(1)$. 
Le secteur de Higgs est très étendu et fait intervenir de
larges multiplets, de dimension 10, 16, 45, 120, 126 selon le modèle. C'est un
secteur extrêmement important pour la phénoménologie.

\subsubsection{Le modèle minimal $SU(5)$ supersymétrique}

Une fois la supersymétrie introduite pour résoudre les problèmes de hiérarchie
des échelles (chapitres 1 et 2), nous avons vu au début de ce chapitre
qu'elle permettait en plus une unification beaucoup plus précise des couplages
de jauge, la rendant presque indispensable pour les modèles Grand-Unifiés. 
Dans un modèle minimal $SU(5)$ supersymétrique, les multiplets vont être remplacés
par des supermultiplets. Les représentations utilisées sont
($\bf{10}+\bf{\bar{5}}$) pour les champs de matière et leurs sfermions, $\bf{24}$
pour les bosons et leurs partenaires fermioniques les "bosinos". Pour les même 
raisons qu'au chapitre 2, nous devons introduire 2 supermultiplets,
$\bf{5}$ et $\bf{\bar{5}}$, pour les Higgs et les Higgsinos, et enfin un
supermultiplet~\footnote{Il n'y en a qu'un seul, car il est auto-conjugué.} 
$\bf{24}$ de Higgs (et Higgsinos) pour briser $SU(5)$. 
L'existence de scalaires supplémentaires permet d'autres
interactions et notamment d'autres modes de désintégration pour le proton ({\it
via} échange de Higgsinos par exemple, ou de squarks si la $R$-parité n'est pas
conservée)
 
\subsection{Les conséquences typiques des GUTs}
\subsubsection{La désintégration du proton}

\begin{figure}[htbp!]
\begin{center}
\begin{picture}(350,100)(-20,-20)
\Photon(0,0)(-30,0){4}{4}
\ArrowLine(20,-20)(0,0)
\ArrowLine(0,0)(20,20)
\Text(27,25)[]{$e^+$}
\Text(25,-25)[]{$d$}
\Text(-36,-3)[]{$X$}

\Photon(100,0)(70,0){4}{4}
\ArrowLine(120,-20)(100,0)
\ArrowLine(100,0)(120,20)
\Text(127,25)[]{$u$}
\Text(125,-25)[]{$\bar{u}$}
\Text(64,-3)[]{$X$}

\Photon(200,0)(170,0){4}{4}
\ArrowLine(220,-20)(200,0)
\ArrowLine(200,0)(220,20)
\Text(227,25)[]{$\bar{\nu}$}
\Text(225,-25)[]{$d$}
\Text(164,-3)[]{$Y$}

\Photon(300,0)(270,0){4}{4}
\ArrowLine(320,-20)(300,0)
\ArrowLine(300,0)(320,20)
\Text(327,25)[]{$u$}
\Text(325,-25)[]{$\bar{d}$}
\Text(264,-3)[]{$Y$}

\end{picture} 
\end{center}
\caption{Diagrammes de Feynman montrant quelques nouveaux vertex permis dans les
GUT.}
\label{feynman}
\end{figure}
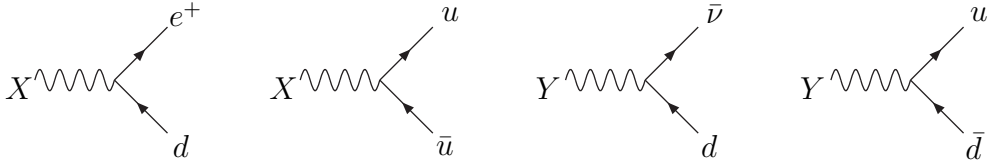

Une conséquence directe de l'existence d'une Grande Unification est la 
désintégration du proton.
En effet, les bosons $X$, $Y$ et les triplets de Higgs couplent les indices de couleurs de
$SU(3)$ avec les indices d'isospin faible de $SU(2)$. Nous avons donc des 
interactions possibles du genre
$q\to l+X$, $q+q\to Y \to q+l$,..., comme montré par les diagrammes de
Feynman de la figure~(\ref{feynman}). 

Autrement dit, dans les modèles GU il y a des interactions telles que $\Delta
B=\Delta L=1$, \underline{le nombre baryonique n'est pas conservé}. Rien n'empêche alors le 
proton de se désintégrer. Les modes de désintégration possibles incluent: 
\beq 
p\to e^+\, \pi^0,\ \ \mu^+\, K^0,\ \ \bar{\nu}\,K^+, \ \ \bar{\nu}\,\pi^+,...
\eeq
Pour évaluer le temps de vie du proton nous pouvons considérer une
interaction effective à basse énergie entre 4 fermions (qui représente par
exemple un processus baryon$\to$ lepton + méson). L'interaction est médiée
par l'échange d'un boson $X$ ou $Y$ très lourd (la GU est supposée se
réaliser à $\simeq 10^{16}$ GeV et $m_X\simeq m_Y\simeq
\mathcal{O}(M_{GU})$) donc l'interaction peut être considérée comme
ponctuelle. L'amplitude est proportionnelle à~\footnote{Par analogie avec
la constante de Fermi des interactions faibles, $G_F=\frac{g^2}{8 
m^2_W}$.}
$G_X=\displaystyle\frac{g_X^2}{8 m^2_X}$,
et prend la forme suivante dans le modèle $SU(5)$ minimal:
\begin{eqnarray}
&&\left(\epsilon_{ijk} u_{R_k} \gamma_\mu u_{L_j}\right)~G_X
~ \left(2 e_R
~\gamma^\mu ~d_{L_i} + e_L~\gamma^\mu~d_{R_i} \right)~, \nonumber \\   
&&\left(\epsilon_{ijk} u_{R_k} \gamma_\mu d_{L_j}\right)~G_X~ 
\left(\nu_L ~\gamma^\mu ~d_{R_i}\right)~,
\label{fourtwentyfive}
\end{eqnarray}
o\`u nous avons explicité les chiralités des fermions et nous négligeons 
les éventuels mélanges entre les différentes 
générations de quarks et leptons. Une interaction de la forme 
(\ref{fourtwentyfive}) donnerait la forme suivante du taux de 
désintégration:
\beq \Gamma_B=C\, G_X^2 m_p^5. \eeq 
Le facteur $m_p^5$ ($m_p$ est la masse du proton) vient de considérations 
dimensionnelles~\footnote{Nous rappelons qu'une largeur de désintégration 
$\Gamma$ a la dimension d'une masse.} et $C$ est un facteur qui dépend du 
modèle de la Grande Unification et de la dynamique hadronique
du proton (qui est un ensemble complexe de 3 quarks confinés, ce qui amène des
complications au calcul). Les calculs
détaillés donnent dans le cas $SU(5)$ non-supersymétrique un temps de vie du
proton entre $2.5\times 10^{28}$ et $1.6\times 10^{30}$ ans~\cite{Mohapatra}. 
Ceci est en désaccord avec les mesures de SuperKamiokande:
\beq \tau_p(p\to e^+ \pi^0)>1.6\times 10^{33}\ \mathrm{ans}. \eeq   

De ce point de vue là aussi les prédictions du modèle non-supersymétrique ne
sont donc pas en accord avec les données expérimentales. En revanche, les modèles
supersymétriques ne sont généralement pas exclus par cette limite sur la 
désintégration du proton, et ceci est notamment dû à une
échelle d'unification plus haute (donc à des masses des $X$ ou $Y$ plus élevées).
Typiquement, $\tau_p\sim 10^{35}$ ans dans les modèles supersymétriques. Les futurs projets
expérimentaux (Fréjus, Hyperkamiokande,...) devraient atteindre cette limite.  
L'observation de la désintégration du proton étant un des tests les plus
importants pour les GUTs, elle est de ce fait très attendue.     

\subsubsection{Les masses des quarks et leptons}
Les masses des quarks et leptons sont données, comme dans le MS, à la brisure de
la symétrie électrofaible $SU(2)\otimes U(1)$. Quarks et leptons se situent dans 
les mêmes représentations ce qui implique des relations particulières entre leurs masses.  
Dans $SU(5)$ par exemple, on a les
multiplets de fermions $\bf \bar{5}$ et $\bf 10$ qui contiennent respectivement
$d_L^c$, $e_L$ et $d_L$, $e^c_L$, $u_L$, $u_L^c$ (pour la première famille).
Un terme de masse est donc proportionnel 
à~\footnote{Nous rappelons que $\bar{\psi}\psi=\bar{\psi}_R\psi_L +
\bar{\psi}_L\psi_R=-\psi^{\mathrm{T}}_L C \psi^c_L + h.c.$.} 
$({\bf \bar{5}})\otimes({\bf
10})$ et pour former un scalaire on peut coupler ce terme à un Higgs dans une
représentation $\bf 5$ ce qui nous donne :
\beq 
\mathcal{Y}\, (\psi^i[{\bf \bar{5}}])^{\mathrm{T}}_L C(\chi^{ij}[{\bf 10}])_L (H_j[{\bf
5}])^{\dag}. 
\eeq
Après la brisure, $H[{\bf 5}]\to <0|H[{\bf 5}]|0>=v_H$, on a le terme de masse
\beq 
\mathcal{Y} v_H \left( \sum^3_{a=1} \bar{d_a}d_a +\bar{e}e
\right). 
\eeq
La masse de l'électron est donc la même que celle du quark down, ainsi que pour
les deux autres familles, à l'échelle $M_{GU}$ :
\beqn 
m_d=m_e, \\ m_s=m_{\mu}, \\ m_{\tau}=m_b. 
\eeqn
Mais il nous faut encore \underline{renormaliser} jusqu'à basse énergie ou nous connaissons 
la valeur des masses. Pour la troisième famille, nous avons
approximativement~\cite{JE-Beyond} : 
\beq
\frac{m_b}{m_{\tau}}\simeq
\left[\frac{\alpha_3(m_b)}{\alpha_3(M_{GU})}\right]^{4/7}\simeq 3, 
\eeq
ce qui est (à peu près) observé 
expérimentalement~\footnote{Pour $m_b\simeq4.1-4.9$ GeV et 
$m_\tau\simeq 1.78$ GeV nous trouvons que 
le rapport expérimental est autour de 2.5.~\cite{PDG04}.}. En incluant les 
corrections radiatives
d'ordres supérieurs, la supersymétrie, etc, il y a toujours accord entre la
relation prédite et l'expérience. Malheureusement, la renormalisation 
appliquée aux 2 premières familles donnent des résultats en contraction avec les
mesures. Ceci veut dire qu'il faut construire un modèle non-minimal et y inclure 
d'autres multiplets de Higgs (des {\bf 45} par exemple). Mais d'une façon 
générale, l'organisation nouvelle des
particules proposée par les modèles Grand-Unifiés implique des relations particulières
entre les masses des quarks et leptons. Beaucoup de travaux ont donc pour but
d'expliquer dans ce type de modèle le spectre de masses observé.

\subsection{Discussion}
Résumons cette section sur les modèles Grand-Unifiés en passant en revue les
divers avantages et faiblesses de ce type de modèles.

$\circ$ L'organisation des fermions d'une famille dans les représentations
plus grandes 
($\bf{10}+\bf{\bar{5}}$ dans le cas de $SU(5)$) rend la structure des quarks et leptons beaucoup plus
simple que dans le MS. 

$\circ$ Les quarks et leptons sont dans les mêmes multiplets ce qui explique la 
charge électrique fractionnaire (multiple de 1/3) des quarks et donc la
quantification de la charge. 

$\circ$ Les masses des quarks et leptons sont reliées et certains rapports 
prédits (par exemple $m_b/m_{\tau}$
Cependant, une relation viable pour les deux premières 
générations demanderait, par exemple, un secteur de Higgs 
non-minimal ou des opérateurs de dimensions 5.

$\circ$ Ces modèles réalisent pleinement l'unification des interactions (une
seule constante de couplage pour la partie non-gravitationnelle) et à haute
énergie ce qui est en accord avec la stabilité observé du proton.

$\circ$ Ces modèles sont aussi justifiés par l'unification
précise des couplages suggérée expérimentalement (c'est-à-dire à la prédiction précise de 
$\sin^2\theta_W$ ou de $\alpha_S$ à partir d'une unification postulée)

$\diamond$ Malheureusement, en général les GUTs ne donnent pas
d'explication sur l'origine du nombre de familles ni sur l'origine de la 
saveur (matrices CKM et MNS,...)
 
$\diamond$  Une unification à si haute énergie implique un grand "désert" de
$10^{12-13}$ GeV. Qu'y a t'il vraiment dans cette gamme d'énergie, est-ce 
vraiment "désert" ?

$\diamond$  La plupart des GUT phénologiquement viables ont un secteur de 
Higgs compliqué incluant beaucoup de multiplets et de représentations de 
dimensions élevées.

\par\hfill\par
Les modèles Grand Unifiés tels que présentés ici n'ont de toute façon pas la
prétention de tout expliquer mais il semble très intéressant et attirant d'inclure cette
idée et la plupart des ingrédients dans des modèles réalistes décrivant la
physique jusqu'à de très hautes échelles d'énergie. Pour tester ces idées, il
faudrait en particulier pouvoir observer la désintégration du proton (qui contraint en
particulier le secteur de Higgs) et la supersymétrie qui est quasi-indispensable aux GUTs, et d'enrichir
notre connaissance de la physique du neutrino (qui en quelque sorte "sonde" des
énergies proches des GUTs).


\newpage

\section{La brisure de la supersymétrie}


La brisure de la supersymétrie est nécessaire. En effet, nous n'observons
pas de particules supersymétriques de même masses que leurs partenaires
usuels: $m_e \ne m_{\tilde e}\,,\, m_\gamma \ne m_{\tilde\gamma}$. Or,
l'algèbre de supersymétrie implique que deux états liés par une
transformation de supersymétrie ont les mêmes valeurs propres pour
l'opérateur $P^2$ donc la même masse. Il faut donc briser la
supersymétrie. La question est de savoir si la brisure est explicite
c'est-à-dire présente dans le Lagrangien sous-jacent à la théorie ou bien,
si la brisure est spontanée, c'est-à-dire induite par un état de vide non
supersymétrique. Plusieurs raisons jouent en défaveur d'une brisure
explicite. C'est d'une part non esthétique, ensuite ce n'est pas une
manière analogue à celle avec laquelle sont brisées les symétries de
jauge, enfin, cela conduirait à des inconsistances dans les théories de
supergravité. Ainsi, les théoriciens se sont concentrés sur la brisure
spontanée de la supersymétrie.

Si le vide est non supersymétrique, il existe un état fermionique $\chi$
qui est couplé au vide par l'intermédiaire de l'opérateur fermionique
$Q$ (qui correspond à la charge supersymétrique) :

\begin{equation}
< 0 | Q | \chi > \; = \; f_\chi^2 \; \not= \; 0.
\end{equation}
Le fermion $\chi$ est l'équivalent du boson de Goldstone dans les
symmetries bosoniques spontanément brisées. Le champ $\chi$ étant un
champ fermionique, on l'appelle donc fermion de Goldstone ou Goldstino.

Jusqu'à présent, nous n'avons parlé de supersymétrie que dans le sens
d'une symétrie globale, c'est-à-dire dont les transformations ne dépendent
pas de l'espace-temps, donc n'incluant pas la gravité. C'est dans ce cadre
que l'on se place pour le moment : ainsi ce que l'on brise à ce stade est
une symétrie globale. Or, un problème apparaît lorsqu'on brise la
supersymétrie globale: l'énergie du vide est positive. Pour le constater,
il suffit de voir la valeur dans le vide de l'anticommutateur des charges
$Q$:
\begin{equation}
\{ Q, Q \} \propto \gamma_\mu P^\mu.
\end{equation}
et donc en introduisant la relation de fermeture $|\chi > < \chi | = 1$ :

\begin{equation}
< 0 | \{ Q, Q \} | 0 > \; = \; | < 0 | Q | \chi > |^2 = f_\chi^4 \; 
\propto \; < 0 | P_0 | 0 > = E_0,
\end{equation}
nous constatons ainsi que la brisure de la supersymétrie globale entraîne:

\begin{equation}
E_0 \; = \; f_\chi^4 \; > \; 0.
\end{equation}
On obtient une valeur de l'énergie du vide strictement positive.

\subsection{Terme F et terme D}

Afin de voir comment on peut obtenir une valeur de l'énergie du vide non
nulle, reprenons le potentiel effectif d'une théorie supersymétrique
globale:
\begin{equation}
V \; = \; \Sigma_i | \frac{\partial W}{\partial \phi^i} |^2 \; 
+ \; \frac{1}{2} \Sigma_\alpha g^2_\alpha | \phi^* T^\alpha \phi |^2.
\end{equation}
Rappelons que le premier terme est appelé terme F, et le second est le
terme D. Ainsi pour obtenir une valeur de l'énergie du vide non nulle, il
faut soit que le terme F soit défini positif (valeur moyenne dans le vide
non nulle) soit que le terme D soit défini positif.

\subsubsection{Terme D}

L'option $D > 0$ implique de construire un modèle avec un groupe de
symétrie de jauge $U(1)$. L'exemple le plus simple consiste à prendre un
supermultiplet chiral avec une charge unité pour lequel le potentiel
effectif est:
\begin{equation}
V_D \; = \; \frac{1}{2} ( \xi + g \phi^* \phi )^2.
\end{equation} 
Le terme supplémentaire $\xi$ n'est pas permis dans une théorie
non-abélienne c'est la raison pour laquelle il nous faut utiliser une
théorie qui possède un groupe de jauge $U(1)$. Le minimum du potentiel
effectif est atteint pour $< 0 | \phi | 0 > = 0$ et alors $V_D = 1/2 \xi^2
> 0$ et la supersymétrie est spontanément brisée. Dans cet état de vide,
on trouve que:
\begin{equation}
m_\phi \; = \; g \xi, \; m_\psi \; = \; 0, \; m_V \; = \; m_{\tilde V} \;
= \; 0.
\end{equation}
On distingue nettement dans cet exemple la différence de masse entre boson
et fermion qui correspondent au supermultiplet $(\phi, \psi)$.
Malheureusement, on ne peut pas utiliser le groupe de jauge $U(1)$ de
l'électromagnétisme dans le modèle standard. En effet, dans le Modèle
Standard, il y a des champs qui ont des signes différents pour
l'hypercharge permettant au potentiel effectif de s'annuler. On doit
rajouter un nouveau groupe de jauge $U(1)$ mais aussi de nombreux nouveaux
champs pour annuler les anomalies triangulaires supplémentaires qui
apparaissent à cause de ce nouveau groupe de jauge. Ainsi, le modèle de
brisure de supersymétrie avec un terme D n'a pas suscité énormément
d'intérêt bien qu'aujourd'hui il connaisse un renouveau dans le cadre des
modèles dérivés des théories de cordes.

\subsubsection {Terme F}

L'option $F > 0$ implique des champs chiraux supplémentaires avec des
couplages artificiels: ceux du Modèle Standard ne suffisent pas; l'exemple
le plus simple consiste à prendre trois supermultiplets chiraux $A,B,C$
avec le superpotentiel suivant:

\begin{equation}
W \; = \; \alpha A B^2 \; + \; \beta C (B^2 - m^2).
\end{equation} 
On trouve les termes F correspondants :

\begin{equation}
F_A \; = \; \alpha B^2, \; F_B \; = \; 2 B (\alpha A + \beta C), \; F_C \; 
= \; \beta (B^2 - m^2),
\end{equation}
et le potentiel effectif se réécrit:

\begin{equation}
V_F \; = \; \Sigma_i | F_i |^2 \; =
\; 4 |B(\alpha A + \beta C)|^2 \; + \; | \alpha B^2|^2 \; + 
\; | \beta (B^2 - m^2) |^2.
\end{equation}
On vérifie que les trois termes de l'équation précédente ne peuvent pas
s'annuler simultanément. Ainsi, on obtient nécessairement $V_F > 0$ et la
supersymétrie est brisée.

Pour conclure cette section, nous pouvons dire qu'il n'existe pas de moyen
satisfaisant de briser la supersymétrie globale car d'une façon ou d'une
autre, on est amené soit à rajouter de nouveaux champs de matière et des
couplages artificielles, soit à introduire un nouveau groupe de jauge et
d'autres champs. Dans les années $1980$, des tentatives ont été faites
pour "dissimuler" les champs supplémentaires dans un secteur caché mais il
s'est révélé qu'il était très compliqué d'obtenir un modèle
phénoménologiquement viable. De plus, la brisure de la supersymétrie
globale implique nécessairement une énergie du vide non nulle: à première
vue, cela n'est pas une mauvaise chose; en effet, les observations
actuelles tendent à prouver que l'énergie du vide n'est pas nulle. Le
problème fondamental vient du fait que la mesure de cette énergie du vide
est complétement en désaccord avec la prédiction de l'énergie du vide de
la supersymétrie globale: la valeur mesurée de l'énergie du vide est:
\begin{equation}
\Lambda \sim 10^{-123} m^4_P,
\end{equation}
alors que l'énergie du vide de la supersymétrie bris\'ee prédit une
constante cosmologique:

\begin{equation}
\Lambda \sim (1 \; TeV)^4 \sim 10^{-64} \, m^4_P.
\end{equation}
Nous obtenons une différence de 60 ordres de grandeur! Pour discuter de la
brisure de la supersymétrie de manière plus satisfaisante, il faut une
théorie supersymétrique incluant la gravité: c'est l'objectif du prochain
chapitre.

\subsection{Supersymétrie locale ou Supergravité}

Jusqu'à présent, nous avons considéré les transformations globales de
supersymétrie dans lesquelles le spineur des transformations
infinitésimales $E$ est indépendant de l'espace-temps. Nous allons
maintenant considérer un spineur de transformations dépendant de
l'espace-temps $E(x)$ par analogie avec les symmétries bosoniques qui, une
fois rendu locales, donnent naissance aux théories de jauge.

De plus, le fait de rendre la supersymétrie locale conduit à un mécanisme
analogue au mécanisme de Higgs pour briser les symétries bosoniques: le
mécanisme de super-Higgs qui permettra de briser de manière élégante la
supersymétrie. En prime, une théorie locale de la supersymétrie contient
nécessairement, comme nous le verrons, la gravité (c'est la raison pour
laquelle cette théorie se nomme la Supergravité)  et ouvre la perspective
d'unifier toutes les interactions des particules et les champs de matière
avec des transformations de supersymétrie étendue:

\begin{equation}
G (J = 2 ) \; \to \; {\tilde G} ( J = 3/2 ) \; \to \; V ( J = 1 ) \; \to 
\; q, \ell ( J = 
1/2 ) \; \to \; H ( J = 0 ),
\end{equation}
où $G$ est le graviton de spin 2 et ${\tilde G}$ est le gravitino, son
partenaire supersymétrique de spin-3/2, qui s'inscrivent tous deux dans le
supermultiplet de la gravitation:

\beq
\left( \begin{array}{c} G \\ \tilde{G} \end{array}\right) = 
 \left( \begin{array}{c} 2 \\  \frac{3}{2}\end{array}\right).
\eeq
La Supergravité est un ingrédient essentiel pour la compréhension des
interactions gravitationnelles des particules supersymétriques, et donc
nécessaire à une discussion cohérente de l'\'energie du vide.

\subsection{Supergravité}

Pour comprendre pourquoi le fait de rendre la supersymétrie locale
implique une présence de la gravité dans la théorie, considérons ce qui
arrive si l'on applique à un supermultiplet chiral les deux
transformations de supersymétrie suivantes:
\begin{eqnarray}
\delta_i \phi \; & = & \; \sqrt{2} {\bar E_i} \psi + \dots, \\
\delta_j \psi \; & = & \; - i \sqrt{2} \gamma_\mu \partial^\mu \phi E_j + 
\dots
\end{eqnarray}
Si l'on construit le commutateur sur les champs $\phi$ ou $\psi$, on obtient:

\begin{equation}
[\delta_i, \delta_j] (\phi, \psi) = - 2 ({\bar E_j} \gamma_\mu E_j) i 
\partial_\mu (\phi, \psi).
\end{equation}
On observe que l'effet sur les champs est équivalent à une translation
d'espace-temps, puisque $i \partial_\mu \leftrightarrow P_\mu$.

Si les transformations spinorielles infinitésimales $E_{i,j}$ sont
indépendantes de $x$, la translation est globale et la théorie est
invariante par translation globale. Mais si les $E_{i,j}$ dépendent de la
position dans l'espace-temps, on obtient que le commutateur appliqué au
champ est équivalent à un changement de coordonnées locales sur les champs
et donc la théorie est invariante par changement de coordonnées locales.
Or, nous savons qu'une théorie invariante par changement de coordonnées
locales contient nécessairement la gravité : l'invariance par changement
de coordonnées locales est un des points de départ de la construction de
la Relativité Générale qui est la théorie de la gravité.

\subsubsection{Analogie avec les théories de jauge}

Dans cette section nous verrons que le gravitino émerge naturellement
comme champ de jauge de la supersymétrie. Considérons, dans un premier
temps, la variation du terme cinétique d'un fermion sous une
transformation de jauge : $\delta ( i {\bar \psi}\gamma^\mu \partial_\mu
\psi )$. Dans une transformation de jauge, le champ fermionique devient :
\begin{equation}
\psi (x) \; \to \; e^{i \epsilon (x)} \psi (x), 
\end{equation}
o\`u $\epsilon(x)$ est une variation de phase dépendante de la position
dans l'espace-temps. Ainsi, un terme supplémentaire, par rapport à une
théorie où la phase est indépendante de l'espace-temps, apparaît dans la
variation du terme cinétique:
\begin{equation}
- {\bar \psi} \gamma_\mu \psi \partial^\mu \epsilon (x).
\end{equation}
Pour annuler cette variation et maintenir l'invariance du lagrangien sous
les transformations de jauge, on introduit un champ dit champ de jauge:
$A^\mu (x)$. Dans une symétrie de jauge abélienne, il intervient dans le
Lagrangien sous cette forme:
\begin{equation}
{\bar \psi} (x) \gamma_\mu \psi (x) A^\mu (x)
\end{equation}
et se transforme ainsi :
\begin{equation}
\delta A^\mu (x) \; = \; 
\partial^\mu \epsilon (x).
\end{equation}
Dans le cas de transformations locales de supersymétrie, la variation du
champ fermionique est:
\begin{equation}
\delta \psi (x) = - i \gamma_\mu \partial^\mu ( \phi (x) E 
(x)) + \dots, 
\end{equation}
et ainsi, la variation du terme cinétique fermionique contient un terme:
\begin{equation}
\propto \; {\bar \psi} \gamma_\mu \gamma_\nu \partial^\nu \phi \partial^\mu E(x)
\end{equation}
qui est compensé par l'introduction d'un terme contenant un nouveau champ
$\psi_\mu$ :
\begin{equation}
\kappa {\bar \psi} \gamma_\mu \gamma_\nu \partial^\nu \phi \psi_\mu (x),
\end{equation}
qui se transforme comme ceci sous la supersymétrie:
\begin{equation}
\delta \psi_\mu (x) \; = \; - \frac{2}{\kappa} \partial_\mu E (x).
\end{equation}
Le nouveau champ $\psi_\mu$ est un spineur avec un indice de Lorentz :
donc il poss\`ede spin 3/2. Il est le champ de jauge de la supersymétrie :
c'est un fermion de jauge. Comme il a un spin 3/2, il ne peut être que le
partenaire supersymétrique du graviton de spin 2: c'est donc le gravitino.

\subsubsection{Le Lagrangien de pure Supergravité}

Considérons le plus simple Lagrangien pour un gravitino et un graviton. Il
consiste en la somme du Lagrangien de Einstein-Hilbert pour la Relativité
Générale (description du graviton) et celui de Rarita-Schwinger pour un
champ de spin 3/2 (descrption du gravitino). Bien entendu, il faut rendre
invariant le Lagrangien de Rarita-Schwinger sous transformations générales
de coordonnées en covariantisant la dérivée:

\begin{equation}
L \; = \; - \frac{1}{2 \kappa^2} \sqrt{-g} R \; - \; \frac{1}{2} 
\epsilon^{\mu \nu \rho \sigma} {\bar \psi_\mu} \gamma_5 \gamma_\nu {\cal 
D}_\rho \psi_\sigma,
\end{equation}
où $g = det(g_{\mu \nu})$ et $g_{\mu \nu}$ est le tenseur métrique:

\begin{equation}
g_{\mu \nu} \; \equiv \; \epsilon^m_\mu \epsilon^n_\nu \eta_{mn}
\end{equation}
où $\epsilon^m_\mu$ est le vierbein (qui décrit, entre autres, le champ du
graviton) et
\begin{equation}
{\cal D}_\rho \; \equiv \; \partial_\rho + \frac{1}{4} \omega^{mn}_\rho 
[\gamma_m, \gamma_n],
\end{equation}
est la dérivée covariante avec $\omega^{mn}_\rho$ la connexion de spin. Ce
Lagrangien est naturellement invariant sous transformations locales de
supersymétrie. Les champs se transforment de la manière suivante:
\begin{eqnarray}
\delta \epsilon^m_\mu \; & = & \; {\bar E} (x) \gamma^m \psi_\mu (x), \\
\delta \omega^{mn}_\mu \; & = & \; 0, \\
\delta \psi_\mu \; & = & \; \frac{1}{\kappa} {\cal D}_\mu E (x).
\end{eqnarray}

\subsection{Le mécanisme de Super-Higgs}

Dans le Lagrangien précédent, nous n'avons pas inclus le couplage à la
matière, dont nous parlerons plus tard. Cependant, nous pouvons déjà
parler du mécanisme de brisure spontanée de la supersymétrie locale: le
mécanisme de super-Higgs. En effet, nous venons d'introduire le gravitino
qui se trouve au centre de ce mécanisme. Rappelons que dans le mécanisme
de Higgs conventionnel, un boson de Goldstone sans masse (de spin 0 donc
avec une seule polarisation) est absorbé par un boson de jauge sans masse
(d'hélicité + ou -1 donc avec deux \'etats de polarisation) qui va
acquérir par ce processus les degrés de liberté du boson de Goldstone et
donc posséder trois états de polarisations qui lui permettent d'acquérir
une masse:
\begin{equation}
2 \times (V_{m = 0}) \; + \; 1 \times (GB) \; = \; 3 \times (V_{m \ne 0}).
\end{equation}
Dans une théorie localement supersymétrique, les deux états de
polarisation (hélicité + ou - 1/2) du fermion de Goldstone sans masse sont
absorbés par un gravitino sans masse d'hélicité + ou - 3/2, donc avec deux
états de polarisation, pour lui donner les quatres états de polarisations
nécessaire à l'obtention d'une masse:
\begin{equation}
2 \times (\psi^\mu_{m = 0}) \; + \; 2 \times (GF) \; = \; 4 \times 
(\psi^\mu_{m \ne 0}).
\end{equation}
Ce processus brise la supersymétrie locale, puisqu'il donne une masse au
gravitino alors que le graviton n'en a pas: $m_G = 0 \ne m_{\tilde G} \ne
0$. C'est le seul moyen consistent pour briser la supersymétrie. De plus,
il peut se faire en donnant une énergie nulle au vide:
\begin{equation}
< 0 | V | 0 > \; = \; 0 \; \leftrightarrow \; \Lambda \; = \; 0.
\end{equation}
Ainsi, on peut obtenir à la fois une brisure de la supersymétrie locale et
une constante cosmologique nulle, ce qui n'était pas le cas de la brisure
globale de supersymétrie.

\subsection{Couplage de la Supergravité à la matière}

Le Lagrangien complet de la Supergravité incluant les multiplets vecteurs
et chiraux peut être obtenu par la méthode de calcul tensoriel locale.
C'est un travail laborieux. Nous ne donnerons que quelques éléments clés
sans écrire de preuves. Parmi ces expressions nouvelles par rapport à la
supersymétrie globale, on trouve principalement le potentiel de Kähler. Le
potentiel de Kähler est une fonction des champs scalaires (contenus dans
les supermultiplets chiraux) : $G(\phi, \phi^*)$. Le superpotentiel est
relié au potentiel de Kähler. Ce dernier est aussi appelé variété de
Kähler, car il décrit une géométrie interne qui influe directement sur les
termes cinétiques des champs fermioniques et scalaires et qui détermine la
masse des particules après la brisure de supersymétrie. C'est aussi un
paramètre d'ordre pour la brisure de supersymétrie:
\begin{equation}
m_{\tilde G} \; \equiv \; m_{3/2} = e^{G_0\over 2}\,|<W>|,
\end{equation}
où $G_0$ est la valeur dans le vide de $G$ et $\ <W>$ est la valeur dans le vide
du superpotentiel $W$\footnote{Où $W\,=W_{cach}+W_{obs}$ et $<W_{obs}>=0$,
mais $<W_{cach}>\not=0$.}.

Comme nous l'avons dit, $G$ détermine les termes cinétiques des champs
fermioniques et des champs scalaires. Dans le cas des champs scalaires:
\begin{equation}
L_K \; = \; G^j_i \partial^\mu \phi^*_j \partial_\mu \phi^i ,
\end{equation}
où $G^j_i$ est la métrique de Kähler:
\begin{equation}
G^j_i \; \equiv \; \frac{\partial^2 G}{\partial \phi^i \partial \phi^*_j},
\end{equation}
Le potentiel effectif sans la présence des termes $D$ est aussi déterminé
par le potentiel de Kähler:
\begin{equation}
V \; = \; e^G [ G_i (G^{i}_j)^{-1} G^j - 3] \,avec \; G_i \; \equiv \; 
\frac{\partial G}{\partial \phi^i}.
\end{equation}
On remarque que ce potentiel peut s'annuler sans empêcher la brisure de
supersymétrie, ce qui n'était pas le cas dans la supersymétrie globale. En
effet, on peut avoir simultanément $m_{\tilde G}^2 = e^G \not= 0$ et
$V=0$.

L'état de vide de la théorie doit correspondre à un minimum de $V$. Or, il
se trouve que pour des formes générales de $G$, et pour certaines valeurs
des champs, le potentiel $V$ devient négatif: ceci constitue une
catastrophe pour la cosmologie, puisque cela signifierait que la constante
cosmologique est négative et donc que l'Univers s'effondre sur lui-même.
Heureusement, il existe une classe particulière de potentiels de Kähler,
non n\'gatifs. Il se trouve que c'est la classe de potentiels qui émerge
des théories de cordes.

De même qu'il existe un potentiel de Kähler qui détermine la géométrie et
la cinétique des champs chiraux, il existe une fonction appelée fonction
cinétique notée $f(\phi)$ qui détermine les termes cinétiques des
supermultiplets vecteurs.

\subsection{Théorie effective aux basses énergies} 

La théorie basse-énergie trouve son origine dans une théorie haute-énergie
: la théorie la plus générale de supergravité n'est pas renormalisable,
mais les termes non renormalisables ne sont pas importants \`a basses
\'energies. Donc, en premi\`ere approximation on peut ne garder que des
termes renormalisables dans le potentiel effectif de la supergravité à
basses énergies. Pour des choix génériques du potentiel de Kähler, on
obtient des masses issues de la brisure de la supersymétrie non nulles
pour les gauginos:
\begin{equation}
m_{1/2} \; \propto \; m_{\tilde G} \; \equiv \; m_{3/2}.
\end{equation}
L'universalité des masses des jauginos correspondant aux différents
groupes de jauge $SU(3), SU(2)$ et $U(1)$ n'est pas systématique, mais
cela émerge naturellement si la géométrie n'est pas trop compliquée, par
exemple si la fonction cinétique $f(\phi)$ est un singulet. En développant
le potentiel effectif, on trouve des termes proportionnels à $|\phi|^2$,
qui sont interprétés comme les masses des scalaires issues de la brisure
de la supersymétrie:
\begin{equation}
m_{0} \; \propto \; m_{\tilde G} \; \equiv \; m_{3/2}.
\end{equation}
Dans ce cas, il n'y a aucune raison théorique d'avoir universalité des
masses: les modèles issus des théories de cordes brisent souvent cette
universalité. Par contre, il y a de bonnes raisons phénoménologiques pour
penser que les masses des scalaires avec la même charge sont identiques.
En effet, dans le cas contraire, on aurait des interactions véhiculées par
des Sparticules virtuelles qui provoqueraient des changements de saveurs.

Des formes génériques du potentiel effectif conduisent aussi \`a la
présence de termes d'interaction trilinéaires issues de la brisure de
supersymétrie entre particules scalaires :
\begin{equation}
A_\lambda \lambda \phi^3: \; \; A_\lambda \; \propto \; m_{\tilde G} \; 
\equiv \; m_{3/2},
\end{equation}
Encore une fois, l'universalité n'a pas de raison fondamentale. Si la
théorie supersymétrique inclut aussi des termes d'interaction bilinéaires
comme c'est le cas dans l'extension minimale du Modèle Standard à la
supersymétrie (MSSM) avec des termes $\mu \phi^2$ où le scalaire en
question est le Higgs, on s'attend à trouver aussi des termes $B_\mu \mu
\phi^2$ dans le potentiel effectif.

La forme finale du Lagrangien de brisure 'explicite' de la théorie basse
énergie de la supergravité suggérée par la brisure spontanée de la
supersymétrie locale est:
\begin{equation}
- \frac{1}{2} \Sigma_a m_{1/2_a} {\tilde V}_a {\tilde 
V}_a - \Sigma_i m^2_{0_i} |\phi^i|^2 - (\Sigma_\lambda A_\lambda 
\lambda \phi^3 + \Sigma_\mu B_\mu \mu \phi^2 + {\rm Herm. Conj.}), 
\end{equation} 
qui contient de nombreux paramètres libres. La brisure de la supersymétrie
est explicite dans la théorie basse énergie mais, comme nous l'avons déjà
vu, elle est néanmoins douce car la renormalisation des paramètres
$m_{1/2_a}, m_{0_i}, A_\lambda$ et $B_\mu$ est logarithmique, sans
corrections quadratiques. Il faut souligner que ces paramètres ne sont pas
fondamentaux, et que le mécanisme de brisure de la supersymétrie
sous-jacent est spontanée.

La renormalisation logarithmique des paramètres signifie que l'on peut
calculer leurs valeurs de basse-énergie à partir de ses valeurs à hautes
énergies provenant d'une théorie des supercordes ou de supergravité. Pour
le cas de la masse basse-énergie des gauginos $M_a$, on a :
\begin{equation}
\frac{M_a}{m_{1/2_a}} \; = \; \frac{\alpha_a}{\alpha_{GUT}}
\end{equation}
à une boucle, où $\alpha_a$ est le couplage de jauge et où l'on a supposé
l'unification des couplages à l'échelle de la supergravité.

Pour le cas des scalaires, on a :
\begin{equation}
\frac{ \partial m^2_{0_i}}{\partial t} \; = \; \frac{1}{16 \pi^2} [ 
\lambda^2 (m_0^2 + A_\lambda^2) - g_a^2 M_a^2]
\label{m0}
\end{equation}
avec $t = {\rm ln} ( Q^2 / m_{GUT}^2)$ et où les coefficients liés au
groupe de jauge ont été supprim\'es. Dans le cas des deux premières
générations, le premier terme dans la partie droite de l'équation peut
être omis, car les couplages de Yukawa sont faibles. On obtient ainsi :
\begin{equation}
m^2_{0_i} \; = \; m_0^2 + C_i m_{1/2}^2,
\end{equation}
dans le cas o\`u les masses initiales sont identiques pour les scalaires,
et les coefficients $C_i$ sont calculables dans tous les modèles. Le
premier terme dans la partie droite de l'équation (\ref{m0}) est important
pour la troisième génération et pour les bosons de Higgs du MSSM.

En effet, le signe du premier terme est positif et celui du second est
négatif. Cela signifie que le dernier terme augmente $m^2_{0_i}$ lorsque
l'échelle de renormalisation $Q$ diminue, alors que le terme positif fait
diminuer $m^2_{0_i}$ lorsque $Q$ diminue. Dans le cas du boson de Higgs
$H_u$, le terme positif n'est pas négligeable car $H_u$ a un couplage de
Yukawa élevé avec le quark top: $\lambda_t \sim g_{2,3}$. Une observation
très intéressante est que la brisure de symétrie spontanée électrofaible
est expliquable. En effet, on retrouve la forme du potentiel de brisure
électrofaible de manière naturelle puisque $m_{H_u}^2$ devient négatif \`a
basse \'energie.  Ainsi, la brisure spontanée de supersymétrie entraine la
brisure spontanée de la symétrie électrofaible. Cela se produit à une
échelle d'énergie exponentiellement plus petite que l'échelle de la
supergravité:
\begin{equation}
\frac{m_W}{m_P} \; = \; exp( \frac{ - {\cal O}(1)}{\alpha_t}): \; \; 
\alpha_t \equiv \frac{\lambda_t^2}{4 \pi}.
\end{equation}
Les calculs d'évolution permettent de d\'emontrer que $m_W \sim 100$~GeV
émerge naturellement si $m_t \sim 60$ to $200$~GeV, ce qui a \'et\'e
confirmé ensuite par des exp\'eriences.

\subsection{Conclusion sur la brisure de la supersymétrie locale}

Comme nous l'avons vu, la brisure de supersymétrie locale serait induite
par le mécanisme de super-Higgs. Cependant, le détail du mécanisme de
brisure reste un grand mystère. Pour que les sparticules soient plus
lourdes que les particules, tout en restant à une masse inférieure à 10
Tev (pour permettre la stabilisation des bosons de Higgs), une méthode
consiste à ajouter des champs non présents dans la théorie. On appelle ce
secteur, le secteur caché. Dans certains modèles, le secteur caché
communique gravitationnellement avec le secteur observable. Dans d'autres
modèles, les messagers sont des bosons de jauge. Dans le secteur caché, on
suppose l'existence d'une interaction forte qui permet la brisure
dynamique de la supersymétrie \`a une \'echelle \'elev\'ee. Dans le cas de
l'interaction forte usuelle, le passage de la zone perturbative à la zone
non pertubative de la théorie à basse énergie provoque la condensation
et le confinement des quarks. Par analogie, on attend un condensat de
jauginos\cite{Nilles:1998sx} dans le secteur caché, ce qui briserait la supersymétrie. Par
exemple, dans la théorie des cordes hétérotiques dont le groupe de jauge
est $E_8 \times E_8$, le premier groupe exceptionnel contient le modèle
standard alors que le second facteur décrit le secteur caché, dans lequel
apparait le condensat de jauginos.

Les théories de cordes donnent naissance à des théories sp\'ecifiques de
supergravité dites no-scale. Dans ces modèles, la masse du gravitino est
indéterminée au niveau de l'arbre car le potentiel effectif est constant.
Dans un tel modèle, il n'y a qu'une seule \'echelle de masse au départ:
celle de Planck. La masse du gravitino provient alors des corrections
radiatives au potentiel effectif. La forme de ce potentiel au niveau de
l'arbre ne fournit pas de termes de masse provenant de la brisure de
supersymétrie pour les scalaires. Pourtant, il existe des moyens de
communiquer la brisure de la supersymétrie\cite{Randall:1998uk} au secteur observable. Par
exemple, dans ces théories de "no-scale" supergravité provenant des
théories de cordes, il existe souvent un second champ scalaire du secteur
caché qui paramétrise la fonction cinétique des bosons de jauge et permet
aux gauginos d'obtenir une masse. Les masses des particules scalaires sont
alors données par les corrections radiatives lorsque l'on amène les
paramètres de la théorie vers l'échelle électrofaible.

Pour conclure ce paragraphe, il est intéressant de regarder ce que devient
l'énergie du vide ou la constante cosmologique après renormalisation. On
parle bien évidemment du vide de la théorie supersymétrique sous-jacente.
Les corrections à boucle sur l'énergie du vide sont quadratiquement
divergentes dans une théorie g\'en\'erique de supergravité, ce qui suggère
une contribution \`a l'\'energie du vide de l'ordre de $m^2_{3/2} m^2_P$
et donc $O(10^{-32})m^4_P$. Cette contribution est supprim\'ee dans
certains modèles, qui ont une correction à une boucle de l'ordre de $
O(10^{-64})m^2_P$. Pourtant, il faudrait sans doute une symétrie
supplémentaire pour amener la constante cosmologique au niveau requis,
soit $10^{-123} m^4_P$. Ceci nous conduit maintenant \`a discuter de la
théorie des cordes, qui est notre meilleur candidat pour une Théorie de
Tout incluant la gravité.

\newpage


\section{Vers une "Théorie de Tout"}


\subsection{Les problèmes de la gravité quantique}

Un des éléments fondamentaux qui manquent à notre compréhension de
l'Univers et des interactions fondamentales est l'unification des deux
grandes théories du XXème siècle : la relativité générale et la mécanique
quantique. Ecrire une théorie unifiée est un des enjeux majeurs de notre
siècle. La solution du problème de la constante cosmologique devra se
situer dans le cadre d'une telle "Théorie de Tout". 

La gravitation échappe à la théorie quantique surtout car des infinis
incontrôlables et non renormalisables apparaissent lorsque l'on veut
calculer des diagrammes contenant des boucles avec des gravitons. Ces
corrections sont des puissances qui divergent de plus en plus rapidement
lorsqu'on augmente l'ordre du calcul perturbatif, parce que la constante
de couplage de la gravitation possède une dimension.

Il existe aussi des problèmes non perturbatifs qui émergent lorsque l'on
cherche à quantifier la gravitation. Ces problèmes sont pour la première
fois apparus lorsque les physiciens se sont intéressés aux trous noirs. Le
trou noir est une solution non perturbative des équations de la Relativité
Générale dans laquelle la courbure de l'espace-temps li\'ee aux forces
gravitationnelles devient tr\`es importante : aucune particule ne peut
sortir de l'horizon qui l'entoure. L'existence de cet horizon est li\'ee
aux questions d'entropie et de temp\'erature des trous noirs. En
effet, la masse d'un trou noir est proportionnelle à la surface de son
horizon, qui se situe au rayon de Schwarzschild: $r_s= \frac{G\,m}{c^2}$ :
\begin{equation}
S\;=\frac{1}{4}\;A. 
\end{equation}
où $A$ est la surface et $S$ l'entropie.

Lorsque la masse d'un trou noir augmente, sa surface augmente, et aussi
son entropie. Proche de l'horizon du trou noir, les effets quantiques
cr\'eent des paires de particules, dont une est virtuelle et se trouve \`a
l'int\'erieur du trou noir tandis que l'autre est r\'eelle et se trouve
\`a l'ext\'erieur de l'horizon. Cette derni\`ere est rayonn\'ee par le
trou noir : c'est que l'on appelle le rayonnement d'Hawking. Cette
radiation est stochastique, avec tous les charact\'eristiques d'une
radiation thermique \'emise par un corps noir. Cet effet est \`a rapprocher
de l'effet Unruh, selon lequel un observateur dans un r\'ef\'erentiel
acc\'el\'er\'e detecte une radiation thermique. Dans le cas d'un trou
noir, sa température est li\'ee \`a sa masse :
\begin{equation}
T\;=\frac{M}{2\pi}. 
\end{equation}

L'entropie du trou noir refl\`ete la perte d'information \`a travers son
horizon, et l'\'etat thermique de la radiation de Hawking, d'une nature
stochastique, en est l'expression. Pour la d\'ecrire, il faut utiliser un
\'etat quantique mixte. Or, on peut imaginer la pr\'eparation d'un trou
noir \`a partir d'un \'etat pur. Donc il faut envisager une transition
entre un \'etat mixte et un \'etat pur, ce qui n'est pas admis par la
m\'ecanique quantique habituelle. On peut voir ce probl\`eme d\'ej\`a au
niveau de la radiation de Hawking : considérons un état quantique pur
compos\'e de deux particules $A, B$ dans une superposition d'\'etats
individuels : $|A,B> \equiv \sum_i c_i |A_i > |B_i>$. Si la particule $A$
tombe dans le trou noir et $B$ s'échappe, l'information tenue par la
particule $|A>$ est perdue :
\begin{equation}
\sum_i \, c_i |A_i B_i ><A_i B_i| \to \sum_i |c_i|^2 |B_i > <B_i|
\end{equation}
et $B$ émerge dans un état mixte! Or, la mécanique quantique ne permet pas
l'évolution d'un état pur en état mixte.

Cette discussion met en évidence un conflit entre la théorie quantique et
la Relativité Générale. Au moins un de ces deux piliers de la physique du
XX\`eme si\`ecle doit être modifié. Les physiciens des particules
préfèrent modifier la Relativité Générale en l'élevant à la théorie des
cordes. Nous continuons avec cette théorie.

\subsection{Introduction à la théorie des cordes}

Comme nous avons d\'ej\`a vu, un des problèmes majeurs de la gravité
quantique est la présence d'un nombre illimit\'e de divergences. Ces
divergences peuvent être reliées à l'absence de cut-off (coupure) à
courtes distances dans les théories de champs usuelles où les particules
sont ponctuelles. En effet, on peut rapprocher de manière infinie proche
des objets ponctuels, donnant ainsi naissance à des interactions infinies:
\begin{equation}
\int^{\Lambda \to \infty} \, d^4k \left (\frac{1}{k^2} \right )
\leftrightarrow \int_{1/ \Lambda \to 0} \, d^4x
\left ( \frac{1}{x^6} \right ) \sim \Lambda^2 \to \infty.
\end{equation}
On peut adoucir voire effacer ces divergences si l'on impose un cut-off
naturel.

\begin{figure}
\centerline{\includegraphics[height=3in]{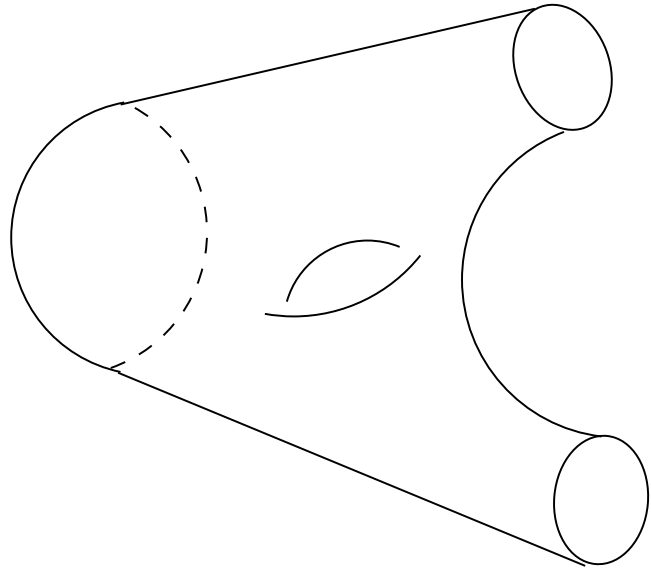}}
\caption[]{Une représentation d'une interaction en théorie des cordes.}
\label{fig:ws}
\end{figure}

Pour faire ceci, il suffit de considérer des objets étendus plutôt que
ponctuels pour réduire voire supprimer ces divergences. L'option la plus
simple est de remplacer les particules ponctuelles par des objets
unidimensionnels : les cordes. Les lignes d'univers associées à ces objets
étendus deviennent des surfaces d'Univers (world-sheets). Si les cordes
sont fermées, on obtient des tubes dans l'espace-temps, qui forment des
circuits de plomberie lorsqu'ils interagissent entre eux, comme indique la
figure~\ref{fig:ws}. On peut imaginer étendre ce principe a des objets
avec davantage de dimensions comme des membranes, dont les lignes
d'Univers se transforment en "volumes d'Univers". Nous reviendrons à ces
objets lorsque nous parlerons des aspects non perturbatifs de la théorie.

Historiquement les théories de cordes furent introduites pour décrire les
interactions fortes avant que ne soit développée la QCD. Les états liés
l'étaient par l'intermédiaire de cordes poss\`edant une certaine tension.
Les amplitudes des processus lors desquels deux particules donnent $n$
particules pouvaient être directement dérivées d'une théorie quantique des
cordes. Avec l'avènement de la QCD, la théorie des cordes fut presque
oubliée comme théorie des interactions fortes~\footnote{Quelques
théoriciens continuent \`a travailler sur ce sujet.}. En plus, il a
\'et\'e reconnu que l'unitarité n\'ecessitait que les cordes fermées
soient présentes dans la théorie.

Il apparaissait que le spectre des états quantiques de la corde fermée
incluerait une particule de spin-2 et sans masse, ce qui était un
inconvénient pour une th\'eorie de l'interaction forte. Pourtant, cela
fournit l'idée que les cordes pouvait être une Théorie de Tout: en effet,
une particule de spin 2 sans masse peut être interprétée comme un graviton,
et dans ce cas la tension de la corde deviendrait beaucoup plus \'elev\'ee
: $\mu = O(m^2_P)$.

Le spectre d'excitation des cordes fournit un nombre infini de différentes
particules. Puisque les cordes se propagent sur une surface d'Univers, le
formalisme est bidimensionnel. Les vibrations des cordes peuvent être
décrites en terme d'ondes qui propagent dans les deux sens autour de la
corde ferm\'ee, vers la gauche ou vers la droite :
\begin{equation}
\phi (r,t) \to \phi_L (r-t), \, \phi_R (r+t).
\end{equation}
Si la corde est fermée, les ondes gauches et droites sont indépendantes.
Après quantification, on peut réécrire la théorie comme une théorie de
champs à deux dimensions, car on travaille sur des surfaces d'Univers.
Comparée à une théorie à quatre dimensions, il est relativement simple
d'obtenir une théorie finie. Dans ce cas, la théorie possède une symétrie
conforme qui est décrite par un groupe de symétrie de dimension infinie à
deux dimensions. Cette symétrie classique ne doit pas être brisée par des
anomalies lorsqu'on rend la théorie quantique. Annuler l'anomalie conforme
implique que dans un espace-temps plat les ondes gauches et droites sont
toutes les deux équivalentes à $26$ bosons si la théorie n'a pas de
supersymétrie et à $10$ supermultiplets de bosons et fermions si la
théorie poss\`ede une supersymétrie N=1 sur la surface d'Univers.

\subsection{Les grandes classes de théories de cordes}

Parmi les modèles de théories de cordes consistantes, on trouve la {\it
corde bosonique}, une théorie en $26$ dimensions qui ne contient pas de
fermions. La corde bosonique a un vide instable, c'est-à-dire qu'un
espace-temps plat est instable.

On a aussi des modèles de {\it supercordes} consistants en $10$
dimensions, qui contiennent des fermions et possèdent des vides stables,
donc des espaces-temps plats stables, mais n'ayant pas de fermions
chiraux.

Le modèle de {\it corde hétérotique en $10$ dimensions} est aussi un
modèle supersymétrique, mais il possède en plus du modèle précédent des
fermions chiraux, car la violation de la parité a été intégrée puisque les
ondes gauches et droites sont traitées différemment. Cette théorie est
appelée hétérotique car elle est contruite à partir d'une théorie de
supercordes pour la partie fermionique et d'une théorie de cordes
bosoniques à $26$ dimensions pour la partie bosonique. Les $16$ dimensions
supplémentaires qui sont autant de degrés de liberté supplémentaires sont
perçues comme des champs supplémentaires dans la théorie à $10$
dimensions.

Enfin, on peut citer les modèles de {\it cordes hétérotiques en $4$
dimensions} qui sont obtenus en compactifiant les six dimensions
suppl\'ementaires à des rayons de l'ordre de la longueur de Planck, ou
alors en travaillant directement à $4$ dimensions en remplaçant les
dimensions manquantes par d'autres degrés de liberté internes comme des
fermions ou des groupes de symétrie. De cette manière, il a été possible
d'incorporer un groupe de jauge unifi\'e et même des modèles ressemblant
au Modèle Standard.

\subsubsection{Les théories de supercordes}

Comme nous l'avons dit, le modèle de corde bosonique possède beaucoup plus
de désavantages que les autres modèles: il a $26$ dimensions, pas de
fermions et un vide instable! Les physiciens se sont donc concentrés sur
des modèles supersymétriques donc de supercordes. Or, il existe cinq
théories de supercordes :

- le type IIA qui se réduit à basse énergie à une supergravité non chirale
$N=2$ à $d=10$ dimensions ;

- le type IIB qui se réduit à basse énergie à une supergravité chirale
$N=2$ à $d=10$ dimensions ;

- la théorie hétérotique $E(8)\times E(8)$ qui se réduit à basse énergie à
une supergravité $N=1$ à $d=10$ couplée à une théorie de Yang-Mills avec
le groupe de jauge $E(8)\times E(8)$ ;

- la théorie hétérotique $SO(32)$ qui se réduit à basse énergie à une
supergravité $N=1$ à $d=10$ couplée à une théorie de Yang-Mills de groupe
de jauge $SO(32)$ ;

- le type I qui contient à la fois des cordes ouvertes et fermées et qui
se réduit à une supergravité $N=1$ à $d=10$ couplée à une théorie de
Yang-Mills de groupe de jauge $SO(32)$.

Chacune de ces théories est différente de l'autre. La théorie type I est
la seule qui contienne à la fois des cordes ouvertes et fermées alors que
les autres ne contiennent que des cordes fermées. De plus, la structure de
jauge des cinq théories est sensiblement différente dans leur structure à
basse énergie.  Nous sommes donc en présence de cinq théories qui
permettent apparemment de décrire la gravité comme une force quantique et
qui sont différentes. Comment comprendre cela ? Existe-t-il un lien entre
les différentes théories?

\subsection{La structure non-perturbative de la théorie}

Dans la construction des théories de cordes, les physiciens ont commencé
par travailler sur une théorie de première quantification, c'est-à-dire
une théorie où l'on quantifie la position et l'impulsion. Dans cette
théorie, les interactions sont introduites à la main et ne peuvent pas
dériver d'une seule action. L'unitarité doit être vérifiée pas à pas. La
formulation est sur couche de masse. La formulation de la théorie est
perturbative.

La théorie de seconde quantification (ou théorie des champs) des cordes
quantifie les champs. Les interactions y sont explicites dans l'action,
l'unitarité est en principe garantie si le Hamiltonien est hermitien, la
théorie peut être formellement écrite perturbativement ou non
perturbativement, et la théorie possède une formulation hors couche de
masse. Le grand problème de la théorie des champs de cordes est que, bien
que sa formulation soit indépendante d'une théorie de perturbation, elle
est actuellement trop difficile pour être résolue dans une région non
perturbative (forts couplages).

L'idée clé pour sonder et comprendre la structure non perturbative de la
théorie est d'utiliser la dualité. La notion de dualité peut être comprise
facilement en considérant la théorie du champ magnétique et du champ
électrique. Les équations de Maxwell sont invariantes par les
transformations de dualité suivantes : l'on change $E$ en $-B$ et $B$ en
$-E$ et l'on change $e$ en $g$, où $g$ est la charge d'un monopole
magnétique. Or, dans le th\'eorie quantique $e\,g=2\pi\,n$ où $n$ est un
entier. On conclut donc que la théorie du champ électrique défini dans une
région à fort couplage ($e$ grand) est équivalente à une théorie du champ
magnétique où le couplage est faible ($g$ petit) et vice-versa. On voit
sur cet exemple que l'on peut sonder la théorie électrodynamique à fort
couplage (donc difficile à résoudre) en étudiant une théorie de monopoles
magnétiques à faible couplage.

C'est exactement le même principe que l'on applique aux théories de
cordes. Les différentes théories de cordes sont reliées par un réseau de
relations de dualité qui permet de comprendre la région non perturbative
d'une théorie en utilisant la zone perturbative d'une autre théorie.
D'autre part, ces relations de dualité laisse supposer qu'une théorie mère
gouverne toutes les théories de cordes comme dans le cas de
l'électromagnétisme: on a la théorie du champ magnétique reliée à la
théorie du champ électrique par dualité, cela implique l'existence d'une
théorie regroupant les deux théories : l'électromagnétisme. La théorie
mère de ces théories des supercordes est appelée la {\it M-théorie}. La
lettre $M$ pourrait avoir le sens de {\it m\`ere}, ou peut signifier le mot
{\it membrane}, ou bien {\it magique} ou {\it myst\'erieuse}, et certains y entendent le
mot $matrice$, d'apr\`es une autre formulation de la théorie.

Mais tout d'abord, citons quelques conséquences de l'existence de la
dualité. Des objets étendues de dimensions supérieure à un, appelées
membranes, apparaissent lorsqu'on étudie des solutions non perturbatives
des théories de supercordes, et aussi de supergravité en dimension $10$ ou
$11$. Ce sont des solutions classiques des équations du mouvement :
$solitons$ qui ont des masses:
\begin{equation}
m \propto {1\over g_s}.
\end{equation}   
Il est clair que ces membranes deviennent des états légers lorsqu'on
augmente le couplage $g_s$. On peut, par dualité, étudier le régime de
couplage fort de ces objets dans une certaine théorie en étudiant un objet
dual dans une autre théorie où le couplage est faible et où les calculs
sont (en principe!) faisables puisque perturbatifs.

Parmi ces solitons , solutions non perturbatives des théories de
supercordes, on trouve une classe particulière appelée membranes de
Dirichlet ou D-membranes: les extrémités de cordes ouvertes y sont
rattachées. Un exemple de l'utilisation de ces membranes est le calcul de
l'entropie d'un trou noir dont nous avons discuté préalablement. Le trou
noir est décrit par une D-membrane sur laquelle sont agglutinées des
cordes. On parle d'ailleurs de boules de cordes. Pour calculer l'entropie
d'un trou noir, il suffit donc de calculer le nombre d'états de la
D-membrane.

La dualité simplifie le calcul. L'existence d'un trou noir implique un
couplage fort, mais on peut se ramener au calcul des différents états
d'une corde à couplage faible (ce qui est bien connu) et par dualité
revenir aux différents états de la membrane à couplage fort. Pour certains
types de trous noirs et pour certaines théories de cordes, on retrouve la
formule d'entropie macroscopique de Hawking et Bekenstein. Enfin, le
paradoxe dont nous parlions à propos de l'état pur qui devient mixte est
maintenant réglé. En effet, le syst\`eme de deux particules $A, B$ qui
était dans un état pur et qui semblait évoluer vers un état mixte peut
être maintenant perçu comme un état intriqué avec l' état du trou noir.
Or, l'état quantique de ce dernier n'est pas facilement accessible
physiquement puisque l'on ne peut pas sortir d'informations d'un trou
noir.

\subsubsection{La M-théorie}

Les relations de dualité entre les cinq théories de supercordes impliquent
que ces cinq théories ne sont pas autre chose que des solutions
différentes d'une seule théorie, la M-théorie. Dans cette approche, les
théories de supercordes ne sont rien d'autre que des développements autour
des différents vides (un pour chaque théorie) d'une même théorie. La
M-théorie vit, quant à elle, dans un espace-temps à $11$ dimensions: ce
serait la limite à couplage fort de la théorie IIA et de l'hétérotique
$E(8)\times E(8)$. En compactifiant la M-théorie sur un cercle on retrouve
la théorie type IIA , en la compactifiant sur un cercle divisée par $Z_2$,
on retrouve la théorie hétérotique $E(8)\times E(8)$. On ne connait pas
l'action de cette M-théorie, mais on sait que la théorie effective aux
basses énergies de la M-théorie est la supergravité $d=11$ avec $N=1$.

\begin{figure}
\centerline{\includegraphics[height=3in]{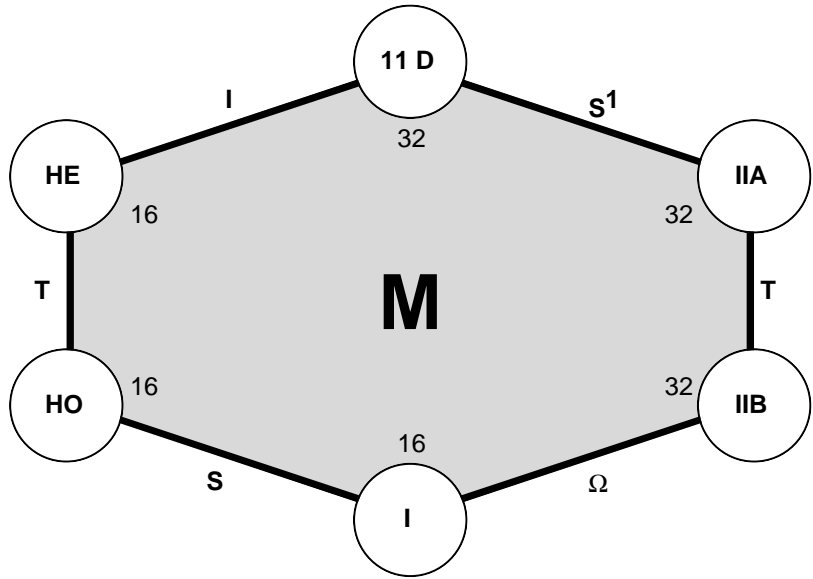}}
\caption[]{Les différents vides de la M-théorie reliés par les différentes
relations de dualité. Les nombres $16$ et $32$ sont le nombre de
composantes d'un spineur dans la théorie.}
\end{figure}

La théorie IIA et la M-théorie sont reliées par la S-dualité qui associe
une observable $f(g_s)$ de la théorie IIA à une observable $f\prime(g_s)$
de la M-théorie de sorte que $f(g_s)=f\prime(\frac{1}{g_s})$. Le résultat
important qui ressort de la S-dualité appliquée à la théorie type IIA est
l'existence de la M-théorie en 11 dimensions avec un couplage directement
relié à la dimension de la onzième dimension. Ainsi, on obtient :
\begin{equation}   
R_{11}\,=\,(g_s)^{\frac{2}{3}}.
\end{equation}
La onzième dimension n'apparaît visible que dans la limite de couplage
fort. C'est la raison pour laquelle cette dimension supplémentaire n'avait
pas été perçue avant l'avènement de la dualité: en n'ayant accès qu'au
domaine perturbatif, il était impossible de la déceler. Comme la taille de
la onzième dimension est liée inversement au couplage, l'unification des
couplages est fonction de la taille de cette onzième dimension. Or, il se
trouve que pour que les couplages s'unifient, la taille de cette onzième
dimension doit être supérieure à celles des six autres dimensions
supplémentaires. En compactifiant les six dimensions à une échelle de
l'ordre de $M_{GUT}$ donc $10^{16}$ GeV , on obtient une taille de l'ordre
de $(10^{12}\,GeV)^{-1}$ à $(10^{13}\,GeV)^{-1}$ pour la onzième dimension
(appelée $R_{11}$) pour que l'unification des quatre couplages ait lieu à
$M_{GUT}$. Ainsi, à des énergies comprises entre $10^{12}$ Gev et $10^{16}
$ GeV, le monde apparaît de dimension cinq. On peut représenter l'étape
intermédiaire à cinq dimensions par deux plateaux de dimensions $4$ mais
qui, à haute énergie, ont chacun $10$ dimensions: les six autres ont été
compactifiées. Les deux plateaux ou membranes sont séparées par la onzième
dimension.

Dans une autre approche, on unifie d'abord les trois couplages des
groupes de jauge à $M_{GUT}$, et puis le couplage gravitationnel rejoint
les trois autres couplages à une échelle d'énergie plus haute $M_{11}$.
Dans ce cas, on trouve que $M_{11}$ est environ deux fois plus grand que
$M_{GUT}$ et que $R_{11}$ est de l'ordre de $(10^{14}\,GeV)^{-1}$ à
$(10^{15}\,GeV)^{-1}$.

Un lien peut maintenant être fait avec la brisure de supersymétrie dans un
secteur caché : notre monde et tous ses champs sont contraints de rester
sur une des deux membranes, et le secteur caché est sur l'autre membrane.
Seuls les champs gravitationnels (nous parlons au pluriel car on trouve le
graviton, le gravitino et le dilaton) peuvent se propager dans le bulk,
c'est-à-dire dans la dimension suppl\'ementaire, et la supersymétrie est
brisée dans le secteur caché, c'est-à-dire sur la membrane cachée. La
brisure est ensuite transmise au secteur visible par les champs
gravitationnels. Ainsi, des mécanismes fondés sur des modèles de
supergravité $d=11$ couplée à une théorie de Yang-Mills avec groupe de
jauge $E(8)\times E(8)$ sont proposés pour décrire la brisure de
supersymétrie. Le secteur caché correspond à un groupe de jauge $E(8)$ et
le secteur visible à l'autre. On peut citer le mécanisme de condensation
de gauginos qui brise la supersymétrie dans le secteur caché: lorsqu'on
diminue l'énergie et qu'on passe une certaine valeur, une interaction
forte du secteur caché fait condenser deux gauginos et brise la
supersymétrie.

Nous conclurons ce paragraphe par citer ce qu'il reste à accomplir. Le
plus grand travail sera d'écrire l'action de la M-théorie et quantifier la
théorie. On suppose, à cause de sa théorie basse énergie - la supergravité
à $11$ dimensions - que la théorie contient des 2-membranes et des
5-membranes qui apparaissent comme des cordes fermées à plus basses
dimensions, mais actuellement on ne sait pas quantifier l'action d'une
membrane avec les techniques connues. D'autre part, on ne sait pas
clairement comment ces membranes interagissent bien que, grâce à la
dualité, on puisse donner les excitations des membranes en termes de
cordes. La M-théorie ne peut pas dans l'état actuel de son développement
prédire la valeur de la constante cosmologique ou expliquer pourquoi la
supersymétrie est brisée. Enfin, la M-théorie devra une fois écrite
permettre de déterminer l'énergie d'unification sans avoir à ajuster la
taille de la onzième dimension: elle devra donc permettre de prédire à la
fois la taille de cette dimension mais aussi la valeur de l'énergie
d'unification des couplages.

\subsection{Compactifications, prédictions et limites des théories de
supercordes}

Une prédiction importante de la théorie perturbative des cordes est la
valeur de l'énergie d'unification:
\begin{equation}
M_{GUT} = O(g) \times \frac{m_P}{\sqrt{8 \pi}} \simeq\mbox{few} \times 10^{17} 
\mbox{GeV}.
\label{topdown}
\end{equation}
Cette valeur est environ $20$ fois plus grande que la valeur calcul\'ee
par une approche qui monte depuis les basses énergies vers les hautes
énergies ... Pourtant, il est impressionnant que l'estimation
(\ref{topdown}) soit si proche d'un calcul qui traverse $14$ ordres de
grandeur des basses énergies vers les hautes! Comme nous l'avons déjà fait
remarquer, dans la M-theorie, la onzième dimension peut être réglée de
manière à ce que tous les couplages s'unifient à $10^{16}$ GeV. Il est
cependant important de constater que deux approches différentes, la
théorie quantique des champs et sa renormalisation d'une part et la
théorie des cordes d'autre part conduisent, à des valeurs de l'énergie
d'unification qui sont si proche.

Parmi les limitations de la théorie des cordes, on trouve le problème de
la compactification donc, finalement, le lien avec notre monde à $4$
dimensions.

Les moyens de compactifier la théorie, c'est-\`a-dire, rendre invisibles
les dimensions suppl\'ementaires, sont innombrables. On a beaucoup
d'options pour retrouver le modèle standard plus d'autres interactions non
observées. Comment choisir parmi toutes les possibilités pour
compactifier? Est-ce qu'il faut choisir une vari\'et\'e, ou bien un espace
avec une g\'eom\'etrie g\'en\'eralis\'ee, un orbifold, par exemple? La
premi\`ere option \'etudi\'ee en d\'etail afin de maintenir la
supersym\'etrie \'etait de compactifier sur une vari\'et\'e de Calabi-Yau.
Or, toute compactification sur un espace Calabi-Yau est paramétrisée par
les champs de moduli qui d\'ecrivent les possibles déformations de ces
espaces : {\it a priori}, les champs de moduli n'ont pas de valeur
déterminée.

Cela revient à se poser la question de savoir quel est l'état de vide de
la théorie à $4$ dimensions. L'énergie du vide à $4$ dimensions devrait
correspondre correspondre à la constante cosmologique. Or, il y a un
nombre énorme d'états de vide possibles après compactification à $4$
dimensions qui pourraient correspondre à cette énergie: le problème de
la dégénérescence du vide reste \`a resoudre.


\section{Les dimensions supplémentaires}


\subsection{Pourquoi (pas) des dimensions supplémentaires?}

Les dimensions supplémentaires sont présentes dans les théories de cordes
comme nous l'avons vu. Pourtant, l'idée d'un espace-temps à plus de quatre
dimensions date des années $1920$: Kaluza et Klein furent les premiers à
introduire une dimension supplémentaire dans le but d'unifier
électromagnétisme et gravitation. L'électromagnétisme pouvait s'expliquait
géométriquement comme la gravitation si l'on ajoutait une dimension
supplémentaire (une ligne et une colonne de plus à la matrice représentant
le tenseur métrique). En compactifiant cette dimension, ils montrèrent que
l'on retrouvait les lois de l'élecromagnétisme telles qu'elles sont dans
notre monde à quatre dimensions.

Les théories modernes proviennent plus ou moins de la M-théorie où deux
membranes sont séparées par une dimension supplémentaire. Pourtant, ces
théories s'en éloignent plus ou moins fortement. En effet, la M-théorie et
sa version basse énergie - la supergravité en dimension $11$ couplée à une
théorie de Yang-Mills de groupe de jauge $E(8)\times E(8)$ - semble
n'autoriser que les champs gravitationnels à se propager dans toutes les
dimensions et contraint les champs qui font notre monde sur une membrane
et les champs du secteur caché sur l'autre membrane; de plus, la dimension
de la onzième dimension est déterminée par l'unification des couplages.
Alors que dans les théories de Kaluza et Klein, on peut trouver
différentes tailles pour la (ou les) dimension(s) supplémentaire(s), et
toutes les variantes possibles pour la localisation des champs, y comprise
la propagation des champs de jauge dans des dimensions supplémentaires.

\subsection{Les différents types de modèles}

Il existe de nombreux modèles différents. Les deux classes principales
sont celle où la dimension supplémentaire est {\it plate} et celle où elle
est {\it courbe}.

Dans le premier cas, on trouve des modèles avec dimension supplémentaire
plate grande, c'est-à-dire pouvant aller jusqu'à $0.1$ mm qui est la
limite haute de taille de dimension supplémentaire d'après des expériences
de type Cavendish. La taille de la ou des dimensions supplémentaires
affecte directement la masse de Planck $M_\star$ dans l'espace à $4+n$
dimensions: 
\begin{equation}
M_p\,=\,M_\star^{2+n}\,R^n ,
\end{equation}
où R est la taille de la dimension supplémentaire. Par exemple, dans le
cas où $n=2$ et si l'on se fonde sur une théorie de cordes pour écrire que
\begin{equation}
M_\star^{2+n}\,=\,\frac{1}{g_s\,g_{ym}^2}M_s^{2+n} ,
\end{equation}
avec $ M_s$ de l'ordre du TeV, on obtient des limites sur la taille des
dimensions supplémentaires par la cosmologie qui donnerait $R\leq
10^{-8}$cm si seule la gravité se propage dans les dimensions
supplémentaires. Dans chacun des modèles, on peut choisir d'autoriser
certains champs à se propager dans toutes les dimensions et d'autres à
rester localis\'es sur les membranes.

Parmi les modèles avec une dimension supplémentaire courbe, citons le
modèle de Randall-Sundrum, o\`u il y a deux membranes. Les champs du
Modèle Standard sont localisés sur une membrane et la gravitation se
propage partout. Randall et Sundrum ont montré\cite{Randall:1999ee} que l'on pouvait redéfinir
le problème de hiérarchie entre la masse de Planck et l'échelle
électrofaible sans avoir à faire appel à la supersymétrie. Il existe
cependant des modèles supersymétriques de ce type où la supersymétrie
est brisée par des anomalies.

\subsection{Signatures expérimentales}

Il est possible de déceler la présence de dimensions supplémentaires à
l'échelle du TeV au LHC. Pour cela, il faut bien évidemment que les
particules en jeux puissent sentir la présence de la (ou des)dimension(s)
supplémentaire(s). Les signatures typiques sont des excitations dites de
Kaluza-Klein ; des résonnances dans l'évolution des sections efficaces des
réactions en fonction de l'énergie peuvent apparaitre. Elles sont dues au
"tours" d'excitations de Kaluza-Klein. On peut comprendre ces "tours" en
faisant une analogie avec une particule qui rencontre un puit de potentiel
en mécanique quantique: son énergie va être quantifiée dans le puit
fournissant une tour d'excitations. Dans notre cas, la dimension
supplémentaire joue le rôle du puit.

D'autres signatures pourraient être fournies par la création d'un micro
trou noir. En effet, tout objet peut se transformer en trou noir s'il est
contracté en deça de son rayon de Schwarschild qui est défini par
$\frac{G\,m}{c^2}$. Bien évidemment, plus la masse est grande, plus l'on
peut obtenir facilement le rayon de Schwarschild. Or, comme on l'a vu, les
dimensions supplémentaires peuvent augmenter la valeur de $G$. Donc s'il y
a des dimensions supplémentaires de taille suffisante, on peut espérer
avec les énergies en jeu au LHC passer en deça du rayon de Schwarschild et
créer des micro trous noirs. Ceux-ci devraient s'évaporer rapidemment,
puisque le rayonnement de Hawking implique que le trou noir rayonne de
manière inversement proportionnelle à sa masse. Des \'etudes par les
collaborations ATLAS et CMS ont d\'emontr\'e que ce rayonnement de Hawking
serait sans doute visible au LHC.

\subsection{Des dimensions en moins à haute énergie?}

Tandis que ce n'est pas pr\'evu dans le th\'eorie des cordes, il est
possible qu'au moins quelques-unes des dimensions de l'espace-temps
paraissent discr\`etes à hautes énergies. Le principe est d'imposer un
cut-off non par la dimension donnée aux particules comme c'est le cas en
théorie des cordes, mais par la structure interne de l'espace-temps. Comme
dans les th\'eories de champs sur le réseau, le mouvement est décomposé en
une succession de bonds d'un point discret de l'espace à un autre. Aux
basses \'energies, la r\'esolution exp\'erimentale n'est pas suffisante pour
d\'etecter ces bonds, et une dimension discr\`ete ressemble \`a une
dimension continue. Pourtant, aux hautes \'energies, cette dimension
dispara\^it, et la dimensionalit\'e de l'espace-temps diminue!


\section{L'énergie noire}


Nous terminons ce cours avec un des plus grands mystères de la Physique
contemporaine: la mystérieuse énergie noire. Commençons par nous rappeler
des équations de la Relativité Générale:
\begin{equation}
R_{\mu\nu}-\frac{1}{2}R g_{\mu\nu}\,=\,k T_{\mu \nu} ,
\end{equation}
o\`u $R_{\mu\nu}$ est le tenseur de Ricci (ou de courbure) , $R$ sa trace
et $T_{\mu\nu}$ est le tenseur d'énergie-impulsion. Ces équations relient
la géométrie de l'espace-temps à la présence de matière (ou d'énergie).
Pourtant, cette forme n'est pas la forme la plus générale. En effet, cette
équation vient de l'intégration d'une équation. On a la conservation du
tenseur énergie-impulsion: $\nabla_{\mu} T^{\mu\nu}=0$ où $\nabla_{\mu}$
est la dérivée covariante et on a aussi $\nabla_{\mu}
(R^{\mu\nu}-\frac{1}{2}R g^{\mu\nu})=0$. En intégrant cette dernière
équation, on constate qu'il est possible de rajouter un terme
proportionnel au tenseur métrique $g^{\mu\nu}$. En effet, on a toujours
$\nabla_{\nu}g^{\mu\nu}=0$. La forme la plus générale des équations de la
Relativité Générale est donc:
\begin{equation}
R_{\mu\nu}-\frac{1}{2}R g_{\mu\nu}\,+\Lambda g_{\mu\nu}\,=\,k T_{\mu \nu} ,
\end{equation}
où $\Lambda$ est la constante cosmologique. Les mesures actuelles des
Supernovae lointaines tendent à prouver que l'\'energie dans le vide n'est
pas nulle, et pourrait correspondre \`a une constante cosmologique. En
effet, ces mesures et d'autres montrent que l'Univers est dans une phase
d'expansion qui est accélérée.

Or, si l'on reprend les équations de la Relativité Générale, et qu'on
place la partie contenant $\Lambda$ du côté du tenseur énergie-impulsion
et qu'on interprête le terme $-\Lambda g_{\mu\nu}$ comme un terme
d'énergie-impulsion homogène et isotrope, on trouve qu'il correspond à un
fluide très particulier qui a une densité d'énergie positive et une
pression négative. Ce fluide qui remplit tout l'Univers de manière
homogène permet d'expliquer la phase d'accélération de l'Univers. La
pression négative permet l'expansion accéléree de l'Univers. Ce fluide a
reçu, à cause de sa propriété inconnue jusqu'alors de pression négative,
le nom de quintessence ou cinquième élément. C'est aussi ce que l'on nomme
l'énergie noire.

Si $\Lambda$ est une vraie constante, le rapport pression sur densité du
fluide de quintessence devrait être égale à $-1$. Il se trouve que les
mesures actuelles donnent un rapport pression sur densité de $ < -0.8$.

Les mesures réalisées d'après le CMB montre que l'Univers est plat
c'est-à-dire que $$\Omega\,=\frac{\rho}{\rho_c}=1$$ où $\rho$ est la
densité totale d'énergie-matière de l'Univers et $\rho_c$ est la densité
critique qui correspond à un Univers plat. Les mesures sur les Supernovae
combinées au fait que l'on sait que l'Univers est plat indique que $70\%$
de la densité d'énergie-matière de l'Univers est de l'énergie noire. Les
$30\%$ restant sont les $25\%$ de matière noire non baryonique et les
$5\%$ de matière baryonique (il est à noter que sur ces $5\%$, la matière
visible sous forme d'étoiles ne représente qu'un dixième!).

Qu'est-ce que l'énergie noire? Si on la décrit par un champs scalaire, à
quel particule correspond-il? Cela reste mystérieux. De plus, comme nous
l'avons déjà dit dans le chapitre sur la brisure de supersymétrie, si
cette énergie correspond à l'énergie du vide, comment obtenir le bon vide?
A l'heure actuelle, aucune théorie ne peut prédire la valeur de la
constante\footnote{On utilise le mot constante par habitude mais comme on
l'a vu il se pourrait que cette constante n'en soit pas une.}
cosmologique, donc aucune théorie ne possède le véritable état de vide.
C'est le grand défi contemporain. La "Théorie du Tout" ne pourra être que
la théorie qui prédit la valeur de l'énergie noire.



\newpage
\noindent {\Huge \bf Bibliographie générale }
\par\hfill\par
\noindent \underline{Modèle Standard et brisure électrofaible}

$\bullet$ Le cours à cette école de Gif 2004 de D.Treille : {\it Le Modèle
Standard}.

$\bullet$ S.~Weinberg,
  ``{\it The quantum theory of fields. Vol. 2: Modern applications}'',
(Cambridge Univ. Press, Cambridge, UK) (1996).

$\bullet$ J.M.Frère, ``{\it La théorie électrofaible et au-delà} '', 
22ème école Joliot-Curie de Physique nucléaire 2003, 
"Interaction faible et noyau : l'histoire continue....",
p27.

$\bullet$ C.~T.~Hill and E.~H.~Simmons,
  ``{\it Strong dynamics and electroweak symmetry breaking}'', 
  Phys.\ Rept.\  {\bf 381} (2003) 235
  [Erratum-ibid.\  {\bf 390} (2004) 553]
  [arXiv:hep-ph/0203079].

$\bullet$ E.~Witten,
  ``{\it When symmetry breaks down}'', 
  Nature {\bf 429} (2004) 507.

\par\hfill\par
\noindent \underline{Supersymétrie :}

$\bullet$ S.~P.~Martin,
  ``{\it A supersymmetry primer}'',
  arXiv:hep-ph/9709356.

$\bullet$ G.~L.~Kane,
  ``{\it Weak scale supersymmetry: A top-motivated-bottom-up approach}'',
  arXiv:hep-ph/0202185.

$\bullet$ M.~F.~Sohnius,
  ``{\it Introducing Supersymmetry,}''
  Phys.\ Rept.\  {\bf 128} (1985) 39.

$\bullet$ A.Deandrea, ``{\it Interactions électrofaibles et introduction à la
supersymétrie}'',

disponible à l'adresse : http://deandrea.home.cern.ch/deandrea/seminars/ew.ps

$\bullet$ J. R. Ellis,``{\it Supersymmetry for alp hikers}'',
~\cite{JE-SUSY}.

\par\hfill\par
\noindent \underline{Physique des neutrinos :}

$\bullet$ W.~Buchmuller,
  ``{\it Neutrinos, grand unification and leptogenesis}'',
  arXiv:hep-ph/0204288.

$\bullet$ J.~R.~Ellis,
  ``{\it Particle physics and cosmology}'',
  arXiv:astro-ph/0305038.

$\bullet$ J.R.Ellis, ``{\it Limits of the standard model}'',
~\cite{JE-NU}.

$\bullet$ S.~F.~King,
  ``{\it Neutrino mass models}'',
  Rept.\ Prog.\ Phys.\  {\bf 67} (2004) 107
  [arXiv:hep-ph/0310204].

$\bullet$ M.~C.~Gonzalez-Garcia and Y.~Nir,
  ``{\it Developments in neutrino physics}'',
  Rev.\ Mod.\ Phys.\  {\bf 75} (2003) 345
  [arXiv:hep-ph/0202058].

$\bullet$ D.Vignaud, ``{\it Comment la masse vint aux neutrinos}'',
22ème école Joliot-Curie de Physique nucléaire 2003, 
Interaction faible et noyau : l'histoire continue...., 
p289.

\par\hfill\par
\noindent \underline{Grande Unification, supergravité, théories des cordes}

$\bullet$ T.~P.~Cheng and L.~F.~Li,
  ``{\it Gauge Theory Of Elementary Particle Physics}'',
(Oxford Science Publications, Clarendon Press, Oxford, UK) (1984).

$\bullet$ G.~G.~Ross,
  ``{\it Grand Unified Theories}'',
(Frontiers In Physics, Benjamin/Cummings, Reading, USA) (1984).

$\bullet$ R.N.Mohapatra, ``{\it Unification and supersymmetry}'',
~\cite{Mohapatra}.

$\bullet$ J. R. Ellis, ``{\it  Beyond the standard model for
hillwalkers}'',
~\cite{JE-Beyond}

$\bullet$ J.~R.~Ellis,
 ``{\it Supersymmetry for Alp hikers}'',~\cite{JE-SUSY}.
  
$\bullet$ M.~Kaku,
  ``{\it Introduction to superstrings and M-theory}'',
(Springer, New York, USA) (1999).

$\bullet$ J.~Wess and J.~Bagger,
  ``{\it Supersymmetry and supergravity},''
(Princeton Univ. Press, Princeton, USA) (1992).

$\bullet$ S.~Weinberg,
  ``{\it The quantum theory of fields.  Vol. 3: Supersymmetry},''
(Cambridge Univ. Press, Cambridge, UK) (2000) 419.
  
$\bullet$ D.~Bailin and A.~Love,
  ``{\it Supersymmetric gauge field theory and string theory},''
(Graduate student series in physics, IOP, Bristol, UK) (1994).

\end{document}